\documentstyle[12pt]{article}
\newcounter{lyter}[equation]
\title{General Superfield Quantization Method. \protect \\
III. Construction of Quantization Scheme}
\author{A.A. Reshetnyak\thanks{E-mail: reshetnyak@ssti.ru}}
\date{\it Department of Mathematics, Seversk State Technological Institute,
\protect \\
Seversk, {\rm 636036}, Russia}
\begin{document}
\maketitle
\begin{abstract}
Quantization procedure in the framework of general superfield quantization
method (GSQM), based on the Lagrangian [1] and Hamiltonian
[2] formulations for general superfield theory of fields,
for irreducible  gauge theories in Lagrangian formalism
is proposed.

Extension procedure for supermanifold ${\cal M}_{cl}$ of
superfields ${\cal A}^{\imath}(\theta)$, ghost number construction are
considered. Classical and $\hbar$-deformed
 generating (master) equations,
existence theorems for their solutions are formulated in
$T^{\ast}_{odd}{\cal M}_{min}$, $T^{\ast}_{odd}{\cal M}_{ext}$.
Analogous scheme is
realized for BV similar generating equations. Master equations versions
for GSQM and BV similar scheme are deformed in powers
of superfields ${\stackrel{\circ}{\Gamma}}{}^p(\theta)$ =
$\bigl({\stackrel{\circ}{\Phi}}{}^B(\theta)$,
${\stackrel{\circ}{\Phi}}{}^{\ast}_B(\theta)\bigr)$ into supermanifold
$T_{odd}(T^{\ast}_{odd}{\cal M}_{ext})$. Arbitrariness in a choice of
solutions for these equations is described.

Investigation of formal Hamiltonian systems for  II class theories
[2] defined via corresponding   master equations solutions is
conducted. Gauge fixing for those theories is described by two ways.

Functional integral of superfunctions on $T_{odd}(T^{\ast}_{odd}{\cal
M}_{ext})$ is defined. Properties for  generating functionals of
Green's  superfunctions are studied.

$\theta$-component  quantization formulation, connection with BV method and
superfield quantization [3] are established. Quantization scheme realization
is demonstrated on  6 models.
\end{abstract}
{\large PACS codes: 03.50.-z, 11.10.Ef, 11.15.-q, 12.90.+b \protect \\
Keywords: Lagrangian superfield quantization, Gauge theory,
Superfields.}
\renewcommand{\thesection}{\Roman{section}}
\renewcommand{\thesubsection}{\thesection.\arabic{subsection}}
\section{Introduction}
\renewcommand{\theequation}{\arabic{section}.\arabic{equation}}
\renewcommand{\thelyter}{\alph{lyter}}

Lagrangian and Hamiltonian formulations for general superfield theory of
fields (GSTF) formulated in papers [1,2]
directly appear by the basis for creation of GSQM rules  for
gauge theories in Lagrangian (in the usual sense) formalism
generalizing in a natural way the rules for BV quantization method [4] in the
superfield form. In this case a procedure for GSQM is realized
for irreducible superfield (on $\theta$) gauge model in the Hamiltonian (or
not always equivalently in Lagrangian) formulation for GSTF. The last as it
had been noticed in [2] contains the all information, among them, on an usual
(relativistic) field theory model. Not repeating the motivations for creation
of GSTF and GSQM stated in detail in [1] let us point out that by the paper's
aim is the construction of complete and noncontradictory rules for GSQM
in the Lagrangian formalism in the framework of GSTF.

In Sec.II having partially preparatory nature it is considered an algorithm,
connected with gauge invariance, of  extension of the supermanifold
${\cal M}_{cl}$ and superalgebras defined on ${\cal M}_{cl}$ including
operatorial ones.

Group-theoretic construction of the ghost number concept is
realized in Sec.III. Sec.IV is devoted to formulation  of a GSQM generating
equation  in a minimal sector $T^{\ast}_{odd}{\cal M}_{min}$ together
with its consequences. In the next section the analogous problems are solved
in an
extended sector of supervariables $T^{\ast}_{odd}{\cal M}_{ext}$ together
with analysis of possible representations for master equations induced by
transformations of the operators from superalgebras
{\boldmath${\cal B}_{min}$} and {\boldmath${\cal B}_{ext}$}. Quantum master
equation as a deformation
of GSQM classical master equation  in $T^{\ast}_{odd}{\cal M}_{ext}$
in powers of $\hbar$ is formulated  in Sec.VI. Here it is
considered its properties and another BV similar quantum (in the sense of
deformation on $\hbar$) master equation and its consequences as well.

In Sec.VII  a so-called global nontrivial arbitrariness in a
choice of solutions of above-mentioned variants for master equations  is
defined.

The deformation procedure in powers of superfields
${\stackrel{\circ}{\Gamma}}{}^p(\theta)$  for quantum master equations from
Sec.VI with superfunction
$S_{H}(\Gamma(\theta), {\stackrel{\circ}{\Gamma}}(\theta),\hbar)$ and
derivation of their consequences are fulfilled in Sec.VIII.
The concept of a trivial global arbitrariness in a choice of
solutions of master equations in $T_{odd}(T^{\ast}_{odd}{\cal M}_{ext})$
is introduced here as well.

The problem of gauge fixing associated with a choice of the
required Lagrangian
surface in $T^{\ast}_{odd}{\cal M}_{ext}$,
a restriction of the master equation solution  on which possesses by
nondegenerate supermatrix of the 2nd partial superfield derivatives with
respect to superfields parametrizing this surface, is carefully considered
in Sec.IX by means of arbitrariness operation in solutions and phase
anticanonical transformation. The same procedure is solved for GSTF model
satisfying both to  GSQM and  BV similar master equations in
$T_{odd}(T^{\ast}_{odd}{\cal M}_{ext})$.

By one of the most main objects of the paper it appears the functional
integral of superfunctions whose definition and properties  are
examined in Sec.X. All generating
functionals (superfunctions) of Green's (super)functions
$Z\bigl({\stackrel{\circ}{
\Gamma}}(\theta),\Phi^{\ast}(\theta), J^{\ast}(\theta)\bigr)$,
$W\bigl({\stackrel{\circ}{
\Gamma}}(\theta),\Phi^{\ast}(\theta), J^{\ast}(\theta)\bigr)$ and
effective action $\mbox{\boldmath$\Gamma$}\bigl({\stackrel{\circ}{
\Gamma}}(\theta),\langle\Phi(\theta)\rangle,
\Phi^{\ast}(\theta)\bigr)$
being intensively used in the quantum  gauge fields theory are introduced.
Properties of these superfunctions including Ward identities are established.

Component (on $\theta$) formulation for GSQM rules   is proposed
in Sec.XI. The interconnection of GSQM with BV method [4] and superfield
Lagrangian quantization of the work [3] is entirely discovered in
Sec.XII.

The application of general scheme of GSQM rules
is demonstrated in Sec.XIII on 6
GSTF models  considered for the first time in Refs.[1,2] in the
framework of Lagrangian and Hamiltonian formulations for GSTF. These examples
describe free and interacting massive complex spinless scalar superfields,
massive spinor superfields of spin $\frac{1}{2}$ and free massless and massive
vector superfields for arbitrary
$D\geq 2$ and represent the basic field-theoretic models for
construction, for instance, interacting superfield (on $\theta$) (non-)Abelian
Yang-Mills type theories not having the analogs in usual relativistic gauge
fields theories on a basis of the gauge principle [5] realization.

At last, in conclusion it is summarized to the definite results of the
work. The all assumptions in the framework of which the paper is made in
fact had been pointed out in the introduction of Ref.[1].

In paper it is used unless otherwise stated the system of notations
suggested in work [1]. In particular  all derivatives  including
superfield variational, partial superfield and component variational (in the
usual sense) with respect to superantifields $\Phi^{\ast}_A(\theta)$,
${\stackrel{\circ}{\Phi}}{}^{\ast}_A(\theta)$ and their components are
understood as left  and corresponding ones with
respect to superfields $\Phi^A(\theta)$, ${\stackrel{\circ}{\Phi}}{}^A(
\theta)$ and their components as right according to failure to mention.
In opposite case  the signs $"l"$ for the latter derivatives and $"r"$ for
former ones (for instance,
$\displaystyle\frac{\partial_l\phantom{xxxx}}{\partial\Phi^A(\theta)}$, $
\displaystyle\frac{\delta_r\phantom{xxx}}{\delta\Phi^{\ast}_A(\theta)}$)
are added.
\section{Extension of ${\cal M}_{cl}$ Induced by  Gauge Invariance}

By the principal problem in
quantum theory of gauge fields one appears the construction of the generating
functional of Green's functions for gauge theory of general type (GThGT) or
for one of special type (GThST) in the Lagrangian [1] and Hamiltonian [2]
formulations for GSTF
belonging both to the I class theories and to the II class ones. In
Hamiltonian formalism that theory is defined
by superfunction $S_{H}(\Gamma(\theta),\theta)$ belonging to
$C^{k}(T^{*}_{odd}{\cal M}_{cl}\times \{\theta\})$, $k\leq \infty$ being by
superalgebra of $k$-times differentiable superfunctions given on the extended
by coordinate $\theta$ supermanifold $T^{*}_{odd}{\cal M}_{cl}\times\{
\theta\}$ with coordinates
$\Gamma^p(\theta)$=$({\cal A}^{\imath}(\theta), {\cal A}^{\ast}_{
\imath}(\theta))$, $\theta$ ($p=1,\ldots,2n,\, \imath=1,\ldots,n=(n_{+},
n_{-})$)
appearing by classical superfields, superantifields and odd time respectively
with following values of the Grassmann parities $\varepsilon_{P},
\varepsilon_{\bar{J}}, \varepsilon$ $(\varepsilon$ = $\varepsilon_{P}$ +
$\varepsilon_{\bar{J}})$ [1,2]
$$
(\varepsilon_{P}, \varepsilon_{\bar{J}}, \varepsilon){\cal A}^{\imath}(
\theta) = (0,\varepsilon_{\imath}, \varepsilon_{\imath}),\
(\varepsilon_{P}, \varepsilon_{\bar{J}}, \varepsilon){\cal A}^{\ast}_{
\imath}(\theta) = (1, \varepsilon_{\imath}, \varepsilon_{\imath} + 1),\
(\varepsilon_{P}, \varepsilon_{\bar{J}}, \varepsilon)\theta=(1,0,1)\,.
$$

${\cal A}^{\imath}(\theta), {\cal A}^{\ast}_{\imath}(\theta))$ are the
elements of Berezin superalgebra
$\tilde{\Lambda}_{D\vert Nc + 1}(z^{a},\theta;{\bf K})$, ${\bf K}$ =
(${\bf R}$ or ${\bf C}$) defined on
superalgebra ${\Lambda}_{D\vert Nc + 1}(z^{a},\theta;{\bf K})$ [1,2].
Both superalgebras consist of elements being transformed with respect to
superfield (ir)reducible representations of supergroup $J=\bar{J}\times P$
with one-parametric supergroup $P$ corresponding to nilpotent variable
$\theta$.
The superfield representation $T$ of
supergroup $J$ is given on the superalgebra $\tilde{\Lambda}_{D\vert Nc +
1}(z^{a},\theta;{\bf K})$ [1,2]. In its turn elements from
$\tilde{\Lambda}_{D\vert Nc + 1}(z^{a},\theta;{\bf K})$,
${\Lambda}_{D\vert Nc + 1}(z^{a},\theta;{\bf K})$ appear by
superfunctions on ${\cal M}= \tilde{\cal M} \times \tilde{P}$ parametrized
by coordinates $(z^{a},\theta)$, $z^{a} \in \tilde{\cal M}$, where $\tilde{
\cal M}$ may be by usual quotient space of supergroup $\bar{J}$ over its
subsupergroup $\bar{J}_{\tilde{A}}$ being by a some set
of the internal automorphisms for $\bar{J}$ [1].

As it had been shown in [1]
one can choose, in particular case,  $\bar{J}$ in the form of Poincare
supergroup acting on the superspace
\begin{eqnarray}
\tilde{\cal M} = {\bf R}^{1,D - 1\mid Nc},\;
c=2^{[D/2]}, z^{a}=(x^{\mu}, \theta^{Aj}), A=1,\ldots,2^{[D/2]},\;j=
\overline{1,N}\;, 
\end{eqnarray}
where $D = \dim{\bf R}^{1,D - 1}$ is the dimension of Minkowski space,
$N$
is the number of supersymmetries, $x^{\mu}$ and $\theta^{Aj}$ denote the
Lorentz vector and spinor coordinates on $\tilde{\cal M}$. In the writing
${\Lambda}_{D\vert Nc + 1}(z^{a},\theta;{\bf K})$ the signs $D$, $Nc$ and
1 denote the number of even with respect to $\varepsilon_{\bar{J}}$,
$\varepsilon$ gradings, odd ($Nc$) with respect to $\varepsilon_{\bar{J}}$,
$\varepsilon$ gradings generating elements $z^a$ and odd with respect to
$\varepsilon_P$, $\varepsilon$ generating element $\theta$ [1].

The fact of belonging of a GThGT (GThST) with $S_{H}(\Gamma(\theta),\theta)$
to the I class theories  [2] means a possibility of fulfilment for this model,
defined in Hamiltonian formulation of GSTF, of the following equation
(on solutions ${}^1\overline{\Gamma}{}^{p}(\theta)$  for Hamiltonian systems
(HS) of corresponding different types) written with help of antibracket [2]
\begin{eqnarray}
{(S_H(\Gamma(\theta),\theta), S_H(\Gamma(\theta),\theta))_{\theta}}_{\mid
 {}^1\overline{\Gamma}(\theta)} = 0\;. 
\end{eqnarray}
Belonging of a superfield (on $\theta$) model to the II class theories means
the necessary fulfilment of Eq.(2.2) in the whole $T^{\ast}_{odd}{\cal M}_{cl
}$ for any configuration $\Gamma^p(\theta)$.
Moreover, the superfunction $S_{H}(\Gamma(\theta))$ does not
explicitly depend upon $\theta$ in the last case.

Generalized Hamiltonian system (GHS) describing dynamics in $T^{*}_{odd}{\cal
M}_{cl}\times\{\theta\}$ at least
for the I class theories contains the so-called generalized constraints in
Hamiltonian formalism (GCHF) $\Theta_{\imath}^{H}(\Gamma(\theta),\theta)$
in addition to one's own HS [2]. GCHF restrict possible solutions of HS
subsystem in GHS. In realizing of the condition
\begin{eqnarray}
{\rm deg}_{{\cal A}^{\ast}(\theta)}\Theta_{\imath}^{H}(\Gamma(\theta),\theta)
=0 
\end{eqnarray}
\sloppy
GCHF appear by holonomic constraints in Hamiltonian formalism (HCHF) [2]. In
the Lagrangian formalism the role being analogous to GCHF (HCHF) plays the
differential constraints in Lagrangian formalism (DCLF)
(holonomic constraints in Lagrangian formalism
(HCLF) being singled out by the relation in brackets) [1]
$$
\Theta_{\imath}\bigl({\cal A}(\theta),
{\stackrel{\ \circ}{\cal A}}(\theta),\theta\bigr) =
\Theta_{\imath}^{H}(\Gamma(\theta),\theta)_{\mid
{\stackrel{\ \circ}{\cal A}}(\theta)= {\stackrel{\ \circ}{\cal
A}}(\Gamma(\theta),\theta)}=0\
\Bigl({\rm deg}_{{\stackrel{\ \circ}{\cal
A}}(\theta)}\Theta_{\imath}\bigl({\cal A}(\theta),
{\stackrel{\ \circ}{\cal A}}(\theta),\theta\bigr)=0\Bigr)\,.
$$
DCLF above are
expressed through constraints in Hamiltonian formalism via
Legendre transform of superfunction $S_{L}(\theta) \equiv$
$S_{L}\bigl({\cal A}(\theta), {\stackrel{\ \circ}{\cal A}}(\theta),
\theta\bigr)$ $\in$ $C^k(T_{odd}{\cal M}_{cl}\times\{\theta\})$ with respect
to odd velocities ${\stackrel{\ \circ}{\cal A}}{}^{\imath}(\theta)$ under
assumption of invertibility for supermatrix of the 2nd partial superfield
derivatives of $S_{L}(\theta)$ with respect to superfields
${\stackrel{\ \circ}{\cal A}}{}^{\imath}(\theta)$ [2].   From basic
theorems of papers [1,2] analyzing the problems of functional
independence of DCLF (HCLF) and GCHF (HCHF) it follows, that among them the
identities may exist  whose number  is equal to $m=(m_{+},m_{-}),\;0\le m
< n$.

It is the presence of above identities serves in general by obstacle to
construct the generating functional of Green's functions in the Lagrangian
formalism (in the usual sense) without violation of locality and covariance
with
respect to indices of representation $T$ restriction  onto subsupergroup
$\bar{J}$: $T_{\vert\bar{J}}$.

In order to pass over the above obstacle it is necessary to extend the
supermanifold ${\cal M}_{cl}$ of superfields ${\cal A}^{\imath}(\theta)$
and therefore $T^{\ast}_{odd}{\cal M}_{cl}$ by additional supervariables, for
instance, by analogy with BV method [4].
The latter means that GThGT (GThST) in Hamiltonian formulation for GSTF
belonging to the I class with superfunction $S_{H}(\Gamma(\theta),\theta)$
restricted from $T^{*}_{odd}{\cal M}_{cl}\times\{\theta\}$ to ${\cal M}_{cl}$
must be continued  to superfunction $\tilde{S}_{H}(\theta)$
defined on the type $"T^{*}_{odd}{\cal M}_{cl}"$ supermanifold  with
additional coordinates and belonging already at least for $\hbar=0$ to the II
class GThGT.

Thus, by the central feature of the quantization procedure itself
are a generating equation of the form (2.2) for arbitrary configuration of
the superfields (coordinates) of extended supermanifold (for $\hbar=0$) being
considered for $\tilde{S}_{H}(\theta)$. In
correspondence with Statement 5.4 from Ref.[2]  the corresponding
HS is built with respect to the solution $\tilde{S}_{H}(\theta)$ of the last
equation, so that $\tilde{S}_{H}(\theta)$ is its integral.  In this
case $\tilde{S}_{H}(\theta)$ must satisfy according to Statement 5.3 [2]
to condition to be "proper". It means the rank value for supermatrix of the
2nd partial superfield derivatives of $\tilde{S}_{H}(\theta)$ with respect to
coordinates of $"T^{*}_{odd}{\cal M}_{cl}"$ type supermanifold calculated
on the special supersurface
is equal to half of the above supermanifold special dimension. As a
result the problem announced in beginning of the section may
be practically decided with use of $\tilde{S}_{H}(\theta)$ in the extended
space.

Not opening of explicit geometric matter of the extension procedure for ${\cal
M}_{cl}$ and $T^{*}_{odd}{\cal M}_{cl}$ (in fact using the  smooth
vector bundles theory on the basis of supermathematics) let us present the
functional-differential way for continuation of ${\cal
M}_{cl}$ being based on a certain interpretation of
gauge transformations of general  type (GTGT)  and ones of special
type (GTST) in Lagrangian and Hamiltonian formulations for GSTF [1,2]. These
transformations  defined in corresponding
supermanifolds in infinitesimal form [1,2] are given by the formulae
\begin{eqnarray}
{}\hspace{-1em} 1){} & \hspace{1em}
{\cal A}^{\imath}(\theta) \mapsto {\cal A}'^{\imath
}(\theta) = {\cal A}^{\imath}(\theta) + \delta_g{\cal A}^{\imath}(\theta);\
\delta_g{\cal A}^{\imath}(\theta) = \displaystyle\int d\theta'
\hat{\cal R}^{\imath}_{\alpha}\bigl({\cal A}(\theta),
{\stackrel{\ \circ}{\cal A}}(\theta),\theta;\theta'\bigr)\xi^{\alpha}(\theta')
\,;{} & \\ 
{}\hspace{-1em} 2){} &   \hspace{1em}
{\cal A}^{\imath}(\theta) \mapsto {\cal A}'^{\imath}(\theta) =
{\cal A}^{\imath}(\theta) + \delta{\cal A}^{\imath}(\theta);\
\delta{\cal A}^{\imath}(\theta) =
{\cal R}_0{}^{\imath}_{\alpha}({\cal A}(\theta),\theta)\xi_0^{\alpha}(
\theta)\,; \\ 
{}\hspace{-1em} 3){} &         \hspace{1em}
 (\Gamma^{p}(\theta), D^{\imath}(\theta))
\mapsto  (\Gamma^{p}(\theta),
D^{\imath}(\theta) + \delta D^{\imath}(\theta));\
\delta D^{\imath}(\theta) = \displaystyle\int d\theta'
\hat{{\cal R}}^{\imath}_{H{}\alpha}(\Gamma(\theta),\theta;\theta')
\xi^{\alpha}(\theta')\,; {} & \\ 
{}\hspace{-1em} 4){} & \hspace{1em}
{\cal A}^{\imath}(\theta) \mapsto {\cal A}'^{\imath}(\theta) =
{\cal A}^{\imath}(\theta) + \delta{\cal A}^{\imath}(\theta);\
\delta{\cal A}^{\imath}(\theta) = {\cal R}^{\imath}_{H{}\alpha}({\cal
A}(\theta),\theta)\xi_0^{\alpha}(\theta)  
\end{eqnarray}
with arbitrary superfunctions
$\xi^{\alpha}(\theta')$, $\xi^{\alpha}_{0}(\theta)$, $\alpha
=1,\ldots,m$.

\begin{sloppypar}
The ge\-ne\-ra\-tors of GTGT (GGTGT)
$\hat{\cal R}^{\imath}_{\alpha}\bigl({\cal A}(\theta),
{\stackrel{\ \circ}{\cal A}}(\theta),\theta;\theta'\bigr)$ in Lag\-ran\-gi\-an
and
GGTGT $\hat{\cal R}^{\imath}_{H{}\alpha}(\Gamma(\theta), \theta; \theta')$
in Ha\-mil\-to\-ni\-an for\-mu\-la\-ti\-ons for GSTF ha\-ve lo\-cal on
$\theta$ re\-pre\-sen\-ta\-ti\-on in fact con\-nec\-ting them with GGTST
${\cal R}_{0}{}^{\imath}_{\alpha}({\cal A}(\theta),\theta)$  and
${\cal R}^{\imath}_{H{}\alpha}({\cal A}(\theta),\theta)$ by me\-ans of the
re\-la\-ti\-ons [1,2]
\end{sloppypar}
\begin{eqnarray}
{} & {} & \hat{\cal R}^{\imath}_{\alpha}(\theta;\theta') \equiv
\hat{\cal R}^{\imath}_{\alpha}\bigl({\cal A}(\theta),
{\stackrel{\ \circ}{\cal A}}(\theta),\theta;\theta'\bigr) =
\displaystyle\sum\limits_{k=0}^{1} \left(\left(\displaystyle\frac{d}{d\theta}
\right)^k
\delta(\theta - \theta')\right)
{\hat{ {\cal R}}}_{k}{}^{\imath}_{\alpha}
\bigl( {\cal A}(\theta), {\stackrel{\ \circ}{\cal A}}(\theta), \theta\bigr)\;
, \\
{} & {} & \hat{\cal R}^{\imath}_{H{}\alpha}(
\theta;\theta')\equiv \hat{\cal R}^{\imath}_{H{}\alpha}(\Gamma(\theta),\theta;
\theta')
= \displaystyle\sum\limits_{k=0}^1
\displaystyle\left(\left(\frac{d}{d\theta}\right)^k \delta(\theta - \theta')
\right){\cal R}^{\imath}_{H{}k\alpha}(\Gamma(\theta),\theta)\;.
\end{eqnarray}
Let us remind that GGTST in (2.5), (2.7) one can consider by coinciding just
as $\hat{\cal R}^{\imath}_{\alpha}(\theta;\theta')$,
$\hat{\cal R}^{\imath}_{H{}\alpha}(\theta; \theta')$ after substitution into
formers the superfields ${\stackrel{\ \circ}{\cal A}}{}^{\imath}(\theta)$
expressed through $\Gamma^p(\theta)$, $\theta$ as a result of transition from
Lagrangian
formulation for GSTF to Hamiltonian one [2] (at least ${\cal R}^{\imath}_{
H{}\alpha}(\theta)$ and ${\cal R}_{0}{}^{\imath}_{\alpha}(\theta)$,
$\hat{\cal R}^{\imath}_{\alpha}(\theta;\theta')$ and
$\hat{\cal R}^{\imath}_{H{}\alpha}(\theta;\theta')$ are connected by
corresponding equivalence
transformations after their expression in a some definite formulation for
GSTF).

Grassmann parities for GGTGT, GGTST, superfields $\xi^{\alpha}(\theta')$,
$\xi^{\alpha}_{0}(\theta)$, $D^{\imath}(\theta)$ (being auxiliary Lagrange
multipliers for GCHF) are defined by means of the table
\begin{eqnarray}
\begin{array}{lcccccccl}
{} & \hat{\cal R}^{\imath}_{\alpha}(\theta;\theta') &
\hat{\cal R}^{\imath}_{H{}\alpha}(\theta; \theta') &
{\hat{\cal R}}_{k}{}^{\imath}_{\alpha}(\theta) &
{\cal R}^{\imath}_{H{}k\alpha}(\theta) & \xi^{\alpha}(\theta')&
\xi^{\alpha}_0(\theta) & D^{\imath}(\theta) &{} \\
\varepsilon_P & 1 & 1 &  \delta_{1k} & \delta_{1k} & 0 & 0 & 0 & {} \\
\varepsilon_{\bar{J}} & \varepsilon_{\imath}+\varepsilon_{\alpha} &
\varepsilon_{\imath}+\varepsilon_{\alpha} & \varepsilon_{\imath}+
\varepsilon_{\alpha} & \varepsilon_{\imath}+\varepsilon_{\alpha} &
\varepsilon_{\alpha} &\varepsilon_{\alpha} & \varepsilon_{\imath} & {}\\
\varepsilon & \varepsilon_{\imath}+\varepsilon_{\alpha}+1 &
\varepsilon_{\imath}+\varepsilon_{\alpha}+1 & \varepsilon_{\imath}+
\varepsilon_{\alpha}+\delta_{1k}  & \varepsilon_{\imath}+\varepsilon_{\alpha}
+ \delta_{1k} &
\varepsilon_{\alpha} &\varepsilon_{\alpha} & \varepsilon_{\imath} &
\hspace{-1em}.
\end{array} 
\end{eqnarray}
GTGT (2.4), (2.6) are  the transformations of invariance for superfunctionals
respectively
\begin{eqnarray}
{} & {} & Z[{\cal A}] = \displaystyle\int d\theta S_L(\theta)\;, \\ 
{} & {} & Z^{(1)}_H[\Gamma,D] = \displaystyle\int d\theta\left(
{\stackrel{\ \circ}{\cal A}}{}^{\imath}(\theta){\cal A}^{\ast}_{\imath
}(\theta) - S_H(\Gamma(\theta),\theta) +
D^{\imath}(\theta)\Theta_{\imath}^{H}(\Gamma(\theta),\theta)\right),
\end{eqnarray}
whereas the only superfunction $S({\cal A}(\theta),\theta)$ from
Corollary 2.2 of the Theorem 2 from Ref.[1] in Lagrangian formulation
for GSTF (equivalently from Corollary 1.2 of  the Theorem 1 from Ref.[2]
in Hamiltonian one) is invariant with respect to GTST (2.5), (2.7).
$S({\cal A}(\theta),\theta)$ appears by the potential term in the almost
natural system determined in both formulations by superfunctions
$S_{L}(\theta)$ and $S_{H}(\theta)$ respectively
\begin{eqnarray}
{} & S_L\bigl({\cal A}(\theta),
{\stackrel{\ \circ}{\cal A}}(\theta),\theta\bigr)=
T\bigl({\cal A}(\theta), {\stackrel{\ \circ}{\cal A}}(\theta)\bigr) -
S({\cal A}(\theta),\theta)\;, & {} \\ 
{} & \hspace{-1em}
S_{H}(\Gamma(\theta),\theta) =
\left({\stackrel{\ \circ}{\cal A}}{}^{\imath}(\theta){\cal A}^{\ast}_{\imath
}(\theta) -
S_L\bigl({\cal A}(\theta),
{\stackrel{\ \circ}{\cal A}}(\theta),\theta\bigr)
\right)_{\mid{\stackrel{\ \circ}{\cal A}}(\theta) = {\stackrel{\ \circ}{\cal
A}}(\Gamma(\theta),\theta)} \hspace{-1em}= T(\Gamma(\theta)) +
S({\cal A}(\theta),\theta)\,. & {} 
\end{eqnarray}
The invariance of the superfunctionals (2.11), (2.12) and
$S({\cal A}(\theta),\theta)$  with respect to GTGT (2.4), (2.6) and GTST
(2.5) or (2.7) respectively is expressed in the presence of the
corresponding identities
\begin{eqnarray}
\int\hspace{-0.3em}
d\theta {}\frac{\delta_r Z[{\cal A} ]}{\delta{\cal A}^{\imath
}(\theta)}{}{\hat{{\cal R}}}^{\imath}_{\alpha}(\theta; {\theta}') = 0,\
\hspace{-0.2em}\int \hspace{-0.2em}d\theta
\frac{\delta_r Z_H^{(1)}[\Gamma,D]}{\delta D^{\imath}(\theta)\phantom{xxxx}}
\hat{\cal R}^{\imath}_{H{}\alpha}(\theta;\theta') = 0,\
  {S,}_{\imath}({\cal A}(\theta),\theta){\cal R}_{0}{}^{\imath}_{
\alpha}(\theta) = 0\,.  
\end{eqnarray}
At last, it should be noted that  sets of generators
$\hat{\cal R}^{\imath}_{H{}\alpha}(\theta;\theta')$,
$\hat{\cal R}^{\imath}_{H{}\alpha}(\theta;\theta')$,
${\cal R}_{0}{}^{\imath}_{\alpha}({\cal A}(\theta),\theta)$  are complete and
functionally (for ${\cal R}_{0}{}^{\imath}_{\alpha}({
\cal A}(\theta),\theta)$  linearly) independent correspondingly [1,2].

Further interpretation of GTST (2.5), (2.7) consists in
change in the indicated formulae of arbitrary superfields $\xi^{\alpha}_{0
}(\theta)$ $\in$ $\tilde{\Lambda}_{D\vert Nc + 1}(z^{a},\theta;{\bf K})$ onto
new arbitrary superfunctions being by a  standard set of ghost
superfields $C^{\alpha}(\theta)$ $\in$
$\tilde{\Lambda}_{D\vert Nc +1}(z^{a},\theta;{\bf K})$ by the formal rule
\begin{eqnarray}
\xi^{\alpha}_0(\theta) = C^{\alpha}(\theta)\nu,\ \nu\in{}^1\Lambda_1(\theta)
\;.
\end{eqnarray}

Grassmann gradings for $C^{\alpha}(\theta)$ and their derivatives
${\stackrel{\,\circ}{C}}{}^{\alpha}(\theta)$ are determined from ones for
$\xi^{\alpha}_0(\theta)$ (2.10) and $\nu$
($(\varepsilon_{P}, \varepsilon_{\bar{J}}, \varepsilon)\nu
= (1,0,1)$)
\begin{eqnarray}
(\varepsilon_{P}, \varepsilon_{\bar{J}}, \varepsilon)C^{\alpha}(\theta) =
(\varepsilon_{P}({\stackrel{\,\circ}{C}}{}^{\alpha}(\theta))+1,
\varepsilon_{\bar{J}}({\stackrel{\,\circ}{C}}{}^{\alpha}(\theta)),
\varepsilon({\stackrel{\,\circ}{C}}{}^{\alpha}(\theta))+1) =
(1, \varepsilon_{\alpha}, \varepsilon_{\alpha} + 1)\,. 
\end{eqnarray}
Represent the variations of superfields ${\cal A}^{\imath}(\theta)$
in (2.5), (2.7) in the form of differentials formally
having written to this end in these formulae instead of arbitrary superfields
$\xi^{\alpha}_{0}(\theta)$ the differentials of
arbitrary ones  $\tilde{\xi}^{\alpha}_{0}(\theta)$ $\in$ $\tilde{\Lambda}_{
D\vert Nc + 1}(z^{a},\theta;{\bf K})$.
For instance, for (2.7) we have
\begin{eqnarray}
d{\cal A}^{\imath}(\theta) = {\cal R}^{\imath}_{H{}\alpha}({\cal A}(\theta),
\theta)d\tilde{\xi}^{\alpha}_{0}(\theta),\ \xi^{\alpha}_{0}(\theta) =
d\tilde{\xi}^{\alpha}_{0}(\theta)\;. 
\end{eqnarray}
In its turn, express the differentials $d\tilde{\xi}^{\alpha}_{0}(\theta)$
themselves according to formula (2.16) in the form
\begin{eqnarray}
d\tilde{\xi}^{\alpha}_{0}(\theta) = C^{\alpha}(\theta)d\theta\;. 
\end{eqnarray}
It follows from (2.18), (2.19) that variations of ${\cal A}^{\imath}(\theta)$
(2.5), (2.7) may be represented as the 1st order
system of differential equations with respect to derivatives on $\theta$ in
the normal form respectively
\begin{eqnarray}
{} & {} & \displaystyle\frac{d_r {\cal A}^{\imath}(\theta)}{d\theta\phantom{
xxxx}} = {\cal R}_0{}^{\imath}_{\alpha}({\cal A}(\theta),{\theta})C^{\alpha
}(\theta),\ {\cal R}_0{}^{\imath}_{\alpha}({\cal A}(\theta),{\theta}) \equiv
\hat{\cal R}_0{}^{\imath}_{\alpha}({\cal A}(\theta),{\theta})\;
 , \\ 
{} & {} & \displaystyle\frac{d_r{\cal A}^{\imath}(\theta)}{d\theta\phantom{
xxxx}} = {\cal R}^{\imath}_{H{}\alpha}({\cal A}(\theta),{\theta})C^{\alpha
}(\theta),\ {\cal R}^{\imath}_{H{}\alpha}({\cal A}(\theta),{\theta}) =
\hat{\cal R}_0{}^{\imath}_{\alpha}({\cal A}(\theta),{\theta})\;.  
\end{eqnarray}

The  transformation for the GTGT (2.4), (2.6) being analogous to the above
procedure for GTST realized in (2.16), (2.18)--(2.21) is impossible without
principal change of the initial supergroup $J$, superspace ${\cal M}$,
superalgebras  ${\Lambda}_{D\vert Nc +1}(z^{a},\theta;{\bf K})$,
$\tilde{\Lambda}_{D\vert Nc +1}(z^{a},\theta;{\bf K})$ structure by means of
the extension of subsupergroup $P$, subsuperspace $\tilde{P}$ to
corresponding two-parametric sets with help of additional to
$\theta$ Grassmann coordinate $\theta^{(1)}$ ($(\theta^{(1)})^2=0$). Leaving
the exact algorithm construction  for the interpretation of GTGT (2.4),
(2.6) out the paper's scope I only note that in  formulae (2.4), (2.6) it is
necessary to change the all superfields
$({\cal A}^{\imath}(\theta), {\cal A}^{\ast}_{\imath}(\theta),
\xi^{\alpha}(\theta))$ onto ones depending upon two-dimensional odd time
$(\theta, \theta^{(1)})$:  $(\Gamma^p(\theta, \theta^{(1)}),
\xi^{\alpha}(\theta, \theta^{(1)}))$.

Relationship (2.16)  realizes in the formula form a projection on the 1st
argument  of the following mapping image
\begin{eqnarray}
\xi^{\alpha}_0(\theta) \mapsto \bigl({C}^{\alpha}(\theta), \overline{C}{}^{
\alpha}(\theta), B^{\alpha}(\theta)\bigr)\;,  
\end{eqnarray}
in which
the new arbitrary superfunctions being by standard sets of antighost
 $\overline{C}{}^{\alpha }(\theta)$ and auxiliary
$B^{\alpha}(\theta)$ superfields ($\overline{C}{}^{\alpha}( \theta)$,
$B^{\alpha}(\theta)$ $\in$ $\tilde{\Lambda}_{D\vert Nc + 1}(z^{a},\theta;{\bf
K})$) [4] are associated to arbitrary superfields $\xi^{\alpha}_0(\theta)$  in
addition to ${C}^{\alpha}(\theta)$. The table of Grassmann parities for
additional superfields and their derivatives
${\stackrel{\,\circ}{\overline{C}}}{}^{\alpha}(\theta)$,
${\stackrel{\,\circ}{B}}{}^{\alpha}(\theta)$ is written down in the form
\begin{eqnarray}
\begin{array}{lccccl}
{} & \overline{C}{}^{\alpha}(\theta) & B^{\alpha}(\theta) &
{\stackrel{\,\circ}{\overline{C}}}{}^{\alpha}(\theta) &
{\stackrel{\,\circ}{B}}{}^{\alpha}(\theta) & {}\\
\varepsilon_{P} & 1 & 0 & 0 & 1 & {} \\
\varepsilon_{\bar{J}} & \varepsilon_{\alpha} & \varepsilon_{\alpha} &
\varepsilon_{\alpha} & \varepsilon_{\alpha} & {} \\
\varepsilon & \varepsilon_{\alpha} + 1 & \varepsilon_{\alpha} &
\varepsilon_{\alpha} & \varepsilon_{\alpha} + 1 & .
\end{array}   
\end{eqnarray}
Let us designate the joint sets composed from superfields $\{
{\cal A}^{\imath}(\theta)$, $C^{\alpha}(\theta)$, $\overline{C}{}^{\alpha}(
\theta)$, $B^{\alpha}(\theta)\}$ for coinciding continual components of
indices $\imath, \alpha$  through composite superfields correspondingly
\renewcommand{\theequation}{\arabic{section}.\arabic{equation}\alph{lyter}}
\begin{eqnarray}
\setcounter{lyter}{1}
{} & \Phi^{A}_{min}(\theta) \equiv ({\cal A}^{\imath}(\theta),
C^{\alpha}(\theta)),\ A=(\imath,\alpha),\
{}\footnotemark[1][A]=n+m\;, {} &
\\
\setcounter{equation}{24}
\setcounter{lyter}{2}
{} & \Phi^{B}(\theta) \equiv (\Phi^{A}_{min}(\theta),\Phi^{C}_{aux
}(\theta)) \equiv (\Phi^{A}_{min}(\theta), \overline{C}{}^{\alpha}(\theta),
B^{\alpha}(\theta)) \equiv ({\cal A}^{\imath}(\theta), \Phi^{D}_{add}(
\theta))\,, {}&
\nonumber \\
{} &    B=(\imath,\alpha,\alpha,\alpha)\equiv (A,C) \equiv
(\imath, D),\  [B]=n+3m,\ [C]=2m,\ [D]=3m\,. {} & 
\end{eqnarray}
\footnotetext[1]{symbol $[A]$ in (2.24) denotes a number of discrete indices
entering in condensed index $A$ in contrast to  the sign $[{D}/{2}]$ in (2.1)
denoting the integer part of number $D/2 \in {\bf R}$}Composite
superfields $\Phi^{A}_{min}(\theta),\, \Phi^{C}_{aux}(\theta),\,
\Phi^{D}_{add}(\theta), \Phi^{B}(\theta)$ parametrize the sets (in changing
of ${\cal A}^{\imath}(\theta)$, $\xi^{\alpha}_0(\theta)$ according to (2.16),
(2.22) the superfields $\Phi^{B}(\theta)$ are changed as well) having the
complicated geometric-differential structure (in what follows formally,
without explicit definitions and descriptions being called, in general, by
supermanifolds(!))
and being designated correspondingly
\begin{eqnarray}
\setcounter{lyter}{1}
{} & {} & {\cal M}_{min} = \{\Phi^{A}_{min}(\theta) \vert
{\cal A}^{\imath}(\theta) \in {\cal M}_{cl}, C^{\alpha}(\theta)\in{\cal
M}_{C}\}\;, \\
\setcounter{equation}{25}
\setcounter{lyter}{2}
{} & {} & {\cal M}_{aux} = \{\Phi^{C}_{aux}(\theta) \vert
\overline{C}{}^{\alpha}(\theta) \in {\cal M}_{\overline{C}}, B^{\alpha
}(\theta)\in{\cal M}_B\}\;, \\ 
\setcounter{equation}{25}
\setcounter{lyter}{3}
{} & {} & {\cal M}_{add} = \{\Phi^{D}_{add}(\theta) \vert
{C}^{\alpha}(\theta) \in {\cal M}_{C}, \Phi^{C}_{aux}(\theta)\in{\cal M}_{aux
}\}\;, \\ 
\setcounter{equation}{25}
\setcounter{lyter}{4}
{} & {} & {\cal M}_{(min,\overline{C})} = \{(\Phi^{A}_{min}(\theta),
\overline{C}{}^{\alpha}(\theta)) \vert
\Phi^{A}_{min}(\theta) \in {\cal M}_{min}, \overline{C}{}^{\alpha}(\theta)
\in{\cal M}_{\overline{C}}\}\;, \\ 
\setcounter{equation}{25}
\setcounter{lyter}{5}
{} & {} & {\cal M}_{ext} = \{\Phi^{B}(\theta) \vert
\Phi^{A}_{min}(\theta) \in {\cal M}_{min}, \Phi^{C}_{aux}(\theta)\in{\cal
M}_{aux}\}\;, 
\end{eqnarray}
where ${\cal M}_{t}, t\in \{ C,\overline{C},B\}$ are the sets parametrized
by $C^{\alpha}(\theta)$, $\overline{C}{}^{\alpha }(\theta )$,
$B^{\alpha }(\theta)$ respectively.

The Grassmann gradings of the composite superfields (2.24) according to
(2.17), (2.23) are defined by the formulae
\begin{eqnarray}
\setcounter{lyter}{1}
{}&
\begin{array}{lccccl}
{} & \Phi^{A}_{min}(\theta)  & \Phi^{C}_{aux}(\theta) & \Phi^{D}_{add}(\theta)
 & \Phi^{B}(\theta) & {}\\
\varepsilon_{P} & (\varepsilon_{P})_A & (\varepsilon_{P})_C &
(\varepsilon_{P})_D & (\varepsilon_{P})_B & {} \\
\varepsilon_{\bar{J}} & (\varepsilon_{\bar{J}})_A & (\varepsilon_{\bar{J}})_C
& (\varepsilon_{\bar{J}})_D  & (\varepsilon_{\bar{J}})_B & {} \\
\varepsilon & \varepsilon_A & \varepsilon_C &
\varepsilon_D & \varepsilon_B & ,
\end{array} & {}
 \\ 
\setcounter{equation}{26}
\setcounter{lyter}{2}
{} & \varepsilon_A \equiv (\varepsilon_{\imath},\varepsilon_{\alpha} + 1),\
\varepsilon_C \equiv (\varepsilon_{\alpha} + 1,\varepsilon_{\alpha}),\
\varepsilon_D \equiv (\varepsilon_{\imath},\varepsilon_C),\
\varepsilon_B \equiv (\varepsilon_{\imath},\varepsilon_D)
\equiv (\varepsilon_A,\varepsilon_C)\;. & {} 
\end{eqnarray}
The derivatives on $\theta$ of  superfields (2.24)
${\stackrel{\circ}{\Phi}}{}^{A}_{min}(\theta)$,
${\stackrel{\circ}{\Phi}}{}^C_{aux}(\theta)$,
${\stackrel{\circ}{\Phi}}{}^D_{add}(\theta)$,
${\stackrel{\circ}{\Phi}}{}^B(\theta)$
possess by shifted (on 1) $\varepsilon_{P}$, $\varepsilon$ gradings.

Just as $T_{odd}{\cal M}_{cl}$ [1]  the supermanifolds $T_{odd}{\cal M}_{t}$,
$t \in \{C,\overline{C},B\}$ are formally introduced, parametrized by pairs of
superfields ($t^{\alpha}(\theta)$, ${\stackrel{\,\circ}{t}}{}^{
\alpha}(\theta)$) respectively. As a consequence of
$T_{odd}{\cal M}_{t}$ construction above the following supermanifolds are
defined as well
\renewcommand{\theequation}{\arabic{section}.\arabic{equation}}
\begin{eqnarray}
T_{odd}{\cal M}_{s},\ s\in\{min, aux,  (min,\overline{C}), add, ext\}\,,
\end{eqnarray}
parametrized by coordinates
$\bigl(\Phi^{A}_{min
}(\theta)$, ${\stackrel{\circ}{\Phi}}{}^{A}_{min}(\theta)\bigr)$,
$\bigl(\Phi^{C}_{aux}(\theta)$, ${\stackrel{\circ}{\Phi}}{}^C_{aux}(\theta)
\bigr)$,
$\bigl(\Phi^{(A,\alpha)}_{(min,\overline{C})}(\theta)$,
${\stackrel{\circ}{\Phi}}{}^{(A,\alpha)}_{(min,\overline{C})}(\theta)\bigr)$,
$\bigl(\Phi^{D}_{add}(\theta)$, ${\stackrel{\circ}{\Phi}}{}^D_{add}(\theta)
\bigr)$,
$\bigl(\Phi^{B}(\theta)$, ${\stackrel{\circ}{
\Phi}}{}^B(\theta)\bigr)$ respectively. By analogy with
juxtaposition to any ${\cal A}^{\imath}(\theta)$ of ${\cal A}^{\ast}_{
\imath}(\theta)$ let us associate with any superfield from
$\Phi^{D}_{add}(\theta)$ the following superantifields
\renewcommand{\theequation}{\arabic{section}.\arabic{equation}\alph{lyter}}
\begin{eqnarray}
\setcounter{lyter}{1}
{} & \Phi^{\ast}_{D{}add}(\theta)= (C^{\ast}_{\alpha}(\theta),
\Phi^{\ast}_{C{}aux}(\theta)) =  (C^{\ast}_{\alpha}(\theta),
\overline{C}{}^{\ast}_{\alpha}(\theta), B^{\ast}_{\alpha}(\theta))\;, &{} \\
\setcounter{equation}{28}
\setcounter{lyter}{2}
{} & \Phi^{\ast}_{A{}min}(\theta) =
({\cal A}^{\ast}_{\imath}(\theta), C^{\ast}_{\alpha}(\theta)),\
\Phi^{\ast}_B(\theta) = (\Phi^{\ast}_{A{}min}(\theta),
\Phi^{\ast}_{C{}aux}(\theta)) = ({\cal A}^{\ast}_{\imath}(\theta),
\Phi^{\ast}_{D{}add}(\theta))\;.&{} 
\end{eqnarray}
Let us formally define the supermanifolds $T^{\ast}_{odd}{\cal M}_{t}$, $t
\in \{C,\overline{C},B\}$ by the relationships
\renewcommand{\theequation}{\arabic{section}.\arabic{equation}}
\begin{eqnarray}
T^{\ast}_{odd}{\cal M}_{t}
= \{(t^{\alpha}(\theta),
t^{\ast}_{\alpha}(\theta))\vert t^{\alpha}(\theta) \in {\cal M}_t\},\
t\in \{C, \overline{C}, B\}\;. 
\end{eqnarray}
Objects $T^{*}_{odd}{\cal M}_{min}$, $T^{*}_{odd}{\cal M}_{aux}$,
$T^{*}_{odd}{\cal M}_{add}$, $T^{*}_{odd}{\cal M}_{(min,\overline{C})}$,
$T^{\ast}_{odd}{\cal M}_{ext}$ are formally determined from relations
(2.29) analogously to (2.27) with corresponding parametrization.

At last, one can define the supermanifolds of the form
\begin{eqnarray}
T_{odd}(T^{\ast}_{odd}{\cal M}_{s}), s \in C_0 = \{C,
\overline{C}, B, min, aux, add, (min,\overline{c}), ext\}\,, 
\end{eqnarray}
parametrized by superfields
$\bigl(\Phi^{A_s}_{s}(\theta)$, ${\stackrel{\circ}{\Phi}}{}^{A_s}_{s}(
\theta)$, $\Phi^{\ast}_{A_s,s}(\theta)$,
${\stackrel{\circ}{\Phi}}{}^{\ast}_{A_s,s}(\theta)\bigr)$,
$A_s$ = $\alpha$, $\alpha$, $\alpha$, $A$, $C$,
$D$, $(A,\alpha)$, $B$ for $s$ being by the
corresponding element from ordered set $C_0$ in (2.30). Introduce on
$T^{*}_{odd}{\cal M}_{s}$ the local coordinates which ignore the
superfield-superantifield structure
\begin{eqnarray}
\Gamma^{p_s}_{s}(\theta) = (\Phi^{A_s}_{s}(\theta),
\Phi^{\ast}_{A_s,s}(\theta)),\ p_s = 1,\ldots,2[A_s],\; s\in C_0\;.
\end{eqnarray}
Sets of the pairs
$\bigl(\Gamma^{p_s}_{s}(\theta), {\stackrel{\circ}{\Gamma}}{}^{p_s}_{s
}(\theta)\bigr)$ appear by the coordinates on
$T_{odd}(T^{*}_{odd}{\cal M}_{s})$ according to (2.31).  Because the
superfields $({\cal A}^{\imath}(\theta), {\cal A}^{\ast}_{ \imath}(\theta))$
$\equiv$ $\Gamma^{p}_{cl}(\theta)$ are transformed
with respect to supergroup $J$  superfield representation $T\oplus T^{*}$ $
\equiv$ $T^{cl}\oplus T^{cl{}*}$
[1,2] then taking into consideration that superfields $\xi^{\alpha}_0(\theta)$
($\tilde{\xi}^{\alpha}_0(\theta)$) are the elements of a some superfield
representation $T_{\xi}$ space of  supergroup $J$  as well, the
formulae (2.16), (2.19), (2.22) in fact mean the superfields
$C^{\alpha}(\theta)$, $\overline{C}{}^{\alpha}(\theta)$, $B^{\alpha}(\theta)$
are transformed with respect to the same representation $T_{\xi}$
and their superantifields $\Phi^{\ast}_{D{}add}(\theta)$ with respect to
conjugate one $T^{\ast}_{\xi}$ (as $T^{\ast}$ and $T$ in Ref.[2]).
Therefore the transformation laws for
${\stackrel{\circ}{\Phi}}{}^D_{add}(\theta)$,
${\stackrel{\circ}{\Phi}}{}^{\ast}_{D{}add}(\theta)$ are known
too.

Gradings $\varepsilon_{P}, \varepsilon_{\bar{J}}, \varepsilon$   for
superantifields $\Phi^{\ast}_{A_s,s}(\theta)$, $s \in C_0$ read as follows
according to (2.26)
\begin{eqnarray}
(\varepsilon_{P}, \varepsilon_{\bar{J}}, \varepsilon)\Phi^{\ast}_{A_s,s
}(\theta)
= ((\varepsilon_{P})_{A_s} + 1, (\varepsilon_{\bar{J}})_{A_s},
\varepsilon_{A_s}+1)\;.
\end{eqnarray}
For ${\stackrel{\circ}{\Phi}}{}^{\ast}_{A_s,s}(\theta)$
the $\varepsilon_{\bar{J}}$ grading is the same as in (2.32) for $\Phi^{
\ast}_{A_s,s}(\theta)$ but $\varepsilon_{P}$ and $\varepsilon$ ones are
shifted on 1.  Besides it is convenient to introduce the Grassmann parities
values for  $\Gamma^{p_s}_{s}(\theta)$, ${\stackrel{\circ}{\Gamma}}{}^{p_s}_{s
}(\theta)$ considered as the whole superfields
\begin{eqnarray}
(\varepsilon_{P}, \varepsilon_{\bar{J}}, \varepsilon)\Gamma^{p_s}_{s}(\theta)
= ((\varepsilon_{P})_{p_s}, (\varepsilon_{\bar{J}})_{p_s},
\varepsilon_{p_s}),\
(\varepsilon_{P}, \varepsilon_{\bar{J}}, \varepsilon){\stackrel{\circ}{
\Gamma}}{}^{p_s}_{s}(\theta) = ((\varepsilon_{P})_{p_s} +1,
(\varepsilon_{\bar{J}})_{p_s}, \varepsilon_{p_s}+ 1)\,.
\end{eqnarray}
Eqs.(2.20), (2.21) in terms of new geometric objects
are the ordinary differential equations (ODE) of the 1st order on $\theta$ in
${\cal M}_{min}\times\{\theta\}$.

\begin{sloppypar}
The variations of arbitrary superfunctions
${\cal F}(\theta)$ $\equiv$
${\cal F}\bigl(\Gamma_{cl}(\theta)$, ${\stackrel{\circ}{\Gamma}}_{cl}(
\theta)$, $\theta\bigr)$ $\in$
$C^{k}\bigl(T_{odd}(T^{\ast}_{odd}
{\cal M}_{cl})$ $\times$ $\{\theta\}\bigr)$ $\equiv$ $D_{cl}^{k}$, $k\leq
\infty$
with respect to GTGT (2.4) and
${\cal J}(\theta)$ $\equiv$
${\cal J}({\cal A}(\theta), \theta)$ $\in$
$C^{k}\bigl({\cal M}_{cl}\times\{\theta\}\bigr)$ with respect to GTST (2.5),
(2.7) may be represented by means of the formulae respectively
\end{sloppypar}
\renewcommand{\theequation}{\arabic{section}.\arabic{equation}\alph{lyter}}
\begin{eqnarray}
\setcounter{lyter}{1}
{} & \delta{\cal F}(\theta) =
\displaystyle\int d\theta'\displaystyle\frac{\delta {\cal F}(\theta)
\phantom{x}}{\delta{\cal A}^{\imath
}(\theta')}\int d\theta_1
{\hat{{\cal R}}}^{\imath}_{\alpha}(\theta'; {\theta}_1)\xi^{\alpha}(\theta_1)
\;, {} &\\
\setcounter{equation}{34}
\setcounter{lyter}{2}
{} & \delta{\cal J}(\theta) = {{\cal J},}_{\imath}(\theta)
{\cal R}_0{}^{\imath}_{\alpha}({\cal A}(\theta),\theta)\xi_0^{\alpha}(\theta),
\ \delta{\cal J}(\theta) = {{\cal J},}_{\imath}(\theta)
{\cal R}^{\imath}_{H{}\alpha}({\cal A}(\theta),\theta)\xi_0^{\alpha}(\theta)
\;. 
\end{eqnarray}
Under correspondences being given by (2.16), (2.19) for the above indicated
arbitrary superfunctions ${\cal F}(\theta)$, ${\cal J}(\theta)$ not explicitly
depending upon $\theta$  the only variations (2.34b) for GTST can be
represented almost equivalently
in the form of translation along $\theta$ coordinate on a parameter $\mu
\in {}^1\Lambda_1(\theta)$  induced by variation of
superfields ${\cal A}^{\imath}(\theta)$ (or
equivalently onto solutions $\bar{\cal A}^{\imath}(\theta)$ for corresponding
Eqs.(2.20), (2.21))
\renewcommand{\theequation}{\arabic{section}.\arabic{equation}}
\begin{eqnarray}
\delta{\cal J}({\cal A}^{\imath}(\theta))
\equiv \delta_{\mu}{\cal J}(\theta)_{\mid\bar{\cal
A}} \equiv \displaystyle\frac{d_r{\cal J}(\theta)}{d\theta\phantom{xxxx}}_{
\mid\bar{\cal A}}\mu = {\cal J},_{\imath}(\theta)
\frac{d_r{\cal A}^{\imath}(\theta)}{d\theta\phantom{xxxx}}_{
\mid\bar{\cal A}}\mu \;. 
\end{eqnarray}
Formula (2.35) is suitable in equal form
for both variations (2.34b). The absence in the right-hand side of
variations (2.35), in view of nilpotent operator $\frac{d}{d
\theta}$ presence, the projector $P_{0}(\theta)$ from the system of projectors
$P_{a}(\theta)$, $a=0,1$ acting in $D^{k}_{cl}(\theta)$
[1,2]
is possible only in realizing of the
solvability conditions for systems (2.20), (2.21) according to the
Statement 3.6 from Ref.[2], which (i.e. solvability conditions) for the above
systems are not in general fulfilled.
The accounting of explicit dependence upon $\theta$ for ${\cal J}(\theta)$
under interpretation of its variation (2.34b) by means of the type (2.35)
formula is
possible only  with help of the above additional variable $\theta^{(1)}$
introduction. Remind  that
concept of solvability for ODE of the 0,1,2 orders with respect to
$\theta$ [1,2] means the fulfilment on  solutions of corresponding system
of all its  differential consequences being obtained by differentiation on
$\theta$ of this system.
The last remarks mean a notion of
"almost equivalence" for formulae (2.34b), (2.35).

Let us continue the process of extension of algebraic structures earlier
given on ${\cal M}_{cl}$, $T_{odd}{\cal M}_{cl}$, $T^{*}_{odd}{\cal M}_{cl}$
and etc. [1,2] onto corresponding  geometric objects built on a basis of
${\cal M}_{s}$, $s \in C_0$. Superalgebras
${\bf K}[T_{odd}{\cal M}_{cl}\times
\{\theta\}]$, ${\bf K}[[T_{odd}{\cal M}_{cl}\times\{\theta\}]]$,
$C^{k}(T_{odd}{\cal M}_{cl}\times\{\theta\})$,
${\bf K}[T^{\ast}_{odd}{\cal M}_{cl}\times
\{\theta\}]$, ${\bf K}[[T^{\ast}_{odd}{\cal M}_{cl}\times\{\theta\}]]$,
$C^{k}(T^{\ast}_{odd}{\cal M}_{cl}\times\{\theta\})$,
${\bf K}[T_{odd}(T^{\ast}_{odd}{\cal M}_{cl})\times
\{\theta\}]$, ${\bf K}[[T_{odd}(T^{\ast}_{odd}{\cal M}_{cl})\times\{
\theta\}]]$, $D^{k}_{cl}$ [1,2]
are successively continued to the corresponding sets of
superfunctions under formal change of  index $"cl"$ on arbitrary element
$s \in C_0$ (2.30).
Here under ${\bf K}[\ \;]$, ${\bf K}[[\ \;]]$ and $C^{k}(\ \;)$ we understand
respectively the
superalgebras over number field ${\bf K}$ of superfunctions being by finite
polynomials on expanding in series in powers of  generating
elements--coordinates on $T_{odd}{\cal M}_{s}\times\{\theta\}$, $T^{*}_{odd
}{\cal M}_{s}\times\{\theta\}, \ldots$,  i.e. superfields $\bigl(\Gamma^{
p_s}_{s}(\theta)$, ${\stackrel{\circ}{\Gamma}}{}^{p_s}_{s}(\theta)\bigr)$,
by formal power series
(${\bf K}[[\ \;]]$) and by superalgebra $C^{k}(\ )$, $k \le \infty$ with
partial superfield differentiation of its elements for which the
expansions in Taylor's series are valid [1,2]. The last superalgebra
consists of $k$-times differentiable superfunctions. In addition it is
naturally to regard that with respect to part of generating elements, for
instance, $\Phi^{D}_{add}(\theta)$ and $\Phi^{\ast}_{B}(\theta)$ the
superalgebra $C^{k}(T^{*}_{odd}{\cal M}_{ext}\times\{\theta\})$, $k \le
\infty$ consists of superfunctions being considered as (finite) power series
with respect to indicated variables.
For example, for ${\cal F}\bigl(C(\theta), {\stackrel{\,\circ}{C}}(\theta),
(\theta)\bigr)$ $\in$ ${\bf K}[[T_{odd}{\cal M}_{C}
\times\{\theta\}]]$    the representation in the
form of series with respect to $C^{\alpha}(\theta)$ and (or) ${
\stackrel{\,\circ}{C}}{}^{\alpha}(\theta)$, to be finite for polynomials
corresponding in the such case for local superfunctions relatively
to $C^{\alpha}(\theta)$
and (or) ${\stackrel{\,\circ}{C}}{}^{\alpha}(\theta)$, is valid
according to Ref.[1] with notations being analogous to ones suggested in
[1,2]
\renewcommand{\theequation}{\arabic{section}.\arabic{equation}}
\begin{eqnarray}
 \hspace{-1em} {\cal F}\bigl( C(\theta),
{\stackrel{\circ}{C}}(\theta), \theta \bigr) =
\sum\limits_{l=0}\frac{1}{l!}
{\cal F}_{(\alpha)_l} \bigl({\stackrel{\,\circ}{C}}(\theta),\theta\bigr)
\vec{C}^{(\alpha)_l}(\theta) = \sum\limits_{k,l=0}
\frac{1}{k!l!}{\cal F}_{(\alpha)_l(\beta)_k}(\theta)
\vec{\stackrel{\,\circ}{C}}{}^{(\beta)_k}(\theta)
\vec{C}^{(\alpha)_l}(\theta)\,, {} &  
\end{eqnarray}
\vspace{-3ex}
\renewcommand{\theequation}{\arabic{section}.\arabic{equation}\alph{lyter}}
\begin{eqnarray}
\setcounter{lyter}{1}
{} & \vec{\stackrel{\,\circ}{C}}{}^{
(\beta)_k}(\theta) =
\displaystyle\prod\limits_{p=1}^k {\stackrel{\,\circ}{C}}{}^{
\beta_p}(\theta),\
\vec{C}^{(\alpha)_l}(\theta) =
\displaystyle\prod\limits_{p=1}^l C^{\alpha_p}(\theta)\;, & {} \\
\setcounter{equation}{37}
\setcounter{lyter}{2}
{} & {\cal F}_{(\alpha)_l} \bigl({\stackrel{\,\circ}{C}}(\theta),\theta\bigr)
\equiv {\cal F}_{\alpha_1\ldots\alpha_l} \bigl({\stackrel{\,\circ}{C}}(
\theta), \theta\bigr),\  {\cal F}_{(\alpha)_l(\beta)_k}(\theta) \equiv
{\cal F}_{\alpha_1\ldots\alpha_l\beta_1\ldots\beta_k}(\theta)\;. & {}
\end{eqnarray}
Series coefficients ${\cal F}_{(\alpha)_l}\bigl({\stackrel{\,\circ}{C}}(
\theta),\theta\bigr)$ and ${\cal F}_{(\alpha)_l(\beta)_k}(\theta)$
possess by  the following symmetry properties written, for instance, for the
latters
\renewcommand{\theequation}{\arabic{section}.\arabic{equation}}
\begin{eqnarray}
{} & {\cal F}_{(\alpha)_l(\beta)_k}(\theta) = (-1)^{(\varepsilon_{\alpha_s}
+1)( \varepsilon_{\alpha_{s-1}} +1)} {\cal
F}_{\alpha_1\ldots\alpha_s\alpha_{s-1}\ldots\alpha_l(\beta)_k}(\theta) = & {}
\nonumber \\
{} & (-1)^{\varepsilon_{\beta_r}\varepsilon_{\beta_{r-1}}}
{\cal F}_{(\alpha)_l\beta_1\ldots\beta_r\beta_{r-1}\ldots\beta_k}(\theta),\
s=\overline{2,l}, r= \overline{2,k}\;. & {}
\end{eqnarray}
For ${\cal F}\bigl(C(\theta), {\stackrel{\,\circ}{C}}(\theta), \theta\bigr)
\in
C^{k}(T_{odd}{\cal M}_{C}\times\{\theta\})$, $k \le \infty$ we assume
the expansion in formal functional Taylor's series in powers of
$\delta C^{\alpha}(\theta)$=$(C^{\alpha}(\theta)$ $-$ $C_0^{\alpha}(\theta))$,
$\delta{\stackrel{\,\circ}{C}}{}^{\alpha}(\theta)$=$\bigl({\stackrel{\,
\circ}{C}}{}^{\alpha}(\theta)$ $-$
${\stackrel{\,\circ}{C}}{}^{\alpha}_0(\theta)\bigr)$ in a some neighbourhood
(possibly in the whole $T_{odd}{\cal M}_{C}\times\{\theta\})$ of a point
$\bigl(C^{\alpha}_0(\theta)$, ${\stackrel{\,\circ}{C}}{}^{\alpha}_0(\theta)
\bigr)$ $\in$ $T_{odd}{\cal M}_{C}$ being valid. Moreover we presuppose the
$C^{k}(T_{odd}{\cal M}_{C}\times\{\theta\})$ is equipped by  structure of a
norm and a convergence of the series with respect to that norm
\begin{eqnarray}
{} & {\cal F}\bigl( C(\theta), {\stackrel{\,\circ}{C}}(\theta), \theta \bigr)
= \displaystyle\sum\limits_{
k,l=0}\displaystyle\frac{1}{k!l!}
{\cal F}_{(\alpha)_l(\beta)_k}\bigl( C_0(\theta), {\stackrel{\,\circ}{C}}_0(
\theta), \theta \bigr)
{\delta\vec{\stackrel{\,\circ}{C}}{}^{(\beta)_k}}(\theta)
{\delta\vec{C}^{(\alpha)_l}}(\theta)\equiv & {}
\nonumber \\
{} & \displaystyle\sum\limits_{k,l=0}\displaystyle\frac{1}{k!l!}
\left(\prod\limits_{p=0}^{l-1}
\displaystyle\frac{{\partial}\phantom{xxxxxx}}{\partial C^{\alpha_{l-p}
}(\theta)} \prod\limits_{q=0}^{k-1}
\displaystyle\frac{{\partial}
\phantom{xxxxxx}}{{\partial{\stackrel{\,\circ}{C}}{}^{\beta_{k-q}}(\theta)}}{
\cal F}\bigl(C(\theta),
{\stackrel{\,\circ}{C}}(\theta), \theta \bigr) \right)_{ \mid C_{0}(\theta),
{\stackrel{\circ}{C}}_{0}(\theta)}
\hspace{-1.5em}{\delta\vec{\stackrel{\,\circ}{
C}}{}^{(\beta)_k}}(\theta) {\delta\vec{C}^{(\alpha)_l}}(\theta)\;, &{} 
\end{eqnarray}
with decomposition coefficients satisfying to the properties (2.38). It is not
complicated to write down the analogs of formulae (2.36), (2.39) for any
${\cal F}(\theta)$ $\in$ ${\bf K}[[T_{odd}(T^{*}_{odd}{\cal M}_{s})\times\{
\theta\}]]$ and ${\cal F}(\theta)$ $\in$ $D^{k}_{s}$, $s\in C_0$ respectively.
It is sufficient in this case to change  in the above-mentioned formulae
the superfields $C^{\alpha}(\theta)$, ${\stackrel{\,\circ}{C}}{}^{\alpha
}(\theta)$
onto $\Gamma^{p_s}_{s}(\theta)$ and
${\stackrel{\circ}{\Gamma}}{}^{p_s}_{s}(\theta)$, $s\in C_0$ respectively.

An action of projector systems
$\{P_{a}(\theta)\}$, $a=0,1$ and $\{\tilde{P}_{b}(\theta)$, $U(\theta)$,
$V(\theta)$ (${\cal W}(\theta)$ = $(U+V)(\theta))\}$, $b=0,1$ defined in [1,2]
are naturally continued from $D^{k}_{cl}$ to $D^{k}_{s}$, $k \le \infty$
with following from this fact  all consequences for
representation of $D^{k}_{s}$. Besides, the special involution $*$ [1,2] is
naturally continued onto $T_{odd}(T^{*}_{odd}{\cal M}_{s})$ $\times$
$\{\theta\}$ and $D^{k}_{s}$, $s \in C_0$ as well so that
$P_{0}(T_{odd}(T^{\ast}_{odd}{\cal
M}_{s}))$ and subsuperalgebra $P_{0}D^{k}_{s}$ = $C^{k}(P_{0}(T_{odd}(
T_{odd}{\cal M}_{s})))$ parametrized by $P_0(\theta)$-components of
superfields $\bigl(\Gamma^{p_s}_{s}(\theta)$,
${\stackrel{\circ}{\Gamma}}{}^{p_s}_{s}(\theta)\bigr)$, i.e. by
$\bigl(P_0(\theta)\Gamma^{p_s}_{s}(\theta)$,
${\stackrel{\circ}{\Gamma}}{}^{p_s}_{s}(\theta)\bigr)$, are
invariant with respect to action of involution $*$.

The classes of regular over ${\bf K}$ superfunctionals  $C_{F,cl}$ and
$C_{FH,cl}$ are continued to ones
$C_{F,s}$ and $C_{FH,s}$ $s \in C_0$ consisting of the continued
superfunctionals defined
on $T_{odd}{\cal M}_{s}$ $\times$ $\{\theta\}$ and $T_{odd}(T^{*}_{odd}{\cal
M}_{s})$ $\times$ $\{\theta\}$ correspondingly by means of the formulae
\renewcommand{\theequation}{\arabic{section}.\arabic{equation}\alph{lyter}}
\begin{eqnarray}
\setcounter{lyter}{1}
{} & {} & F[{\cal A}_s] = \displaystyle\int d\theta{\cal F}\bigl({\cal
A}_s(\theta),
{\stackrel{\ \circ}{\cal A}}_{s}(\theta),\theta\bigr),\ {\cal F}(\theta)\in
C^k(T_{odd}{\cal M}_{s} \times \{\theta\})\;,  \\
\setcounter{equation}{40}
\setcounter{lyter}{2}
{} & {} & F_H[{\Gamma}_s] = \displaystyle\int d\theta{\cal J}\bigl({\Gamma
}_s(\theta), {\stackrel{\circ}{\Gamma}}_{s}(\theta),\theta\bigr),\
{\cal J}(\theta)\in D^k_s\;. 
\end{eqnarray}
The formulae of the variational calculus [1,2] are valid
for these superfunctionals under
formal change of $\Gamma^{p}_{cl}(\theta)$
onto $\Gamma^{p_s}_{s}(\theta)$ and ${\stackrel{\circ}{\Gamma}}{}^{p}_{cl}(
\theta)$ onto ${\stackrel{\circ}{\Gamma}}{}^{p_s}_{s}(\theta)$.

For an arbitrary superfunction from $D^{k}_{s}$ it is not difficult to
produce the transformation laws under action  of the superfield
representation $T^{s}$, $s\in C_0$ operators constructed with respect to $T$,
$T^{*}$, $T_{\xi}$, $T^{*}_{\xi}$.

At last, the superalgebra of the 1st order differential operators (with
respect to differentiation on ${\cal A}^{\imath}(\theta))$
{\boldmath${\cal A}_{cl}$} over $C^{k}(T_{odd}{\cal M}_{cl}\times\{\theta\})$
[1] and one of the 1st and 2nd orders operators
{\boldmath${\cal B}_{cl}$} over $C^{k}(T^{*}_{odd}{\cal M}_{cl}$ $\times$
$\{\theta\})$ and $D^{k}_{cl}$ [2]
 in a natural and literal way
are continued to {\boldmath${\cal A}_{s}$} and {\boldmath${\cal B}_{s}$}
acting on $C^{k}(T_{odd}{\cal M}_{s}\times\{\theta\})$ and
$C^{k}(T^{\ast}_{odd}{\cal M}_{s}\times\{\theta\})$, $D^{k}_s$, $s \in C_0$
respectively with help of continued projector system $\{P_{a}(\theta)\}$ and
involution $*$. Let us indicate the corresponding formulae for the basis
elements $U^{s}_{a}(\theta)$, ${\stackrel{\circ}{U}}{}^{s}_{a}(\theta)$ from
{\boldmath${\cal A}_{s}$} and
$U^{s}_{a}(\theta)$, ${\stackrel{\circ}{U}}{}^{s}_{a}(\theta)$,
$V^{s}_{a}(\theta)$, ${\stackrel{\circ}{V}}{}^{s}_{a}(\theta)$,
${\cal W}^{s}_{a}(\theta)$, ${\stackrel{\circ}{\cal W}}{}^{s}_{a}(\theta)$,
$\Delta^{s}_{ab}(\theta)$, $a,b=0,1$ from {\boldmath${\cal B}_{s}$}, $s \in
C_0$
\begin{eqnarray}
\setcounter{lyter}{1}
{} & U^{s}_{a}(\theta) =
P_1(\theta)\Phi^{A_{s}}_s(\theta)
\displaystyle\frac{\partial_l
\phantom{xxxxxxxx}}{\partial P_a(\theta)\Phi^{A_{s}}_s(\theta)},\
{\stackrel{\circ}{U}}{}^{s}_{a}(\theta) =
{\stackrel{\circ}{\Phi}}{}^{A_s}_{s}(\theta)\displaystyle\frac{\partial_l
\phantom{xxxxxxxx}}{\partial P_a(\theta)\Phi^{A_{s}}_s(\theta)}\;, & {}
\\
\setcounter{equation}{41}
\setcounter{lyter}{2}
{} & V^{s}_{a}(\theta)=
P_1(\theta)\Phi^{\ast}_{A_{s},s}(\theta)
\displaystyle\frac{\partial\phantom{xx
xxxxxxx}}{\partial P_a(\theta)\Phi^{\ast}_{A_{s},s}(\theta)},\
{\stackrel{\circ}{V}}{}^{s}_{a}(\theta) =
{\stackrel{\circ}{\Phi}}{}^{\ast}_{A_s,s}(\theta)
\displaystyle\frac{\partial\phantom{xxxxxxxxx}}{\partial
P_a(\theta)\Phi^{\ast}_{A_{s},s}(\theta)}\;, & {}
\\ 
\setcounter{equation}{41}
\setcounter{lyter}{3}
{} & \hspace{-2em}{\cal W}^{s}_{a}(\theta)= (U^{s}_{a} + V^{s}_{a})(\theta)=
P_1(\theta)\Gamma^{p_s}_s(\theta)
\displaystyle\frac{\partial_l\phantom{xxxxxxxx}}{\partial
P_a(\theta)\Gamma^{p_s}_s(\theta)},\
{\stackrel{\circ}{\cal W}}{}^{s}_{a}(\theta)=
{\stackrel{\circ}{\Gamma}}{}^{p_s}_{s}(\theta)
\displaystyle\frac{\partial_l\phantom{xxxxxxxx}}{\partial
P_a(\theta)\Gamma^{p_s}_s(\theta)}\;, & {}
\\ 
\setcounter{equation}{41}
\setcounter{lyter}{4}
{} & \hspace{-1.5em}
\Delta^{s}_{ab}(\theta) = (-1)^{\varepsilon_{A_s}}
\displaystyle\frac{\partial_l \phantom{xxxxxx}}{\partial
P_a\Phi^{A_{s}}_s(\theta)}
\displaystyle\frac{\partial\phantom{xxxxxxx}}{\partial
P_b\Phi^{\ast}_{A_{s},s}(\theta)} =
\displaystyle\frac{1}{2}
\displaystyle\frac{\partial_l\phantom{xxxxxxxx}}{\partial
P_a(\theta)\Gamma^{p_s}_s(\theta)}
\omega^{p_sq_s}_s(\theta)
\displaystyle\frac{\partial_l\phantom{xxxxxxxx}}{\partial
P_b(\theta)\Gamma^{q_s}_s(\theta)}(-1)^{\varepsilon_{p_s}},  & {}
\nonumber \\
{} & \hspace{-2em}
\omega^{p_sq_s}_s(\theta) = - (-1)^{(\varepsilon_{p_s} + 1)
(\varepsilon_{q_s} + 1)}\omega^{q_sp_s}_s(\theta) =
P_0(\theta)\omega^{p_sq_s}_s(\theta),\
\left\|\omega^{p_sq_s}_s(\theta)\right\| = \left\|
\begin{array}{cc}
0_{[A_s]} &  1_{[A_s]} \\
-1_{[A_s]}  & 0_{[A_s]}
\end{array}
\right\|.
\end{eqnarray}
Grassmann gradings $\varepsilon_P, \varepsilon_{\bar{J}}, \varepsilon$ for
operators $U^{s}_{a}(\theta)$, $V^{s}_{a}(\theta)$, ${\cal
W}^{s}_{a}(\theta)$ are equal to (0,0,0), whereas for
${\stackrel{\circ}{U}}{}^{s}_{a}(\theta)$,
${\stackrel{\circ}{V}}{}^{s}_{a}(\theta)$,
${\stackrel{\circ}{\cal W}}{}^{s}_{a}(\theta)$, $\Delta^s_{ab}(\theta)$ to
(1,0,1) respectively.

Involution $*$ acting on $T_{odd}(T^{*}_{odd}{\cal M}_{s})$ $\times$ $\{
\theta\}$ by the rule
\renewcommand{\theequation}{\arabic{section}.\arabic{equation}}
\begin{eqnarray}
{} & \left(\Gamma^{p_s}_{s}(\theta)\right)^{\ast} =
\Gamma^{p_s}_{s}(\overline{\theta}) \equiv
\overline{\Gamma^{p_s}_{s}(\theta)} \equiv \Gamma^{p_s}_{s}(-\theta),\
\left({\stackrel{\circ}{\Gamma}}{}^{p_s}_{s}(\theta)\right)^{\ast} =
{\stackrel{\circ}{\Gamma}}{}^{p_s}_{s}(\theta),\ \theta^{\ast}= - \theta
\end{eqnarray}
transfers the superfield $\Gamma^{p_s}_{s}(\theta)$
into $*$-conjugate one $\overline{\Gamma^{p_s}_{s}(\theta)}$ being
transformed with respect to supergroup $J$ $*$-conjugate representation if
$\Gamma^{p_s}_{s}(\theta)$ are transformed on $T^{s}\otimes T^{s{}\ast}$
representation. By means of (2.41), (2.42) one can write down the other
elements of {\boldmath${\cal A}_{s}$} and {\boldmath${\cal B}_{s}$} as well
in the almost superfield form
\renewcommand{\theequation}{\arabic{section}.\arabic{equation}\alph{lyter}}
\begin{eqnarray}
\setcounter{lyter}{1}
{} &
U^{s}_{+}(\theta) = (U^{s}_{0} + U^{s}_{1})(\theta) =
P_1(\theta)\Phi^{A_{s}}_s(\theta)
\displaystyle\frac{\partial_l
\phantom{xxxx}}{\partial \Phi^{A_{s}}_s(\theta)},\ \;
{\stackrel{\circ}{U}}{}^{s}_{+}(\theta) =
{\stackrel{\circ}{\Phi}}{}^{A_s}_{s}(\theta)\displaystyle\frac{\partial_l
\phantom{xxxx}}{\partial \Phi^{A_{s}}_s(\theta)}\;, & {}
\\
\setcounter{equation}{43}
\setcounter{lyter}{2}
{} &
U^{s}_{-}(\theta) = (U^{s}_{0} - U^{s}_{1})(\theta) =
P_1(\theta)\Phi^{A_{s}}_s(\theta)
\displaystyle\frac{\partial_l
\phantom{xxxx}}{\partial \overline{\Phi^{A_{s}}_s(\theta)}},\
{\stackrel{\circ}{U}}{}^{s}_{-}(\theta) =
{\stackrel{\circ}{\Phi}}{}^{A_s}_{s}(\theta)\displaystyle\frac{\partial_l
\phantom{xxxx}}{\partial \overline{\Phi^{A_{s}}_s(\theta)}}\;, & {}
 \\
\setcounter{equation}{43}
\setcounter{lyter}{3}
{} & V^{s}_{+}(\theta)= (V^{s}_{0} + V^{s}_{1})(\theta) =
P_1(\theta)\Phi^{\ast}_{A_{s},s}(\theta)
\displaystyle\frac{\partial\phantom{xx
xxx}}{\partial \Phi^{\ast}_{A_{s},s}(\theta)},\
{\stackrel{\circ}{V}}{}^{s}_{+}(\theta) =
{\stackrel{\circ}{\Phi}}{}^{\ast}_{A_s,s}(\theta)
\displaystyle\frac{\partial\phantom{xxxxx}}{\partial
\Phi^{\ast}_{A_{s},s}(\theta)}\;, & {}
 \\
\setcounter{equation}{43}
\setcounter{lyter}{4}
{} & V^{s}_{-}(\theta)= (V^{s}_{0} - V^{s}_{1})(\theta) =
P_1(\theta)\Phi^{\ast}_{A_{s},s}(\theta)
\displaystyle\frac{\partial\phantom{xx
xxx}}{\partial \overline{\Phi^{\ast}_{A_{s},s}(\theta)}},\
{\stackrel{\circ}{V}}{}^{s}_{-}(\theta) =
{\stackrel{\circ}{\Phi}}{}^{\ast}_{A_s,s}(\theta)
\displaystyle\frac{\partial\phantom{xxxxx}}{\partial
\overline{\Phi^{\ast}_{A_{s},s}(\theta)}}\;, & {}
 \\
\setcounter{equation}{43}
\setcounter{lyter}{5}
{} & {\cal W}^{s}_{+}(\theta)= ({\cal W}^{s}_{0} + {\cal W}^{s}_{1})(\theta)=
P_1(\theta)\Gamma^{p_s}_s(\theta)
\displaystyle\frac{\partial_l\phantom{xxxx}}{\partial
\Gamma^{p_s}_s(\theta)},\
{\stackrel{\circ}{\cal W}}{}^{s}_{+}(\theta)=
{\stackrel{\circ}{\Gamma}}{}^{p_s}_{s}(\theta)
\displaystyle\frac{\partial_l\phantom{xxxx}}{\partial
\Gamma^{p_s}_s(\theta)}\;, & {}
 \\
\setcounter{equation}{43}
\setcounter{lyter}{6}
{} & {\cal W}^{s}_{-}(\theta)= ({\cal W}^{s}_{0} - {\cal W}^{s}_{1})(\theta)=
P_1(\theta)\Gamma^{p_s}_s(\theta)
\displaystyle\frac{\partial_l\phantom{xxxx}}{\partial
\overline{\Gamma^{p_s}_s(\theta)}},\
{\stackrel{\circ}{\cal W}}{}^{s}_{-}(\theta)=
{\stackrel{\circ}{\Gamma}}{}^{p_s}_{s}(\theta)
\displaystyle\frac{\partial_l\phantom{xxxx}}{\partial
\overline{\Gamma^{p_s}_s(\theta)}}\;, & {}
  \\
\setcounter{lyter}{1}
{} & \hspace{-2.0em}\Delta^s(\theta) =
\sum\limits_{a,b}\Delta^s_{ab}\hspace{-0.1em}(\theta)= (-1)^{\varepsilon_{
A_s}}\displaystyle\frac{\partial_l \phantom{xxxx}}{\partial\Phi^{A_{s}}_s
(\theta)}
\displaystyle\frac{\partial\phantom{xxxxx}}{\partial
\Phi^{\ast}_{A_{s},s}(\theta)} = \displaystyle\frac{1}{2}
\displaystyle\frac{\partial_l\phantom{xxxx}}{\partial\Gamma^{p_s}_s(\theta)}
\omega^{p_sq_s}_s(\theta)
\displaystyle\frac{\partial_l\phantom{xxxx}}{\partial
\Gamma^{q_s}_s(\theta)}(-1)^{\varepsilon_{p_s}},  & {}
\\
\setcounter{equation}{44}
\setcounter{lyter}{2}
{} & \Delta^s_{+-}(\theta) =
\sum\limits_{a,b}(-1)^{\varepsilon_{A_s}}
\displaystyle\frac{\partial_l
\phantom{xxxxxxxx}}{\partial P_a(\theta)\Phi^{A_{s}}_s(\theta)}
\displaystyle\frac{\partial\phantom{xxxxxxxxx}}{\partial
\overline{P_b(\theta)\Phi^{\ast}_{A_{s},s}(\theta)}} = (-1)^{\varepsilon_{
A_s}}
\displaystyle\frac{\partial_l \phantom{xxxx}}{\partial\Phi^{A_{s}}_s(\theta)}
\displaystyle\frac{\partial\phantom{xxxxx}}{\partial
\overline{\Phi^{\ast}_{A_{s},s}(\theta)}}\;,& {}
 \\
\setcounter{equation}{44}
\setcounter{lyter}{3}
{} & \Delta^s_{-+}(\theta) =
\sum\limits_{a,b}(-1)^{\varepsilon_{A_s}}
\displaystyle\frac{\partial_l
\phantom{xxxxxxxx}}{\partial \overline{P_a(\theta)\Phi^{A_{s}}_s(\theta)}}
\displaystyle\frac{\partial\phantom{xxxxxxxxx}}{\partial
{P_b(\theta)\Phi^{\ast}_{A_{s},s}(\theta)}} = (-1)^{\varepsilon_{A_s}}
\displaystyle\frac{\partial_l
\phantom{xxxx}}{\partial\overline{\Phi^{A_{s}}_s(\theta)}}
\displaystyle\frac{\partial\phantom{xxxxx}}{\partial
{\Phi^{\ast}_{A_{s},s}(\theta)}}\;,& {}
 \\
\setcounter{equation}{44}
\setcounter{lyter}{4}
{} & \Delta^s_{--}(\theta) =
\sum\limits_{a,b}(-1)^{\varepsilon_{A_s}}
\displaystyle\frac{\partial_l
\phantom{xxxxxxxx}}{\partial \overline{P_a(\theta)\Phi^{A_{s}}_s(\theta)}}
\displaystyle\frac{\partial\phantom{xxxxxxxxx}}{\partial
\overline{P_b(\theta)\Phi^{\ast}_{A_{s},s}(\theta)}} = (-1)^{\varepsilon_{
A_s}}
\displaystyle\frac{\partial_l
\phantom{xxxx}}{\partial\overline{\Phi^{A_{s}}_s(\theta)}}
\displaystyle\frac{\partial\phantom{xxxxx}}{\partial
\overline{\Phi^{\ast}_{A_{s},s}(\theta)}}\;.& {}
\end{eqnarray}
In writing of (2.43), (2.44) it is taken into consideration according to
[1,2] that by relationship
\renewcommand{\theequation}{\arabic{section}.\arabic{equation}}
\begin{eqnarray}
\displaystyle\frac{\partial_l\phantom{xxxx}}{\partial\overline{\Gamma^{p_s
}_s(\theta)}} =
\displaystyle\frac{\partial_l\phantom{xxxxxxxx}}{\partial
P_0(\theta)\Gamma^{p_s}_s(\theta)} -
\displaystyle\frac{\partial_l\phantom{xxxxxxxx}}{\partial
P_1(\theta)\Gamma^{p_s}_s(\theta)} \equiv
\left(\displaystyle\frac{\partial_l\phantom{xxxx}}{\partial{\Gamma^{p_s
}_s(\theta)}}\right)^{\ast}
\end{eqnarray}
involution $*$ is continued onto {\boldmath${\cal B}_{s}$}, $s\in C_0$.

All algebraic
properties for operators from {\boldmath${\cal A}_{cl}$}, {\boldmath${\cal
B}_{cl}$} and conclusions of the papers [1,2]
without changes are transferred onto {\boldmath${\cal A}_{s}$}, {\boldmath${
\cal B}_{s}$}. The  operators $\Delta^s_{ab}(\theta)$ action on the
product
of any superfunctions ${\cal F}(\theta)$, ${\cal J}(\theta)$  yields the
following antibrackets $(\ ,\ )^s_{ab}$, $s \in C_0$, $a,b$=$0,1$
\begin{eqnarray}
({\cal F}(\theta),{\cal J}(\theta))^s_{ab} =
\frac{\partial{\cal F}(\theta)\phantom{xxxxx}}{\partial
P_a(\theta)\Phi^{A_{s}}_s(\theta)}
\frac{\partial{\cal J}(\theta)\phantom{xxxxxx}}{\partial
{P_b(\theta)\Phi^{\ast}_{A_{s},s}(\theta)}} -
\frac{\partial_r{\cal F}(\theta)\phantom{xxxxx}}{\partial
{P_b(\theta)\Phi^{\ast}_{A_{s},s}(\theta)}}
\frac{\partial_l{\cal J}(\theta)\phantom{xxxx}}{\partial
P_a(\theta)\Phi^{A_{s}}_s(\theta)}\;.
\end{eqnarray}
In its turn  the operators (2.44) action on the such product generates the other
antibrackets in the superfield form ($(\ ,\ )^{s}_{++}$, $(\ ,\ )^{s}_{--}$,
$(\ ,\ )^{s}_{\pm\mp})$. For example, for $(\ ,\ )^{s}_{++}$ we have
\begin{eqnarray}
{} & ({\cal F}(\theta),{\cal J}(\theta))^s_{++} = \sum\limits_{a,b}
({\cal F}(\theta),{\cal J}(\theta))^s_{ab} \equiv
({\cal F}(\theta),{\cal J}(\theta))^s_{\theta} =
\displaystyle\frac{\partial{\cal F}(\theta)\phantom{x}}{\partial
\Phi^{A_{s}}_s(\theta)}
\displaystyle\frac{\partial{\cal J}(\theta)\phantom{xx}}{\partial
\Phi^{\ast}_{A_{s},s}(\theta)} - & {} \nonumber \\
{} & - (-1)^{(\varepsilon({\cal F}) + 1)(\varepsilon({\cal J}) + 1)}
({\cal F} \longleftrightarrow {\cal J}) =
\displaystyle\frac{\partial_r{\cal F}(\theta)\phantom{x}}{\partial
\Gamma^{p_s}_s(\theta)}\omega^{p_sq_s}_s(\theta)
\displaystyle\frac{\partial_l{\cal J}(\theta)\phantom{x}}{\partial
\Gamma^{q_s}_s(\theta)}\;.  & {}
\end{eqnarray}
All antibrackets satisfy to the standard properties for odd Poisson bracket
[2], i.e. to an generalized antisymmetry, Leibnitz rule, Jacobi identity.
In addition
their $\varepsilon_P$, $\varepsilon_{\bar{J}}$, $\varepsilon$ gradings are
identical to ones for antibrackets given on $D^k_{cl}$ [2] and, for instance,
for $(\ ,\ )^{s}_{\theta}$ reads as follows
\begin{eqnarray}
(\varepsilon_P, \varepsilon_{\bar{J}}, \varepsilon)
({\cal F}(\theta),{\cal J}(\theta))^s_{\theta}=
(\varepsilon_P({\cal F}) + \varepsilon_P({\cal J}) +1,
\varepsilon_{\bar{J}}({\cal F}) + \varepsilon_{\bar{J}}({\cal J}),
\varepsilon({\cal F}) + \varepsilon({\cal J}) +1)\;. 
\end{eqnarray}
For any operator $B^{s}(\theta)$ $\in$ $\bigl\{
{\stackrel{\circ}{U}}{}^{s}_{+}(\theta)$,
${\stackrel{\circ}{V}}{}^{s}_{+}(\theta)$,
${\stackrel{\circ}{\cal W}}{}^{s}_{+}(\theta)$, $\Delta^s(\theta)\bigr\}$
and for $(\ ,\ )^{s}_{\theta}$ the relationship holds
\begin{eqnarray}
B^{s}(\theta)({\cal F}(\theta),{\cal J}(\theta))^s_{\theta} =
(B^{s}(\theta){\cal F}(\theta),{\cal J}(\theta))^s_{\theta} +
(-1)^{\varepsilon({\cal F}) +1}({\cal F}(\theta),B^{s}(\theta){\cal
J}(\theta))^s_{\theta}\;. 
\end{eqnarray}
The so-called transformations of operators from {\boldmath${\cal B}_{cl}$} [2]
are continued onto {\boldmath${\cal B}_{s}$}, $s\in C_0$.
Let us confine ourselves by the case of operators $B^s(\theta)$ defined
before (2.49) having required of the following transformation rule
fulfilment
\begin{eqnarray}
B^{s}(\theta) \mapsto B'^{s}(\theta) = B^{s}(\theta) + ({\cal F}(\theta),\
\;)^s_{\theta} \equiv B^{s}(\theta) + {}^s{\rm ad}_{{\cal F}(\theta)},\
(\varepsilon_P, \varepsilon_{\bar{J}}, \varepsilon){\cal F}(\theta)=(0,0,0)\;
. 
\end{eqnarray}
The condition of nilpotency preservation  for $B'^{s}(\theta)$ leads to
equation on  ${\cal F}(\theta)$ $\in$ $D^{k}_{s}$ of the form
\begin{eqnarray}
B^{s}(\theta){\cal F}(\theta) + \frac{1}{2}({\cal F}(\theta),{\cal
F}(\theta))^s_{\theta} = f({\stackrel{\circ\ }{\Gamma_s}}(\theta),\theta),\
(\varepsilon_P, \varepsilon_{\bar{J}}, \varepsilon)f(\theta)=(1,0,1)\;.
\end{eqnarray}
In particular the restriction of $f({\stackrel{\circ}{\Gamma}}_s(\theta),
\theta)$ onto $T^{*}_{odd}{\cal M}_{s}$  vanishes.
Transformation (2.50) permits to obtain the vanishing for one from
$B'^{s}(\theta)$, at least for
${\stackrel{\circ}{U}}{}'^{s}_{+}(\theta)$ or
${\stackrel{\circ}{V}}{}'^{s}_{+}(\theta)$,
with help of a choice for ${\cal F}(\theta)$ (see Ref.[2] for details).
\section{Ghost Number}
\setcounter{equation}{0}

In order to single out the definite subsets in ${\cal M}_{ext}$ and then in
$D^{k}_{ext}$ it is convenient to introduce a special group structure.

By group ${\bf Gh}$ of homogeneous in the extended sense (scaled or ghost)
transformations of the superspace ${\cal M}$ let us call the one-parametric
group
of linear transformations of ${\cal M}$ defined by its action on coordinates
$(z^{a}, \theta)$
\begin{eqnarray}
\forall g^{\tau}\in {\bf Gh}: g^{\tau}(z^{a},\theta)=
(g^{\tau}z^{a},g^{\tau}\theta) = (e^{\gamma\tau}z^a,e^{\delta\tau}\theta),\
\gamma,\delta\in {\bf Z},\;\tau\in{\bf R}\;. 
\end{eqnarray}
The weights of $z^{a}$ and $\theta$ (being usually called by the ghost
numbers) are equal respectively
\begin{eqnarray}
\gamma \equiv {\rm gh}(z^a)=0,\ \delta \equiv {\rm gh}(\theta)= - 1\;. 
\end{eqnarray}
Formally defining the elements $dz^{a}$ and $d\theta$ on dual to $T_{(z,
\theta)}{\cal M}$
(being by tangent space over point $(z^{a},\theta) \in {\cal M}$
with basis $\{\frac{d\phantom{x}}{dz^{a}}$, $\frac{d}{d\theta}\}$)
cotangent space $T^{*}_{(z,\theta)}{\cal M}$ we put
\begin{eqnarray}
{\rm gh}(dz^a) = \left({\rm gh}\textstyle\frac{d\phantom{x}}{d z^a}\right)=0,\
{\rm gh}(d\theta) = -1= - {\rm gh}\left(\textstyle\frac{d}{d\theta
}\right)\footnotemark[2].  
\end{eqnarray}
\footnotetext[2]{for such definition of ghost number the sign of integral with
respect to variable $\theta$ ($\int_{\theta}$) must formally possess by ghost
number to be equal to $+2$}Representation $U$ of the group ${\bf Gh}$ in
${\cal M}_{ext}$
\begin{eqnarray}
U: {{\bf Gh}} \to U({{\bf Gh}}) \subset GL({\cal M}_{ext})
\end{eqnarray}
is realized by means of relations on its coordinates
\begin{eqnarray}
{} & U(g^{\tau})\Phi^B(\theta) = \tilde{g}{}^{\tau}\Phi^B(\theta)=
\bigl(e^{\alpha_1\tau}{\cal A}^{\imath}(\theta), e^{\alpha_2\tau}C^{\alpha
}(\theta), e^{\alpha_3\tau}\overline{C}{}^{\alpha}(\theta),
e^{\alpha_4\tau}B^{\alpha}(\theta)\bigr)\;, & {}
\nonumber \\
{} & (\alpha_1,\alpha_2,\alpha_3,\alpha_4) \equiv ({\rm gh}({\cal
A}^{\imath}(\theta)), {\rm gh}({C}^{\alpha}(\theta)),
{\rm gh}(\overline{C}{}^{\alpha}(\theta)), {\rm gh}(B^{\alpha}(\theta))) =
(0, 1, -1, 0)\;. & {} 
\end{eqnarray}
Representation $U$ induces in fibres over point with coordinates $\Phi^B(
\theta) \in {\cal M}_{ext}$: $T^{\Phi}_{odd}{\cal M}_{ext}$,
 $T^{*{}\Phi}_{odd}{\cal M}_{ext}$,
$T^{(\Phi,\Phi^{\ast})}_{odd}(T^{\ast}_{odd}{\cal M}_{ext})$\footnote[3]{we
assume $T_{odd}{\cal M}_{ext}$,
$T^{*}_{odd}{\cal M}_{ext}$, $T_{odd}(T^{*}_{odd}{\cal M}_{ext})$
are the bundles over base
${\cal M}_{ext}$ (without presentation of their explicit construction)}
correspondingly the representations
\renewcommand{\theequation}{\arabic{section}.\arabic{equation}\alph{lyter}}
\begin{eqnarray}
\setcounter{lyter}{1}
{} & {} & TU: {\bf  Gh} \to TU({\bf Gh}) \subset GL\bigl(T_{odd}{\cal
M}_{ext}\bigr)\;,
\\
\setcounter{equation}{6}
\setcounter{lyter}{2}
{} & {} & T^{\ast}U: {{\bf Gh}} \to T^{\ast}U({{\bf Gh}}) \subset
GL\bigl(T^{\ast}_{odd}{\cal M}_{ext}\bigr)\;,\\
\setcounter{equation}{6}
\setcounter{lyter}{3}
{} & {} & T(T^{\ast}U): {{\bf Gh}} \to T(T^{\ast}U)({{\bf Gh}}) \subset
GL\bigl(T_{odd}(T^{\ast}_{odd}{\cal M}_{ext})\bigr)
\end{eqnarray}
being realized on the coordinates of these supermanifolds by the formulae
\renewcommand{\theequation}{\arabic{section}.\arabic{equation}}
\begin{eqnarray}
{} & TU(g^{\tau})\bigl(\Phi^{B}(\theta),
{\stackrel{\circ}{\Phi}}{}^B(\theta)\bigr) =
\bigl(\tilde{g}{}^{\tau}\Phi^B(\theta),
{\stackrel{\,\circ}{g}}{}^{\tau}{\stackrel{\circ}{\Phi}}{}^B(\theta)\bigr) =
\bigl(\tilde{g}{}^{\tau}\Phi^B(\theta),
\bigl(e^{(\alpha_1+1)\tau}{\stackrel{\ \circ}{\cal A}}{}^{\imath}(\theta),
e^{(\alpha_2+1)\tau}{\stackrel{\,\circ}{C}}{}^{\alpha}(\theta), &{}
\nonumber \\
{} &
e^{(\alpha_3+1)\tau} {\stackrel{\,\circ}{\overline{C}}}{}^{\alpha}(\theta),
e^{(\alpha_4+1)\tau}{\stackrel{\,\circ}{B}}{}^{\alpha}(\theta)\bigr)\bigr),\
{\rm gh}\bigl({\stackrel{\circ}{\Phi}}{}^B(\theta)\bigr)=
{\rm gh}(\Phi^B(\theta)) + 1 = (1,2,0,1)\;, {} &
\\
{} & \hspace{-2.0em}T^{\ast}U(g^{\tau})(\Phi^B(\theta),\Phi^{\ast}_B(\theta))
\hspace{-0.2em} = \hspace{-0.2em}
\bigl(\tilde{g}{}^{\tau}\Phi^B(\theta), g^{\ast\tau}\Phi^{\ast}_B(\theta)
\bigr) \hspace{-0.2em}= \hspace{-0.2em}\bigl(\tilde{g}{}^{\tau}\Phi^B(\theta),
\bigl(e^{-(\alpha_1+1)\tau}\hspace{-0.2em}{\cal A}^{\ast}_{\imath}(\theta),
e^{-(\alpha_2+1)\tau}\hspace{-0.2em}C^{\ast}_{\alpha}(\theta), &{} \nonumber \\
{} & e^{-(\alpha_3+1)\tau
}\overline{C}{}^{\ast}_{\alpha}(\theta),
e^{-(\alpha_4+1)\tau}B^{\ast}_{\alpha}(\theta)\bigr)\bigr),\
{\rm gh}(\Phi^{\ast}_B(\theta)) = -1 - {\rm gh}(\Phi^B(\theta))=
(-1,-2,0,-1)\;, & {}
\\
{} & T(T^{\ast}U)(g^{\tau})\bigl(\Gamma^{p}_{ext}(\theta),
{\stackrel{\circ}{\Gamma}}{}^{p}_{ext}(\theta)\bigr)
= \bigl(T^{\ast}U(g^{\tau})\Gamma^{p}_{ext}(\theta),
\bigl({\stackrel{\,\circ}{g}}{}^{\tau}{\stackrel{\circ}{\Phi}}{}^B(\theta),
{\stackrel{\circ}{g}}{}^{\ast\tau}{\stackrel{\circ}{\Phi}}{}^{\ast}_B(\theta)
\bigr)\bigr),\
{\stackrel{\,\circ}{g}}{}^{\ast\tau}{\stackrel{\circ}{\Phi}}{}^{\ast
}_B(\theta) = & {} \nonumber \\
{} & \hspace{-2.0em}
\bigl(e^{-\alpha_1\tau}\hspace{-0.2em}
{\stackrel{\ \circ}{\cal A}}{}^{\ast}_{\imath}(\theta),
e^{-\alpha_2\tau}\hspace{-0.2em}{\stackrel{\,\circ}{C}}{}^{\ast}_{\alpha
}(\theta),
e^{-\alpha_3\tau}\hspace{-0.2em}{\stackrel{\,\circ}{\overline{C}}}{}^{\ast
}_{\alpha}(\theta),e^{-\alpha_4\tau}\hspace{-0.2em}{\stackrel{\,\circ}{B}}{}^{
\ast}_{\alpha}(\theta)\bigr),\; {\rm gh}\bigl({\stackrel{\circ}{\Phi}}{}^{
\ast}_B(\theta)\bigr)\hspace{-0.2em}=\hspace{-0.2em}- {\rm gh}({\Phi^B
}(\theta))\hspace{-0.2em}=\hspace{-0.2em} (0,-1,1,0)\,.
{} & 
\end{eqnarray}
It should be noted the
nontrivial ($ \ne 0$) value of ghost number on ${\cal M}$, ${\cal M}_{ext}$,
etc. is connected with nontrivial   $\varepsilon_P$ grading
($\varepsilon_P = {\rm gh}(mod\,2)$).

Composing the linear span of all possible tensor products of elements from
${\cal M}_s$, $T_{odd}{\cal M}_s$, $T^{\ast}_{odd}{\cal M}_s$,
$T_{odd}(T^{\ast}_{odd}{\cal M}_s)$
one can obtain the group ${\bf  Gh}$ representation $\hat{T}^s$ realization,
connected with representation  $U$ (3.4), on the superalgebra $D^k_s$, $s\in
C_0$
\begin{eqnarray}
\hat{T}^s: {{\bf Gh}} \to \hat{T}^s({{\bf Gh}}) \subset
GL(D^k_{s})\,.
\end{eqnarray}
Representation $\hat{T}^s$ on the coordinates of elements from $D^k_{s}$
considered as the superalgebra acts by the rule taking formulae (3.5),
(3.7)--(3.9) into account
\begin{eqnarray}
{} & \hat{T}^s(g^{\tau})\left(
\vec{\Phi}{}^{(A)_l}_s(\theta)  \vec{\stackrel{
\circ}{\Phi}}{}^{(A)_f}_s(\theta)
\vec{\Phi}{}^{\ast}_{s{}(A)_h}(\theta)
\vec{\stackrel{\circ}{\Phi}}{}^{\ast}_{s{}(A)_d}(\theta)\right) =
\bigl({\stackrel{\mbox{\boldmath$\longrightarrow$}}{\tilde{g}{}^{\tau}\Phi_s}
 }\bigr)^{(A)_l}(\theta)
\Bigl({\stackrel{\mbox{\boldmath$\longrightarrow$}}{{\stackrel{\,\circ}{
g}}{}^{\tau}{\stackrel{\circ}{\Phi}}_s}}\Bigr)^{(A)_f}(\theta)
\times & {} \nonumber \\
{} & \bigl({\stackrel{\mbox{\boldmath$\longrightarrow$}}{g^{\ast{}\tau}\Phi^{
\ast}_s}}\bigr)_{(A)_h}(\theta)
\Bigl({\stackrel{\mbox{\boldmath$\longrightarrow$}}{{\stackrel{\,\circ}{g}}{
}^{
\ast\tau}{\stackrel{\circ}{\Phi}}{}^{\ast}_s}}\Bigr)_{(A)_d}(\theta) =
\exp\Bigl(\sum\limits_{i=1}^l{\rm gh}(\Phi^{(A)_i}_s) +
\sum\limits_{j=1}^f\bigl({\rm gh}(\Phi^{(A)_j}_s) + 1\bigr) -
\sum\limits_{k=1}^h\bigl({\rm gh}(\Phi^{(A)_k}_s) + 1\bigr)
 & {} \nonumber \\
{} &
- \sum\limits_{p=1}^d{\rm gh}(\Phi^{(A)_p}_s)\Bigr)
\vec{\Phi}{}^{(A)_l}_s(\theta)
\vec{\stackrel{\circ}{\Phi}}{}^{(A)_f}_s(\theta)
\vec{\Phi}{}^{\ast}_{s{}(A)_h}(\theta)
\vec{\stackrel{\circ}{\Phi}}{}^{\ast}_{s{}(A)_d}(\theta),\
 l,f,h,d \in \mbox{\boldmath$N_0$}\,, & {}
\end{eqnarray}
so that
\begin{eqnarray}
{} & {\rm gh}\left(\vec{\Phi}{}^{(A)_l}_s(\theta)
\vec{\stackrel{\circ}{\Phi}}{}^{(A)_f}_s(\theta)
\vec{\Phi}{}^{\ast}_{s{}(A)_h}(\theta)
\vec{\stackrel{\circ}{\Phi}}{}^{\ast}_{s{}(A)_d}(\theta)\right) =
\sum\limits_{i=1}^l{\rm gh}(\Phi^{(A)_i}_s) +
\sum\limits_{j=1}^f\bigl({\rm gh}(\Phi^{(A)_j}_s) + 1\bigr) -
& {} \nonumber\\
{} & \sum\limits_{k=1}^h\bigl({\rm gh}(\Phi^{(A)_k}_s) + 1\bigr)
- \sum\limits_{p=1}^d{\rm gh}(\Phi^{(A)_p}_s)\;. & {} 
\end{eqnarray}

According to definition, by homogeneous superfunction ${\cal F}(\theta)$ in
the extended sense
(superfunctional $F_{H,s}$) on $D^{k}_{s}$ ($C_{FH,s}$) of degree
$g={\rm gh}({\cal F}(\theta))$ call the superfunction
${\cal F}\bigl(\Gamma_{
s}(\theta)$, ${\stackrel{\circ}{\Gamma}}_{s}(\theta), \theta\bigr)$
$\equiv$ ${\cal F}(\theta)$
$\in D^{k}_{s}$ $\subset D^{k}_{ext}$ satisfying to equation on value of $g$
\begin{eqnarray}
{\cal F}\bigl(\tilde{g}{}^{\tau}\Phi_s(\theta), {\stackrel{\,\circ}{g}}{}^{
\tau}{
\stackrel{\circ}{\Phi}}_s(\theta), g^{\ast{}\tau}\Phi^{\ast}_s(\theta),
{\stackrel{\,\circ}{g}}{}^{\ast\tau}{\stackrel{\circ}{\Phi}}{}^{\ast}_s(
\theta), g^{\tau}\theta\bigr) =
e^{{\rm gh}({\cal F})}{\cal F}\bigl(\Gamma_{
s}(\theta), {\stackrel{\circ}{\Gamma}}_{s}(\theta), \theta\bigr),\;
{\rm gh}({\cal F}) \in {\bf Z}\;. 
\end{eqnarray}
For arbitrary homogeneous superfunction in the extended sense ${\cal
F}(\theta) \in D^{k}_{s}$ with definite ${\rm gh}({\cal F})$  the Euler's
theorem holds
\begin{eqnarray}
\hat{N}_s\left(\frac{\partial_l\phantom{xxx}}{\partial\Gamma_s(\theta)},
\frac{\partial_l\phantom{xxx}}{\partial{\stackrel{\circ}{\Gamma}}_{s
}(\theta)}, \frac{\partial\phantom{}}{\partial\theta},
\Gamma_{s}(\theta), {\stackrel{\circ}{\Gamma}}_{s}(\theta),
\theta\right){\cal F}(\theta) = {\rm gh}({\cal F}){\cal F}(\theta)\;, 
\end{eqnarray}
where the linear operator $N_s$ (ghost number operator) is defined by the
formula
\begin{eqnarray}
{} & \hspace{-2em}
\hat{N}_s(\theta) = {\rm gh}(\Phi^{A_s}_s)\Phi^{A_s}_s(\theta)
\displaystyle\frac{\partial_l\phantom{xxxx}}{\partial\Phi^{A_s}_s(\theta)} +
(1+{\rm gh}(\Phi^{A_s}_s)){\stackrel{\circ}{\Phi}}{}^{A_s}_s(\theta)
\displaystyle\frac{\partial_l\phantom{xxxxx}}{\partial{\stackrel{\circ}{
\Phi}}{}^{A_s}_s(\theta)} - (1+{\rm gh}(\Phi^{A_s}_s))\times
{} & \nonumber \\
{} & \hspace{-2em}
\Phi^{\ast}_{A_s,s}(\theta)
\displaystyle\frac{\partial\phantom{xxxxxx}}{\partial\Phi^{\ast}_{A_s,s}(
\theta)}
- {\rm gh}(\Phi^{A_s}_s){\stackrel{\circ}{\Phi}}{}^{\ast}_{A_s,s}(\theta)
\displaystyle\frac{\partial\phantom{xxxxxx}}{\partial{\stackrel{\circ}{\Phi}}{
}^{\ast}_{A_s,s}(\theta)} -
\theta\frac{\partial\phantom{}}{\partial\theta},\ \;
(\varepsilon_P,\varepsilon_{\bar{J}},\varepsilon)\hat{N}_s(\theta) =
(0,0,0)\;. {} & 
\end{eqnarray}
All elements of superalgebra {\boldmath${
\cal B}_{ext}$} with $\varepsilon_P \ne 0$ and antibrackets (2.46), (2.47)
shift the ghost number values of homogeneous
with respect to "${\rm gh}$" components of superfunctions from $D^{k}_{s}$, on
which the mentioned objects act, on $+1$
\renewcommand{\theequation}{\arabic{section}.\arabic{equation}\alph{lyter}}
\begin{eqnarray}
\setcounter{lyter}{1}
{} & {\rm gh}\left(({\cal F}(\theta), {\cal J}(\theta))_{ij}^{ext}\right)
= {\rm gh}({\cal F}) + {\rm gh}({\cal J}) + 1,\ \;i,j\in\{0,1\},\ i,j
\in\{+,-\}\;, & {}
\\
\setcounter{equation}{16}
\setcounter{lyter}{2}
{} & {\rm gh}(B(\theta){\cal F}(\theta))= {\rm gh}({\cal F}(\theta)) + 1,\ \;
(B(\theta) \in \mbox{\boldmath${\cal B}_{ext}$}, \varepsilon_P(B)=1)\,. & {}
\end{eqnarray}
Let us formally ascribe to  all such operators from {\boldmath${\cal B}_{
ext}$}
and to antibrackets $(\ ,\ )_{ij}$ the ghost number being equal to 1
\renewcommand{\theequation}{\arabic{section}.\arabic{equation}}
\begin{eqnarray}
{} & {\rm gh}(B(\theta)) = {\rm gh}\left((\ \;,\ \;)_{ij}^{ext}\right) = 1\;,
& {} 
\end{eqnarray}
whereas to other ones with $\varepsilon_P=0$ the zero ghost number.
The requirement of the ghost number conservation  in transforming of the
operators
$B^{s}(\theta)$ (2.50) leads to the following additional restriction on the
structure of superfunction ${\cal F}(\theta)$ and therefore on
$f\bigl({\stackrel{\circ}{\Gamma}}_{s}(\theta)$, $\theta\bigr)$ in (2.51)
\begin{eqnarray}
{\rm gh}({\cal F}(\theta)) = {\rm gh}(f) - 1=0\;.
\end{eqnarray}
\section{Classical Master Equation in  Minimal Sector}
\setcounter{equation}{0}

The restriction of
superfunction $S_{H}(\Gamma_{cl}(\theta),\theta)$ $\in$
$C^{k}(T^{\ast}_{odd}{\cal M}_{cl}\times\{\theta\}$) [2] with corresponding GHS

\renewcommand{\theequation}{\arabic{section}.\arabic{equation}\alph{lyter}}
\begin{eqnarray}
\setcounter{lyter}{1}
{} & {} & \displaystyle\frac{d_r\Gamma^p_{cl}(\theta)}{d\theta\phantom{xxxx}}=
\left(\Gamma^p_{cl}(\theta), S_H(\Gamma_{cl}(\theta),\theta)\right)^{s}_{
\theta}
\,,\\ 
\setcounter{equation}{1}
\setcounter{lyter}{2}
{} & {} & \Theta^H_{\imath}(\Gamma_{cl}(\theta), \theta) = ({\cal A}^{\ast}_{
\imath}(\theta),
S_H(\Gamma_{cl}(\theta),\theta))^{s}_{\theta} - (-1)^{\varepsilon_{\imath}}
\left({{S''_H}}^{-1}
\right)_{\imath k}(\theta)\Biggl[\frac{\partial}{\partial\theta}
({\cal A}^k(\theta),S_H(\theta))^{s}_{\theta}(-1)^{\varepsilon_{k}}
\nonumber \\
{} & {} & \hspace{3em} - (-1)^{\varepsilon_{\jmath}}({\cal A}^k(\theta),({\cal A}^{\ast}_{
\jmath}(\theta),S_H(\Gamma_{cl}(\theta),
\theta))^{s}_{\theta})^{s}_{\theta}({\cal A}^{\jmath}(\theta),S_H(\Gamma_{cl}(
\theta),\theta))^{s}_{\theta}\Biggl] = 0,\;s=cl\;,
\\ 
\renewcommand{\theequation}{\arabic{section}.\arabic{equation}}
{} & {} & ({S''_H})^{\imath\jmath}(\theta) =
\frac{\partial\phantom{xxxx}}{\partial{\cal A}^{\ast}_{\imath}(\theta)
}\frac{\partial S_H(\Gamma_{cl}(\theta),\theta)}{\partial{\cal A}^{\ast}_{
\jmath}(\theta)\phantom{xxxxx}},\ \;
{\rm sdet}\left\|({S''_H})^{\imath\jmath}(\theta)\right\| \neq 0\,,
\end{eqnarray}
determining the I class GThGT  in Hamiltonian formalism for GSTF by means
of requirements that $S_{H}(\Gamma_{cl}(\theta),\theta)$ would not
explicitly depend upon $\theta$ and would appear by eigenfunction for
ghost number operator $\hat{N}_{cl}(\theta)$ (3.15) with zero ghost number,
is defined by the equations
\renewcommand{\theequation}{\arabic{section}.\arabic{equation}\alph{lyter}}
\begin{eqnarray}
\setcounter{lyter}{1}
{} & \tilde{P}_1(\theta)S_H(\Gamma_{cl}(\theta),\theta) = 0
\Longleftrightarrow
S_H(\Gamma_{cl}(\theta),\theta)= S_H(\Gamma_{cl}(\theta))\;, & {} \\ 
\setcounter{equation}{3}
\setcounter{lyter}{2}
{} & \hat{N}_{cl}(\theta)S_H(\Gamma_{cl}(\theta),\theta) = 0
\Longrightarrow S_H(\Gamma_{cl}(\theta)) = {\cal S}_0({\cal A}(\theta))\;,
& {} 
\end{eqnarray}
according to (3.2), (3.5), (3.8). The nonvanishing of superfunction
${\cal S}_{0}({\cal A}(\theta))$ will be guaranteed by condition of potential
term existence for $S_{H}(\Gamma_{cl}(\theta),\theta)$ of the form
(2.14). Really the following expressions are valid
\renewcommand{\theequation}{\arabic{section}.\arabic{equation}}
\begin{eqnarray}
S({\cal A}(\theta),0) = {\cal S}_{0}({\cal A}(\theta)),\ \;
{\rm gh}(T({\cal A}(\theta), {\cal A}^{\ast}(\theta))) \leq -1,\ \;
{\rm deg}_{{\cal A}(\theta)}{\cal S}_{0}({\cal A}(\theta)) \geq 2\;,   
\end{eqnarray}
where the integer-valued functions of degree ${\rm deg}_{
{\cal A}(\theta)}$,
${\rm deg}_{{\cal A}^{\ast}(\theta)}$ (2.3),
${\rm deg}_{d}$
and of least degree $\min{\rm deg}_d$, $d\in D=\{
\Gamma_s(\theta)$, ${\stackrel{\circ}{\Gamma}}_s(\theta)$,
$\Gamma_s(\theta){\stackrel{\circ}{\Gamma}}_{s'}(\theta)\}$, $s,s'\in C_0$
had been defined in Ref.[1] for arbitrary superfunction with respect
to any generating element $d \in D$.

Restricted GHS
corresponding to  ${\cal S}_{0}({\cal A}(\theta))$ arises from (4.1) already
with HCHF $\Theta_{\imath}^{H}({\cal A}(\theta))$
\renewcommand{\theequation}{\arabic{section}.\arabic{equation}\alph{lyter}}
\begin{eqnarray}
\setcounter{lyter}{1}
{} & \displaystyle\frac{d_r{\cal A}^{\imath}(\theta)}{d\theta\phantom{
xxxx}}=0,\ \;
\displaystyle\frac{d_r{\cal A}^{\ast}_{\imath}(\theta)}{
d\theta\phantom{xxxx}}= - (-1)^{\varepsilon_{\imath}}{{\cal S}_{0},}_{
\imath}({\cal A}(\theta))\;, {} & \\ 
\setcounter{equation}{5}
\setcounter{lyter}{2}
{} & \Theta_{\imath}^{H}({\cal A}(\theta))=
- (-1)^{\varepsilon_{\imath}}{{\cal S}_{0},}_{\imath}({\cal A}(\theta))=0\;.
\end{eqnarray}
The  equivalence of GHS (4.1) to one's own HS subsystem (4.1a) by virtue of
Statement 3.3 from Ref.[2] is lost for the case of GHS (4.5) in spite of the
restricted dynamical equations presence [2]
\renewcommand{\theequation}{\arabic{section}.\arabic{equation}}
\begin{eqnarray}
\Theta_{\imath}^{H}({\cal A}(\theta))= - \left(
\displaystyle\frac{d_r{\cal A}^{\ast}_{\imath}(\theta)}{
d\theta\phantom{xxxx}} + (-1)^{\varepsilon_{\imath}}{{\cal S}_{0},}_{
\imath}({\cal A}(\theta))\right). 
\end{eqnarray}
That restricted by means of Eqs.(4.3) superfield theory with superfunction
${\cal S}_0({\cal A}(\theta))$  determines the II class GThST on $T^{\ast
}_{odd}{\cal M}_{cl}$, i.e. ${\cal S}_0({\cal A}(\theta))$ appears by not
only integral for HS (4.5a) but satisfies to Eq.(2.2) for any
${\cal A}^{\imath}(\theta) \in {\cal M}_{cl}$ as well. It means [2] both GHS
(4.5) and HS (4.5a) are solvable. The supermatrix rank
of the 2nd partial superfield derivatives with respect to
${\cal A}^{\imath}(\theta)$, ${\cal A}^{\jmath}(\theta)$ of superfunction
${\cal S}_{0}({\cal A}(\theta))$ by virtue of initial postulate
$2_{H}$, written in Ref.[2], on the structure of $S_{H}(\Gamma_{cl}(\theta),
\theta)$ is equal on a stationary surface $\Sigma$ to $(n-m)$, where $m$ is
the number of identities among
$\Theta_{\imath}^{H}({\cal A}(\theta))$
\begin{eqnarray}
{\rm rank}_{\varepsilon}\left\|{{\cal S}_{0},}_{\imath\jmath}({\cal
A}(\theta))\right\|_{\mid \Sigma} =  n-m = (n_{+}+ m_{+}, n_{-} + m_{-}),\
\Sigma \subset {\cal M}_{cl},\ {\rm dim}\Sigma = m\,.   
\end{eqnarray}
In addition to Eq.(2.2), its consequence is trivially fulfilled  in the
whole $T^{\ast}_{odd}{\cal M}_{cl}$ (Statement 5.5 of work [2])
\begin{eqnarray}
\Delta^{cl}(\theta){\cal S}_{0}({\cal A}(\theta)) = 0\;. 
\end{eqnarray}
For obtained GThST of the II class by virtue of corollary 2.2 for Theorem 2
in Ref.[1] and corollary 1.2 for Theorem 1 in [2] the identities, following
from the 3rd group equations in (2.15) with allowance made for algebraic
properties for projectors $\{\tilde{P}_a(\theta), U(\theta)\}$ applied for
decomposition of mentioned equations, are valid
\begin{eqnarray}
{{\cal S}_{0},}_{\imath}({\cal A}(\theta)){\cal R}_{0}{}^{\imath
}_{\alpha}({\cal A}(\theta)) = 0,\
{\cal R}_{0}{}^{\imath}_{\alpha}({\cal A}(\theta))={\cal R}_{0}{}^{
\imath}_{\alpha}({\cal A}(\theta),0) 
\end{eqnarray}
in identifying of ${\cal R}_{0}{}^{\imath}_{\alpha}({\cal A}(\theta),\theta)$
$\equiv$ ${\cal R}^{\imath}_{H{}\alpha}({\cal A}(\theta),\theta)$.
GTST presented in the form of n ODE system (2.21) can be written as
follows
\renewcommand{\theequation}{\arabic{section}.\arabic{equation}\alph{lyter}}
\begin{eqnarray}
\setcounter{lyter}{1}
{} & {} & \displaystyle\frac{d_r{\cal A}^{\imath}(\theta)}{d\theta\phantom{
xxxx}}= ({\cal A}^{\imath}(\theta),
S_1(\Gamma_{cl}(\theta),C(\theta))^{cl}_{\theta}\;
,\\ 
\setcounter{equation}{10}
\setcounter{lyter}{2}
{} & {} & S_1(\theta) \equiv
S_1(\Gamma_{cl}(\theta),C(\theta)) =
{\cal A}^{\ast}_{\imath}(\theta){\cal R}_{0}{}^{\imath}_{\alpha}({\cal A
}(\theta))C^{\alpha}(\theta)\;, \nonumber \\
{} & {} & (\varepsilon_P,\varepsilon_{\bar{J}},
\varepsilon)S_1(\theta)=(0,0,0),\ S_1(\theta)\in
C^k((T^{\ast}_{odd}{\cal M}_{cl})\times{\cal M}_C)\;. 
\end{eqnarray}
The extension  of n ODE system (4.10a) to the 1st order on $\theta$
2n ODE system in the normal form with respect to derivatives ${\stackrel{
\circ}{\Gamma}}_{cl}(\theta)$ reads as follows
\renewcommand{\theequation}{\arabic{section}.\arabic{equation}}
\begin{eqnarray}
\frac{d_r\Gamma^p_{cl}(\theta)}{d\theta\phantom{xxxx}}=
\left(\Gamma^p_{cl}(\theta), S_1(\theta)\right)^{cl}_{\theta}\,. 
\end{eqnarray}
In (4.11) the dependence upon superfields $C^{\alpha}(\theta)$ is parametric
one. System (4.11) itself does not satisfy to solvability condition.
The supermatrix rank of the 2nd partial superfield derivatives with respect to
$\Gamma^{p}_{cl}(\theta)$, $\Gamma^{q}_{cl}(\theta)$ of $S_{1}(\theta)$
calculated in $T^{\ast}_{odd}{\cal M}_{min}\times\{\theta\}$ for linearly
independent GGTST ${\cal R}_{0}{}^{\imath}_{\alpha}({\cal A}(\theta),\theta)$
is equal (with respect to $\varepsilon$ grading) on the supersurface $\Sigma
\cup \{{\cal A}^{\ast}_{0{}\imath}(\theta), C^{\alpha}_0(\theta)\}$ $\subset
T^{\ast}_{odd}{\cal M}_{min}$
\begin{eqnarray}
{\rm rank}_{\varepsilon}\left\|
\frac{\partial_r\phantom{xxxxx}}{\partial\Gamma^p_{min}(\theta)}
\frac{\partial_lS_1(\theta)\phantom{x}}{\partial\Gamma^q_{min}(\theta)}
\right\|_{\mid \Sigma\cup\{{\cal A}^{\ast}_{0{}\imath}(\theta),
C^{\alpha}_0(\theta)\}} \hspace{
-1.0em} = 2(m_{-},m_{+}),\ ({\cal A}^{\ast}_{0{}\imath}(\theta),
C^{\alpha}_0(\theta))=(0,0)\,. 
\end{eqnarray}
A distribution of rank values for supermatrix of the form (4.7), (4.12) with
respect to enumeration of the columns and rows on parities
$\varepsilon_{\bar{J}}$ and $\varepsilon$ is differred in accordance with
whether the parities of
indices for supermatrix elements coincide with parities of corresponding
generating elements or are opposite to them. So, the rank values (4.7) are
identical for $\varepsilon_{\bar{J}}$ and $\varepsilon$  by virtue of
${\cal A}^{\imath}(\theta)$, ${\cal S}_0({\cal A}(\theta))$ gradings but for
expression (4.12) are distinct with respect to change of boson and fermion
values for rank of supermatrix (4.12). It is natural to denote the ranks with
respect to the mentioned gradings by additional sign $"\varepsilon"$,
$"\varepsilon_{\bar{J}}"$: ${\rm rank}_{\varepsilon}$, ${\rm rank}_{
\varepsilon_{\bar{J}}}$.
In this case, the rank with respect to $\varepsilon $
(${\rm rank}_\varepsilon$) have been written in (4.12) and for the case of
${\rm rank}_{\varepsilon_{\bar{J}}}$  the value $2m=2(m_{+}, m_{-})$ must
stand in the right-hand side of formula (4.12).

In view  of $\varepsilon_{P}$,
$\varepsilon_{\bar{J}}$, $\varepsilon$ gradings specification for the first
time introduced in [1] that principal difference had been earlier ignored.
For instance, in Refs.[4,6,7] where for fields and
antifields the only $\varepsilon$ grading  in fact was introduced,
the corresponding rank had been calculated, under omission, with respect to
$\varepsilon_{\bar{J}}$ one!

Note the dimensions of ${\cal M}_{cl}$ with respect to
$\varepsilon$,  $\varepsilon_{\bar{J}}$ gradings over number field
${\bf K}$ are equal to
\begin{eqnarray}
{\rm dim}_{{\bf K},\varepsilon_{\bar{J}}}{\cal M}_{cl} =
{\rm dim}_{{\bf K},\varepsilon}{\cal M}_{cl} =
\left({\rm dim}_{+{\bf K},\varepsilon},
{\rm dim}_{-{\bf K},\varepsilon}\right){\cal M}_{cl} = n\,,
\end{eqnarray}
whereas the  dimensions of $T_{odd}{\cal M}_{cl}$,
$T^{\ast}_{odd}{\cal M}_{cl}$ with respect to the same parities are differred
\renewcommand{\theequation}{\arabic{section}.\arabic{equation}\alph{lyter}}
\begin{eqnarray}
\setcounter{lyter}{1}
{} & {\rm dim}_{{\bf K},\varepsilon_{\bar{J}}}T_{odd}{\cal M}_{cl} =
{\rm dim}_{{\bf K},\varepsilon_{\bar{J}}}T^{\ast}_{odd}{\cal M}_{cl} = 2n
\,, & {} \\ 
\setcounter{equation}{14}
\setcounter{lyter}{2}
{} & \hspace{-1em}{\rm dim}_{{\bf K},\varepsilon}T_{odd}{\cal M}_{cl} =
{\rm dim}_{{\bf K},\varepsilon}T^{\ast}_{odd}{\cal M}_{cl} = n + n^P= (n_{+}
+ n_-,n_{-} + n_+),\;n^P = (n_-,n_+),{} &  
\end{eqnarray}
from where it is seen that the structures of the even (odd) subsupermanifolds
on $\varepsilon$ and $\varepsilon_{\bar{J}}$ parities  are different. In what
follows we will work in GSQM mathematical means, in general, with notion of
${\rm rank}_{{\bf K},\varepsilon}$ and  by omitting
under rank notion  it will be understood namely the last type of rank for
even supermatrix just as in calculating of  the  geometric
objects dimensions we use the operation ${\rm dim}_{{\bf K},\varepsilon}$. For instance,
the dimensions of superspaces ${\cal M}_{k}$, $T_{odd}{\cal M}_{k}$,
$T^{\ast}_{odd}{\cal M}_{k}$, $k\in\{C,B,min\}$ are given by the following
table
\begin{eqnarray}
{} &
\setcounter{lyter}{1}
\begin{array}{lccccccl}
{} & {\cal M}_C & {\cal M}_B & {\cal M}_{min} & T_{odd}{\cal M}_C &
T_{odd}{\cal M}_B & T_{odd}{\cal M}_{min} & {} \\
{\rm dim}_{{\bf K},\varepsilon} & m^P & m & n + m^P & m+m^P &
m+m^P & n + n^P + m + m^P & {} \\
{\rm dim}_{{\bf K},\varepsilon_{\bar{J}}} & m & m & n+m & 2m & 2m & 2(n+m) & ,
\end{array} & {} \\ 
\setcounter{equation}{15}
\setcounter{lyter}{2}
{} &
\left({\rm dim}_{{\bf K},\varepsilon},
{\rm dim}_{{\bf K},\varepsilon_{\bar{J}}}\right)T_{odd}{\cal M}_{s} =
\left({\rm dim}_{{\bf K},\varepsilon},
{\rm dim}_{{\bf K},\varepsilon_{\bar{J}}}\right)T^{\ast}_{odd}{\cal M}_{s},\
s\in C_0\;. & {} 
\end{eqnarray}
It is easy to find from (4.14), (4.15) the corresponding
dimensions of the other geometric structures ${\cal M}_{s}$, $T_{odd}{\cal
M}_{s}$, $T^{\ast}_{odd}{\cal M}_{s}$, $T_{odd}(T^{\ast}_{odd}{\cal M}_{s})$,
$s \in C_0$ (2.30).
These concepts for dimensions and ranks should be transferred onto
corresponding objects in Refs.[1,2]. From the notations in (4.13), (4.15b)
it is
easy to obtain ${\rm dim}_{+{}{\bf K},\varepsilon(\varepsilon_{\bar{J}})}$
and ${\rm dim}_{-{}{\bf K},\varepsilon (\varepsilon_{\bar{J}})}$
being respectively by designations  for even and
odd dimensions of the sets in (4.14), (4.15) with respect to
$\varepsilon(\varepsilon_{\bar{J}})$ parity.

Next, for construction  the II class GThGT  let us combine the systems
(4.5), (4.11) by means of addition of the corresponding vector fields
components  standing from the right in Eqs.(4.5a), (4.11). As a result obtain
the following
almost Hamiltonian system  of the 1st order on $\theta$ 2n ODE in normal
form given on $T^{\ast}_{odd}{\cal M}_{min}\times\{\theta\}$
\renewcommand{\theequation}{\arabic{section}.\arabic{equation}}
\begin{eqnarray}
\frac{d_r\Gamma^p_{cl}(\theta)}{d\theta\phantom{xxxx}}=
\left(\Gamma^p_{cl}(\theta), S_{[1]}(\Gamma_{cl}(\theta),C(\theta))
\right)^{cl}_{\theta},\ p=\overline{1,2n} 
\end{eqnarray}
with superfunction $S_{[1]}(\theta)$ defined by equality
\begin{eqnarray}
S_{[1]}(\theta)\equiv
S_{[1]}(\Gamma_{cl}(\theta),C(\theta)) = {\cal S}_0({\cal A}(\theta))
+ S_{1}(\Gamma_{cl}(\theta),C(\theta))\;. 
\end{eqnarray}
Conditions (4.3a,b) are obviously fulfilled for superfunction
$S_{[1]}(\theta)$
\begin{eqnarray}
(\hat{N}_{min}(\theta),\tilde{P}_1(\theta))S_{[1]}(\theta) =(\hat{N}_{min}(
\theta),,\tilde{P}_1(\theta))S_{1}(\theta) = (0,0)
\Longrightarrow {\rm gh}(S_{[1]}(\theta)) = 0\,. 
\end{eqnarray}
At last extend the system (4.16) to HS of the form
\begin{eqnarray}
\frac{d_r\Gamma^{p_{min}}_{min}(\theta)}{d\theta\phantom{xxx
xxx}}=
\left(\Gamma^{p_{min}}_{min}(\theta), S_{[1]}(\Gamma_{min}(\theta))
\right)^{min}_{\theta},\ p_{min}=\overline{1,2(n+m)},\;
S_{[1]}(\Gamma_{min}(\theta)) \equiv S_{[1]}(\theta). 
\end{eqnarray}
Theory of superfields $\Gamma_{min}^{p_{min}}(\theta)$ being described by
superfunction $S_{[1]}(\theta)$ leads to HS (4.19) not satisfying
to solvability condition for GThGT [not forgetting on presence of HCHF
(4.5b) and formally having  continued the relation (4.6) into
$T^{\ast}_{odd}{\cal M}_{min}$: $\Theta_{\imath}(\Gamma_{min}(\theta))$ =
$- \Bigl(\frac{d_r {\cal A}_{\imath}^{\ast}(\theta)}{d\theta\phantom{xxxx}}$
$+$
$(- 1)^{\varepsilon_{\imath}}S_{[1]},_{\imath}(\Gamma_{\min}(\theta))\Bigr)$
we already obtain the GCHF and therefore from initial GThST the
\mbox{GThGT},
so that $\Theta_{\imath}(\Gamma_{\min}(\theta))_{\mid C(\theta)=0}$ =
$\Theta_{\imath}^{H}({\cal A}(\theta))$]. The unsolvability for Eq.(4.19)
follows from the
nonfulfilment of Eq.(2.2) for $S_{[1]}(\theta)$ in the whole
$T^{\ast}_{odd}{\cal M}_{min}$ (see Statement 5.4 in [2])
\renewcommand{\theequation}{\arabic{section}.\arabic{equation}\alph{lyter}}
\begin{eqnarray}
\setcounter{lyter}{1}
{} & (S_{[1]}(\theta),S_{[1]}(\theta))^{min}_{\theta} = \frac{1}{2}
{\cal A}_{\imath}^{\ast}(\theta)\left(
{{\cal R}_{0}{}^{\imath}_{\alpha},}_{\jmath}({\cal A}(\theta))
{\cal R}_{0}{}^{\jmath}_{\beta}({\cal A}(\theta)) - (-1)^{\varepsilon_{
\alpha}\varepsilon_{\beta}}(\alpha \longleftrightarrow \beta)\right)\times
 & {}\nonumber \\
{} & C^{\beta}(\theta)C^{\alpha}(\theta)(-1)^{\varepsilon_{\alpha}+1}\equiv
{\cal F}_2(\Gamma_{cl}(\theta),C(\theta)) \neq 0\;, {} & \\ 
\setcounter{equation}{20}
\setcounter{lyter}{2}
{} & ({\rm min}{\rm deg}_{C}, {\rm deg}_{C}, {\rm gh},\varepsilon_P,\varepsilon_{\bar{J}},
\varepsilon){\cal F}_2(\theta) = (2,2,1,1,0,1)\;. {} & 
\end{eqnarray}
Thus the equation
\renewcommand{\theequation}{\arabic{section}.\arabic{equation}}
\begin{eqnarray}
(S_{[1]}(\theta),S_{[1]}(\theta))^{min}_{\theta} = 0 
\end{eqnarray}
is valid only with accuracy up to the 1st degree with respect to
superfields $C^{\alpha}(\theta)$. Moreover, the necessary condition of an
ordinary solvability for Eq.(4.21) in correspondence with Statement 5.3 [2]
holds.
Really, the rank (${\rm rank}_{\varepsilon}$) of Hesse supermatrix according
to assumptions $1_H$, $2_{H}$, Corollary 1.2 from Ref.[2] and
taking (4.7), (4.12) into account is given by relations
\renewcommand{\theequation}{\arabic{section}.\arabic{equation}\alph{lyter}}
\begin{eqnarray}
\setcounter{lyter}{1}
{} &
{\rm rank}_{\varepsilon}\left\|
\displaystyle\frac{\partial_r\phantom{xxxxx}}{\partial\Gamma^p_{min}(\theta)}
\displaystyle\frac{\partial_lS_{[1]}(\theta)\phantom{}
}{\partial\Gamma^q_{min}(\theta)} \right\|_{\mid
\Sigma\cup\{{\cal A}^{\ast}_{0{}\imath}(\theta),
C^{\alpha}_0(\theta)\}}\hspace{ -1.0em} =
{\rm rank}_{\varepsilon}\left\|
\displaystyle\frac{\partial_r\phantom{xxxx}}{\partial\varphi^a_{min}(\theta)}
\displaystyle\frac{\partial_lS_{[1]}(\theta)\phantom{}
}{\partial\varphi^b_{min}(\theta)}\right\|_{\mid
\Sigma\cup\{{\cal A}^{\ast}_{0{}\imath}(\theta),
C^{\alpha}_0(\theta)\}}\hspace{ -1.0em} = {} & \nonumber \\
{} & n + 2m^P-m \leq
{\rm dim}_{{\bf K},\varepsilon}T^{\ast}_{odd}{\cal M}_{min}  \;, {} &
\\ 
\setcounter{equation}{22}
\setcounter{lyter}{2}
{} & \varphi^a_{min}(\theta) = ({\cal A}^A(\theta), -{\cal A}^{\ast}_{
\alpha}(\theta), C^{\alpha}(\theta)),
\varphi^{\ast}_{a{}min}(\theta) = ({\cal A}^{\ast}_A(\theta), {\cal A}^{
\alpha}(\theta), C^{\ast}_{\alpha}(\theta))\;, {} & \nonumber \\
{} &  a,b=((A,\alpha,\alpha),(B,\beta,\beta)),\;
A,B=\overline{1,n-m},\;\alpha,\beta=\overline{1,m}\;.{} & 
\end{eqnarray}
This fact is seen especially simple in the point from $T^{\ast}_{odd}{\cal
M}_{min}$ with coordinates
\renewcommand{\theequation}{\arabic{section}.\arabic{equation}}
\begin{eqnarray}
\Gamma^{p_{min}}_{0{}min}(\theta) = (\Phi_0^{A_{min}}(\theta),
\Phi^{\ast}_{0{}A_{min}}(\theta))=(({\cal A}^{\imath}_0(\theta),0),(0,0))\;,
\end{eqnarray}
in which supermatrix (4.22a) has the form
\begin{eqnarray}
{} &
\left\|
\displaystyle\frac{\partial_r\phantom{xxxxx}}{\partial\Gamma^p_{min}(\theta)}
\displaystyle\frac{\partial_lS_{[1]}(\theta)\phantom{}
}{\partial\Gamma^q_{min}(\theta)} \right\|_{\mid \Gamma^{p_{min}}_{0{}min}(
\theta)}\hspace{ -1.0em} =
\begin{array}{c}
\begin{array}{lcccccc}
\vspace{-3ex}
{} & \scriptstyle{{\cal A}^B} & \hspace{-1em}\scriptstyle{\cal A}^{\beta} &
\hspace{-1em}\scriptstyle{\cal A}^{\ast}_B & \hspace{-1em}\scriptstyle{\cal
A}^{\ast}_{
\beta} & \hspace{-1em}\scriptstyle{C^{\beta}} & \hspace{-1em}
\scriptstyle{C^{\ast}_{\beta}} \\
\phantom{\scriptstyle{\cal A}^A} & \phantom{{{\cal S}_0,}_{AB}({\cal
A}_0)(-1)^{\varepsilon_A}} &
\hspace{-1em}\phantom{K^{12}_{A\beta}} & \hspace{-0.8em}\phantom{\ast} &
\hspace{-0.8em}\phantom{{\cal R}_0{}^{\alpha}_{\beta}({\cal A}_0)} &
\hspace{-0.8em}\phantom{{\cal R}_0{}^{\alpha}_{\beta}({\cal A}_0)}
& \hspace{-0.8em}\phantom{\ast }
\end{array}\\
\begin{array}{l||cccccc||}
\scriptstyle{\cal A}^A & {{\cal S}_0,}_{AB}({\cal A}_0)(-1)^{\varepsilon_A} &
\hspace{-1em}K^{12}_{A\beta} & \hspace{-0.8em}\ast & \hspace{-0.8em}\ast &
\hspace{-0.8em}\ast & \hspace{-0.8em}\ast \\
\scriptstyle{\cal A}^{\alpha} &
K^{21}_{\alpha B} & \hspace{-1em}K^{22}_{\alpha\beta}
& \hspace{-0.8em}\ast & \hspace{-0.8em}\ast &
\hspace{-0.8em}\ast &
\hspace{-0.8em}\ast \\
\scriptstyle{\cal A}^{\ast}_A & \ast & \hspace{-1em}\ast &\hspace{-0.8em}\ast
&\hspace{-0.8em}\ast & \hspace{-0.8em}\ast & \hspace{-0.8em}\ast  \\
\scriptstyle{\cal A}^{\ast}_{\alpha} & \ast & \hspace{-1em}\ast &
\hspace{-0.8em}\ast &\hspace{-0.8em}\ast &
\hspace{-0.8em}{\cal R}_0{}^{\alpha}_{\beta}({\cal A}_0) & \hspace{-0.8em}\ast
\\
\scriptstyle{C^{\alpha}} & \ast & \hspace{-1em}\ast &\hspace{-0.8em}\ast &
\hspace{-0.8em}{\cal R}_0{}^{\beta}_{\alpha}({\cal A}_0)
& \hspace{-0.8em}\ast & \hspace{-0.8em}\ast  \\
\scriptstyle{C^{\ast}_{\alpha}} & \ast & \hspace{-1em}\ast &
\hspace{-0.8em}\ast
& \hspace{-0.8em}\ast & \hspace{-0.8em}\ast & \hspace{-0.8em}\ast
\end{array}
\end{array}
= {} &\nonumber \\
{} & \hspace{-1em}
\begin{array}{c}
\begin{array}{lcccc}
\vspace{-3ex}
\phantom{\scriptstyle{\cal A}^{\imath}} & \scriptstyle{\cal A}^{\jmath} &
\hspace{-1.0em}\scriptstyle{\cal A}^{\ast}_{\jmath}
 & \hspace{-0.8em}\scriptstyle{C^{\beta}} & \hspace{-0.8em}
\scriptstyle{C^{\ast}_{\beta}} \\
\phantom{\scriptstyle{\cal A}^{\jmath}} & \phantom{{{\cal S}_0,}_{\imath
\jmath}({\cal A}_0)(-1)^{\varepsilon_{\imath}}} &
\hspace{-1.0em}\phantom{{\cal R}_0{}^{\imath}_{\beta}({\cal A}_0)} &
\hspace{-0.8em}\phantom{{\cal R}_0{}^{\imath}_{\beta}({\cal A}_0)}
& \hspace{-0.8em}\phantom{\scriptstyle{C^{\ast}_{\alpha}}}
\end{array}\\
\begin{array}{l||cccc||}
\scriptstyle{\cal A}^{\imath} & {{\cal S}_0,}_{\imath\jmath}({
\cal A}_0)(-1)^{
\varepsilon_{\imath}} &  \hspace{-1.0em}\ast & \hspace{-0.8em}\ast &
\hspace{-0.8em}\ast \\
\scriptstyle{\cal A}^{\ast}_{\imath} & \ast & \hspace{-1.0em}\ast &
\hspace{-0.8em}{\cal R}_0{}^{\imath}_{\beta}({\cal A}_0) & \hspace{-0.8em}
\ast  \\
\scriptstyle{C^{\alpha}} & \ast & \hspace{-1.0em}{\cal R}_0{}^{\jmath}_{
\alpha}({\cal A}_0)
& \hspace{-0.8em}\ast & \hspace{-0.8em}\ast  \\
\scriptstyle{C^{\ast}_{\alpha}} & \ast & \hspace{-1.0em}\ast &
\hspace{-0.8em}\ast & \hspace{-0.8em}\ast
\end{array}
\end{array},
\begin{array}{l}
\Gamma^p_{min}(\theta) =
({\cal A}^{\jmath}(\theta), {\cal A}^{\ast}_{\jmath}(\theta), C^{\beta
}(\theta), C^{\ast}_{\beta}(\theta)), \\
\Gamma^q_{min}(\theta) =
({\cal A}^{\imath}(\theta), {\cal A}^{\ast}_{\imath}(\theta),
C^{\alpha}(\theta), C^{\ast}_{\alpha}(\theta)), \\
\imath=(A,\alpha), \jmath=(B,\beta), \\
A,B=\overline{1,n-m},\alpha,\beta=\overline{1,m}.
\end{array} {} &
\end{eqnarray}
Under sign "$\ast$" in the above supermatrices it should be
understood  the zero values. The signs of ${\cal A}^{\imath}(\theta)$,
${\cal A}^{\jmath}(\theta),\ldots$
point out the  enumeration of the columns and rows for these supermatrices.
$\left\|{\cal
R}_{0}{}^{\alpha}_{\beta}({\cal A}_0({\theta}))\right\|$ appears by
nondegenerate supermatrix with value for its rank with respect to $
\varepsilon_{\bar{J}}$ $(\varepsilon)$
being equal to $m$ $(m^P)$ by virtue of linear
independence of ${\cal R}_{0}{}^{\imath}_{\alpha}({\cal A}(\theta))$ and
identities (4.9). At last the subsupermatrices $K^{12}_{A\beta}$,
$K^{21}_{\alpha B}$, $K^{22}_{\alpha\beta}$ appear by linear combinations of
the  corresponding rows and columns of the nondegenerate supermatrix
${{\cal S}_0,}_{AB}({\cal A}_0)$ and therefore do not give a nonvanishing
contribution into rank value of supermatrix (4.24).

To fulfill the Eq.(4.21) together with
condition (4.22) it is necessary to modify $S_{[1]}(\theta)$ by means
of special superfunction $S_{2}(\Gamma_{min}(\theta))$ to be now quadratic
with respect to $C^{\alpha}(\theta)$. Moreover the modified Eq.(4.21)
for superfunction
\begin{eqnarray}
S_{[2]}(\theta) \equiv
S_{[2]}(\Gamma_{min}(\theta)) = S_{[1]}(\Gamma_{min}(\theta)) +
S_{2}(\Gamma_{min}(\theta))   
\end{eqnarray}
in general case having the form
\renewcommand{\theequation}{\arabic{section}.\arabic{equation}\alph{lyter}}
\begin{eqnarray}
\setcounter{lyter}{1}
{} & (S_{[2]}(\theta),S_{[2]}(\theta))^{min}_{\theta} =
{\cal F}_3(\Gamma_{min}(\theta)) + {\cal F}_4(\Gamma_{min}(\theta)) \;, {} &
\\ 
\setcounter{equation}{26}
\setcounter{lyter}{2}
{} &
({\rm min}{\rm deg}_{C},{\rm deg}_{C}, {\rm gh},\varepsilon_P,
\varepsilon_{\bar{J}},\varepsilon){\cal F}_k(\theta) = (k,k,1,1,0,1),\
k=3,4\;. {} &   
\end{eqnarray}
will be satisfied with accuracy up to $O(C^{2}(\theta))$. Naturally in
(4.26b) it is taken into account the continuation of the ghost number and
Grassmann gradings
properties from $S_{[1]}(\theta)$ onto $S_{[2]}(\theta)$
\renewcommand{\theequation}{\arabic{section}.\arabic{equation}}
\begin{eqnarray}
\hat{N}_{min}(\theta)S_{2}(\theta) = 0,\
(\varepsilon_P,\varepsilon_{\bar{J}},\varepsilon)S_2(\theta) = (0,0,0)\;.
\end{eqnarray}

It does not arise from (4.27), in general, but from the construction of
$S_{[2]}(\theta)$ itself with respect to right-hand side of Eq.(4.20a) it
follows that $S_{2}(\theta)$ may be chosen in the form not explicitly
depending upon $\theta$.
Thus, conditions (4.27) determine $S_{2}(\theta)$ as more than quadratic
addition  to $S_{[1]}(\theta)$, and therefore $S_{2}(\theta)$ does not
influence on value of rank (4.22) written for $S_{[2]}(\theta)$ in its
calculating on the supersurface
$\Sigma\cup\{{\cal A}^{\ast}_{0{}\imath}(\theta),
C^{\alpha}_0(\theta),C^{\ast}_{0{}\alpha}(\theta)=0\}$ in view of the
expression
\begin{eqnarray}
({\rm min}{\rm deg}_{CC^{\ast}},{\rm min}{\rm deg}_{{\cal A}^{\ast}C})
S_2(\theta) \in \{(3,4),(3,0),(0,4)\}\;. 
\end{eqnarray}
One can construct the HS with respect to $S_{[2]}(\theta)$ (4.25) being by
deformation for HS (4.19) in powers of $C^{\alpha}(\theta)$ having replaced
to this end the $S_{[1]}(\theta)$ on $S_{[2]}(\theta)$ in the last equations.
Obtained HS will not be solvable again by virtue of the fact that
${\cal F}_{k}(\theta) \neq 0,$ $k=3,4$ in (4.26a).

The deformation process of $S_{[2]}(\theta)$ in powers of
$C^{\alpha}(\theta)$ by means  of special superfunction $S_{3}(\theta)$
addition, constructed with respect to ${\cal F}_{3}(\theta)$ in order
to the superfunction $S_{[3]}(\theta)$ = $(S_{[2]}$ + $S_3)(\theta)$ would
satisfy to Eq.(4.26a) with accuracy up to  $O(C^{3}(\theta))$ and so on
may be continued in the framework, for instance, of
polynomial theory (i.e. ${\cal S}_{0}(\theta)$, ${\cal R}_{0}{}^{\imath}_{
\alpha}(\theta)$ are the polynomials on ${\cal A}^{\imath}(\theta))$ with
respect to $\Gamma_{min}(\theta)$ and on a definite stage of
$S_{\infty}(\theta)$ construction  will be broken  satisfying to
Eq.(4.26a) for $S_{[\infty]}(\theta)$  in all powers of $C^{\alpha}(\theta)$.
Thus, the procedure of superfunction $S_{[1]}(\theta)$ deformation in
superfield form  by components of higher powers of
$C^{\alpha}(\theta)$ in order to satisfy to Eq.(4.20a) and
therefore, in agreement with Statement 5.4 from paper [2], to solvability
of corresponding HS of the type (4.19) is described.

The construction of superfunction $S_{H{}s}(\Gamma_{s}(\theta))$, $s=min$
taking values in $\Lambda_1(\theta;{\bf R})$
\begin{eqnarray}
 S_{H{}s}(\Gamma_{s}(\theta)) \equiv S_{[\infty]}(\Gamma_{s}(\theta)) =
{\cal S}_0(\theta) + \sum\limits_{k\geq 1} S_k(\Gamma_{s}(\theta)),\
({\rm min}{\rm deg}_{C},{\rm deg}_{C})S_k(\Gamma_{s}(\theta)) = (k,k)\,,
\end{eqnarray}
satisfying to the equation (usually called the master equation or generating
equation of gauge algebra [4,6,7])
\renewcommand{\theequation}{\arabic{section}.\arabic{equation}}
\begin{eqnarray}
 (S_{H{}s}(\Gamma_{s}(\theta)),S_{H{}s}(\Gamma_{s}(\theta)))^s_{\theta}=
0,\ \forall\Gamma_s(\theta)\in T^{\ast}_{odd}{\cal M}_s,\ s=min\;, 
\end{eqnarray}
to the conditions (4.22) without change of the Lagrangian surface,
parametrized by $\varphi^a_{min}(\theta)$, and
to the following equations for $s=min$
\begin{eqnarray}
(\hat{N}_s(\theta), \tilde{P}_1(\theta),\varepsilon_P,\varepsilon_{\bar{J}},
\varepsilon)S_{H{}s}(\Gamma_{s}(\theta))=(0,0,0,0,0),
\end{eqnarray}
appears by the result of the produced procedure. Corresponding solvable HS has
the form
\begin{eqnarray}
\frac{d_r\Gamma^{p_s}_{s}(\theta)}{d\theta\phantom{xxxx}}=
\left(\Gamma^{p_s}_{s}(\theta), S_{H{}s}(\Gamma_s(\theta))\right)^{s
}_{\theta},\ s=min
\end{eqnarray}
and possesses by $S_{H{}{min}}(\theta)$ as  one's integral.
Besides HCHF(4.5b) are the additional conditions to HS (4.32).

The following equation is represented by principal one appearing by virtue of
validity of the Statement 5.5 from Ref.[2] for the II class GThGT, being
described by $S_{H{}{min}}(\theta)$, by  the consequence of the Eq.(4.30)
\begin{eqnarray}
\Delta^s(\theta)S_{H{}s}(\Gamma_{s}(\theta))=0,\ \forall\Gamma_s(\theta)\in
T^{\ast}_{odd}{\cal M}_s, s=min\;. 
\end{eqnarray}
Briefly the algorithm of construction of the II class GThGT given  on
$T^{\ast}_{odd}{\cal M}_{min}$, being defined by
superfunction $S_{H{}{min}}(\Gamma_{min}(\theta))$, starting from
the I class original GThGT  given on $T^{\ast}_{odd}{\cal M}_{cl}$ with
$S_H(\Gamma_{cl}(\theta),\theta)$, in general, not appearing by integral for
HS determined with respect to the latter superfunction
(therefore being unsolvable), may describe in the form
of following operations sequence realization:

\noindent
{\bf 1)} The restriction by means of ghost number and operator $\tilde{
P}_1(\theta)$ of the I class  original GThGT
determined on  $T^{\ast}_{odd}{\cal M}_{cl}$ $\times$ $\{\theta\}$ with
$S_H(\Gamma_{cl}(\theta),\theta)$ to the II class GThST  given
on $T^{\ast}_{odd}{\cal M}_{cl}$ with
superfunction ${\cal S}_0({\cal A}(\theta))$ $\in$ $C^{k}({\cal M}_{
cl})$ (with additional HCHF $\Theta_{\imath}^{H}({\cal A}(\theta))$).
This stage is described by relations (4.1)--(4.5);

\noindent
{\bf 2)} Join of HS for the II class GThST  with ${\cal S}_{0}({\cal
A}(\theta))$ and system of the 1st order on $\theta $
 n ODE  in  the normal form, describing GTST for
${\cal S}_0({\cal A}(\theta))$  in
$T_{odd}{\cal M}_{cl}$ $\times$ ${\cal M}_{C}$ and obtaining the
superfunction $S_{[1]}(\Gamma_{cl}(\theta), C(\theta))$ $\in$ $C^k\bigl((T^{
\ast}_{odd}{\cal M}_{cl})$ $\times {\cal M}_C\bigr)$.
That step is described by the formulae (4.10a)--(4.12), (4.16)--(4.18);

\noindent
{\bf 3)} The extension of the almost HS from the 2nd stage  to HS in
$T^{\ast}_{odd}{\cal M}_{cl}$
defined by means of $S_{[1]}(\theta)$ and obtaining of GThGT but not yet of
the II class. The set of formulae (4.19)--(4.24) corresponds to this step;

\noindent
{\bf 4)} Deformation of $S_{[1]}(\theta)$ in powers of $C^{\alpha}(\theta)$
with obtaining of $S_{H{}{min}}(\Gamma_{min}(\theta))$ determining
the II class GThGT. The mathematical description starting from formula (4.25)
to Eq.(4.33) inclusively corresponds to that stage.

Definite moments of $S_{H{}{min}}(\theta)$ construction beginning from
${\cal S}_{0}({\cal A}(\theta))$, in particular the possibility of
$S_{k}(\Gamma_{min}(\theta))$ construction with respect to the right-hand side of
the preceding equation for $S_{[k -
1]}(\theta)$ for the $k$th degree on $C^{\alpha}(\theta)$, require the
demonstrative verification.

Therefore formulate the obtained results in the form of theorem (without its
proof here)

\noindent
\underline{\bf Theorem 1} (Existence of a solution for the classical
master equation in $C^k\bigl(T^{\ast}_{odd}{\cal M}_{min})$)

\noindent
A solution for equation (4.30) in fulfilling of the conditions (4.31),
boundary condition
\begin{eqnarray}
{}\hspace{-10.1em}{\rm a)}\hspace{10.1em}S_{H{}s}(\Gamma_{s}(\theta)){
\hspace{-0.5em}\phantom{\Bigl(}}_{\mid
\Phi^{
\ast}(\theta)=0} = {\cal S}_0({\cal A}(\theta)),\ s=min\;, 
\end{eqnarray}
b) an existence of the Lagrangian surface  $\Lambda_{min}$ $\subset$
$T^{\ast}_{odd}{\cal M}_{min}$ parametrized by coordinates
$\varphi^{a_s}_s(\theta)$ ($(\varphi^{a_s}_s(\theta),\varphi^{b_s}_s(
\theta))^{(\Gamma_s)}_{\theta}$ = $0$, $a_s$ = $1,\ldots,n_{s}$, $n_{min}$ =
$[A_{min}] = n + m$) so that  on the supersurface $\tilde{\Sigma}$ $\equiv$
$\Sigma \cup \{{\cal A}^{\ast}_{0{}\imath}(\theta),
C^{\alpha}_0(\theta), C^{\ast}_{0{}\alpha}(\theta)\}$ the
formula holds
\begin{eqnarray}
{} &
{\rm rank}_{\varepsilon}\left\|
\displaystyle\frac{\partial_r\phantom{xxxx}}{\partial\Gamma^{p_{s}
}_s(\theta)}
\displaystyle\frac{\partial_lS_{H{}s}(\theta)\phantom{}
}{\partial\Gamma^{q_{s}}_s(\theta)\phantom{x}} \right\|_{\mid\tilde{\Sigma}}
 = {\rm rank}_{\varepsilon}\left\|
\displaystyle\frac{\partial_r\phantom{xxx}}{\partial\varphi^{a_{s}}_s(\theta)}
\displaystyle\frac{\partial_lS_{H{}s}(\theta)\phantom{}
}{\partial\varphi^{b_{s}}_s(\theta)}\right\|_{\mid\tilde{\Sigma}} = {} &
\nonumber \\
{} & n - m + 2m^P \leq
{\rm dim}_{{\bf K},\varepsilon}T^{\ast}_{odd}{\cal M}_{min},\
p_s,q_s=\overline{1,2n_s},\ s=min {} &
\end{eqnarray}
and $\Gamma^{p_{min}}_{0{}min}(\theta)$ $\in$ $\Lambda_{min}$
($\Gamma^{p_{min}}_{0{}min}(\theta)$ is defined in (4.23)), in class of
superfunctions $S_{H{}{min}}(\theta)$ $\in$ $C^k\bigl(T^{\ast}_{odd}{\cal
M}_{min})$ with representation (4.29) exists.
\vspace{1ex}

Having realized the 1st step of the procedure for constructing of the II
class GThGT, connected with transition from $S_{H}(\Gamma_{cl}(\theta),\theta)$
to ${\cal S}_{0}({\cal A}(\theta))$, for  the obtained II class GThST
with HCHF (4.5b) the identities (4.9)  are written equivalently
in terms of vector fields $\hat{\cal R}_{0{}\alpha}(\theta)$ over ${\cal
M}_{cl}$ with basis $\{\frac{\partial\phantom{xxxx}}{\partial{\cal A}^{
\imath}(\theta)}\}$  in the form
\begin{eqnarray}
\hat{\cal R}_{0{}\alpha}(\theta){\cal S}_{0}({\cal A}(\theta)) =0,\ \
\hat{\cal R}_{0{}\alpha}(\theta) =
{\cal R}_{0}{}^{\imath}_{\alpha}({\cal A}(\theta))
\frac{\partial\phantom{xxx}}{\partial{\cal A}^{
\imath}(\theta)}\,.
\end{eqnarray}
The supercommutator of two arbitrary operators $\hat{\cal R}_{0{}\alpha}(\theta)$,
$\hat{\cal R}_{0{}\beta}(\theta)$ naturally defined in the linear space of
vector fields must vanish in calculating on ${\cal S}_{0}({\cal
A}(\theta))$ just as for $\hat{\cal R}_{0{}\alpha}(\theta)$ themselves
\begin{eqnarray}
[\hat{\cal R}_{0{}\alpha}(\theta),\hat{\cal R}_{0{}\beta}(\theta)]_s
{\cal S}_{0}({\cal A}(\theta)) =0\;. 
\end{eqnarray}
The last fact by virtue of completeness for the set of GGTST
${\cal R}_{0}{}^{\imath}_{\alpha}({\cal A}(\theta))$
means that the coordinates of the vector field $\bigl[\hat{\cal R}_{0{}
\alpha}(\theta), \hat{\cal R}_{0{}\beta}(\theta)\bigr]_s$
have the following general form
\begin{eqnarray}
{} & {{\cal R}_{0}{}^{\imath}_{\alpha},}_{\jmath}(\theta)
{\cal R}_{0}{}^{\jmath}_{\beta}(\theta) - (-1)^{\varepsilon_{
\alpha}\varepsilon_{\beta}}(\alpha \leftrightarrow \beta)= -
{\cal R}_{0}{}^{\imath}_{\gamma}(\theta)
{\cal F}_{0}{}^{\gamma}_{\alpha\beta}({\cal A}(\theta)) -
{{\cal S}_{0},}_{\jmath}(\theta)
{\cal M}_{0}{}^{\imath\jmath}_{\alpha\beta}({\cal A}(\theta))\;,
\end{eqnarray}
with superfunctions ${\cal F}_{0}{}^{\gamma}_{\alpha\beta}({\cal A}(\theta))$,
${\cal M}_{0}{}^{\imath\jmath}_{\alpha\beta}({\cal A}(\theta))$ $\in$ $C^k({
\cal M}_{cl})$  possessing by vanishing ghost numbers values and following
$\varepsilon_{P}$, $\varepsilon_{\bar{J}}$, $\varepsilon$ parities and
symmetry properties as well
\renewcommand{\theequation}{\arabic{section}.\arabic{equation}\alph{lyter}}
\begin{eqnarray}
\setcounter{lyter}{1}
{} & \begin{array}{l|cccl}
{}& \varepsilon_P & \varepsilon_{\bar{J}} & \varepsilon &{}\\\hline
{\cal F}_{0}{}^{\gamma}_{\alpha\beta}(\theta) & 0    &
\varepsilon_{\gamma} + \varepsilon_{\alpha}+ \varepsilon_{\beta} &
\varepsilon_{\gamma} + \varepsilon_{\alpha}+ \varepsilon_{\beta} & {}\\
{\cal M}_{0}{}^{\imath\jmath}_{\alpha\beta}(\theta) & 0 &
\varepsilon_{\imath} + \varepsilon_{\jmath} + \varepsilon_{\alpha}+
\varepsilon_{\beta} &
\varepsilon_{\imath} + \varepsilon_{\jmath} + \varepsilon_{\alpha}+
\varepsilon_{\beta} & ,
\end{array} {} &
\\ 
\setcounter{equation}{39}
\setcounter{lyter}{2}
{} & \hspace{-1.5em} {\cal F}_{0}{}^{\gamma}_{\alpha\beta}(\theta) = -
(-1)^{\varepsilon_{\alpha} \varepsilon_{\beta}}{\cal F}_{0}{}^{\gamma}_{
\beta\alpha}(\theta),\
{\cal M}_{0}{}^{\imath\jmath}_{\alpha\beta}(\theta) = -
(-1)^{\varepsilon_{\imath} \varepsilon_{\jmath}}
{\cal M}_{0}{}^{\jmath\imath}_{\alpha\beta}(\theta) = -
(-1)^{\varepsilon_{\alpha} \varepsilon_{\beta}}
{\cal M}_{0}{}^{\imath\jmath}_{\beta\alpha}(\theta).{} & 
\end{eqnarray}
Relationships (4.36), (4.38) partially define the gauge algebra of GTST with
its generators ${\cal R}_{0}{}^{\imath}_{\alpha}({\cal A}(\theta))$.
The comparison of (4.38) with the form of ${\cal F}_{2}(\theta)$ in (4.20a)
permits to transform ${\cal F}_{2}(\theta)$ and then to obtain the explicit
form for $S_{2}(\Gamma_{min}(\theta))$
\begin{eqnarray}
\setcounter{lyter}{1}
{} & \hspace{-2em}
 {\cal F}_{2}(\Gamma_{cl}(\theta),C(\theta)) = \frac{1}{2}
{\cal A}^{\ast}_{\imath}(\theta)\left[{\cal R}_{0}{}^{\imath}_{\gamma}(\theta)
{\cal F}_{0}{}^{\gamma}_{\alpha\beta}(\theta) +
{{\cal S}_{0},}_{\jmath}(\theta)
{\cal M}_{0}{}^{\imath\jmath}_{\alpha\beta}(\theta)\right]
C^{\beta}(\theta)C^{\alpha}(\theta)(-1)^{\varepsilon_{\alpha}},
{} & \\ 
\setcounter{equation}{40}
\setcounter{lyter}{2}
{} & \hspace{-2em}
S_{2}(\Gamma_{min}(\theta)) = - \left[\frac{1}{2}C^{\ast}_{
\gamma}(\theta){\cal F}_{0}{}^{\gamma}_{\alpha\beta}(\theta) -
\frac{1}{4}{\cal A}^{\ast}_{\imath}(\theta){\cal A}^{\ast}_{
\jmath}(\theta){\cal M}_{0}{}^{\imath\jmath}_{\alpha\beta}(\theta)(-1)^{
\varepsilon_{\imath}}\right]
C^{\beta}(\theta)C^{\alpha}(\theta)(-1)^{\varepsilon_{\alpha}}. & {} 
\end{eqnarray}
Thus $Q({\cal S}_{0})$ $\equiv$ ${\rm Ker}\{{\cal S}_{0},_{
\imath}(\theta)\}$ is not only by affine $C^k({\cal M}_{cl})$-module but
appears
by a special form superalgebra with respect to supercommutator $[\ ,\ ]_{s}$.
\section{Classical Master Equation in $T^{\ast}_{odd}{\cal M}_{ext}$}
\setcounter{equation}{0}

In principle, superfunction $S_{H{}min}(\theta)$ is already suitable to
construct the generating functional of Green's functions in view of the
properness (4.35) for solution of Eq.(4.30) in the terminology of works
[4,6,7].
As it have been already  mentioned it is usually impossible to do the such
step in the Lagrangian quantization procedure
without violation of locality and irreducibility with respect to indices of
$T_{\mid\bar{J}}$ representation.

Let us extend $S_{H{}s}(\theta) \in C^k(T_{odd}^{\ast}{\cal M}_{s})$ from
value of $s=min$ to $s=ext$ in a such way that the latter superfunction
leads to the II class GThGT in $T_{odd}^{\ast}{\cal M}_{ext}$.
By boundary condition for that extension it appears the following restriction
for $S_{H{}ext}(\theta)$ $\equiv$ $S_{H{}ext}(\Gamma_{ext}(\theta))$
\renewcommand{\theequation}{\arabic{section}.\arabic{equation}}
\begin{eqnarray}
S_{H{}ext}(\Gamma_{ext}(\theta)){\hspace{-0.5em}\phantom{\Bigl(}}_{
\mid\Gamma_{aux}(\theta)=0}=
S_{H{}min}(\Gamma_{min}(\theta))\;. 
\end{eqnarray}
All other properties of $S_{H{}min}(\theta)$ either are literally
inherited by $S_{H{}ext}(\theta)$ or are specially modified.

\noindent
\underline{\bf Theorem 2} (Existence of a solution for the classical master
equation in $C^k(T_{odd}^{\ast}{\cal M}_{ext})$)

\noindent
A solution of the equation
\begin{eqnarray}
(S_{H{}s}(\Gamma_{s}(\theta)),S_{H{}s}(\Gamma_{s}(\theta)))^s_{\theta}=0,\ \;
\forall\Gamma_s(\theta)\in T_{odd}^{\ast}{\cal M}_{s},\ s=ext 
\end{eqnarray}
under validity of the system
\begin{eqnarray}
(\tilde{P}_1(\theta), {\rm gh}, \varepsilon_P, \varepsilon_{\bar{J}},
\varepsilon)S_{H{}s}(\Gamma_{s}(\theta)) = (0,0,0,0,0),\; s=ext
\end{eqnarray}
and in fulfilling of the additional conditions
\begin{eqnarray}
{} \hspace{-10.3em}{\rm a)}\hspace{10.3em}S_{H{}ext}(\Gamma_{ext}(
\theta)){\hspace{-0.5em}\phantom{\Bigl(}}_{
\mid\Gamma_{aux}(\theta)=C(\theta)=0}={\cal S}_0({\cal A}(\theta))\;, 
\end{eqnarray}
b) an  existence of the Lagrangian surface $\Lambda_s\subset T^{\ast}_{odd}{
\cal M}_s$
parametrized by coordinates $\varphi^{a_s}_s(\theta)$, $a_s=1,\ldots,n_s$
($n_{ext} = n + 3m$) so that on the supersurface $\tilde{\Sigma}_{ext}$ $
\equiv$ $\tilde{\Sigma}\cup\{\Gamma^{p_{aux}}_{0{}aux}(\theta)\}$ $\subset$
$T^{\ast}_{odd}{\cal M}_s$ the relations hold
\begin{eqnarray}
{} &
{\rm rank}_{\varepsilon}\left\|
\displaystyle\frac{\partial_r\phantom{xxxx}}{\partial\Gamma^{p_{s}
}_s(\theta)}
\displaystyle\frac{\partial_lS_{H{}s}(\theta)\phantom{}
}{\partial\Gamma^{q_{s}}_s(\theta)\phantom{x}} \right\|_{
\mid\tilde{\Sigma}_{ext}}
 = {\rm rank}_{\varepsilon}\left\|
\displaystyle\frac{\partial_r\phantom{xxx}}{\partial\varphi^{a_{s}}_s(\theta)}
\displaystyle\frac{\partial_lS_{H{}s}(\theta)
}{\partial\varphi^{b_{s}}_s(\theta)\phantom{x}}\right\|_{
\mid\tilde{\Sigma}_{ext}} = {} & \nonumber \\
{} & n + 2m^P + m \leq
{\rm dim}_{{\bf K},\varepsilon}T^{\ast}_{odd}{\cal M}_{ext}=n+n^P+3(m+m^P),\
p_s,q_s=\overline{1,2n_s},\ s=ext\;, {} &
\\ 
{} & \Gamma^{p_{ext}}_{0{}ext}(\theta)=(\Gamma^{p_{min}}_{0{}min}(\theta),
\Gamma^{p_{aux}}_{0{}aux}(\theta)) = (\Gamma^{p_{min}}_{0{}min}(\theta),0)
\in \Lambda_{ext}\;, & {} 
\end{eqnarray}
($\Gamma^{p_{min}}_{0{}min}(\theta)$ is given in (4.23)) in class of
superfunctions
$S_{H{}{s}}(\theta)$ $\in$ $C^k\bigl(T^{\ast}_{odd}{\cal M}_s)$, $s=ext$
with representation
\begin{eqnarray}
{} & S_{H{}ext}(\theta) = S_{H{}min}(\theta) + \sum\limits_{k\geq 1}S_{k
}(\Gamma_{ext}(\theta)) \equiv
S_{H{}min}(\theta) + S_{H{}aux}(\Gamma_{ext}(\theta))\;, & {} \nonumber \\
{} & ({\rm min}{\rm deg}_{\Gamma_{aux}},{\rm deg}_{\Gamma_{aux}})
S_k(\Gamma_{ext}(\theta)) = (k,k) & {}
\end{eqnarray}
exists.

\noindent
\underline{\bf Remarks:}

\noindent
{\bf 1)} Solutions  $S_{H{}{s}}(\theta)$, $s=min, ext$
in (4.30),  (5.2) are not uniquely determined in the framework of the
conditions for Theorems 1,2;

\noindent
{\bf 2)} Lagrangian  surfaces $\Lambda_s$, $s=min,ext$ are not uniquely
defined by means of conditions of the above Theorems as well. Fixing of the
quadratic parts for superfunctions $S_{H{}{s}}(\theta)$ =
${}^0S_{H{}{s}}(\theta)$ + $S_{H{}{s}{}{\rm nl}}(\theta)$,
${\rm deg}_{\Gamma_s}{}^0S_{H{}s}(\theta) = 2$ determines $\Lambda_s$,
$s=min$, $ext$ to be essentially  unique in that sense for $\Lambda_s$,
$\Lambda_{0{}s}$
corresponding to $S_{H{}{s}}(\theta)$, ${}^0S_{H{}{s}}(\theta)$ that
under conditions of the Theorems 1,2 the coordinates
$\varphi^{a_s}_s(\theta)$, $\varphi^{a_s}_{0{}s}(\theta)$
parametrizing $\Lambda_s$, $\Lambda_{0{}s}$  respectively
are  connected by the relationships
\begin{eqnarray}
\varphi^{a_s}_s(\theta) = \varphi^{a_s}_{0{}s}(\theta)+
{\cal F}^{a_s}_s(\varphi_{0{}s}(\theta),\varphi^{\ast}_{0{}s}(\theta)),\;
{\rm min}{}{\rm deg}_{\varphi_{0{}s}\varphi^{\ast}_{0{}s}}
{\cal F}^{a_s}_s(\varphi_{0{}s}(\theta),\varphi^{\ast}_{0{}s}(\theta))\geq
2\,.  
\end{eqnarray}
Coordinates
$\bigl(\varphi^{a_s}_s, \varphi^{\ast}_{a_s,s}\bigr)(\theta)$  on
$T^{\ast}_{odd}{\cal M}_s$, $s=min$, $ext$
appear by anticanonically conjugate  pairs. Grassmann parities, ghost numbers
for $\varphi^{a_s}_s(\theta)$, $\varphi^{a_s}_{0{}s}(\theta)$,
${\cal F}^{a_s}_s(\theta)$
can be chosen by coinciding for fixed $a_s$. Thus, fixing of the explicit
form for ${\cal S}_0({\cal A}(\theta))$,
${\cal R}_{0}{}^{\imath}_{\alpha}({\cal A}(\theta))$ determines
$\Lambda_{min}$ in the essentially unique way;

\noindent
{\bf 3)} in fixing of $S_{H{}min}(\theta)$ in series (5.7) by the conditions
of Theorem 1 and under inclusion  $\Lambda_{min}$  $\subset$ $\Lambda_{ext}$
it is possible to represent the Lagrangian surface $\Lambda_{ext}$  in the
form
\begin{eqnarray}
\Lambda_{ext} = \Lambda_{min} \cup \Lambda_{aux},\ \Lambda_{min} \cap
\Lambda_{aux}=0,\
{\rm dim}_{{\bf K},\varepsilon}\Lambda_{aux}=2m\;, 
\end{eqnarray}
where $\Lambda_{aux}$ is defined by means of quadratic part of $S_{H{}aux}(
\Gamma_{ext}(\theta))$ with respect to $\Gamma_{aux}(\theta)$;

\noindent
{\bf 4)} to solve the problem on covariant construction of the generating
functional of Green's functions starting from the II class GThGT  being
defined by $S_{H{}min}(\theta)$
it is sufficient, in general, to deform the superfunction
$S_{H{}min}(\theta)$
to $S_{H{}(min,\overline{C})}(\Gamma_{(min,\overline{C})}(\theta))$ given on
$T^{\ast}_{odd}{\cal M}_{(min,\overline{C})}$. In
this connection  it is necessary to write in Theorem 2 the values
$(min,\overline{C})$,
$n_{(min,\overline{C})}$ = $n+2m$ instead of values $ext$, $n_{ext}$
and the value of rank in (5.5) must be given by the following supernumber
\begin{eqnarray}
n+2m^P\leq {\rm dim}_{{\bf K},\varepsilon}T^{\ast}_{odd}{\cal M}_{(min,
\overline{C})} = n+n^P +2(m+m^P)\;.
\end{eqnarray}
In fact the superfunction
\begin{eqnarray}
{} & S_{H{}(min,\overline{C})}(\Gamma_{(min,\overline{C})}(\theta)) =
S_{H{}min}(\Gamma_{min}(\theta)) + \displaystyle\frac{1}{2}
\sigma^{\alpha\beta}(\theta)\overline{C}{}^{\ast}_{\alpha}(\theta)
\overline{C}{}^{\ast}_{\beta}(\theta)\;, {} & \nonumber \\
{} & \sigma^{\alpha\beta}(\theta) = P_0(\theta)\sigma^{\alpha\beta}(\theta) =
(-1)^{\varepsilon_{\alpha} \varepsilon_{\beta}}\sigma^{\beta\alpha}(\theta),\
{\rm rank}\left\|\sigma^{\beta\alpha}(\theta)\right\|=m {} &
\end{eqnarray}
satisfies to Theorem 2 in this case.
\vspace{1ex}

In the framework of the remark 3) the equation (5.2) with allowance made for
validity of (4.30) leads to fulfilment of the following one
\begin{eqnarray}
2(S_{H{}min}(\theta),S_{H{}aux}(\theta))^{min}_{\theta} +
(S_{H{}aux}(\theta),S_{H{}aux}(\theta))^{ext}_{\theta} =0\;.
\end{eqnarray}
By its particular solution it appears, for instance, the superfunction
\begin{eqnarray}
\tilde{S}_{H{}aux}(\Gamma_{aux}(\theta))=
\overline{C}{}^{\ast}_{\alpha}(\theta)B^{\alpha}(\theta),\
({\rm min}{\rm deg}_{\Gamma_{aux}},{\rm deg}_{\Gamma_{aux}}, {\rm deg}_{
\Gamma_{min}})\tilde{S}_{H{}aux}(\theta)=(2,2,0)\;.
\end{eqnarray}
Thus, the superfunction  $\tilde{S}_{H{}ext}(\theta)$
\begin{eqnarray}
\tilde{S}_{H{}ext}(\Gamma_{ext}(\theta)) = S_{H{}min}(\theta) +
\overline{C}{}^{\ast}_{\alpha}(\theta)B^{\alpha}(\theta)\;,
\end{eqnarray}
satisfies both to the requirements of Theorem 2 with $\varphi^{a_{aux}}_{
aux}(\theta)$ = $(\overline{C}{}^{\ast}_{\alpha}(\theta)$,
$B^{\alpha}(\theta))$ and to conditions (5.9).
Part of $\tilde{S}_{H{}ext}(\Gamma_{ext}(\theta))$ being invariant with
respect to involution $*$ and therefore
having the form $P_0(\theta)\tilde{S}_{H{}ext}(\theta)$
appears by none other than the Batalin-Vilkovisky action
 given for irreducible gauge theories for $\hbar=0$ [4,7]
\begin{eqnarray}
S_{BV}(P_0\Gamma_{ext}(\theta)) = P_0(\theta)\tilde{S}_{H{}ext
}(\Gamma_{ext}(\theta)) = (P_0(\theta)\tilde{S}_{H{}ext}(\theta))^{\ast}
.
\end{eqnarray}
It appears by important  the remark about way of representation for
Eqs.(4.30), (5.2) if to make use the transformation of operators
${\stackrel{\circ}{U}}{}^s_{+}(\theta)$, ${\stackrel{\circ}{V}}{}^s_{+}(
\theta)$ by means of relationships (2.50), (2.51). Namely having chosen
\renewcommand{\theequation}{\arabic{section}.\arabic{equation}\alph{lyter}}
\begin{eqnarray}
\setcounter{lyter}{1}
{} & \hspace{-1.7em}S_{H{}s}(\Gamma_{s}(\theta)) \hspace{-0.1em}=
\hspace{-0.1em}
S_{i{}s}\bigl(\Gamma_{s}(\theta),{\stackrel{\circ}{\Gamma}}_s(\theta)\bigr)
\hspace{-0.1em}- \hspace{-0.1em}{\cal F}_i\bigl(\Gamma_{s}(\theta),
{\stackrel{\circ}{\Gamma}}_s(\theta)
\bigr),\ i\hspace{-0.1em}=\hspace{-0.1em}1,2,\; s\hspace{-0.1em}\in
\hspace{-0.1em} \{cl,min,(min,\overline{C}),ext\}, {} & \\
\setcounter{equation}{16}
\setcounter{lyter}{2}
{} & {\cal F}_1\bigl(\Gamma_{s}(\theta),{\stackrel{\circ}{\Gamma}}_s(\theta)
\bigr)= {\stackrel{\circ}{\Phi}}{}^{A_s}_{s}(\theta)\Phi^{\ast}_{A_s,s
}(\theta),\
{\cal F}_2\bigl(\Gamma_{s}(\theta),{\stackrel{\circ}{\Gamma}}_s(\theta)
\bigr) = -
{\stackrel{\circ}{\Phi}}{}^{\ast}_{A_s,s}(\theta)\Phi^{A_s}_{s
}(\theta)\;, & {} 
\end{eqnarray}
we obtain in substituting of (5.16) into (5.2), for fixed $s$ in (5.16a),
the equations on $S_{i{}s}\bigl(\Gamma_s(\theta)$,
${\stackrel{\circ}{\Gamma}}_s(\theta)\bigr)$ $\equiv$ $S_{i{}s}(\theta)$ in
more difficult for investigation forms
\renewcommand{\theequation}{\arabic{section}.\arabic{equation}}
\begin{eqnarray}
\textstyle\frac{1}{2}(S_{1{}s}(\theta),S_{1{}s}(\theta))^s_{\theta} +
{\stackrel{\circ}{U}}{}^s_{+}(\theta)S_{1{}s}(\theta)=0,\ \
\textstyle\frac{1}{2}(S_{2{}s}(\theta),S_{2{}s}(\theta))^s_{\theta} +
{\stackrel{\circ}{V}}{}^s_{+}(\theta)S_{2{}s}(\theta)=0\,. 
\end{eqnarray}
At last, note that $S_{H{}ext}(\theta)$ in the framework of the Theorem 2
conditions  determines HS (4.32)  and equation of the form
(4.33) for $s=ext$. The last equation is valid in this
case being by consequence for Eq.(5.2) (for $s=ext$). The latter
conclusion appears by important distinction of GSQM from BV method [4,6]
where Eq.(4.33) in general has not taken place.
\section{Quantum Master Equation in $T^{\ast}_{odd}{\cal M}_{ext}$}
\setcounter{equation}{0}

By following stage in procedure of the  GSQM construction one appears the
deformation of the II class GThGT  defined by $S_{H{}ext}(\theta)$ in powers
of $\hbar$ to the II class GThGT  with superfunction
\renewcommand{\theequation}{\arabic{section}.\arabic{equation}\alph{lyter}}
\begin{eqnarray}
\setcounter{lyter}{1}
{} & S_{H{}s}(\Gamma_s(\theta),\hbar) \equiv S_{H{}s}(\theta,\hbar) =
\sum\limits_{n\geq 0}\hbar^n S_{H{}s}^{(n)}(\Gamma_s(\theta)),\
S_{H{}s}^{(0)}(\Gamma_s(\theta)) \equiv S_{H{}s}(\Gamma_s(\theta))\;, & {}
\\
\setcounter{equation}{1}
\setcounter{lyter}{2}
{} & ({\rm min}{\rm deg}_{\hbar},{\rm deg}_{\hbar})S_{H{}s}^{(n)}(\Gamma_s(
\theta))=(0,0),\  S_{H{}s}^{(n)}(\theta)\in C^k(T^{\ast}_{odd}{\cal M}_{s
}),\;s=ext\;. & {} 
\end{eqnarray}
\subsection{Quantum Master Equation for GSQM}

\underline{\bf Theorem 3} (Existence of a solution for the quantum master
equation in $C^k(T^{\ast}_{odd}{\cal M}_{ext})$)

\noindent
A solution of the equation
\renewcommand{\theequation}{\arabic{section}.\arabic{equation}}
\begin{eqnarray}
(S_{H{}s}(\Gamma_s(\theta),\hbar),S_{H{}s}(\Gamma_s(\theta),\hbar))^s_{
\theta}=0,\ s=ext 
\end{eqnarray}
under validity of the relationships
\begin{eqnarray}
(\tilde{P}_1(\theta), \hat{N}_s(\theta), \varepsilon_P, \varepsilon_{\bar{J}},
\varepsilon)S_{H{}s}(\theta,\hbar) = (0,0,0,0,0),\ s=ext
\end{eqnarray}
and fulfilment of the additional conditions [a) boundary condition and b) of
necessity for solution to be "proper"]
\begin{eqnarray}
{}\hspace{-11em} {\rm a)}\hspace{11em}S_{H{}s}(\theta,
\hbar){\hspace{-0.5em}\phantom{\bigl(}}_{\mid\Gamma_{
aux}(\theta) = C(\theta) = \hbar=0}={\cal S}_0({\cal A}(\theta))\;,
\end{eqnarray}
b) an existence of the Lagrangian surface $\Lambda_s(\hbar)$ parametrized by
coordinates $\varphi^{a_s}_s(\theta,\hbar)$ = $\sum\limits_{n\geq 0}
\hbar^n \varphi^{(n)}{}^{a_s}_s(\theta)$, $\varphi^{(n)}{}^{a_s}_s(\theta)$
$\in$ $C^k(T^{\ast}_{odd}{\cal M}_{s})$, $a_s=\overline{1,n_s}$ so that
on the supersurface $\tilde{\Sigma}_{ext}$ the relations are fulfilled for
$s=ext$
\begin{eqnarray}
{} &
{\rm rank}_{\varepsilon}\left\|
\displaystyle\frac{\partial_r\phantom{xxxx}}{\partial\Gamma^{p_{s}
}_s(\theta)}
\displaystyle\frac{\partial_lS_{H{}s}(\theta,\hbar)\phantom{}
}{\partial\Gamma^{q_{s}}_s(\theta)\phantom{xxx}}
\right\|_{\mid\tilde{\Sigma}_{s}}  = {\rm rank}_{\varepsilon}\left\|
\displaystyle\frac{\partial_r\phantom{xxxxx}}{\partial\varphi^{a_{s}}_s(
\theta,\hbar)}
\displaystyle\frac{\partial_lS_{H{}s}(\theta,\hbar)\phantom{}
}{\partial\varphi^{b_{s}}_s(\theta,\hbar)\phantom{xx}}
\right\|_{\mid\tilde{\Sigma}_{s}} = n + 2m^P + m\,,{} &  \\ 
{} & \Gamma^{p_{ext}}_{0{}ext}(\theta)=(\Gamma^{p_{min}}_{0{}min}(\theta),
\Gamma^{p_{aux}}_{0{}aux}(\theta)) \in \Lambda_{ext}(\hbar)\;, {} & 
\end{eqnarray}
in class of superfunctions $S_{H{}ext}(\theta,\hbar)$ $\in$
$C^k(T^{\ast}_{odd}{\cal M}_{ext})$ exists.
\vspace{1ex}

Remarks 1)--4) written after Theorem 2 with corresponding
modifications are also valid for the II class GThGT  with
$S_{H{}ext}(\theta,\hbar)$. In addition one
can require, in a natural way, the fulfilment of the Theorem 3 for $s=min$,
$(min,\overline{C})$ as well. Moreover the following representation holds
\begin{eqnarray}
\Lambda_{s}(\hbar){\hspace{-0.5em}\phantom{\bigl)}}_{\mid\hbar=0} =
\Lambda_s,\
\varphi^{a_s}_s(\theta,\hbar){\hspace{-0.5em}\phantom{\bigl)}}_{\mid\hbar=0}
=\varphi^{a_s}_s(\theta)\;. 
\end{eqnarray}
Coordinates
$\bigl(\varphi^{a_s}_s(\theta,\hbar)$, $\varphi^{\ast}_{a_s,s}(\theta,
\hbar)\bigr)$  are
connected with initial $\Gamma^{p_s}_s(\theta)$ by means of an anticanonical
transformation. Finally, the corresponding
solvable HS has the form
\begin{eqnarray}
\frac{d_r\Gamma^{p_s}_{s}(\theta)}{d\theta\phantom{xxxx}}=
\left(\Gamma^{p_s}_{s}(\theta), S_{H{}s}(\Gamma_s(\theta),\hbar)\right)^{s
}_{\theta},\ s\in C_1 = \{min,(min,\overline{C}),ext\}\;.
\end{eqnarray}
\underline{\bf Remark:} Superfunction $S_{H{}s}(\theta,\hbar)$
both for $\hbar=0$ and  $\hbar\neq 0$ does not correspond to a some
equivalent superfunction $S_{L{}s}(\theta,\hbar)$ $\in$ $C^k(T_{odd}{\cal
M}_s)$
in the Lagrangian formulation for GSTF because of impossibility to realize
the Legendre transform for
$S_{H{}s}(\theta,\hbar)$ with respect to $\Phi^{\ast}_{A_s,s}(\theta)$.
In the framework of the Theorems 2,3 conditions  it leads, in general, to
invalidity of the relationships (4.1b), (4.6) for HS constructed
with respect to $S_{H{}s}(\theta,\hbar)$.
However as it have been already mentioned the HCHF (4.5b) are inherited by
the continued superfield model.
\vspace{1ex}

According to the Statement 5.5 from Ref.[2] the fulfilment of the following
equation appears by the consequence of Eq.(6.2)
\begin{eqnarray}
\Delta^s(\theta)S_{H{}s}(\Gamma_s(\theta),\hbar)=0,\
\forall\Gamma_s(\theta)\in T^{\ast}_{odd}{\cal M}_{s},\ s=ext\;.
\end{eqnarray}
Theorem 3 for $\hbar=0$ has as one's corollaries the Theorems 1,2 for
corresponding values of $s$. Eq.(6.2) being iteratively solved in powers of
$\hbar$ is equivalent to the system of equations
\begin{eqnarray}
({\rm min}{\rm deg}_{\hbar},{\rm deg}_{\hbar}) & = (0,0): &
(S_{H{}s}^{(0)}(\theta),S_{H{}s}^{(0)}(\theta))^s_{\theta}=0\;, \nonumber \\
\phantom{({\rm min}{\rm deg}_{\hbar},{\rm deg}_{\hbar})} & = (1,1): &
(S_{H{}s}^{(0)}(\theta),S_{H{}s}^{(1)}(\theta))^s_{\theta}=0\;, \nonumber \\
{} \hspace{-4em}\ldots\hspace{4em}  & \ldots & \hspace{5em}\ldots \nonumber \\
\phantom{({\rm min}{\rm deg}_{\hbar},{\rm deg}_{\hbar})} & = (n,n): &
\sum\limits_{k=0}^{n} (S_{H{}s}^{(k)}(\theta),S_{H{}s}^{(n-k)}(\theta))^s_{
\theta}=0\;. 
\end{eqnarray}
By necessary and sufficient condition for the existence of a solution for
system
(6.10) appears the resolvability (in the usual sense) of the 1st
equation, i.e. the validity of Theorem 2.
\subsection{BV Type Quantum Master Equation}

A set of solutions for Eq.(6.2) for $s\in C_1$
taking of its consequence (6.9) into account is embedded (but is not
equivalent)
into one of solutions for ordinary (standard [4,6,7]) quantum master equation
\begin{eqnarray}
\textstyle\frac{1}{2}(S_{H{}s}(\theta,\hbar),S_{H{}s}(\theta,\hbar))^s_{
\theta}= \imath\hbar\Delta^s(\theta)S_{H{}s}(\theta,\hbar)
\end{eqnarray}
being equivalently represented in the form of linear differential equation
\begin{eqnarray}
\Delta^s(\theta)
{\rm exp}\left\{\frac{\imath}{\hbar}S_{H{}s}(\theta,\hbar)\right\} = 0\,.
\end{eqnarray}
Eq.(6.12), in general, may contain the solutions with summands to be not
less than
linear in powers of $\hbar$ and not satisfying to Eq.(6.2) (in terms of
the equivalent system (6.10) not satisfying to one's own
subsystem with exception of the 1st equation in (6.10)).

One can formulate the analog of Theorem 3 for Eq.(6.12). To this end it is
sufficient to replace formally the Eq.(6.2) onto Eq.(6.12). The similar
analogs one can write for $s\in C_1$.

In the framework of local field theories the Eq.(6.12) contains with
respect to indices of $\bar{J}$ supergroup the usual $\delta$-function and its
derivatives calculated in 0 ($\delta(0)$, $\partial_{\mu}\delta(0)$) so that
a regularization of the such expressions, for instance, in the form
$\delta(0)$ = $\partial_{\mu}\delta(0)$ = $0$ (as in the dimensional one)
leads
to the fact that solution for Eq.(6.12) appears by solution for (6.2).

The HS defined with respect to $S_{H{}s}(\theta,\hbar)$ being by solution
of Eq.(6.12), but not of Eqs.(6.2), (6.9), does
not satisfy to the solvability conditions [1,2]. In particular, in
correspondence with Statement 5.6 [2] it means both the absence, in general
case, of nilpotency for corresponding translation generator along formal
integral curves for that HS $s_s(\theta,\hbar)$ [2] and the fact that
superfunction $S_{H{}s}(\theta,\hbar)$  does not appear by integral for
mentioned HS and does not vanish under operator $s_s(\theta,\hbar)$ action
defined by following relationship in acting on an arbitrary
${\cal F}(\theta)$ $\in$ $C^k(T^{\ast}_{odd}{\cal M}_s)$
\begin{eqnarray}
s_s(\theta,\hbar){\cal F}(\theta) = {\rm ad}_{S_{H{}s}(\theta,\hbar)}
{\cal F}(\theta)(-1)^{\varepsilon({\cal F})} =
(S_{H{}s}(\theta,\hbar),{\cal F}(\theta))^s_{\theta}
(-1)^{\varepsilon({\cal F})},\ s\in C_1\;. 
\end{eqnarray}
Eqs.(4.30), (5.2), (6.2), their consequences of the form (6.9) and
Eq.(6.11) as well do not depend upon different choices of local coordinates
$\Gamma^{p_s}_s(\theta)$ connected with each other via anticanonical
transformations in view of antibracket $(\ ,\ )^s_{\theta}$ and operator
$\Delta^s(\theta)$ invariance with respect to these transformations.
That fact permits to find the such
coordinates ${\Gamma'}^{p_s}_s(\theta)$ in which the
superfunctions $S_{H{}s}(\theta)$, $S_{H{}s}(\theta,\hbar)$  remaining
invariable on their values will  have the simplest form of dependence
upon ${\Gamma'}^{p_s}_s(\theta)$.

On the other hand  from nonuniqueness of solutions for Eqs.(6.2),
(6.12) it follows the possibility of $S_{H{}s}(\theta,\hbar)$ transformation
not directly connected with anticanonical ones
under which these superfunctions are converted into others, depending on the
same local coordinates ${\Gamma}^{p_s}_s(\theta)$ just as the initial
$S_{H{}s}(\theta,\hbar)$ and satisfying to the same equations.
\section{Nontrivial Arbitrariness in Solutions of Master Equations}
\setcounter{equation}{0}

By global nontrivial arbitrariness (GNA) in solutions of Eq.(6.12) let us
call the following transformation of corresponding superfunctions
$S_{H{}s}(\theta,\hbar)$
\begin{eqnarray}
{\rm exp}\left\{\frac{\imath}{\hbar}S_{H{}s}(\Psi_s;\theta,\hbar)\right\} =
{\rm exp}\left\{-\hat{T}_1^s(\theta)(\Psi_s(\Gamma_s(\theta),\hbar))\right\}
{\rm exp}\left\{\frac{\imath}{\hbar}S_{H{}s}(\theta,\hbar)\right\}
\end{eqnarray}
with the 2nd order  with respect to
$\frac{\partial\phantom{xxxx}}{\partial\Gamma^{p_s}_s(\theta)}$ operator
$\hat{T}_1^s(\theta)$ defined by relation
\begin{eqnarray}
\hat{T}_1^s(\theta)= [\Delta^s(\theta),\ \;]_s,\
({\rm gh}, \varepsilon_P, \varepsilon_{\bar{J}},\varepsilon)
\hat{T}_1^s(\theta)= (1,1,0,1)\;. 
\end{eqnarray}
From the requirement of satisfaction by $S_{H{}s}(\Psi_s;\theta,\hbar)$ to the
conditions of Theorem 3 analog formulated for Eq.(6.12) it follows the
necessary properties for $\Psi_s(\Gamma_s(\theta),\hbar)$ $\in$
$C^k(T^{\ast}_{odd}{\cal M}_s)$ with its representability in the form
of the type (6.1) series in powers of $\hbar$
\begin{eqnarray}
(\tilde{P}_1(\theta), \hat{N}_s(\theta), \varepsilon_P, \varepsilon_{\bar{J}},
\varepsilon)\Psi_{s}(\theta,\hbar) = (0,-1,1,0,1),\ s\in C_1\,. 
\end{eqnarray}
The 1st order operator $\hat{T}_1^s(\theta)\bigl(\Psi_s(\theta,\hbar)\bigr)$
is given by the explicit expression
\begin{eqnarray}
\hat{T}_1^s(\theta)(\Psi_s(\Gamma_s(\theta),\hbar)) =
\left[\Delta^s(\theta),\Psi_s(\Gamma_s(\theta),\hbar)\right]_{+}=
\Delta^s(\theta)(\Psi_s(\Gamma_s(\theta),\hbar)) - {\rm ad}_{
\Psi_s(\Gamma_s(\theta),\hbar)}\;.
\end{eqnarray}
The fact, that $S_{H{}s}(\Psi_s;\theta,\hbar)$
satisfies to Eq.(6.12), if $S_{H{}s}(\theta,\hbar)$ had appeared by the
solution
for the same equation  arises from commutativity of the operator (7.4) and
$\Delta^s(\theta)$
\begin{eqnarray}
\left[\hat{T}_1^s(\theta)(\Psi_s(\Gamma_s(\theta),\hbar)),\Delta^s(\theta)
\right]_{-}=0\;.
\end{eqnarray}
Consider for Eq.(6.2) the GNA in the form (7.1) defined with respect
to superfunction $\hat{\Psi}_s(\Gamma_s(\theta),\hbar)$ $\equiv$ $
\hat{\Psi}_s(\theta,\hbar)$ additionally satisfying to equation
\begin{eqnarray}
 \Delta^s(\theta)\hat{\Psi}_s(\theta,\hbar) = 0\;.
\end{eqnarray}
Then from (7.4) it follows the representation for
$\hat{T}_1^s(\theta)\bigl(\hat{\Psi}_s(\theta,\hbar)\bigr)$
\begin{eqnarray}
\hat{T}_1^s(\theta)\left(\hat{\Psi}_s(\theta,\hbar)\right) =
- {\rm ad}_{\hat{\Psi}_s(\theta,\hbar)}\;.
\end{eqnarray}
One can show the formula (7.1) is transformed to the expression
\begin{eqnarray}
S_{H{}s}(\hat{\Psi}_s;\theta,\hbar)= {\rm exp}\left\{
{\rm ad}_{\hat{\Psi}_s(\theta,\hbar)}\right\}S_{H{}s}(\theta,\hbar)\;.
\end{eqnarray}
From the last relation with allowance made for identity (7.5) applied to the
operator (7.7) it follows that together with (6.12) the equation holds
\begin{eqnarray}
\Delta^s(\theta)S_{H{}s}(\hat{\Psi}_s;\theta,\hbar)= 0\;,
\end{eqnarray}
which means in its turn  the fulfilment of Eq.(6.2) for
$S_{H{}s}(\hat{\Psi}_s;\theta,\hbar)$.

Thus, a subset of GNA transformations for Eq.(6.12) with additional condition
(7.6) on $\hat{\Psi}_s(\theta,\hbar)$ transfers
a solution for Eq.(6.2) into solution again with conservation of the
consequence (7.9) validity  for new solution.
\section{Quantum Master Equations in $T_{odd}(T^{\ast}_{odd}{\cal M}_s)$}
\setcounter{equation}{0}
\subsection{Quantum Master Equation for GSQM in $T_{odd}(T^{\ast}_{odd}{
\cal M}_s)$}

Consider according to paper [3] the deformation of the II class GThGT
with superfunction $S_{H{}s}(\theta,\hbar)$ satisfying to
Eq.(6.2) in powers of superfields
${\stackrel{\circ}{\Gamma}}{}^{p_s}_s(\theta)$ in order to the validity of
Theorem 3 would remain invariable for deformed superfunction
$S_{H{}s}\bigl(\Gamma_s(\theta),{\stackrel{\circ}{\Gamma}}_s(\theta),
\hbar\bigr)$ $\in$ $D^k_s$, $s\in C_1$.

\noindent
\underline{\bf Theorem 4:} (Existence of a solution for the
quantum master equation in  $D^k_s$)

\noindent
A solution of the equation
\begin{eqnarray}
\left(S_{H{}s}\bigl(\Gamma_s(\theta),{\stackrel{\circ}{\Gamma}}_s(\theta),
\hbar\bigr),
S_{H{}s}\bigl(\Gamma_s(\theta),{\stackrel{\circ}{\Gamma}}_s(\theta),
\hbar\bigr)\right)^s_{\theta}=0,\ s\in C_1\;,
\end{eqnarray}
under validity of the Eqs.(6.3), a) boundary condition
\begin{eqnarray}
S_{H{}s}\bigl(\Gamma_s(\theta),{\stackrel{\circ}{\Gamma}}_s(\theta),
\hbar\bigr)_{\mid\Phi^{\ast}_s(\theta)= {\stackrel{\circ}{\Phi}}{}^{\ast
}_s(\theta)=\Phi_{aux}(\theta)={\stackrel{\circ}{\Phi}}_{aux}(\theta)=
\hbar=0}={\cal S}_0({\cal A}(\theta))\;,
\end{eqnarray}
b) an existence of the Lagrangian surface $\Lambda_s(\hbar)\subset
T_{odd}(T^{\ast}_{odd}{\cal M}_s)$ parametrized by coordinates
$\tilde{\varphi}^{a_s}_s\bigl(\Gamma_s(\theta)$, ${\stackrel{\circ}{
\Gamma}}_s(\theta), \hbar\bigr)$ $\in D^k_s$, $a_s = \overline{1,n_s}$, so
that the condition (6.5) holds on the possibly deformed by superfields
${\stackrel{\circ}{\Gamma}}_s(\theta)$ supersurface $\tilde{\Sigma}_s$
$\subset$ $T_{odd}(T^{\ast}_{odd}{\cal M}_s)$ together with $\Gamma^{p_{ext}
}_{0{}ext}(\theta)$ having the form as in (6.6),
in class of superfunctions
$S_{H{}s}\bigl(\Gamma_s(\theta),{\stackrel{\circ}{\Gamma}}_s(\theta),
\hbar\bigr)$ $\in$ $D^k_s$ exists.
\vspace{1ex}

Although a dependence upon ${\stackrel{\circ}{\Gamma}}{}^{p_s}_s(\theta)$
in Eq.(8.1) is parametric one but the properties and interpretation
of the corresponding formal gauge theory appear by essentially different  in
comparison with ones for
$S_{H{}s}(\theta,\hbar)$ $\in$ $C^k(T^{\ast}_{odd}{\cal M}_s)$.

The fulfilment of Eq.(8.1) does not guarantee a realizability of Eq.(6.9)
in $D^k_s$ for ${\stackrel{\circ}{\Gamma}}{}^{p_s}_s(\theta)$ $\neq 0$. This
is related with the fact that Eq.(8.1) itself, antibracket of 2 arbitrary
superfunctions given on $D^k_s$ and operator $\Delta^s(\theta)$ calculated on
${\cal F}(\theta)$ $\in$ $D^k_s$ are not already
invariant with respect to anticanonical transformations. This invariance is
restored only in restricting of $D^k_s$ to $C^k(T^{\ast}_{odd}{\cal M}_s)$.
Therefore Eq.(6.9) for $S_{H{}s}(\theta,\hbar)$ appears by independent
equation with respect to (8.1) and must be considered additionally.

Corresponding system of 2$n_s$ ODE defined with respect to $S_{
H{}s}(\theta,\hbar)$ from Theorem 4 both for $\hbar=0$ and $\hbar \ne 0$ has
the form of the 1st order on $\theta$ system of $2n_s$ ODE
\begin{eqnarray}
\frac{d_r\Gamma^{p_s}_{s}(\theta)}{d\theta\phantom{xxxx}}=
\left(\Gamma^{p_s}_{s}(\theta), S_{H{}s}
\bigl(\Gamma_s(\theta),{\stackrel{\circ}{\Gamma}}_s(\theta),\hbar
\bigr)\right)^{s}_{\theta},\ s\in C_1
\end{eqnarray}
not already being in normal form and therefore  loses a number of properties
for solutions of the real HS. Call the system (8.3) by the
\underline{HS type system}.

Formally the last system by virtue of Eq.(8.1) is solvable but is not
covariant with respect to anticanonical transformations. The generator
$\tilde{s}_s(\theta,\hbar)$  being formally
defined with help of action on arbitrary ${\cal F}(\theta,\hbar)$ $\in$ $D^k_s$
\begin{eqnarray}
\tilde{s}_s(\theta,\hbar){\cal F}\bigl(\Gamma_s(\theta),{\stackrel{\circ}{
\Gamma}}_s(\theta),\theta,\hbar\bigr)= {\rm ad}_{S_{H{}s}
\bigl(\Gamma_s(\theta),{\stackrel{\circ}{\Gamma}}_s(\theta),\hbar
\bigr)}{\cal F}(\theta,\hbar)(-1)^{\varepsilon({\cal F})} + \frac{
\partial_r {\cal F}(\theta,\hbar)}{\partial\theta\phantom{xxxxx}},\ s\in C_1
\end{eqnarray}
is nilpotent and transfers $S_{H{}s}(\theta,\hbar) \in D^k_s$
into zero.
\subsection{BV Type Quantum Master Equation  in
$T_{odd}(T^{\ast}_{odd}{\cal M}_s)$}

Analogously to Theorem 4 one can write down the formulation for Existence
Theorem of a solution for equation of the form (6.11) with
$S_{H{}s}\bigl(\Gamma_s(\theta),{\stackrel{\circ}{\Gamma}}_s(\theta),
\hbar\bigr)$ $\in$ $D^k_s$ representing the deformation
of superfunction $S_{H{}s}(\Gamma_s(\theta),\hbar)$, satisfying to mentioned
equation, in powers of  ${\stackrel{\circ}{\Gamma}}{}^{p_s}_s(\theta)$. To
this end it is sufficient to replace the relation (8.1) in
Theorem 4 onto formula with equation
\begin{eqnarray}
{} & \frac{1}{2}\left(
S_{H{}s}\bigl(\Gamma_s(\theta),{\stackrel{\circ}{\Gamma}}_s(\theta),
\hbar\bigr),
S_{H{}s}\bigl(\Gamma_s(\theta),{\stackrel{\circ}{\Gamma}}_s(\theta),
\hbar\bigr)\right)^s_{\theta}=
\imath\hbar\Delta^s(\theta)
S_{H{}s}\bigl(\Gamma_s(\theta),{\stackrel{\circ}{\Gamma}}_s(\theta),
\hbar\bigr)\;, & {} \nonumber \\
{} & \forall\bigl(\Gamma_s(\theta),{\stackrel{\circ}{\Gamma}}_s(\theta)
\bigr)\in T_{odd}(T^{\ast}_{odd}{\cal M}_s),\ s\in C_1\;. & {} 
\end{eqnarray}
Corresponding HS type system  given in the form (8.3) now is not formally
solvable. Operator $\tilde{s}_s(\theta,
\hbar)$ (8.4) constructed with respect to $S_{H{}s}(\theta,\hbar)$ from
Eq.(8.5) is not nilpotent and does not possess by this superfunction (i.e.
$S_{H{}s}(\theta,\hbar)$) as one's eigenfunction with zero eigenvalue.

By remarkable case it appears the fact that GNA (7.1), with
$\Psi_s\bigl(\Gamma_s(\theta),{\stackrel{\circ}{\Gamma}}_s(\theta),\hbar
\bigr)$ $\in$ $D^k_s$ satisfying to the conditions (7.3), transfers an any
solution for (8.5) into solution $S_{H{}s}(\Psi_s;\theta,\hbar)$ $\in$
$D^k_s$ again. Among such transformations there exists a subset with
superfunction $\hat{\Psi}_s(\theta,\hbar)$ $\in$ $D^k_s$ satisfying to (7.6)
so that GNA (7.1) in the form  (7.8) transfers solutions of
Eqs.(8.1), (6.9) for
$S_{H{}s}\bigl(\Gamma_s(\theta), {\stackrel{\circ}{\Gamma}}_s(\theta), \hbar
\bigr)$ in their solutions
$S_{H{}s}\bigl(\hat{\Psi}_s;\Gamma_s(\theta), {\stackrel{\circ}{\Gamma}
}_s(\theta), \hbar\bigr)$ again.

\underline{\bf Remark:} HS or HS type system  corresponding to
$S_{H{}s}(\Gamma_s(\theta), \hbar)$ $\in$ $C^k(T^{\ast}_{odd}{\cal M}_s)$ or
$S_{H{}s}\bigl(\Gamma_s(\theta)$, ${\stackrel{\circ}{\Gamma}}_s(\theta),
\hbar\bigr)$ $\in$ $D^k_s$, $s\in C_1$ satisfying, for instance, to Eqs.(6.1)
or (8.5) respectively are modified under GNA to
HS and HS type system  defined with respect to
$S_{H{}s}\bigl(\hat{\Psi}_s;\theta, \hbar\bigr)$,
$S_{H{}s}({\Psi}_s;\theta, \hbar)$ from $C^k(T^{\ast}_{odd}{
\cal M}_s)$, $D^k_s$.
\subsection{Trivial Arbitrariness in  Master Equations Solutions in
\protect \\ $T_{odd}(T^{\ast}_{odd}{\cal M}_s)$}

The structure of supermanifolds $T_{odd}(T^{\ast}_{odd}{\cal M}_s)$,
$s\in C_1$ permits to specify the another type of transformations for
solutions of Eqs.(8.1), (8.5) being different from GNA (7.1).

Define the so-called global trivial arbitrariness (GTA) in a choice of
solutions for Eq.(8.5) transferring its solution $S_{H{}s}(\theta,\hbar)$
$\in$ $D^k_s$ into corresponding one of the form ${S}_{H{}s}''\bigl(
\Gamma_s(\theta)$, ${\stackrel{\circ}{\Gamma}}_s(\theta),\hbar\bigr)$ $\in$
$D^k_s$ by the formula
\begin{eqnarray}
{} & {S}_{H{}s}''\bigl(
\Gamma_s(\theta), {\stackrel{\circ}{\Gamma}}_s(\theta),\hbar\bigr)=
{S}_{H{}s}\bigl(\Gamma_s(\theta), {\stackrel{\circ}{\Gamma}}_s(\theta),
\hbar\bigr) + {\stackrel{\circ}{\cal W}}{}^s_{+}(\theta)\Psi_{1{}s}
\bigl(\Gamma_s(\theta), {\stackrel{\circ}{\Gamma}}_s(\theta),
\hbar\bigr)\;,{} & \nonumber \\
{} &
(\tilde{P}_1(\theta), {\rm gh}, \varepsilon_P, \varepsilon_{\bar{J}},
\varepsilon)\Psi_{1{}s}(\theta,\hbar) = (0,-1,1,0,1),\
\Psi_{1{}s}(\theta,\hbar)\in D^k_s,\ s\in C_1\;. 
\end{eqnarray}
In addition the superfunction $\Psi_{1{}s}(\theta,\hbar)$ must satisfy to the
equation
\renewcommand{\theequation}{\arabic{section}.\arabic{equation}\alph{lyter}}
\begin{eqnarray}
\setcounter{lyter}{1}
{} & \frac{1}{2}\left(\tilde{\cal F}_{1{}s}(\theta,\hbar),
\tilde{\cal F}_{1{}s}(\theta,\hbar)\right)^s_{\theta} +
\left(\tilde{\cal F}_{1{}s}(\theta,\hbar),
S_{H{}s}(\theta,\hbar)\right)^s_{\theta} = \imath\hbar\Delta^s(\theta)
\tilde{\cal F}_{1{}s}(\theta,\hbar)\;, {} &
\\
\setcounter{equation}{7}
\setcounter{lyter}{2}
{} & \tilde{\cal F}_{1{}s}(\theta,\hbar) =
{\stackrel{\circ}{\cal W}}{}^s_{+}(\theta)\Psi_{1{}s}
\bigl(\Gamma_s(\theta), {\stackrel{\circ}{\Gamma}}_s(\theta),
\hbar\bigr)\in D^k_s\;.{} & 
\end{eqnarray}
GTA for Eq.(8.1) is described by the same relations (8.6),
(8.7) but without operator $\Delta^s(\theta)$ in (8.7a), whereas for the
equation
\renewcommand{\theequation}{\arabic{section}.\arabic{equation}}
\begin{eqnarray}
\Delta^s(\theta)
{S}_{H{}s}\bigl(\Gamma_s(\theta), {\stackrel{\circ}{\Gamma}}_s(\theta),
\hbar\bigr) = 0 
\end{eqnarray}
is given by formula (8.6) but instead of relationship (8.7) the following one
must be fulfilled
\begin{eqnarray}
\Delta^s(\theta) {\stackrel{\circ}{\cal W}}{}^s_{+}(\theta)\Psi_{1{}s}
\bigl(\Gamma_s(\theta),{\stackrel{\circ}{\Gamma}}_s(\theta),
\hbar\bigr)=0\;. 
\end{eqnarray}
\section{Gauge Fixing}
\subsection{Gauge Fixing Procedure in $T^{\ast}_{odd}{\cal M}_{ext}$}
\setcounter{equation}{0}

Theorems 2,3 together with their BV similar analogs permit to present a
parametrization for $\Lambda_{ext}(\hbar)$ in terms of variables
$\varphi^a_{ext}(\theta)$, $a=1,\ldots,n+3m$, containing both
$\Phi^A_{min}(\theta)$ and $\Phi^{\ast}_{A{}{min}}(\theta)$.

As it had been already mentioned the explicit construction of
$\varphi^a_{ext}(\theta)$ leads to violation of locality and covariance
relative to indices of $T_{\mid\bar{J}}$ representation.  The
standard problem of  $\Lambda_{ext}(\hbar)$ modification, in satisfying of
the Theorems 2,3 conditions,  to $\Lambda_{ext}'(\hbar)$ $\equiv$ $\Lambda_{
\Phi}$ $\subset$ $T^{\ast}_{odd}{\cal M}_{ext}$ arises with
parametrization of $\Lambda_{\Phi}$ only by
superfields $\Phi^B(\theta)$ in changing of local coordinates system
$\Gamma^{p_{ext}}_{ext}(\theta)$ in a such way that its subsystem consisting
from $\Phi^B(\theta)$ remains invariable providing the direct
connection with original superfield classical theory. Resolution of the
problem in question is realized by means of two different ways with respect
to modification of $S_{H{}ext}(\Gamma_{ext}(\theta),\hbar)$:

\noindent
1) with help of phase anticanonical transformation (AT) in the class of
Abelian hypergauges;

\noindent
2) by means of application of GNA operation to solutions of Eqs.(6.2), (6.11)
for $s=ext$.
\vspace{1ex}

\noindent
\underline{\bf Theorem 5} (on modification of $\Lambda_{ext}(\hbar)$ to
$\Lambda_{\Phi}$)

\noindent
A phase AT exists: $\Gamma^{p_s}_s(\theta)$ $\to$
${\Gamma'}^{p_s}_s(\theta)$ = ${\Gamma'}^{p_s}_s(\Gamma_s(\theta))$
$\in$ $C^k(T^{\ast}_{odd}{\cal M}_s)$, $s=ext$ a such that
$S_{H{}s}(\Gamma_{s}(\theta),\hbar)$ satisfying to Eq.(6.2) is transformed
into $S_{H{}s}'(\Gamma_{s}'(\theta),\hbar)$ satisfying to the same equation.
For  latter superfunction the boundary condition (6.4) is true and the
condition to be "proper" (6.5) is fulfilled both for $\Lambda_{ext}(\hbar)$
and for $\Lambda_{\Phi}$ parametrized by $\varphi^a_{ext}(\theta)$
and $\Phi^B(\theta)$ respectively.

\underline{Proof} of the Theorem is based in the presentation of mentioned
transformation by the explicit formulae with additional requirement
 of the $\varepsilon_P, \varepsilon_{\bar{J}}, \varepsilon$ and
ghost gradings conservation for new $\Gamma_{s}'(\theta)$ coordinates in
the same form as for the old $\Gamma_{s}(\theta)$ ones
\renewcommand{\theequation}{\arabic{section}.\arabic{equation}\alph{lyter}}
\begin{eqnarray}
\setcounter{lyter}{1}
{} & {} & \Phi'{}^B(\theta) = {\Phi}^B(\theta),\
{\Phi'}^{\ast}_B(\theta) = {\Phi}^{\ast}_B(\theta) + \Psi_B(\Phi(\theta)),\
B=1,\ldots,n+3m\;,\\
\setcounter{equation}{1}
\setcounter{lyter}{2}
{} & {} &
(\tilde{P}_1(\theta), {\rm gh}, \varepsilon_P, \varepsilon_{\bar{J}},
\varepsilon)\Psi_B(\Phi(\theta)) = (0,-{\rm gh}({\Phi}^B)-1,
(\varepsilon_P)_B + 1, (\varepsilon_{\bar{J}})_B, \varepsilon_B+
1)\;. 
\end{eqnarray}
The nontrivial condition to be anticanonical for the change of variables (9.1)
follows from the potentiality of the expression
\renewcommand{\theequation}{\arabic{section}.\arabic{equation}}
\begin{eqnarray}
({\Phi'}^{\ast}_A(\theta),{\Phi'}^{\ast}_B(\theta))^{(\Gamma)}_{\theta} =
\frac{\partial\Psi_A(\theta)}{\partial\Phi^B(\theta)} - (-1)^{
\varepsilon_A\varepsilon_B}\frac{\partial\Psi_B(\theta)}{\partial
\Phi^A(\theta)}=0\;, 
\end{eqnarray}
in its turn arising from definition of arbitrary AT:
$\Gamma_s(\theta)$ $\to$ $\Gamma_s'(\Gamma_s(\theta))$, $s\in C_0$ by means
of the relationships taking account of (2.41d)
\begin{eqnarray}
\left({\Gamma'}^{p_s}_s(\theta),{\Gamma'}^{q_s}_s(\theta)
\right)^{(\Gamma_s)}_{\theta} = \omega^{p_sq_s}_s(\theta),\ s\in C_0\;. 
\end{eqnarray}
A local solution of Eqs.(9.2) determines  superfunctions
$\Psi_A(\Phi(\theta))$ through the phase of AT $\Psi(\Phi(\theta))$ $\in$
$C^k(T^{\ast}_{odd}{\cal M}_{ext})$ (usually being called by the gauge
fermion for $\theta=0$ [4,6])
\begin{eqnarray}
\Psi_A(\theta) = \frac{\partial\Psi(\Phi(\theta))}{\partial\Phi^A(\theta)
\phantom{x}} +
C_A(\theta),\ ({\rm gh}, \varepsilon_P, \varepsilon_{\bar{J}},
\varepsilon)\Psi(\Phi(\theta)) = (-1,1,0,1)\;, 
\end{eqnarray}
with the choice for arbitrary constants $C_A(\theta)$ to be equal to 0
being consistent with restrictions on $\Psi_A(\Phi(\theta))$.

The definition of AT by means of relations (9.3)
implies  the invariance
of the antibracket calculated on any ${\cal F}(\theta)$, ${\cal J}(\theta)$
$\in$ $C^k(T^{\ast}_{odd}{\cal M}_{ext}\times \{\theta\})$
\begin{eqnarray}
{} & \left({\cal F}(\Gamma_s(\theta),\theta), {\cal J}(\Gamma_s(\theta),
\theta)\right)^{(\Gamma_s)}_{\theta}=
 \left({\cal F}'(\Gamma_s'(\theta),\theta), {\cal J}'(\Gamma_s'(\theta),
\theta)\right)^{(\Gamma_s')}_{\theta}\;, & {} \nonumber \\
{} & \Gamma_s'(\theta) = \Gamma_s'({\Gamma}_s(\theta)),\
{\cal K}(\Gamma_s(\theta),\theta) = {\cal K}'(\Gamma_s'(\theta),\theta),\
{\cal K}\in\{{\cal F}, {\cal J}\}\,.
& {} 
\end{eqnarray}
From formula of $\Delta^s(\theta)$ definition  through antibracket [2]
\begin{eqnarray}
{} & \Delta^s(\theta) =\displaystyle\frac{1}{2}\omega_{q_sp_s}(\theta)
({\Gamma}^{p_s}_s(\theta),({\Gamma}^{q_s}_s(\theta),\ \;
)^{(\Gamma_s)}_{\theta})^{(\Gamma_s)}_{\theta}(-1)^{\varepsilon_{q_s}},\
 \omega_{q_sp_s}(\theta)= -
(-1)^{\varepsilon_{p_s}\varepsilon_{q_s}}\omega_{p_sq_s}(\theta)=\;
& {} \nonumber \\
{} & = P_0(\theta)\omega_{q_sp_s}(\theta),\
\left\|\omega_{q_sp_s}(\theta)\right\|=
\left\|
\begin{array}{cc}
0_{n_s} & -1_{n_s} \\
1_{n_s}  & 0_{n_s}
\end{array} \right\|,\;\omega^{d_sq_s}_s(\theta)\omega_{q_sp_s}(\theta)=
\delta^{d_s}{}_{p_s},\ s\in C_0 & {} 
\end{eqnarray}
it follows the invariance of the form of action for operator $
\Delta^s(\theta)$ on an arbitrary
${\cal F}(\theta)$ $\in$ $C^k(T^{\ast}_{odd}{\cal M}_{ext}\times \{\theta\})$
\begin{eqnarray}
\Delta'^s(\theta){\cal F}'(\Gamma'_s(\theta),\theta) =
\Delta^s(\theta){\cal F}(\Gamma_s(\theta),\theta),\ \;
 \Delta'^s(\theta) = \Delta^s(\theta)\;,
\end{eqnarray}
where in the 2nd expression  it is taken into account that the operator
$\Delta'^s(\theta)$ is given through $\Gamma_s'(\theta)$
according to formula (9.6) and coincides with $\Delta^s(\theta)$ being by the
invariant of AT itself.

Having expressed of $S_{H{}s}(\Gamma_s(\theta),\hbar)$ through the new
coordinates $\Gamma_s'(\theta)$  obtain the following representation for
$S_{H{}s}'(\Gamma_s'(\theta),\hbar)$
\begin{eqnarray}
{} & S_{H{}s}(\Gamma_s(\theta),\hbar) =
S_{H{}s}'(\Gamma_s'(\theta),\hbar) =
S_{H{}s}'\Bigl({\Phi}_s(\theta),{\Phi}^{\ast}_s(\theta)+
\displaystyle\frac{
\partial\Psi_s(\theta)}{\partial\Phi_s(\theta)},\hbar\Bigr) \equiv
{S'}^{\Psi_s}_{H{}s}(\Gamma_s(\theta),\hbar)\;.
\end{eqnarray}
In (9.8) the notation for any
${\cal F}(\Gamma_s(\theta),\hbar)$ $\in$ $C^k(T^{\ast}_{odd}{\cal M}_{s})$
is introduced as follows
\begin{eqnarray}
{\cal F}\Bigl({\Phi}_s(\theta),{\Phi}^{\ast}_s(\theta)+
\frac{\partial\Psi_s(\theta)}{\partial\Phi_s(\theta)},\hbar\Bigr)
= e^{{\rm ad}_{\Psi_s}}{\cal F}(\Gamma_s(\theta),\hbar) \equiv
{\cal F}^{\Psi_s}\bigl(\Gamma_s(\theta), \hbar\bigr)\,.
\end{eqnarray}
It follows from (9.8) taking into account of (7.1), (7.8) that
$S_{H{}s}'({\Gamma}_s(\theta),\hbar)$ has the form
\begin{eqnarray}
e^{-{\rm ad}_{\Psi_s}}{S}_{H{}s}({\Gamma}_s(\theta),\hbar) =
{S}^{(-\Psi_s)}_{H{}s}({\Gamma}_s(\theta),\hbar)=
{S}_{H{}s}({(-\Psi_s)};{\Gamma}_s(\theta),\hbar) =
S_{H{}s}'({\Gamma}_s(\theta),\hbar)\;. 
\end{eqnarray}
Relations (9.10) determine the fact that for new superfunction $S_{
H{}s}'$ given by the formula (under AT (9.1), (9.8))
\renewcommand{\theequation}{\arabic{section}.\arabic{equation}}
\begin{eqnarray}
S_{H{}s}' = {S}_{H{}s}\circ f_s^{-1},\ \Gamma'_s(\theta)=
f_s({\Gamma}_s(\theta)),\;{\Gamma}_s(\theta)= f_s^{-1}(\Gamma'_s(\theta)),\
f_s = \left({\bf 1}_{\Phi_s}, (e^{{\rm ad}_{\Psi_s}})_{\Phi^{\ast}_s}\right),
\end{eqnarray}
the representation written on equality of  coordinates $\Gamma_s'(\theta)$
is true
\begin{eqnarray}
S_{H{}s}'= e^{-{\rm ad}_{\Psi_s}}{S}_{H{}s}={S}_{H{}s}\circ e^{-{\rm ad}_{
\Psi_s}},\ s=ext 
\end{eqnarray}
meaning that $S_{H{}s}'$ is obtained from ${S}_{H{}s}$ by means of GNA
(7.1) with superfunction $\hat{\Psi}_{ext}(\theta,\hbar)$ $\equiv$ $-\Psi(
\Phi(\theta))$ satisfying among them to Eq.(7.6).
$S_{H{}s}'(\Gamma'_s(\theta),\hbar)$ satisfies both to Eq.(6.2) and to
its consequence (6.9).

A choice of $\Psi_s(\Phi(\theta))$ from the nondegeneracy requirement for
supermatrix of the 2nd derivatives of $S_{H{}s}'({\Gamma}_s(\theta),
\hbar)$ with respect to $\Phi(\theta)$  under restriction onto
$\tilde{\Sigma}_{ext} \subset \Lambda_{\Phi}$
\begin{eqnarray}
{\rm rank}_{\varepsilon}\left\|
\frac{\partial\phantom{xxxx}}{\partial\Phi^A(\theta)}
\frac{\partial_l S_{H{}s}'({\Gamma}_s(\theta),\hbar)}{\partial
\Phi^B(\theta)\phantom{xxxxxx}}\right\|_{\mid\tilde{\Sigma}_{ext}} =
n+m+2m^P\;, 
\end{eqnarray}
leads to the following necessary conditions on the gauge fermion
$\Psi_s(\Phi(\theta))$
\begin{eqnarray}
{} &\hspace{-6em} {\rm 1)}\hspace{6em} &
{\rm rank}_{\varepsilon}\left\|
\frac{\partial_l\phantom{xxxx}}{\partial\overline{C}{}^{\alpha}(\theta)}
\frac{\partial\phantom{xxxx}}{\partial
C^{\beta}(\theta)}(\Psi_s(\Phi(\theta)), S_{H{}s}'({\Gamma}_s(\theta),\hbar)
)^{(\Gamma_s)}_{\theta}\right\|_{\mid\tilde{\Sigma}_{ext}}= m^P, \\
{}&\hspace{-6em} {\rm 2)}\hspace{6em} &
{\rm rank}_{\varepsilon}\left\|
\frac{\partial_l{\Psi_s,}_{\imath}(\Phi(\theta))}{\partial\overline{C}{}^{
\alpha}(\theta)\phantom{xxxx}}\right\|_{\mid\tilde{\Sigma}_{ext}}=m^P.
\end{eqnarray}
The contents of the new  $\varphi'^a_{ext}(\theta)$ and old coordinates
$\varphi^a_{ext}(\theta)$, connected through AT (9.1), for original
$\Lambda_{ext}(\hbar)$ described in Theorem 3 is differred, whereas the ones
of new $\Phi'^A(\theta)$ and old superfields $\Phi^A(\theta)$ under the same
AT parametrizing $\Lambda_{\Phi}$ $\equiv$ $\Lambda_{ext}'(\hbar)$
coincide.$_{\textstyle\Box}$

\vspace{1ex}
It appears a so-called minimal  the choice of gauge fermion for
transition to $\Lambda_{\Phi}$ in the form
\begin{eqnarray}
\Psi_s(\Phi(\theta)) =\overline{C}{}^{\alpha}(\theta)\chi_{\alpha}({\cal
A}(\theta))\;,
\end{eqnarray}
which in general permits  to pass to $\Lambda_{\Phi_{(min,\overline{C})
}}$ for superfunction ${S}_{H{}s}({\Gamma}_s(\theta),\hbar)$ $\in$
$C^k(T^{\ast}_{odd}{\cal M}_{s})$ for $s$ = $(min,\overline{C})$. Formula
(9.16) defines the gauge called in BV method for $\theta=0$  by
singular whereas the nonsingular one has the representation for $s=ext$
\begin{eqnarray}
\Psi_s(\Phi(\theta)) =\overline{C}{}^{\alpha}(\theta)\bigl(\chi_{\alpha}({\cal
A}(\theta)) + g_{\alpha\beta}(\theta)B^{\beta}(\theta)\bigr),\
g_{\alpha\beta}(\theta)=P_0(\theta)g_{\alpha\beta}(\theta),\
{\rm sdet}\left\|g_{\alpha\beta}(\theta)\right\|\neq 0\;. 
\end{eqnarray}
From Theorem 5 proof it follows  that phase AT (9.1) is equivalently given
for ${S}_{H{}s}({\Gamma}_s(\theta),\hbar)$ by means of GNA operation (7.1).
The last conclusion is remarkable by the fact that it is sufficient in
${S}_{H{}s}({\Gamma}_s(\theta),\hbar)$ to shift
the argument of superantifields $\Phi^{\ast}_{B}(\theta)$
on $\left(-\frac{\partial\Psi_s(\theta)}{\partial\Phi^B(\theta)}\right)$
to determine  $S_{H{}s}'({\Gamma}_s(\theta),\hbar)$.
It should be noted the restriction of $S_{H{}s}'(\Gamma_s'(\theta),
\hbar)$ (9.8) onto hypersurface defined by equations $\Phi'^{\ast}_s(
\theta)=0$ leads to the following superfunction
\begin{eqnarray}
S_{H{}s}'(\Gamma'_s(\theta),\hbar){\hspace{-0.5em}\phantom{\Bigl)}}_{
\mid \Phi'^{\ast}_s(\theta)=0}=
\hat{S}_{H{}s}'({\Phi}(\theta),\hbar)= S_{H{}s}\Bigl({\Phi}(\theta),
-\frac{\partial\Psi_s(\theta)}{\partial\Phi(\theta)},\hbar\Bigr)
\end{eqnarray}
being nondegenerate on $\Lambda_{\Phi}$ and belonging to $C^k({\cal M}_{
ext})$.

\noindent
\underline{\bf Remarks:}

\noindent
{\bf 1)} In ignoring of the above-mentioned requirements of covariance and
locality but with preservation of the given Grassmann parities and ghost
numbers
distribution for  superfields, superantifields under AT one can describe
with help of Theorem 5 the all essentially different Lagrangian surfaces on
which the condition b) from Theorems 2,3 is realized, in the sense of their
distinct parametrizations in terms of new coordinates ${\Gamma}_s'(\theta)$.
The number of such hypersurfaces including $\Lambda_{ext}(\hbar)$,
$\Lambda_{\Phi}$ is equal to $2^m$;

\noindent
{\bf 2)} the investigation of the correspondence of an arbitrary GNA (7.1)
and AT remains out the present paper's scope;

\noindent
{\bf 3)} the extension of the application field for Theorem 5 relative to
GThGT,
being described by ${S}_{H{}s}({\Gamma}_s(\theta),\hbar)$ satisfying to
Eq.(6.11), is literally realized by replacement of Eq.(6.2) onto (6.11) in
the formulation of that Theorem.

\noindent
\underline{\bf Corollary 5.1} (on HS being defined by
$S_{H{}s}'(\Gamma_s'(\theta),\hbar)$)

\noindent
HS defined by superfunction $S_{H{}s}'(\Gamma_s'(\theta),\hbar)$ =
${S}_{H{}s}({\Gamma}_s(\theta),\hbar)$, $s=ext$ obtained under AT (9.1)
\begin{eqnarray}
\frac{d_r\Gamma'^{p_s}_{s}(\theta)}{d\theta\phantom{xxxxx}}=
\left(\Gamma'^{p_s}_{s}(\theta), S_{H{}s}'(
\Gamma_s'(\theta),\hbar)\right)^{(\Gamma_s')}_{\theta}
\end{eqnarray}
is equivalent to previous one given
by ${S}_{H{}s}({\Gamma}_s(\theta),\hbar)$, $s=ext$
in the old coordinates $\Gamma_s(\theta)$.

\noindent
\underline{Proof:} Taking into consideration  the antibracket (9.5)
invariance under AT  and Leibnitz rule for antibracket we obtain
\begin{eqnarray}
\left({\Gamma}^{p_s}_{s}(\theta), {S}_{H{}s}(
{\Gamma}_s(\theta),\hbar)\right)^{({\Gamma}_s)}_{\theta}=
\frac{\partial_r \Gamma^{p_s}_s(\Gamma_s'(\theta))}{\partial\Gamma'^{
q_s}_s(\theta)\phantom{xxxx}}
\left(\Gamma'^{q_s}_{s}(\theta), S_{H{}s}'(
\Gamma_s'(\theta),\hbar)\right)^{(\Gamma_s')}_{\theta},\ s=ext\;.  
\end{eqnarray}
Quantities $\frac{d_r \Gamma_s(\theta)}{d\theta\phantom{xxxx}}$  are
transformed under AT by the rule
\begin{eqnarray}
\frac{d_r{\Gamma}^{p_s}_{s}(\Gamma_s'(\theta))}{d\theta\phantom{xxx
xxxxx}} =
\frac{\partial_r \Gamma^{p_s}_s(\Gamma_s'(\theta))}{\partial\Gamma'^{
q_s}_s(\theta)\phantom{xxxx}}
\frac{d_r\Gamma'^{q_s}_{s}(\theta)}{d\theta\phantom{xxxxx}},\ s=ext\;.
\end{eqnarray}
Finally the statement of Corollary  directly follows from nondegeneracy
of Jacobi supermatrix $\left\|\frac{\partial_r\Gamma^{p_s}(\theta)}{\partial
\Gamma'^{q_s}(\theta)\phantom{}}\right\|$ in Eqs.(9.20), (9.21).$_{
\textstyle\Box}$

The formal change of $\Gamma_s'(\theta)$ onto initial coordinates ${\Gamma
}_s(\theta)$ in (9.19) leads to nonequivalent HS defined with respect to
$S_{H{}s}'({\Gamma}_s(\theta),\hbar)$
\begin{eqnarray}
\frac{d_r{\Gamma}^{p_s}_{s}(\theta)}{d\theta\phantom{xxxx}}=
\left({\Gamma}^{p_s}_{s}(\theta), {S}_{H{}s}\Bigl(\Phi_s(\theta),
{\Phi}^{\ast}_s(\theta) -
\frac{\partial\Psi_s(\theta)}{\partial\Phi_s(\theta)},\hbar\Bigr)\right)^{(
\Gamma_s)}_{\theta} \hspace{-0.4em} = \left(\Gamma^{p_s}_s(\theta),
S^{(-\Psi_s)}_{H{}s}({\Gamma}_s(\theta),\hbar)\right)^{(\Gamma_s)}_{\theta}
\hspace{-0.4em}.
\end{eqnarray}
In this case in its subsystem corresponding to
${\stackrel{\circ}{\Phi}}{}^A(\theta)$ the only summands in right-hand side
for
\begin{eqnarray}
{\rm deg}_{\Phi_{add}}\left(\frac{\partial S^{(-\Psi_s)}_{H{}s}(\theta,
\hbar)}{
\partial \Phi^{\ast}_A(\theta)\phantom{xxxx}}\right) \le 1 
\end{eqnarray}
coincide with corresponding ones in the right-hand side of analogous HS
constructed with respect to initial ${S}_{H{}s}({\Gamma}_s(\theta),\hbar)$.
Namely, HS (9.22)\footnote[4]{usually instead of expression
$\left(-\frac{\partial\Psi_s}{\partial\Phi\phantom{x}}\right)$ one  writes
$\frac{\partial\Psi_s}{\partial\Phi\phantom{x}}$ that is achieved by
simple redefinition of $\Psi_s$ $\to$ $(-\Psi_s)$} plays the key role in
BV method [4].  This HS possesses by solvability property with corresponding
translation generator with respect to $\theta$ along its integral curve
\begin{eqnarray}
s_s^{(-\Psi_s)}(\theta){\cal F}(\theta) \hspace{-0.1em}=\hspace{-0.1em} \frac{
\partial_r{\cal F}(\theta)}{
\partial\theta\phantom{xxxx}} + \left({\cal F}(\theta),
S^{(-\Psi_s)}_{H{}s}(\theta,\hbar)\right)^{(\Gamma_s)}_{\theta}
\hspace{-0.3em},\;
{\cal F}(\theta)\in C^k(T^{\ast}_{odd}{\cal M}_{s}\times
\{\theta\}), s=ext 
\end{eqnarray}
\setcounter{footnote}{4}
being by nilpotent
and annuling ${S}^{(-\Psi_s)}_{H{}s}({\Gamma}_s(\theta),\hbar)$.

However the projection of solution for HS (9.22) on its subsystem
\begin{eqnarray}
\frac{d_r{\Phi}^{B}(\theta)}{d\theta\phantom{xxxx}}=
\left(\Phi^B(\theta),
S^{(-\Psi_s)}_{H{}s}({\Gamma}_s(\theta),\hbar)\right)^{(\Gamma_s)}_{\theta}
\end{eqnarray}
with trivial other equations leads to the generator of translation with
respect to $\theta$ along projection of solution for HS (9.22) onto
system (9.25) of the form
\begin{eqnarray}
\tilde{s}_s^{(-\Psi_s)}(\theta){\cal F}(\theta) = \frac{\partial_r{\cal F}(
\theta)}{
\partial\theta\phantom{xxx}} + \frac{\partial{\cal F}(\theta)}{\partial
\Phi^B(\theta)}\frac{\partial S^{(-\Psi_s)}_{H{}s}(\theta,\hbar)}{\partial
\Phi^{\ast}_B(\theta)\phantom{xxxx}}
\end{eqnarray}
not being nilpotent but annuling
${S}^{(-\Psi_s)}_{H{}s}({\Gamma}_s(\theta),\hbar)$. The last generator
$\tilde{s}_s^{(-\Psi_s)}(\theta)$ appears for $\theta=0$, $s=ext$ by the
generator of BRST transformations being applied in BV method on
$C^k(P_0(T^{\ast}_{odd}{\cal M}_{ext}))$. The restriction of $\tilde{s}_s^{
(-\Psi_s)}(\theta)$ on $C^k({\cal M}_{ext})$ by the rule
\begin{eqnarray}
\tilde{s}_s^{(-\Psi_s)}(\theta){\cal F}(\Phi(\theta)
){\hspace{-0.5em}\phantom{\Bigl)}}_{\mid\Phi^{\ast}(\theta)=0} =
\frac{\partial{\cal F}(\Phi(\theta))}{\partial
\Phi^B(\theta)\phantom{xx}}\frac{\partial S^{(-\Psi_s)}_{H{}s}(\theta,\hbar)}{
\partial\Phi^{\ast}_B(\theta)\phantom{xxxx}
}{}{\hspace{-0.5em}\phantom{\Bigr)}}_{\mid\Phi^{\ast}(\theta)=0},\ s=ext
\end{eqnarray}
leads to the operator of BRST transformations for $\theta=0$ acting on
$C^k(P_0{\cal M}_{ext})$ with the same properties as for
$\tilde{s}_s^{(-\Psi_s)}(\theta)$.
\subsection{Gauge fixing in $T_{odd}(T^{\ast}_{odd}{\cal M}_{ext})$}

\underline{\bf Theorem 6} (on modification of $\Lambda_{ext}(\hbar)$ to
$\Lambda_{\Phi}$ for $S_{H{}ext}(\theta,\hbar)$ $\in$ $D^k_{ext}$)

\noindent
There exist the GNA transformation for $S_{H{}s}\bigl(\Gamma_s(\theta),
{\stackrel{\circ}{\Gamma}}_s(\theta), \hbar\bigr)$ $\in$ $D^k_{s}$,
$s=ext$ satisfying to Eq.(8.1) (or (8.5)) and transferring $S_{H{}s}(\theta,
\hbar)$ to a superfunction $S_{H{}s}\bigl(\Psi_s(\theta);\Gamma_s(\theta),
{\stackrel{\circ}{\Gamma}}_s(\theta), \hbar\bigr)$ $\in$ $D^k_{s}$,
($\Psi_s\bigl(\Phi_s(\theta),
{\stackrel{\circ}{\Gamma}}_s(\theta), \hbar\bigr)$ $\in$ $D^k_{s}$)
satisfying to Eq.(8.1) (or (8.5)) as well. Moreover the following conditions
must be realized in question

\noindent
a) for $S_{H{}s}\bigl(\Psi_s(\theta);\Gamma_s(\theta),
{\stackrel{\circ}{\Gamma}}_s(\theta), \hbar\bigr)$ the boundary conditions
(8.2) remains  valid;

\noindent
b) condition of properness (point b) in Theorem 4) is fulfilled both for
$\Lambda_{ext}(\hbar)$ and for $\Lambda_{\Phi}$ parametrized by
$\tilde{\varphi}_s^{a_s}(\theta)$, $\Phi^B(\theta)$ respectively.
\vspace{1ex}

\noindent
\underline{Proof:} Formulae (7.8) for Eq.(8.1) and (7.1) for (8.5)
respectively but with $\hat{\Psi}_s\bigl(\Phi_s(\theta)$, $
{\stackrel{\circ}{\Gamma}}_s(\theta),\hbar\bigr)$ and
${\Psi}_s\bigl(\Phi_s(\theta)$,
${\stackrel{\circ}{\Gamma}}_s(\theta),\hbar\bigr)$ from $D^k_s$ satisfying
to conditions (9.14), (9.15) for irreducible GThGT solve the problem
formulated in Theorem 6 with superfunctions
\begin{eqnarray}
S_{H{}s}\bigl(-K_s(\theta);\Gamma_s(\theta),
{\stackrel{\circ}{\Gamma}}_s(\theta),\hbar\bigr) =
S^{(-K_s)}_{H{}s}\bigl(\Gamma_s(\theta),
{\stackrel{\circ}{\Gamma}}_s(\theta),\hbar\bigr),\
{K}_s(\theta) \in \{\hat{\Psi}_s(\theta), \Psi_s(\theta)\}\,._{
\textstyle\Box} 
\end{eqnarray}

In Theorem 6 it is shown  that the role of ${\stackrel{\circ}{\Gamma}}_{ext}(
\theta)$ under GNA transformations is only parametric one.
For GThGT in $D^k_{ext}$ under last Theorem conditions  the remarks
about minimal gauge fermion (9.16), (9.17) and on the number $2^m$ of
essentially different Lagrangian surfaces (on which the property of
properness for $S_{H{}s}^{(-\Psi_s)}\bigl(\Gamma_s(\theta),
{\stackrel{\circ}{\Gamma}}_s(\theta),\hbar\bigr)$ is fulfilled) are valid
as well. At last as the superfunction ${\Psi}\bigl(\Phi_s(\theta),
{\stackrel{\circ}{\Gamma}}_s(\theta),\hbar\bigr)$ in (9.28) one can choose
its  restriction on hypersurface ${\stackrel{\circ}{\Gamma}}_s(\theta)$ =
$0$ that in fact is described in Theorem 5.

The following HS type system  is formally defined with respect to
$S_{H{}s}^{(-\Psi_s)}\bigl(\Gamma_s(\theta),$ $
{\stackrel{\circ}{\Gamma}}_s(\theta),\hbar\bigr)$
\begin{eqnarray}
\frac{d_r{\Gamma}^{p_s}_{s}(\theta)}{d\theta\phantom{xxxx}}=
\left({\Gamma}^{p_s}_{s}(\theta), {S}^{(-\Psi_s)}_{H{}s}\bigl(\Gamma_s(
\theta), {\stackrel{\circ}{\Gamma}}_s(\theta),\hbar\bigr)
\right)^{(\Gamma_s)}_{\theta},\ s=ext 
\end{eqnarray}
not being equivalent to system (8.3) just as for the case of
${\stackrel{\circ}{\Gamma}}_s(\theta)$ = $0$ in Eqs.(9.22). All its
properties have been described above for Eqs.(8.3). In particular, the
generators of translations of the form (9.24), (9.26), (9.27) constructed
from Eqs.(9.29) possess literally the same
properties and having the operators (9.24), (9.26), (9.27) as the one's own
boundary conditions for ${\stackrel{\circ}{\Gamma}}_{ext}(\theta)$ = $0$
respectively.

Systems (9.22) (and (9.19), (6.8) for $\Psi_s=0$ as well), but not (9.29)
(and (8.3) for $\Psi_s=0$ as
well) can be obtained from the variational problem on extremum for the
superfunctional
\begin{eqnarray}
Z_{H{}s}^{(-\Psi_s)}[\Gamma_s,\hbar] = \int d\theta\left(\frac{1}{2}
\Gamma^{p_s}(\theta)\omega_{p_sq_s}(\theta)
\frac{d_r{\Gamma}^{q_s}_{s}(\theta)}{d\theta\phantom{xxxx}} -
S_{H{}s}^{(-\Psi_s)}\bigl(\Gamma_s(\theta),\hbar\bigr)\right).
\end{eqnarray}

GNA transformation for $S_{H{}s}\bigl(
\Gamma_s(\theta),
{\stackrel{\circ}{\Gamma}}_s(\theta),\hbar\bigr)$ $\in D^k_s$ is not
equivalent to a possibility  of arbitrary AT realization even of phase AT
with  $\Psi(\Phi(\theta))$. Really, the invariance of antibracket does not
follow from definition of anticanonicity for
AT: $\Gamma_s(\theta)$ $\to$ $\Gamma_s'(\theta)$ =
$\Gamma_s'(\Gamma(\theta))$, $\forall{\cal F}(\theta)$, ${\cal J}(\theta)$
$\in$ $D^k_s$, $s=ext$
\begin{eqnarray}
{} & \left({\cal F}\bigl(
\Gamma_s(\theta),{\stackrel{\circ}{\Gamma}}_s(\theta),\theta\bigr),
{\cal J}\bigl(\Gamma_s(\theta),{\stackrel{\circ}{\Gamma}}_s(
\theta),\theta\bigr) \right)^{(\Gamma_s)}_{\theta} \neq
 \left({\cal F}'\bigl(
\Gamma'_s(\theta),{\stackrel{\circ}{\Gamma}}{}_s'(\theta),\theta\bigr),
{\cal J}'\bigl(\Gamma'_s(\theta),{\stackrel{\circ}{\Gamma}}{}_s'(
\theta),\theta\bigr) \right)^{(\Gamma'_s)}_{\theta}, & {} \nonumber \\
{} & {\cal K}'\bigl(\Gamma_s'(\theta),{\stackrel{\circ}{\Gamma}}{}_s'(
\theta),\theta\bigr) =
{\cal K}\bigl(\Gamma_s(\Gamma_s'(\theta)),
{\stackrel{\circ}{\Gamma}}_s\bigl(\Gamma_s'(\theta),
{\stackrel{\circ}{\Gamma}}{}_s'(\theta)\bigr),\theta\bigr),\
{\cal K} \in \{{\cal J},{\cal F}\}\;, & {} 
\end{eqnarray}
that is now connected with more than parametric dependence upon
${\stackrel{\circ}{\Gamma}}_s(\theta)$ under ATs of elements from
$D^k_s$  (see (9.21)).
Therefore, not only the generating equations (8.1), (8.5) do not preserve
one's form even under phase AT but the corresponding transformed HS type
system  (8.3) has not the Hamiltonian form (8.3), (9.29) in new coordinates.

Explicitly we have under AT (9.1) for $S_{H{}s}\bigl(\Gamma_s(\theta),
{\stackrel{\circ}{\Gamma}}_s(\theta),\hbar\bigr)$ taking account of (9.31)
\begin{eqnarray}
{} & S_{H{}s}\Bigl(
\Gamma_s(\theta), {\stackrel{\circ}{\Gamma}}_s(\theta),\hbar\Bigr) =
S_{H{}s}'\Bigl(\Phi_s(\theta), \Phi^{\ast}_s(\theta)
+ \displaystyle\frac{\partial\Psi_s(\theta)}{\partial\Phi_s(\theta)\phantom{x
}},
{\stackrel{\circ}{\Phi}}_s(\theta),{\stackrel{\circ}{\Phi}}{}^{\ast}_s(\theta)
+ {\stackrel{\circ}{U}}{}_1^s(\theta)\displaystyle\frac{\partial\Psi_s(
\theta)}{\partial\Phi_s(\theta)\phantom{x}},\hbar\Bigr)   & {} \nonumber \\
{} & ={\rm exp}\left({{\rm ad}_{\Psi_s(\theta)} +
\Biggl({\stackrel{\circ}{U}}{}_1^s(\theta)
\displaystyle\frac{\partial\Psi_s(\theta)}{\partial\Phi^{A_s}(\theta)}\Biggr)
\displaystyle\frac{\partial\phantom{xxxx}}{\partial{\stackrel{\circ}{
\Phi}}{}^{\ast}_{A_s}(\theta)}}\right)S_{H{}s}'\Bigl(
\Gamma_s(\theta),{\stackrel{\circ}{\Gamma}}_s(\theta),\hbar\Bigr)\;. & {}
\end{eqnarray}
The 2nd summand in exponent is not described by means of GNA transformations
and therefore its exponential function does not commute with
$\Delta^s(\theta)$. The only linear AT are the exceptions for above-mentioned
case. In particular, antibracket for any ${\cal F}(\theta)$, ${\cal
J}(\theta) \in D^k_s$ is invariant under linear AT and hence
$S'_{H{}s}\bigl(\Gamma'_s(\theta),
{\stackrel{\circ}{\Gamma}}{}'_s(\theta),\hbar\bigr)$ satisfies to the
equations of the same form and HS type system  constructed with respect to
this superfunction is equivalent to previous one (8.3).

For the case in question the linear phase AT (9.1) with phase
$\Psi(\Phi(\theta))$ being at most the quadratic one with respect to
superfields $\Phi^B(\theta)$ satisfies to this problem. It means for gauge
fermion (9.16), (9.17) the linear choice of the gauge superfunctions
\begin{eqnarray}
\chi_{\alpha}({\cal A}(\theta)) = \kappa_{\alpha\imath}(\theta){\cal A}^{
\imath}(\theta),\ P_0(\theta)\kappa_{\alpha\imath}(\theta)=
\kappa_{\alpha\imath}(\theta)\;. 
\end{eqnarray}
Therefore it is possible to simplify the form of $S_{H{}ext}(\theta,\hbar)$
for ${\stackrel{\circ}{\Gamma}}_{ext}(\theta)\ne 0$ in
superalgebra $D^k_{ext}$ in the main by means of GNA, GTA transformations
without violation  of Eqs.(8.1), (8.5) form.
\section{Functional Integral, Generating Functionals\protect \\
of Green's Functions}
\setcounter{equation}{0}

Let us define for Abelian hypergauge given by means of superfunctions
\renewcommand{\theequation}{\arabic{section}.\arabic{equation}\alph{lyter}}
\begin{eqnarray}
\setcounter{lyter}{1}
{} & {} & G_B(\Gamma(\theta)) = \Phi^{\ast}_B(\theta) +
\displaystyle\frac{\partial\Psi(\Phi(\theta))}{\partial\Phi^B(\theta)
\phantom{xx}} = 0 \;,
\\
\setcounter{equation}{1}
\setcounter{lyter}{2}
{} & {} & \left(G_B(\Gamma(\theta)),G_A(\Gamma(\theta))\right)^{(\Gamma)}_{
\theta}=0,\ A,B=1,\ldots,n+3m 
\end{eqnarray}
in correspondence with (9.1) the generating functional (superfunction) of
Green's functions  by the formula
\begin{eqnarray}
\setcounter{lyter}{1}
{} & Z\bigl({\stackrel{\circ}{\Gamma}}(\theta),
\Phi^{\ast}(\theta), {\cal J}^{\ast}(\theta)\bigr) =
\displaystyle\int
d\Phi(\theta){\rm exp}\left\{\displaystyle\frac{\imath}{\hbar}\left(
{S}^{\Psi}_{H}\bigl({\Gamma}(\theta),
{\stackrel{\circ}{{\Gamma}}}(\theta),\hbar\bigr) + {\cal J}^{\ast
}_A(\theta)\Phi^A(\theta)\right)\right\}, & {}
\\
\setcounter{equation}{2}
\setcounter{lyter}{2}
{} & (\tilde{P}_1(\theta),{\rm gh}, \varepsilon_P, \varepsilon_{\bar{J}},
\varepsilon)Z(\theta)=(0,0,0,0,0)\;. {} & 
\end{eqnarray}
Integration measure $d\Phi(\theta)$ in (10.2a) is defined by
formal relation
\renewcommand{\theequation}{\arabic{section}.\arabic{equation}}
\begin{eqnarray}
d\Phi(\theta) = \prod\limits_A^r d\Phi^A(\theta),\
A=1,\ldots,[r]\ ({\rm for\ fixed}\ \theta) 
\end{eqnarray}
with infinite formal product of measures with respect to all possible values
of index $A=1,\ldots,n+3m$.
According to footnote after (9.23) the sign of $\Psi(\Phi(\theta))$ in
(10.2a) is replaced on opposite one in contrast to (9.18), (9.22).

The new additional to $\Gamma_{ext}(\theta)$,
${\stackrel{\circ}{\Gamma}}_{ext}(\theta)$ superfields ${\cal
J}^{\ast}_A(\theta)$ were introduced in (10.2) defined on
$\tilde{\Lambda}_{D\vert Nc + 1}(z^{a},\theta;{\bf K})$ and being transformed
with respect to supergroup $J$ superfield representation $\tilde{T}$
(being conjugate to $T$ [1,2] with respect to $\varepsilon$-even
nondegenerate bilinear form), playing the role of sources for Green's
functions (superfunctions) and possessing the properties
\begin{eqnarray}
({\rm gh}, \varepsilon_P, \varepsilon_{\bar{J}},\varepsilon){\cal J}^{\ast
}_A(\theta)=(-{\rm gh}(\Phi^A),\varepsilon_P(\Phi^A),\varepsilon_{\bar{J}
}(\Phi^A),\varepsilon_A)\;.  
\end{eqnarray}
Therefore superfunction $Z\bigl({\stackrel{\circ}{\Gamma}}(\theta),
\Phi^{\ast}(\theta), {\cal J}^{\ast}(\theta)\bigr)$ belongs to
$D^{k}_{ext}$ $\times$ $C^{l}({\cal M}_{{\cal J}^{\ast}})$, $l \leq \infty$,
where ${\cal M}_{{\cal J}^{\ast}}$ is a set parametrized by various
configurations of the superfields ${\cal J}^{\ast}_{A}(\theta)$.
In what follows we will call
${\cal J}^{\ast}_{A}(\theta)$ by supersources. All properties of the
superalgebra $D^{k}_{ext}$ are naturally continued onto superalgebra
$D^{k}_{ext}\times C^{l}({\cal M}_{{\cal J}^{\ast}})$.
\subsection{Functional Integral of Superfunctions}

Let us define an functional integral (FI) of superfunctions as
quasi-Gaussian superfunctions on B-algebra [8] (in the framework of
perturbation theory) with given and sufficient set
of generating elements being by superfields which are transformed with
respect to completely definite supergroup $J$ superfield representation
and depend on fixed $\theta$ by scheme described, for instance, in Ref.[8,9]
for $\theta=0$. To this end  introduce for Gaussian superfunction
\begin{eqnarray}
{\cal F}_{G}(\theta) = {\cal F}_{G}(\Phi(\theta), {\cal J}^{\ast}(\theta),
\theta) = {\rm exp}
\left[\frac{\imath}{\hbar}\left(\frac{1}{2}\Phi^A(\theta)S_{AB}(\theta)
\Phi^B(\theta) + {\cal J}^{\ast}_A(\theta)\Phi^A(\theta)\right)\right] 
\end{eqnarray}
the FI of the form
\begin{eqnarray}
{} &
\displaystyle\int d\Phi(\theta)
{\cal F}_{G}(\theta) =
{\rm sdet}^{-\frac{1}{2}}\left\|\hbar^{-1}S^{s_1}_{AB}(\theta)\right\|
{\rm exp}\left(- \displaystyle\frac{\imath}{2\hbar}{\cal J}^{\ast}_A(\theta)
\Lambda^{AB}(\theta){\cal J}^{\ast}_B(\theta)\right), {} &\\ 
{} & \Lambda^{AB}(\theta)=(S^{s_1}(\theta))^{-1{}AB}(-1)^{\varepsilon_B},\
\Lambda^{T_1}(\theta)=\Lambda(\theta){} & 
\end{eqnarray}
with $T_{1}$-symmetric supermatrix $S^{s_1}_{AB}$ defined in the following
way [8]
\begin{eqnarray}
\left\|S^{s_1}_{AB}(\theta)\right\| = \frac{1}{2}\left\|S_{AB}(\theta)+
S^{T_1}_{AB}(\theta)\right\|= \frac{1}{2}
\left\|
\begin{array}{cc}
S_{1{}AB} + S^T_{1{}AB} &  S_{2{}AB}- S^T_{3{}AB}\\
S_{3{}AB} - S^T_{2{}AB} &  S_{4{}AB}- S^T_{4{}AB}
\end{array} \right\|(\theta)
. 
\end{eqnarray}
The nondegeneracy of supermatrix (10.8) is defined by its
$P_{0}(\theta)$-component supermatrix [1].

Let us define the following FI
\begin{eqnarray}
{} & \displaystyle\int d\Phi(\theta)
\vec{\Phi}{}^{(A)_k}(\theta){\cal F}_{G}(\theta) =
\displaystyle\frac{\partial_l^k \phantom{xxxxxxxxxxxxxxxxxxx}}{
\partial\Bigl(\frac{\imath}{\hbar}{\cal J}^{\ast}_{A_1}(\theta)\Bigr)\ldots
\partial\Bigl(\frac{\imath}{\hbar}{\cal J}^{\ast}_{A_k}(\theta)\Bigr)}
\displaystyle\int d\Phi(\theta) {\cal F}_{G}(\theta) = {} & \nonumber \\
{} & \hspace{-0.8em}
{\rm sdet}^{-\frac{1}{2}}\left\|\hbar^{-1}S^{s_1}_{AB}(\theta)\right\|
\displaystyle\frac{\partial_l^k \phantom{xxxxxxxxxxxxxxxxxxx}}{
\partial\left(\frac{\imath}{\hbar}{\cal J}^{\ast}_{A_1}(\theta)\right)\ldots
\partial\left(\frac{\imath}{\hbar}{\cal J}^{\ast}_{A_k}(\theta)\right)}
{\rm exp}\left(- \displaystyle\frac{\imath}{2\hbar}{\cal J}^{\ast
}_A(\theta)\Lambda^{AB}(\theta){\cal J}^{\ast}_B(\theta)\right). {} &
\end{eqnarray}
One can calculate by means of (10.9) the FIs of quasi-Gaussian superfunctions
of the form
\renewcommand{\theequation}{\arabic{section}.\arabic{equation}\alph{lyter}}
\begin{eqnarray}
\setcounter{lyter}{1}
{} & {\cal F}_{qG}(\theta) = {\cal F}_{qG}(\Phi(\theta),{\cal J}^{\ast}(
\theta),\theta) = {\cal F}(\Phi(\theta),\theta)
{\cal F}_{G}(\theta)\;, {} & \\
\setcounter{equation}{10}
\setcounter{lyter}{2}
{} & {\cal F}(\Phi(\theta),\theta) =
\displaystyle\sum\limits_{l\geq 0}{\cal F}_{(A)_l}(\theta)\vec{\Phi}{}^{
(A)_l}(\theta),\ {\cal F}(\Phi(\theta),\theta) \in C^k({\cal M}_{ext}\times
\{\theta\})\;. {} & 
\end{eqnarray}
Taking representation (10.10) into consideration we obtain the basic
relationship to calculate of arbitrary FIs in question
\renewcommand{\theequation}{\arabic{section}.\arabic{equation}}
\begin{eqnarray}
{} &  \displaystyle\int d\Phi(\theta){\cal F}_{qG}(\theta)
= \displaystyle\sum\limits_{k\geq 0}\displaystyle\int d\Phi(\theta)
{\cal F}_{(A)_k}(\theta)
\displaystyle\frac{\partial_l^k \phantom{xxxxxxxxxxxxxxxxxxx}}{
\partial\Bigl(\frac{\imath}{\hbar}{\cal J}^{\ast}_{A_1}(\theta)\Bigr)\ldots
\partial\Bigl(\frac{\imath}{\hbar}{\cal J}^{\ast}_{A_k}(\theta)\Bigr)}
{\cal F}_{G}(\theta)= {} & \nonumber \\
{} & {\cal F}\left(\displaystyle\frac{\partial_l\phantom{xxxxxx}}{
\partial\Bigl(\frac{\imath}{\hbar}{\cal J}^{\ast
}(\theta)\Bigr)},\theta\right)
\displaystyle\int d\Phi(\theta){\cal F}_{G}(\theta) = & {} \nonumber \\
{} &  {\rm sdet}^{-\frac{1}{2}}\left\|\hbar^{-1}S^{s_1}_{AB}(\theta)\right\|
{\cal F}\left(\displaystyle\frac{\partial_l\phantom{xxxxxx}}{
\partial\Bigl(\frac{\imath}{\hbar}{\cal J}^{\ast
}(\theta)\Bigr)},\theta\right)
{\rm exp}\left(- \displaystyle\frac{\imath}{2\hbar}{\cal J}^{\ast
}_A(\theta)\Lambda^{AB}(\theta){\cal J}^{\ast}_B(\theta)\right). {} &
\end{eqnarray}
FI (10.11) satisfies to the standard properties of FI [8,9]. At first the
result
of its calculation does not depend on the integration order. The FI is
invariant under shifts of integration variable and vanishes in calculating
of the total superfield derivative on $\Phi^{B}(\theta)$ of an
arbitrary superfunction over any of the integration trajectories. A derivative
with respect to parametric superfunction in acting on the integrand is
commutative  with the sign $\smallint d\Phi(\theta)$.
The $\theta$-superfield   analog for change of variables
formula in FI is valid. The way to introduce of superfunctional
$\delta$-function has the following integral representation provided by
introduction of new superfields ${\cal P}_{A}(\theta)$
\renewcommand{\theequation}{\arabic{section}.\arabic{equation}\alph{lyter}}
\begin{eqnarray}
\setcounter{lyter}{1}
{} & \delta(\hbar^{-1}\varphi(\Phi(\theta))) = \displaystyle\int d{\cal P}(
\theta)
{\rm exp}\left(\frac{\imath}{\hbar}\varphi^A(\Phi(\theta)){\cal P}_A(\theta)
\right), {} & \\
\setcounter{equation}{12}
\setcounter{lyter}{2}
{} & (\varepsilon_P,\varepsilon_{\bar{J}},\varepsilon){\cal P}_A(\theta)=
(\varepsilon_P,\varepsilon_{\bar{J}},\varepsilon)\varphi^A(\Phi(\theta)),\
\varphi^A(\Phi(\theta)) \in C^k({\cal M}_{ext})\;, & {} 
\end{eqnarray}
where the equations
\renewcommand{\theequation}{\arabic{section}.\arabic{equation}}
\begin{eqnarray}
y^A(\theta)  = \varphi^A(\Phi(\theta)),\  \
\Phi^A(\theta) = \Psi^A(y(\theta))\;,  
\end{eqnarray}
have unique solutions respectively for unknown superfields $\Phi(\theta)$ in
the 1st subsystem in
(10.13) and for unknowns $y(\theta)$ in the 2nd one.
\subsection{Generating Functionals  of Green's Functions}

Let us rewrite FI (10.2a) by means of formula (10.12) in the form
\begin{eqnarray}
{} & Z\bigl({\stackrel{\circ}{\Gamma}}(\theta),\Phi^{\ast
}(\theta),{\cal J}^{\ast}(\theta)\bigr) =\displaystyle\int
d\Phi(\theta)d\lambda(\theta)d{\Phi'}^{\ast}(\theta)
d{\stackrel{\circ}{\Phi}}{}'^{\ast}(\theta)d\pi(\theta)\times
{} & \nonumber \\
{} &  {\rm exp}\Bigl\{\displaystyle\frac{\imath}{\hbar}\Bigl(
{S}_{H}\bigl(\Phi(\theta),\Phi'^{\ast}(\theta),{\stackrel{\circ}{\Phi}
}(\theta),{\stackrel{\circ}{\Phi}}{}'^{\ast}(\theta),\hbar\bigr) + \Bigl({
\stackrel{\circ}{\Phi}}{}'^{\ast}_{A}(\theta) -
{\stackrel{\circ}{\Phi}}{}^{\ast}_{A}(\theta)\Bigr)\pi^A(\theta)+
{} & \nonumber \\
{} &  \displaystyle\frac{\partial\phantom{xxxx}}{\partial\Phi^A(\theta)}[(
\Phi'^{\ast}_{B}(\theta)- {\Phi}^{\ast}_{B}(\theta))\Phi^B(\theta) -
\Psi(\Phi(\theta))]\lambda^A(\theta) + {\cal J}^{\ast
}_A(\theta)\Phi^A(\theta)\Bigr)\Bigr\}, & {}
\end{eqnarray}
with both the new independent set of superantifields $\Phi'^{\ast}_B(\theta)$
together with their derivatives on $\theta$ ${\stackrel{\circ}{\Phi}}{}'^{
\ast}_B(\theta)$ and superfields $\lambda^{B}(\theta)$, $\pi^{B}(\theta)$
being by the new generating elements of corresponding B-algebra [8] and
possessing by the properties
\begin{eqnarray}
{} & ({\rm gh},\varepsilon_P,\varepsilon_{\bar{J}},\varepsilon)\lambda^A(
\theta)= ({\rm gh}(\Phi^A) + 1,\varepsilon_P(\Phi^A) +1,
\varepsilon_{\bar{J}}(\Phi^A),\varepsilon_A + 1)\;,& {} \nonumber \\
{} & ({\rm gh},\varepsilon_P,\varepsilon_{\bar{J}},\varepsilon)\pi^A(
\theta)= ({\rm gh}(\Phi^A),\varepsilon_P(\Phi^A),
\varepsilon_{\bar{J}}(\Phi^A),\varepsilon_A)\;.& {}
\end{eqnarray}
The third summand in the exponent (10.14) can be rewritten with regard for
definition (10.1a) in the form
\begin{eqnarray}
\frac{\partial\phantom{xxxx}}{\partial\Phi^A(\theta)}[(
\Phi'^{\ast}_{B}(\theta)- {\Phi}^{\ast}_{B}(\theta))\Phi^B(\theta) -
\Psi(\Phi(\theta))]\lambda^A(\theta)=
- G_A\Bigl({\Phi}(\theta),{\Phi}^{\ast}(\theta) -
\Phi'^{\ast}(\theta)\Bigr)\lambda^A(\theta). 
\end{eqnarray}
\underline{\bf Remark}: There is not a necessity in integration variables
$\pi^{B}(\theta)$ and ${\stackrel{\circ}{\Phi}}{}'^{\ast}_{B}(\theta)$
for $S_{H}(\Gamma(\theta),\hbar)$ $\in$ $C^{k}(T^{*}_{odd}{\cal M}_{ext})$ in
FI (10.14). In addition the result of integration with respect to above
superfields is trivially equal to ${\rm sdet}\left\|\hbar\delta^A{}_B
\right\|$ in any case.

Introduce the generating functional (superfunction) of connected
correlated Green's functions by the relation
\begin{eqnarray}
W\bigl({\stackrel{\circ}{\Gamma}}(\theta),\Phi^{\ast
}(\theta),{\cal J}^{\ast}(\theta)\bigr) =
\frac{\imath}{\hbar}{\rm ln}Z\bigl({
\stackrel{\circ}{\Gamma}}(\theta),\Phi^{\ast}(\theta),{\cal J}^{\ast}(\theta)
\bigr) 
\end{eqnarray}
and effective action (EA) by means of Legendre transform of $W(\theta)$ with
respect to supersources ${\cal J}^{\ast}_{A}(\theta)$ with help of the
average superfields $\langle\Phi^{A}(\theta)\rangle$ introduction
\renewcommand{\theequation}{\arabic{section}.\arabic{equation}\alph{lyter}}
\begin{eqnarray}
\setcounter{lyter}{1}
{} & \langle\Phi^{A}(\theta)\rangle = \displaystyle\frac{\partial_l
W\bigl({\stackrel{\circ}{\Gamma}}(\theta),\Phi^{\ast
}(\theta),{\cal J}^{\ast}(\theta)\bigr)}{\partial{\cal J}_A^{\ast}(\theta)
\phantom{xxxxxxxxxxxxx}}, & {}    \\
\setcounter{equation}{18}
\setcounter{lyter}{2}
{} & {\mbox{\boldmath$\Gamma$}}\bigl({\stackrel{\circ}{\Gamma}}(\theta),
\langle\Phi(\theta)\rangle,\Phi^{*}(\theta)\bigr) =
W\bigl({\stackrel{\circ}{\Gamma}}(\theta),\Phi^{\ast
}(\theta),{\cal J}^{\ast}(\theta)\bigr) -
{\cal J}_A^{\ast}(\theta)\langle\Phi^A(\theta)\rangle\;, & {} 
\end{eqnarray}
where ${\cal J}^{\ast}_{A}(\theta)$ are expressed as functions of
$\langle\Phi^{A}(\theta)\rangle$ from Eq.(10.18a). From (10.18) it follows
\renewcommand{\theequation}{\arabic{section}.\arabic{equation}}
\begin{eqnarray}
{\cal J}^{\ast}_{A}(\theta) = - \frac{\partial{\mbox{\boldmath$\Gamma$}}
\bigl({\stackrel{\circ}{\Gamma}}(\theta),\langle\Phi(\theta)\rangle,\Phi^{*}(
\theta)
\bigr)}{\partial \langle\Phi^A(\theta)\rangle\phantom{xxxxxxxxxx}}\,.
\end{eqnarray}
\underline{\bf Remark:} The superfunctions $S_{H}(\Gamma(\theta),\hbar)$
$\in$ $C^{k}(T^{*}_{odd}{\cal M}_{ext})$, $Z(\Phi^{\ast}(\theta)$, $
{\cal J}^{\ast}(\theta))$, $W(\Phi^{\ast}(\theta)$, ${\cal J}^{
\ast}(\theta))$ $\in$ $C^{k}(\tilde{T}^{*}_{odd}{\cal M}_{ext})$
and {\boldmath$\Gamma$}($\langle\Phi(\theta)\rangle, \Phi^{*}(\theta))$
$\in$ $C^k(T^{\ast}_{odd}\langle{\cal M}_{ext}\rangle)$,
$\tilde{T}^{\ast}_{odd}{\cal M}_{ext}$ = $\{({\cal J}^{\ast}_{A}(\theta)$,
$\Phi^{\ast}_{A}(\theta))\}$,
$T^{*}_{odd}\langle{\cal M}_{ext}\rangle$ = $\{(\langle\Phi^{A}(
\theta)\rangle$, $\Phi^{\ast}_{
A}(\theta))\}$ present the real interest. All they are obtained by
means of restriction of their analogs on hypersurface
${\stackrel{\circ}{\Gamma}}(\theta)=0$ that corresponds to a parametric
dependence of those superfunctions upon ${\stackrel{\circ}{\Gamma}}(\theta)$.
\subsection{Properties of Generating Functionals}

Let us define a so-called vacuum functional (superfunction)
\begin{eqnarray}
Z_{\Psi}\bigl({\stackrel{\circ}{\Gamma}}(\theta),\Phi^{\ast}(\theta)\bigr)
\equiv Z\bigl({\stackrel{\circ}{\Gamma}}(\theta),\Phi^{\ast}(\theta),0\bigr)
\;. 
\end{eqnarray}
Translation with respect to $\theta $ of arbitrary ${\cal F}(\theta,\hbar)$
$\equiv$ ${\cal F}\bigl(\Gamma(\theta),{\stackrel{\circ}{\Gamma}}(\theta),
\theta,\hbar\bigr)$ $\in$ $D^{k}_{ext}$ along integral curve
$\check{\Gamma}(\theta)$ of the HS type system (9.29) with $S_{H{}s}^{
\Psi_s}\bigl(
\Gamma_{s}(\theta)$, ${\stackrel{\circ}{\Gamma}}_s(\theta), \hbar\bigr)$
$\equiv$ $S_{H}^{\Psi}\bigl(
\Gamma(\theta)$, ${\stackrel{\circ}{\Gamma}}(\theta), \hbar\bigr)$, $s=ext$,
being its integral, on a constant parameter $\mu$  $\in$ ${}^{1}\Lambda_{1
}(\theta)$ has the form\footnote{in what follows in Sec.X the value of sign
s is equal to $ext$ and is omitted unless otherwise said}
\begin{eqnarray}
{} & \delta_{\mu}{\cal F}(\theta,\hbar)_{\mid\check{\Gamma}(\theta)}=
\displaystyle\frac{d_r{\cal F}(\theta,\hbar)}{d\theta\phantom{xxxxxx}
}_{\mid\check{\Gamma}(\theta)}\mu= {} & \nonumber \\
{} & \Biggl[ \displaystyle\frac{\partial_r
{\cal F}\bigl(\check{\Gamma}(\theta),{
\stackrel{\circ}{\check{\Gamma}}}(\theta),\theta,\hbar\bigr)
}{\partial\theta\phantom{xxxxxxxxxxxxx}} +
\left({\cal F}\bigl(\check{\Gamma}(\theta),{
\stackrel{\circ}{\check{\Gamma}}}(\theta),\theta,\hbar\bigr),
S_{H}^{\Psi}(\check{\Gamma}(\theta ),{\stackrel{\circ}{\check{\Gamma}}
}(\theta),\hbar\bigr)\right)_{\theta}\Biggr]\mu {} & 
\end{eqnarray}
and generators (right and left respectively) of this transformation are
defined by relationships
\begin{eqnarray}
\delta_{\mu}{\cal F}(\theta,\hbar)_{\mid\check{\Gamma}(\theta)}
\hspace{-0.1em} = \hspace{-0.1em}
{s^{\Psi}\hspace{-0.1em}(\theta)}{\cal F}\bigl(\check{\Gamma}(\theta),{
\stackrel{\circ}{\check{\Gamma}}}(\theta),\theta,\hbar\bigr)\mu\hspace{-0.1em}
=\hspace{-0.1em}\mu {s_l^{\Psi
}}\hspace{-0.1em}(\theta){\cal F}(\theta,\hbar)_{\mid\check{\Gamma}(\theta)},\
s_l^{\Psi}\hspace{-0.1em}(\theta)\hspace{-0.1em}= \hspace{-0.1em}
\displaystyle\frac{\partial_l}{\partial\theta} \hspace{-0.1em}-\hspace{-0.1em}
{\rm ad}_{S_H^{\Psi}\hspace{-0.1em}(\theta,\hbar)}.  
\end{eqnarray}
The change of variables in $T_{odd}(T^{*}_{odd}{\cal M}_{ext})$:
$\Gamma(\theta)$ $\to$ $\Gamma^{(1)}(\theta)$ = $\Gamma^{(1)}\bigl(\Gamma(
\theta), {\stackrel{\circ}{\Gamma}}(\theta), \hbar\bigr)$ corresponding to HS
type system (9.29)
\begin{eqnarray}
\hspace{-0.8em}\Gamma^{(1){}p}(\theta)\hspace{-0.2em}=
\hspace{-0.2em}{\rm exp}\{\hspace{-0.1em}\mu s_l^{\Psi}\hspace{-0.1em}(\theta)
\hspace{-0.1em}\}
\Gamma^p(\theta) \hspace{-0.2em}= \hspace{-0.2em}{\rm exp}\{\hspace{-0.2em}-
\mu{\rm ad}_{S_H^{\Psi}\hspace{-0.1em}(\theta,\hbar)}\hspace{-0.1em}\}
\Gamma^p(\theta) \hspace{-0.2em}=\hspace{-0.2em} \Gamma^p(\theta) +
\omega^{pq}(\theta)\displaystyle\frac{\partial_l
S_{H}^{\Psi}(\theta,\hbar\bigr)}{\partial\Gamma^q(\theta)\phantom{xx}}\mu
\end{eqnarray}
is given with help of action of the operator $\hat{T}_{1}(\mu S_{H}^{\Psi
}(\theta,\hbar))$ (7.4) and appears by AT (in the sense of definitions (9.3))
in $T_{odd}(T^{*}_{odd}{\cal M}_{ext})$. Really, we obtain by means of all
antibracket's properties the proof of this fact explicitly
\begin{eqnarray}
{} & \left(\Gamma^{(1){}p}(\theta),\Gamma^{(1){}q}(\theta)\right)^{(\Gamma)}_{
\theta} = \omega^{pq}(\theta) - \mu\Bigl[\left(\left(\Gamma^p(\theta),
S_{H}^{\Psi}(\theta,\hbar)\right)^{(\Gamma)}_{\theta},
\Gamma^q(\theta)\right)^{(\Gamma)}_{\theta}(-1)^{\varepsilon_p} - {} &
\nonumber\\
{} & - (-1)^{(\varepsilon_p+1)(\varepsilon_q+1)}(p\longleftrightarrow q)\Bigr]
= \omega^{pq}(\theta) + \left(\omega^{pq}(\theta),S_{H}^{\Psi
}(\theta,\hbar)\right)^{
(\Gamma)}_{\theta}\mu = \omega^{pq}(\theta)\;. {} & 
\end{eqnarray}
With allowance made for antibracket's properties and generating equation (8.1)
for $S_{H}^{\Psi}(\theta,\hbar)$ the following formulae hold as well
\renewcommand{\theequation}{\arabic{section}.\arabic{equation}\alph{lyter}}
\begin{eqnarray}
\setcounter{lyter}{1}
{} & {} &
\displaystyle\frac{d_r \Gamma^{(1){}p}(\theta)}{d\theta\phantom{xxxxx}}=
\displaystyle\frac{d_r \Gamma^{p}(\theta)}{d\theta\phantom{xxxx}} -
\displaystyle\frac{d_r }{d\theta}\Bigl[
\left(\Gamma^{p}(\theta),S_{H}^{\Psi}(\theta,\hbar)\right)^{(\Gamma)}_{
\theta}\Bigr]\mu\;,
\\
\setcounter{equation}{25}
\setcounter{lyter}{2}
{} & {} &
s^{\Psi}(\theta)\displaystyle\frac{d_r \Gamma^{p}(\theta)}{d\theta\phantom{
xxxx}
}_{\mid\check{\Gamma}(\theta)} =  \displaystyle\frac{1}{2}
\left(\Gamma^{p}(\theta),\left(S_{H}^{\Psi}(\theta,\hbar),S_{H}^{\Psi}(\theta,
\hbar)\right)^{(\Gamma)}_{\theta}\right)^{(\Gamma)}_{\theta}=0\;, \\ 
\setcounter{equation}{25}
\setcounter{lyter}{3}
{} & {} & {\stackrel{\circ}{\Gamma}}{}^{(1){}p}(\theta)_{
\mid\check{\Gamma}(\theta)} =
{\stackrel{\circ}{\Gamma}}{}^{p}(\theta)_{\mid\check{\Gamma}(\theta)}\;.
\end{eqnarray}
At last in the case with $S_{H}^{\Psi}(\theta, \hbar)$ $\in C^{k}(T^{*}_{odd
}{\cal M}_{ext})$ satisfying to master equation (6.2) and to its
consequence (6.9) the AT (10.23) not only has  the
change of variables Berezinian  [8] under AT being equal to 1
but also the unit Berezinian  for the corresponding change of variables under
the same AT realized only on the coordinates of
configuration space: $\Phi(\theta)$ $\to
\Phi^{(1)}(\theta)$
\begin{eqnarray}
\setcounter{lyter}{1}
{} &
{\rm Ber}\left\|
\displaystyle\frac{\partial_l\Gamma^{(1){}p}(\theta)}{\partial\Gamma^q(\theta)
\phantom{xx}}\right\|= {\rm exp}\left[{\rm Str}\,{\rm ln}\left\|
\delta_q{}^p(\theta) +
\displaystyle\frac{\partial_l\phantom{xxx}}{\partial\Gamma^q(\theta)
}\left(\Gamma^p(\theta),S_{H}^{\Psi}(\theta,\hbar)\right)^{\Gamma}_{\theta}\mu
\right\|\right] = & {} \nonumber \\
{} & \hspace{-2em} {\rm exp}\left[{\rm Str}\left(
\displaystyle\frac{\partial_l\phantom{xxx}}{\partial\Gamma^q(\theta)
}\left(\Gamma^p(\theta),S_{H}^{\Psi}(\theta,\hbar)\right)^{\Gamma}_{\theta}\mu
\right)\right] =
{\rm exp}\left(- 2\mu\Delta(\theta)S_{H}^{\Psi}(\theta,\hbar)
\right), & {}
\\
\setcounter{equation}{26}
\setcounter{lyter}{2}
{} &
{\rm Ber}\left\|
\displaystyle\frac{\partial_l\Phi^{(1){}A}(\theta)}{\partial\Phi^B(\theta)
\phantom{xx}}\right\|=
{\rm exp}\left(- \mu\Delta(\theta)S_{H}^{\Psi}(\theta,\hbar)
\right). & {}  
\end{eqnarray}
In obtaining of (10.26) the representations for $\Delta(\theta)$ of the form
(2.44a) have been made use.

But if $S_{H}^{\Psi}(\theta, \hbar)$ belongs to $D^{k}$ and ${\rm deg}_{
{\stackrel{\circ}{\Gamma}}(\theta)}S_{H}^{\Psi}$ is more than $0$ then it is
necessary for equality to 1 of the expressions (10.26) to require
the fulfilment of Eq.(8.8) in question in addition to Eq.(8.1).

\noindent
\underline{\bf Theorem 7} (basic properties of $Z_{\Psi}(\Phi^{*}(\theta))$,
$Z_{\Psi}\bigl({\stackrel{\circ}{\Gamma}}(\theta), \Phi^{*}(\theta)\bigr)$)

\noindent
{\bf I.} For the II class GThGT  with $S_{H}^{\Psi}(\Gamma(\theta),\hbar)$,
solvable HS (9.22), superfunction $Z_{\Psi}(\Phi^{*}(\theta))$ $
\equiv$
$Z_{\Psi}\bigl({\stackrel{\circ}{\Gamma}}(\theta), \Phi^{*}(\theta)\bigr)_{
\mid{\stackrel{\circ}{\Gamma}}(\theta) = 0}$ in the representation (10.2a)
the following statements are valid:

\noindent
1) integrand in FI (10.2a) for $Z_{\Psi}(\Phi^{*}(\theta))$ and superfunction
$Z_{\Psi}(\Phi^{*}(\theta))$ itself are invariant with respect
to AT (10.23) for ${\stackrel{\circ}{\Gamma}}(\theta) = 0$, i.e. the
formulae hold
\begin{eqnarray}
\setcounter{lyter}{1}
{} & Z_{\Psi}(\Phi^{*}(\theta))=Z_{\Psi}^{(1)}(\Phi^{(1){}*}(\theta))\;, & {}
\\
\setcounter{equation}{27}
\setcounter{lyter}{2}
{} & \hspace{-2.0em}Z_{\Psi}^{(1)}(\Phi^{(1){}*}(\theta))\hspace{-0.1em}
= \hspace{-0.3em}
\displaystyle\int \hspace{-0.3em} d\Phi^{(1)}(\theta)
{\rm exp}\left\{\hspace{-0.1em}\displaystyle\frac{\imath}{\hbar}
S_H^{(1){}\Psi}(\Gamma^{(1)}(\theta),\hbar)\hspace{-0.1em}\right\}
\hspace{-0.1em},\
S_H^{(1){}\Psi}(\Gamma^{(1)}(\theta),\hbar)\hspace{-0.1em}
= \hspace{-0.1em}S_H^{\Psi}(\Gamma(\theta),
\hbar); & {} 
\end{eqnarray}
2) integrand in FI (10.2a) for $Z_{\Psi}(\Phi^{*}(\theta))$ and
$Z_{\Psi}(\Phi^{*}(\theta))$ itself are invariant with respect to
change of variables (10.23) for ${\stackrel{\circ}{\Gamma}}(\theta) = 0$
\begin{eqnarray}
\setcounter{lyter}{1}
{} & Z_{\Psi}(\Phi^{*}(\theta))=Z_{\Psi}(\Phi^{(1){}*}(\theta))\;, & {}
 \\
\setcounter{equation}{28}
\setcounter{lyter}{2}
{} & Z_{\Psi}(\Phi^{(1){}*}(\theta))=
\displaystyle\int d\Phi^{(1)}(\theta)
{\rm exp}\left\{\displaystyle\frac{\imath}{\hbar}
S_H^{\Psi}(\Gamma^{(1)}(\theta),\hbar)\right\}\;;  
\end{eqnarray}
3) under change of the gauge fermion $\Psi(\Phi(\theta))$ by the rule
\renewcommand{\theequation}{\arabic{section}.\arabic{equation}}
\begin{eqnarray}
\Psi(\Phi(\theta)) \to \Psi^{(1)}(\Phi(\theta)) = \Psi(\Phi(\theta)) +
\delta\Psi(\Phi(\theta))\; 
\end{eqnarray}
the such that $S_{H}^{\Psi + \delta\Psi}(\Gamma(\theta),\hbar)$
satisfies to the conditions of Theorem 5 (in particular, the hypothesis
(9.13) for
$S_{H}^{\Psi + \delta \Psi}(\theta , \hbar)$ is fulfilled on $\Lambda_{\Phi}$
as well) the superfunction $Z_{\Psi}(\Phi^{*}(\theta))$ does not depend upon a
choice of gauge fermion $\Psi(\Phi(\theta))$ (10.29) with accuracy up to the
1st order with respect to $\delta\Psi(\Phi(\theta))$\footnote{the sign
$o(\delta\Psi(\Phi(\theta)))$ is understood as follows
$\lim\limits_{\left\|\delta\Psi(\theta)\right\|\to 0\phantom{xx}}(\frac{
\left\|o(\delta\Psi(\theta))\right\|}{\left\|\delta\Psi(\theta)\right\|
\phantom{x}})$ = 0,  where $\left\|\cdot\right\|$ is  a some norm in
$C^k(T^{\ast}_{odd}{\cal M}_{ext})$}
\begin{eqnarray}
Z_{\Psi+\delta\Psi}(\Phi^{*}(\theta))=
Z_{\Psi}(\Phi^{*}(\theta)) + \langle o(\delta\Psi(\Phi(\theta)))\rangle
Z_{\Psi}(\Phi^{*}(\theta))\;.  
\end{eqnarray}
{\bf II.} For GThGT formally defined with respect to $S_{H}^{\Psi
}\bigl(\Gamma(\theta), {\stackrel{\circ}{\Gamma}}(\theta), \hbar\bigr)$ with
solvable HS type system (9.29) and
$Z_{\Psi}\bigl({\stackrel{\circ}{\Gamma}}(\theta)$, $\Phi^{*}(\theta)\bigr)$
all statements from the 1st part are valid as well:
\begin{eqnarray}
{} \hspace{-10em}{\rm 1)} \hspace{10em}
Z_{\Psi}\bigl({\stackrel{\circ}{\Gamma}}(\theta),\Phi^{*}(\theta)\bigr) =
Z_{\Psi}^{(1)}\bigl({\stackrel{\circ}{\Gamma}}{}^{(1)}(\theta),\Phi^{(1){}*}(
\theta)\bigr)
\end{eqnarray}
under AT (10.23);
\begin{eqnarray}
{} \hspace{-10.2em}{\rm 2)} \hspace{10.2em}
Z_{\Psi}\bigl({\stackrel{\circ}{\Gamma}}(\theta),\Phi^{*}(\theta)\bigr) =
Z_{\Psi}\bigl({\stackrel{\circ}{\Gamma}}{}^{(1)}(\theta),\Phi^{(1){}*}(
\theta)\bigr)   
\end{eqnarray}
under change of variables (10.23);

3) in the framework of hypothesis {\bf I.}3) under formal exchange of
Theorem 5 onto Theorem 6 the following expression is valid
\begin{eqnarray}
Z_{\Psi+\delta\Psi}({\stackrel{\circ}{\Gamma}}(\theta),\Phi^{*}(\theta))=
Z_{\Psi}({\stackrel{\circ}{\Gamma}}(\theta),\Phi^{*}(\theta)) +
\langle o(\delta\Psi(\Phi(\theta)))\rangle Z_{\Psi}({\stackrel{\circ}{
\Gamma}}(\theta), \Phi^{*}(\theta))\;.
\end{eqnarray}
\underline{\bf Remark:} The notation $\langle{\cal F}\bigl(\Gamma(\theta),
{\stackrel{\circ}{\Gamma}}(\theta),\hbar\bigr)\rangle$ for average value of an
arbitrary superfunction ${\cal F}(\theta,\hbar)$ $\in D^{k}$
with respect to
$Z_{\Psi}\bigl({\stackrel{\circ}{\Gamma}}(\theta)$, $\Phi^{*}(\theta),
{\cal J}^{\ast}(\theta)\bigr)$
is introduced in relationships (10.30), (10.33) defined for
${\cal J}^{\ast}(\theta)$ = 0 by the formulae
\begin{eqnarray}
{} & \langle{\cal F}\bigl(\Gamma(\theta),{\stackrel{\circ}{\Gamma}}(\theta),
\hbar\bigr)\rangle=M\left({\cal F}\bigl(\Gamma(\theta),{\stackrel{
\circ}{\Gamma}}(\theta),\hbar\bigr)\right)
Z_{\Psi}^{-1}\bigl({\stackrel{\circ}{\Gamma}}(\theta),\Phi^{*}(\theta)
\bigr)\;, {} &   \\ 
{} &
M\left({\cal F}(\theta,\hbar)\right) =
\displaystyle\int d\Phi(\theta) {\cal F}(\theta,\hbar)
{\rm exp}\left\{\displaystyle\frac{\imath}{\hbar}
S_H^{\Psi}(\Gamma(\theta),{\stackrel{\circ}{\Gamma}}(\theta),\hbar)\right\}\;,
& {}  \\ 
{} & \hspace{-2em}
M\left({\cal F}(\theta,\hbar)\right) =
\left(
{\cal F}\left(\displaystyle\frac{\partial_l\phantom{xxxxxx}}{
\partial\Bigl(\frac{\imath}{\hbar}{\cal J}^{\ast
}(\theta)\Bigr)},\Phi^{\ast}(\theta),{\stackrel{\circ}{\Gamma}}(\theta),
\hbar\right)
Z\bigl({\stackrel{\circ}{\Gamma}}(\theta),\Phi^{*}(\theta),
{\cal J}^{\ast}(\theta)\bigr)\right){\hspace{-0.7em}\phantom{\Bigr)}}_{
\mid{\cal J}^{\ast}(\theta)=0}\,. {} &
\end{eqnarray}
These relations for ${\cal J}^{\ast }(\theta)\ne 0$ are modified to following
ones respectively
\begin{eqnarray}
{} & \langle{\cal F}\bigl(\Gamma(\theta),{\stackrel{\circ}{\Gamma}}(\theta),
\hbar\bigr)\rangle_{{\cal J}^{\ast}(\theta)}=
M\left({\cal F}(\theta,\hbar)\right)_{{\cal J}^{\ast}(\theta)}
Z^{-1}\bigl({\stackrel{\circ}{\Gamma}}(\theta),\Phi^{*}(\theta),
{\cal J}^{\ast}(\theta)\bigr)\;, {} &    \\ 
{} & M\left({\cal F}(\theta,\hbar)\right)_{{\cal J}^{\ast}(\theta)} =
\displaystyle\int d\Phi(\theta) {\cal F}(\theta,\hbar)
{\rm exp}\left\{\displaystyle\frac{\imath}{\hbar}\left(
S_H^{\Psi}(\Gamma(\theta),{\stackrel{\circ}{\Gamma}}(\theta),\hbar) +
{\cal J}^{\ast}_A(\theta)\Phi^A(\theta)\right)\right\},
& {}   
\end{eqnarray}
and in formula (10.36) it is necessary to take away from the right
the restriction ${\cal J}^{\ast}(\theta)=0$.
\vspace{1ex}

\noindent
\underline{Proof:} {\bf I.} From the formula
\begin{eqnarray}
d\Phi^{(1)}(\theta){\rm exp}\left\{\displaystyle\frac{\imath}{\hbar}
S_H^{(1){}\Psi}(\Gamma^{(1)}(\theta),\hbar)\right\}=d\Phi(\theta)
{\rm Ber}\left\|
\displaystyle\frac{\partial_l\Phi^{(1)}(\theta)}{\partial\Phi(\theta)
\phantom{xx}}\right\|{\rm exp}\left\{\displaystyle\frac{\imath}{\hbar}
S_H^{\Psi}(\Gamma(\theta),\hbar)\right\}
\end{eqnarray}
being by definition the rule of change of integration variables in FI,
from definitions (10.27b), formula (10.26b), consequence (6.9) for
Eq.(6.2) written for $S_{H}^{\Psi}(\Gamma(\theta), \hbar)$ it follows the
validity of  statement {\bf I.}1) (10.27a) under AT (10.23);

{\bf I.}2) Formula
\begin{eqnarray}
{} & d\Phi^{(1)}(\theta){\rm exp}\left\{\displaystyle\frac{\imath}{\hbar}
S_H^{\Psi}(\Gamma^{(1)}(\theta),\hbar)\right\} = d\Phi(\theta)
{\rm Ber}\left\|\displaystyle\frac{\partial_l\Phi^{(1)}(\theta)}{\partial\Phi(
\theta)\phantom{xx}}\right\|\times
{} & \nonumber \\ {} &
{\rm exp}\left\{\displaystyle\frac{\imath}{\hbar}\left(
S_H^{\Psi}(\Gamma(\theta),\hbar) + \left(S_H^{\Psi}(\Gamma(\theta),\hbar),
S_H^{\Psi}(\Gamma(\theta),\hbar)\right)^{(\Gamma)}_{\theta}\mu\right)\right\},
& {} 
\end{eqnarray}
arising from the rule for transformation of the measure in FI (10.28b) under
change of variables  (10.23), from expansion in Taylor's series of
$S_{H}^{\Psi}(\Gamma^{(1)}(\theta),\hbar)$ in powers of $\delta_{\mu}
\Gamma(\theta)$ = $\Gamma^{(1)}(\theta) - \Gamma(\theta)$ in a neighbourhood
of $\Gamma^{{p}}(\theta)$ and from Eqs.(6.2), (6.9) written for $S_{H}^{\Psi
}(\Gamma(\theta),\hbar)$ proves the correctness of (10.28a). That
fact  taking into account of specific character of AT (10.23) permits to
write equivalently for $Z_{\Psi}(\Phi^{\ast}(\theta))$ not defined on $T^{*
}_{odd}{\cal M}_{ext}$(!) the formulae
\begin{eqnarray}
{{\stackrel{\circ}{Z}}_{\Psi}(\Phi^{\ast}(\theta))}_{\mid\check{
\Gamma}(\theta)}=0,\ \;
\delta_{\mu}Z_{\Psi}(\Phi^{\ast}(\theta))_{\mid\check{\Gamma}(\theta)}=0\;.
\end{eqnarray}
Really regarding the operator $\frac{d\phantom{x}}{d\theta}$ by
commutative with  sign $\int d\Phi(\theta)$ (remind that $\theta$ is fixed in
FI, i.e. is the parameter) and not acting onto integration variables,
in this case $\Phi^B(\theta)$, we obtain correspondingly
\renewcommand{\theequation}{\arabic{section}.\arabic{equation}\alph{lyter}}
\begin{eqnarray}
\setcounter{lyter}{1}
{} & {\stackrel{\circ}{Z}}{}_{\Psi}(\Phi^{\ast}(\theta))=
\displaystyle\frac{\imath}{\hbar}\langle{\stackrel{\circ}{\Phi}}{}^{\ast}_A(
\theta)
\displaystyle\frac{\partial S_{H}^{\Psi}(\Gamma(\theta),\hbar)}{\partial\Phi^{
\ast}_A(\theta)\phantom{xxxx}}\rangle{Z}_{\Psi}(\Phi^{\ast}(\theta))\;, & {}
\\
\setcounter{equation}{42}
\setcounter{lyter}{2}
{} & {\stackrel{\circ}{Z}}{}_{\Psi}(\Phi^{\ast}(\theta))_{\mid\check{
\Gamma}(\theta)}= - \displaystyle\frac{\imath}{2\hbar}\langle\left(
{S}_{H}^{\Psi}(\Gamma(\theta),\hbar),{S}_{H}^{\Psi}(\Gamma(\theta),\hbar)
\right)_{\theta}\rangle{Z}_{\Psi}(\Phi^{\ast}(\theta))=0\;. & {}  
\end{eqnarray}
Thus one can say on invariance of $Z_{\Psi}(\Phi^{\ast}(\theta))$ under
its translation  with respect to $\theta$ along $\check{\Gamma}(\theta)$;

{\bf I.}3) Taking into account of the relation
\renewcommand{\theequation}{\arabic{section}.\arabic{equation}}
\begin{eqnarray}
{} & {S}_{H}^{\Psi+ \delta\Psi}(\Gamma(\theta),\hbar) = {\rm exp}\{{\rm ad}_{
\delta\Psi(\theta)}\}{S}_{H}^{\Psi}(\Gamma(\theta),\hbar) = {S}_{H}^{\Psi
}\left(\Phi(\theta),\Phi^{\ast}(\theta) + \displaystyle\frac{\partial\delta
\Psi(\theta)}{\partial\Phi(\theta)\phantom{x}},\hbar\right)= {} & \nonumber \\
{} & {S}_{H}^{\Psi}(\Gamma(\theta),\hbar) + \left(\delta\Psi(\Phi(\theta)),
{S}_{H}^{\Psi}(\Gamma(\theta),\hbar)\right)_{\theta} + o(
\delta\Psi(\Phi(\theta))) {} &   
\end{eqnarray}
we obtain for $Z_{\Psi + \delta\Psi}(\Phi^{\ast}(\theta))$ the representation
\begin{eqnarray}
Z_{\Psi + \delta\Psi}(\Phi^{\ast}(\theta))=
\int \hspace{-0.2em} d\Phi(\theta){\rm exp}\left\{\displaystyle\frac{\imath}{
\hbar}\left[{S}_{H}^{\Psi}(\Gamma(\theta),\hbar) + \left(\delta\Psi(\theta),
{S}_{H}^{\Psi}(\Gamma(\theta),\hbar)\right)_{\theta} + o(
\delta\Psi(\theta))\right]\right\}.
\end{eqnarray}
Let us perform the change of variables (10.23) in FI (10.44) now with
superfunction $\mu(\Phi(\theta))$ $\in C^{k}({\cal M}_{ext})$. This
transformation does not appear
by AT by virtue of fulfilment of the inequality following from (10.24)
\begin{eqnarray}
\left(\Gamma^p(\theta),{S}_{H}^{\Psi}(\Gamma(\theta),\hbar)\right)^{(
\Gamma)}_{\theta}
\left(\mu(\Phi(\theta)),\Gamma^q(\theta)\right)^{(\Gamma)}_{\theta} -
(-1)^{(\varepsilon_p+1)(\varepsilon_q+1)}(p
\longleftrightarrow q) \neq 0\;.  
\end{eqnarray}
Properties {\bf I.}1), {\bf I.}2) for $Z_{\Psi}(\Phi^{\ast}(\theta))$ for
arbitrary $\mu(\Phi(\theta))$ are not fulfilled.  Operator $ -
\mu(\Phi(\theta))$ $\times {\rm
ad}_{S_{H}^{\Psi}(\theta,\hbar)}$ in (10.23) not being now represented
in the form of operator $\hat{T}_1(\hat{\Psi}(\theta))$ (7.7) and therefore
does not appear  by GNA operator for $S_{H}^{\Psi}(\Gamma(\theta),\hbar)$.

As a result the successive chain of the equalities holds
\renewcommand{\theequation}{\arabic{section}.\arabic{equation}\alph{lyter}}
\begin{eqnarray}
\setcounter{lyter}{1}
{} &
Z_{\Psi + \delta\Psi}(\Phi^{(1){}\ast}(\theta))=
\displaystyle\int \hspace{-0.2em} d\Phi^{(1)}(\theta){\rm exp}
\Bigl\{\displaystyle\frac{
\imath}{\hbar}\Bigl[{S}_{H}^{\Psi}(\Gamma^{(1)}(\theta),\hbar) +
\left(\delta\Psi(\Phi^{(1)}(\theta)),
{S}_{H}^{\Psi}(\Gamma^{(1)}(\theta),\hbar)\right)^{(\Gamma^{(1)})}_{\theta}
 & {} \nonumber \\
{} & + o(\delta\Psi(\Phi^{(1)}(\theta)))\Bigr]\Bigr\} =
\displaystyle\int d\Phi(\theta)
{\rm Ber}\left\|\displaystyle\frac{\partial_l\Phi^{(1)}(\theta)}{\partial
\Phi(\theta)\phantom{xx}}\right\|
{\rm exp}\Bigl\{\displaystyle\frac{\imath}{\hbar}\Bigl[
S_H^{\Psi}(\Gamma(\theta),\hbar) +
{} &  \\ 
{} & \hspace{-1.0em}\left(S_H^{\Psi}(\Gamma(\theta),\hbar),
S_H^{\Psi}(\Gamma(\theta),\hbar)\right)^{(\Gamma)}_{\theta}\hspace{-0.3em}\mu
 \hspace{-0.1em}+\hspace{-0.2em} \left(\delta\Psi(\Phi(\theta)),
{S}_{H}^{\Psi}(\Gamma(\theta),\hbar)\right)^{(\Gamma)}_{\theta}
\hspace{-0.2em} + \hspace{-0.2em}
o(\delta\Psi(\Phi(\theta))) \hspace{-0.2em}+ \hspace{-0.2em}
{\cal F}(\Gamma(\theta),\hbar)\Bigr]\Bigr\},
{} & \nonumber \\
\setcounter{equation}{46}
\setcounter{lyter}{2}
{} &  {\rm min}{\rm deg}_{\mu(\theta)\delta\Psi(\theta)}
{\cal F}(\Gamma(\theta),\hbar) = 2\;,& {} \\ 
\setcounter{equation}{46}
\setcounter{lyter}{3}
{} & \hspace{-2em}
{\rm Ber}\left\|
\displaystyle\frac{\partial_l\Phi^{(1){}A}(\theta)}{\partial\Phi^B(\theta)
\phantom{xx}}\right\|\hspace{-0.2em}= \hspace{-0.2em}
{\rm exp}\left(-\left(\mu(\theta),S_{H}^{\Psi}(\Gamma(\theta),\hbar)
\right)^{(\Gamma)}_{\theta} \hspace{-0.2em}
- \mu(\theta)\Delta(\theta)S_{H}^{\Psi}(\Gamma(\theta),\hbar)+
o(\mu(\theta))\right). & {}
\end{eqnarray}
Choosing $\mu(\theta)$ in the  form
\renewcommand{\theequation}{\arabic{section}.\arabic{equation}}
\begin{eqnarray}
\mu(\Phi(\theta)) = \frac{\imath}{\hbar}\delta\Psi(\Phi(\theta))\;,  
\end{eqnarray}
and taking into account the properties for $S_{H}^{\Psi}(\Gamma(\theta),
\hbar)$
mentioned in proving of {\bf I.}1) and {\bf I.}2) parts of Theorem we obtain
\begin{eqnarray}
Z_{\Psi + \delta\Psi}(\Phi^{\ast}(\theta))=
Z_{\Psi}(\Phi^{\ast}(\theta)) + \frac{\imath}{\hbar}\int d\Phi(\theta)
o(\delta\Psi(\Phi(\theta)))
{\rm exp}\left\{\frac{\imath}{\hbar}
S_H^{\Psi}(\Gamma^{(1)}(\theta),\hbar)\right\},  
\end{eqnarray}
that with regard of definitions (10.34), (10.35) proves the formula (10.30)
together with the {\bf I} part of the Theorem. The {\bf II} one is fulfilled
by virtue of relations obtained in proving of the {\bf I}  part being
retained as well in the case of superfunctions $Z_{\Psi}(\theta)$ and
$S_{H}^{\Psi}(\theta, \hbar)$ nontrivially depending on
${\stackrel{\circ}{\Gamma}}(\theta)$ and  additionally in force of
formulae (10.25b,c). In particular, the following transformations for
$Z_{\Psi}(\theta)$, $S_{H}^{\Psi}(\theta,\hbar)$ hold  under AT (10.23)
\renewcommand{\theequation}{\arabic{section}.\arabic{equation}\alph{lyter}}
\begin{eqnarray}
\setcounter{lyter}{1}
{} &
Z_{\Psi}\bigl({\stackrel{\circ}{\Gamma}}{}^{(1)}(\theta),\Phi^{(1){}*}(\theta)
\bigr) =
Z_{\Psi}\bigl({\stackrel{\circ}{\Gamma}}(\theta),\Phi^{*}(\theta)\bigr) -
\displaystyle\frac{\imath}{\hbar}\langle\mu
\left(S_H^{\Psi}(\Gamma(\theta),{\stackrel{\circ}{\Gamma}}(\theta),\hbar),
S_H^{\Psi}(\Gamma(\theta),{\stackrel{\circ}{\Gamma}}(\theta),\hbar)\right)_{
\theta}\rangle\times {} & \nonumber \\
{} &\hspace{-2em}
Z_{\Psi}\bigl({\stackrel{\circ}{\Gamma}}(\theta),\Phi^{*}(\theta)\bigr)
+ \displaystyle\frac{\imath}{\hbar}\langle\delta_{\mu}{\stackrel{
\circ}{\Gamma}}{}^p(\theta)_{\mid\check{\Gamma}(\theta)}\displaystyle\frac{
\partial_l
S_{H}^{\Psi}(\theta,\hbar)}{\partial{\stackrel{\circ}{\Gamma}}{}^p(\theta)
\phantom{xxx}}\rangle
Z_{\Psi}\bigl({\stackrel{\circ}{\Gamma}}(\theta),\Phi^{*}(\theta)\bigr)=
Z_{\Psi}\bigl({\stackrel{\circ}{\Gamma}}(\theta),\Phi^{*}(\theta)\bigr),
{} &\\
\setcounter{equation}{49}
\setcounter{lyter}{2}
{} &
S_{H}^{\Psi}\bigl(\Gamma^{(1)}(\theta),{\stackrel{\circ}{\Gamma}}{}^{(1)}(
\theta),\hbar\bigr)=
S_{H}^{\Psi}\bigl(\Gamma(\theta),{\stackrel{\circ}{\Gamma}}(\theta),\hbar
\bigr) +
\left(S_H^{\Psi}(\Gamma(\theta),{\stackrel{\circ}{\Gamma}}(\theta),\hbar),
S_H^{\Psi}(\Gamma(\theta),{\stackrel{\circ}{\Gamma}}(\theta),\hbar)\right)_{
\theta}\mu  {} & \nonumber \\
{} & + \delta_{\mu}{\stackrel{\circ}{\Gamma}}{}^p(\theta)_{\mid\check{
\Gamma}(\theta)}\displaystyle\frac{\partial_l
S_{H}^{\Psi}(\theta,\hbar)}{\partial{\stackrel{\circ}{\Gamma}}{}^p(\theta)
\phantom{xxx}} =
S_{H}^{\Psi}\bigl(\Gamma(\theta),{\stackrel{\circ}{\Gamma}}(\theta),\hbar
\bigr)\;. {} &  
\end{eqnarray}
In proving of the property {\bf II.}3) in  (10.33) note that $S_{H}^{\Psi +
\delta \Psi}\bigl(\Gamma(\theta), {\stackrel{\circ}{\Gamma}}(\theta),
\hbar\bigr)$  is  constructed by means of GNA from
$S_{H}^{\Psi}\bigl(\Gamma (\theta )$,
${\stackrel{\circ}{\Gamma}}(\theta),\hbar\bigr)$.
At last the interpretation for
$Z_{\Psi}\bigl({\stackrel{\circ}{\Gamma}}(\theta),\Phi^{\ast}(\theta)\bigr)$
in the form (10.41), (10.42) as the "integral" of HS type system (9.29)
is valid taking account of the formulae (10.49).$_{\textstyle\Box}$

\noindent
\underline{\bf Corollary 7.1}

The properties 1,2,3 of Theorem 7 for $S_{H}^{\Psi}(\Gamma(\theta), \hbar)$
and $S_{H}^{\Psi}\bigl(\Gamma(\theta)$,
${\stackrel{\circ}{\Gamma}}(\theta),\hbar\bigr)$ from the I and II parts are
retained under restriction of HS (9.22) and HS type system (9.29) to
equations
\renewcommand{\theequation}{\arabic{section}.\arabic{equation}}
\begin{eqnarray}
{} & \displaystyle\frac{d_r{\Phi}^{A}(\theta)}{d\theta\phantom{xxxx}}=
\left(\Phi^A(\theta),
S^{\Psi}_{H}(\theta,\hbar)\right)_{\theta},\ \;
{\stackrel{\circ}{\Phi}}{}^{\ast}_A(\theta) = 0\;, {} & \nonumber \\
{} & S^{\Psi}_{H}(\theta,\hbar) \in \{
S^{\Psi}_{H}(\Gamma(\theta),\hbar), S^{\Psi}_{H}\bigl(\Gamma(\theta),
{\stackrel{\circ}{\Gamma}}(\theta),\hbar\bigr)\}\;. {} & 
\end{eqnarray}

The corollary is proved analogously to proof of the preceding Theorem with
significant simplifications. Corresponding change of variables constructed
via shift transformation with respect to $\theta $ on a constant parameter
$\mu \in {}^{1}\Lambda_{1}(\theta)$ along (projection on ${\cal M}_{ext}$ of
solutions $\check{\Gamma}(\theta)$ for (9.22), (9.29)) integral curves
$\tilde{\Gamma}(\theta)$ for systems (10.50) are realized by the
formulae
\renewcommand{\theequation}{\arabic{section}.\arabic{equation}\alph{lyter}}
\begin{eqnarray}
\setcounter{lyter}{1}
{} & \Phi^{(1){}A}(\theta) = \Phi^{A}(\theta) +
\displaystyle\frac{\partial S_H^{\Psi}(\theta,\hbar)}{\partial\Phi^{\ast
}_A(\theta)\phantom{xx}}\mu = {\rm exp}(\mu\tilde{s}{}_l^{\Psi}(\theta))
\Phi^A(\theta),\ \;
\Phi^{(1){}\ast}_A(\theta) = \Phi^{\ast}_{A}(\theta)\;, {} &\\ 
\setcounter{equation}{51}
\setcounter{lyter}{2}
{} & \tilde{s}{}_l^{\Psi}(\theta) = \displaystyle\frac{\partial_l}{
\partial\theta} +
\displaystyle\frac{\partial_r S_H^{\Psi}(\theta,\hbar)}{\partial\Phi^{\ast
}_A(\theta)\phantom{xxx}}\displaystyle\frac{\partial_l\phantom{xxxx}}{\partial
\Phi^A(\theta)}\;. {} &  
\end{eqnarray}
That transformation not being by AT has the unit Berezinian
(10.26b). The interpretation of the form (10.41), (10.42) for
$Z_{\Psi}(\Phi^{*}(\theta))$ in question is trivial now.
The deficiencies of system (10.50) have been already mentioned (in
particular $\left(\tilde{s}^{\Psi}_l(\theta)\right)^{2} \ne 0$).
It is the system to which the so-called BRST transformations in BV method
are related:
\renewcommand{\theequation}{\arabic{section}.\arabic{equation}}
\begin{eqnarray}
\delta_{\mu}\Gamma^p(\theta)_{\mid\tilde{\Gamma}(\theta)} =
\frac{d_r{\Gamma}^{p}(\theta)}{d\theta\phantom{xxxx}}{\hspace{-0.5em}
\phantom{\Bigr)}}_{\mid\tilde{\Gamma}(
\theta)}\mu = \mu\tilde{s}{}_l^{\Psi}(\theta){\Gamma}^{p}(\theta)\;.  
\end{eqnarray}
If instead of Eq.(6.2) for $S_{H}^{\Psi}(\Gamma(\theta),\hbar)$ and
Eq.(8.1) for $S_{H}^{\Psi}\bigl(\Gamma(\theta),
{\stackrel{\circ}{\Gamma}}(\theta),\hbar\bigr)$ the BV
similar equations (6.11) and (8.5) are fulfilled respectively, then none of
the
properties from Theorem 7 are realized for the corresponding GThGT. In order to restore
the important part of those properties, i.e. the 2nd and the 3rd ones in
(10.28), (10.30) for GThGT with $S_{H}^{\Psi}(\Gamma(\theta), \hbar)$
and in (10.32), (10.33) for formal GThGT with $S_{H}^{\Psi}(\Gamma(\theta),
{\stackrel{\circ}{\Gamma}}(\theta),\hbar\bigr)$
respectively it is necessary to modify the HS (9.22) and HS type system
(9.29) to  corresponding systems of the form (10.50).
\vspace{1ex}

\noindent
\underline{\bf Theorem 8} (basic properties of $Z_{\Psi}(\Phi^{\ast
}(\theta))$,
$Z_{\Psi}\bigl({\stackrel{\circ}{\Gamma}}(\theta), \Phi^{\ast}(\theta)\bigr)$
under validity of Eqs.(6.11), (8.5))

\noindent
Under fulfilment of Eq.(6.11) for $S_{H}^{\Psi}(\Gamma(\theta), \hbar)$
and Eq.(8.5) for  $S_{H}^{\Psi}(\Gamma(\theta),
{\stackrel{\circ}{\Gamma}}(\theta),\hbar\bigr)$ determining the GThGT with
restricted systems (10.50)  the following statements for $Z_{\Psi}(\Phi^{\ast
}(\theta ))$,
$Z_{\Psi}\bigl({\stackrel{\circ}{\Gamma}}(\theta), \Phi^{\ast}(\theta)\bigr)$
are true: \\
1) the integrand in FI (10.2a) determining  $Z_{\Psi}(\Phi^{\ast}(\theta))$,
$Z_{\Psi}\bigl({\stackrel{\circ}{\Gamma}}(\theta), \Phi^{\ast}(\theta)\bigr)$
and the last superfunctions themselves are invariant with respect to
corresponding changes of variables (10.51); \\
2) vacuum functionals do not depend on a choice of the gauge in the sense of
relations (10.30), (10.33) under change of gauge fermion by the rule described
in point {\bf I.}3) of Theorem 7.
\vspace{1ex}

\noindent
\underline{Proof} of the last statement repeats completely the proof of
preceding Theorem and its corollary with following additional corrections:

1) change of variables (10.51) possesses by non-unit Berezinian of the form
(10.26b);

2) instead of the condition that superfunctions $S_{H}^{\Psi}(\theta, \hbar)$
appear by the
integrals for corresponding system (10.50) it is necessary to use the
Eqs.(6.11), (8.5) respectively.$_{\textstyle\Box}$
\vspace{1ex}

\begin{sloppypar}
Besides the properties described in Theorems 7,8 for
$Z(\Phi^{\ast}(\theta), {\cal J}^{\ast}(\theta))$, $Z\bigl({\stackrel{\circ}{
\Gamma}}(\theta), \Phi^{\ast}(\theta), {\cal J}^{\ast
}(\theta)\bigr)$  from generating  equations (6.2), (8.1) written
for $S_{H}^{\Psi}(\Gamma(\theta), \hbar)$ and
$S_{H}^{\Psi}\bigl(\Gamma(\theta)$, ${\stackrel{\circ}{\Gamma}}(\theta),
\hbar\bigr)$ respectively it follows the Ward identities. They can be
obtained by 2 ways.
\end{sloppypar}

{\bf 1)} The first one is based on the functional averaging both of generating
equation
(6.2) for the II class GThGT with $S_{H}^{\Psi}(\Gamma(\theta), \hbar)$
and of Eq.(8.1) for GThGT with
$S_{H}^{\Psi}\bigl(\Gamma(\theta), {\stackrel{\circ}{\Gamma}}(\theta),
\hbar\bigr)$. We have the sequence of equalities  taking the notation for
$S_{H}^{\Psi}(\theta,\hbar)$ in (10.50) into consideration
\renewcommand{\theequation}{\arabic{section}.\arabic{equation}\alph{lyter}}
\begin{eqnarray}
\setcounter{lyter}{1}
{} & \hspace{-0.5em}
M\hspace{-0.2em}\left(\hspace{-0.2em}(S_{H}^{\Psi}(\theta,\hbar),S_{H}^{\Psi}(
\theta,\hbar))_{\theta}
\hspace{-0.2em}\right)_{\hspace{-0.2em}{\cal J}^{\ast}(\theta)}\hspace{-0.4em} =\hspace{-0.2em} 2
\hspace{-0.2em}
\displaystyle\int\hspace{-0.3em} d\Phi(\theta)
\displaystyle\frac{\partial S_H^{\Psi}(\theta,\hbar)}{\partial\Phi_A(\theta)
\phantom{xx}}
\displaystyle\frac{\partial S_H^{\Psi}(\theta,\hbar)}{\partial\Phi^{\ast
}_A(\theta)\phantom{xx}}
{\rm exp}\hspace{-0.1em}\left\{\hspace{-0.1em}\displaystyle\frac{\imath}{
\hbar}\left({S}^{\Psi}_{H}(\theta,\hbar)\hspace{-0.2em} + \hspace{-0.2em}
{\cal J}^{\ast}(\theta)\Phi(\theta)
\hspace{-0.1em}\right)\hspace{-0.1em}\right\} & {} \nonumber \\
{} & =
2\displaystyle\int d\Phi(\theta)
\displaystyle\frac{\partial S_H^{\Psi}(\theta,\hbar)}{\partial\Phi^{\ast
}_A(\theta)\phantom{xx}}
\left(\displaystyle\frac{\hbar}{\imath}
\displaystyle\frac{\partial\phantom{xxxx}}{\partial\Phi^A(\theta)} -
{\cal J}^{\ast}_A(\theta)\right)
{\rm exp}\left\{\displaystyle\frac{\imath}{\hbar}\left(
{S}^{\Psi}_{H}(\theta,\hbar) + {\cal J}^{\ast
}(\theta)\Phi(\theta)\right)\right\}= & {} \nonumber \\
{} & \hspace{-1.8em} -\displaystyle\frac{2\hbar}{\imath} \Biggl[
\displaystyle\int d\Phi(\theta)
\Delta(\theta){S}^{\Psi}_{H}(\theta,\hbar)
{\rm exp}\left\{\displaystyle\frac{\imath}{\hbar}\left(
{S}^{\Psi}_{H}(\theta,\hbar) + {\cal J}^{\ast
}(\theta)\Phi(\theta)\right)\right\} + {\cal J}^{\ast}_A(\theta)
\displaystyle\frac{\partial Z(\theta)\phantom{x}}{\partial\Phi^{\ast
}_A(\theta)}\Biggr]=0, & {}
 \\
\setcounter{equation}{53}
\setcounter{lyter}{2}
{} &
Z(\theta) \in \{
Z(\Phi^{\ast}(\theta),{\cal J}^{\ast}(\theta)), Z\bigl({\stackrel{\circ}{
\Gamma}}(\theta),\Phi^{\ast}(\theta),{\cal J}^{\ast}(\theta)\bigr)\}\;, {}&
\end{eqnarray}
where  the rules of integration by parts in FI and
differentiation with respect to parametric superantifields $\Phi^{\ast }_{A
}(\theta)$ have been made use.  Taking into account of consequence (6.9) for
Eq.(6.2) and  with allowance made for Eq.(8.8) for GThGT defined on
$D^{k}_{ext}$ we obtain Ward identities for $Z(\theta)$
\renewcommand{\theequation}{\arabic{section}.\arabic{equation}}
\begin{eqnarray}
{\cal J}^{\ast}_A(\theta)\frac{\partial Z(\theta)\phantom{x}}{\partial\Phi^{
\ast}_A(\theta)}=0\;.  
\end{eqnarray}
The functional averaging of the Eqs.(6.9) for $S_{H}^{\Psi}(\theta, \hbar)$
as in (10.50) leads to the same result (10.54) not revealing additional
properties for $Z(\theta)$ (10.53b).

Ward identities for GThGT with generating equations (6.11) for $S_{H}^{\Psi
}(\Gamma(\theta), \hbar)$ and (8.5) for
$S_{H}^{\Psi}\bigl(\Gamma(\theta)$, ${\stackrel{\circ}{\Gamma}}(\theta),
\hbar\bigr)$ are deduced by functional averaging of these equations and have
the same form (10.54).

{\bf 2)} The second way is based on the use  of the following system of
$3(n+3m)$ ODE in normal form constructed with respect to HS (9.22)
\begin{eqnarray}
\displaystyle\frac{d_r{\Gamma}^{p}(\theta)}{d\theta\phantom{xxx}}=
\left({\Gamma}^{p}(\theta), {S}_{H}^{\Psi}(\Gamma(\theta),\hbar)\right)_{
\theta},\ \ {\stackrel{\circ}{\cal J}}{}^{\ast}_A(\theta)=0\;. 
\end{eqnarray}
According to Theorem 7 (property {\bf I.}2)  perform a change of
variables of the form (10.23) with fixed ${\cal J}^{\ast}(\theta)$ in the
integrand of FI (10.2a). Obtain  as the consequence  of (10.41), (10.42) the
expression
\renewcommand{\theequation}{\arabic{section}.\arabic{equation}\alph{lyter}}
\begin{eqnarray}
\setcounter{lyter}{1}
{} \hspace{-1.5em}
{Z}(\Phi^{(1){}\ast}(\theta),{\cal J}^{(1){}\ast}(\theta))=
Z(\Phi^{\ast}(\theta),{\cal J}^{\ast}(\theta)) +
\delta_{\mu}Z(\Phi^{\ast}(\theta),{\cal J}^{\ast}(\theta))_{\mid\check{
\Gamma}(\theta)} = Z(\Phi^{\ast}(\theta),{\cal J}^{\ast}(\theta))\,. 
\end{eqnarray}
On the other hand the formula  being consistent with relations
(10.38), (10.42) is formally valid taking the result of expression (10.53a)
into account
\begin{eqnarray}
\setcounter{equation}{56}
\setcounter{lyter}{2}
\hspace{-1.0em}{\stackrel{\circ}{Z}}(\Phi^{\ast}(\theta),{\cal J}^{\ast}(
\theta)){\hspace{-0.5em}\phantom{\Bigl)}}_{\mid\check{\Gamma}(\theta)}
\hspace{-0.5em}=\hspace{-0.15em} - \frac{\imath}{2\hbar}
M\left((S_{H}^{\Psi}(\theta,\hbar),S_{H}^{\Psi}(\theta,\hbar))_{\theta}
\right)_{\hspace{-0.2em}{\cal J}^{\ast}} \hspace{-0.5em}=\hspace{-0.15em}
{\cal J}^{\ast}_A(\theta)\frac{\partial Z(\Phi^{\ast}(\theta),{\cal J}^{
\ast}(\theta))}{\partial\Phi^{
\ast}_A(\theta)\phantom{xxxxxxxx}}=0.   
\end{eqnarray}
Thus Ward identities for $Z(\Phi^{\ast}(\theta), {\cal J}^{\ast}(\theta))$
means the invariance of superfunction $Z(\theta)$ with respect to
translation on $\theta $ along integral curve of system (10.55).

Obtained derivations and interpretations of Ward identities are valid as well
for formal GThGT with $S_{H}^{\Psi}\bigl(\Gamma(\theta)$,
${\stackrel{\circ}{\Gamma}}(\theta), \hbar\bigr)$,
$Z\bigl({\stackrel{\circ}{\Gamma}}(\theta)$, $\Phi^{\ast}(\theta), {\cal
J}^{\ast}(\theta)\bigr)$  with Eq.(8.1).

There is not  of the similar interpretation for $Z(\theta)$ (10.53b),
for formal GThGTs described in the hypotheses of Theorem 8 with restricted
system (10.50) added by the 2nd subsystem in Eqs.(10.55), as it was made in
(10.56b). However the
Ward identities in this case are deduced by means of the 2nd way under
change of variables (10.51) in FI (10.2a).

For generating functional $W(\theta)$ $\in$ $\{W(\Phi^{\ast}(\theta),
{\cal J}^{\ast}(\theta))$,
$W\bigl({\stackrel{\circ}{\Gamma}}(\theta), \Phi^{\ast}(\theta), {\cal
J}^{\ast}(\theta)\bigr)\}$ the Ward identities have the
same form as for $Z(\theta)$
\renewcommand{\theequation}{\arabic{section}.\arabic{equation}}
\begin{eqnarray}
{\cal J}^{\ast}_A(\theta)\frac{\partial W(\theta)\phantom{x}}{\partial\Phi^{
\ast}_A(\theta)}=0\;.  
\end{eqnarray}
For GThGT and formal GThGT described in Theorem 7 the interpretation of the
form (10.56) is valid for $W(\theta)$ as well
\begin{eqnarray}
{\stackrel{\circ}{W}}(\theta){\hspace{-0.4em}\phantom{\bigl)}}_{
\mid\check{\Gamma}(\theta)} =
{\cal J}^{\ast}_A(\theta)\frac{\partial W(\theta)\phantom{x}}{\partial\Phi^{
\ast}_A(\theta)}=0\;.  
\end{eqnarray}
At last the Ward identities for EA
{\boldmath$\Gamma$}$\bigl({\stackrel{\circ}{\Gamma}}(\theta),
\langle\Phi(\theta)\rangle,
\Phi^{\ast}(\theta)\bigr)$  with allowance made for (10.19) take the form
\begin{eqnarray}
\left(\mbox{\boldmath$\Gamma$}\bigl({\stackrel{\circ}{\Gamma}}(\theta),
\langle\Phi(\theta
)\rangle,\Phi^{\ast}(\theta)\bigr),
\mbox{\boldmath$\Gamma$}\bigl({\stackrel{\circ}{\Gamma}}(\theta),
\langle\Phi(\theta
)\rangle,\Phi^{\ast}(\theta)\bigr)\right)^{(\langle\Phi\rangle,
\Phi^{\ast})}_{\theta} =0\;.
\end{eqnarray}
The HS type system defined with respect to superfunction
{\boldmath$\Gamma$}$\bigl({\stackrel{\circ}{\Gamma}}(\theta),
\langle\Phi(\theta)\rangle,
\Phi^{\ast}(\theta)\bigr)$
(and for ${\stackrel{\circ}{\Gamma}}(\theta)=0$ being by the real HS),
taking into account the fact that superantifields
$\Phi^{\ast}_A(\theta)$ coincide with corresponding  average ones,
\begin{eqnarray}
\frac{d_r\langle{\Gamma}^{p}(\theta)\rangle}{d\theta\phantom{xxxxx}}=
\left(\langle\Gamma^p(\theta)\rangle,
\mbox{\boldmath$\Gamma$}\bigl({\stackrel{\circ}{\Gamma}}(\theta),
\langle\Gamma(\theta)\rangle\bigr)\right)^{(\langle\Gamma\rangle)}_{\theta}
\end{eqnarray}
is  solvable. Ward
identities (10.59) reflect the fact that
{\boldmath$\Gamma$}$(\theta)$ is the integral for system (10.60).

The functional averaging of HS (9.22) and HS type system (9.29) by the
formula (10.37) leads to corresponding systems in terms of superfunctions
{\boldmath$\Gamma$}$(\theta)$ $\in$ $\{{\mbox{\boldmath$\Gamma$}}\bigl({
\stackrel{\circ}{\Gamma}}(\theta),\langle\Gamma(\theta)\rangle\bigr)$,
{\boldmath$\Gamma$}$(\langle\Gamma(\theta)\rangle)\}$, $Z(\theta)$,
$W(\theta)$  and $S_{H}^{\Psi}(\theta,\hbar)$
\begin{eqnarray}
{} & {} & \hspace{-3.5em}
\langle\hspace{-0.1em}\displaystyle\frac{d_r{\Phi}^{A}(\theta)}{d\theta
\phantom{xxxx}}
\hspace{-0.1em}\rangle\hspace{-0.2em} = \hspace{-0.2em}
\langle\hspace{-0.1em}\displaystyle\frac{\partial S_H^{\Psi}(\theta,\hbar)}{
\partial\Phi^{\ast
}_A(\theta)\phantom{xx}}\hspace{-0.1em}\rangle\hspace{-0.2em} =\hspace{-0.2em}
\displaystyle\frac{\hbar}{\imath}Z^{-1}(\theta)
\displaystyle\frac{\partial Z(\theta)\phantom{x}}{\partial\Phi^{\ast
}_A(\theta)}
,\
\langle\hspace{-0.1em}\displaystyle\frac{d_r{\Phi}^{\ast}_{A}(\theta)}{
d\theta\phantom{xxxx}}
\hspace{-0.1em}\rangle\hspace{-0.2em} =\hspace{-0.2em}  - \hspace{-0.0em}
\langle\hspace{-0.1em}\displaystyle\frac{\partial_l S_H^{\Psi}(\theta,
\hbar)}{\partial\Phi^A(\theta)\phantom{xx}}\hspace{-0.1em}\rangle
\hspace{-0.2em} = \hspace{-0.2em} (\hspace{-0.1em} -1\hspace{-0.1em} )^{
\varepsilon_A}\hspace{-0.1em}{\cal J}^{\ast}_A(\theta); \\ 
{} & {} & \hspace{-3.5em}
\langle\displaystyle\frac{d_r{\Phi}^{A}(\theta)}{d\theta\phantom{xxxx}}
\rangle=
\displaystyle\frac{\partial W(\theta)\phantom{}}{\partial\Phi^{\ast
}_A(\theta)},\ \;
\langle\displaystyle\frac{d_r{\Phi}^{\ast}_{A}(\theta)}{d\theta\phantom{xxxx}}
\rangle = (-1)^{\varepsilon_A}{\cal J}^{\ast}_A(\theta)\;; \\ 
{} & {} & \hspace{-3.5em}
\langle\displaystyle\frac{d_r{\Gamma}^{p}(\theta)}{d\theta\phantom{xxxx}}
\rangle=
\omega^{pq}(\theta)
\displaystyle\frac{\partial_l \mbox{\boldmath$\Gamma$}(\theta)\phantom{x}
}{\partial\langle\Gamma^q(\theta)\rangle}\;. 
\end{eqnarray}
The comparison  of Eqs.(10.60) with (10.63) leads to equality of their
left-hand sides
\begin{eqnarray}
\langle\displaystyle\frac{d_r{\Gamma}^{p}(\theta)}{d\theta\phantom{xxxx}}
\rangle=
\frac{d_r\langle{\Gamma}^{p}(\theta)\rangle}{d\theta\phantom{xxxxx}
}{\hspace{-0.5em}\phantom{\Bigr)}}_{\mid\check{\Gamma}(\theta)}\;,  
\end{eqnarray}
indicating on the permutability of averaging  sign with operator
$\frac{d\phantom{x}}{d\theta}$ on the coordinates of
$T^{\ast}_{odd}{\cal M}_{ext}$ along $\check{\Gamma}(\theta)$.  Further
investigation of systems (10.60), (10.63) and the connection of corresponding
generators $s^{\Psi}(\theta)$ and $\langle s^{\Psi}(\theta)\rangle$  appear
by the separate problems.

Finally, relation (10.58) for  EA  taking the   Legendre
transform properties (10.18), (10.19) into consideration has the form
\begin{eqnarray}
{\stackrel{\circ}{\mbox{\boldmath$\Gamma$}}}(\theta){\hspace{-0.4em}\phantom{
\bigl(}}_{\mid\check{\Gamma}(\theta)}
= - 2 \frac{\partial \mbox{\boldmath$\Gamma$}(\theta)\phantom{xx}
}{\partial\langle\Phi^A(\theta)\rangle}\frac{\partial \mbox{
\boldmath$\Gamma$}(
\theta)\phantom{xx}}{\partial\langle\Phi^{\ast}_A(\theta)\rangle} = -
\left(\mbox{\boldmath$\Gamma$}(\theta),
\mbox{\boldmath$\Gamma$}(\theta)\right)^{(\langle\Gamma\rangle)}_{\theta}=0\;.
\end{eqnarray}
Thus, Ward  identities for {\boldmath$\Gamma$}$(\theta)$ mean the invariance
of EA, in the sense of conventions (10.42), with respect to translation on
$\theta $ along integral curve $\check{\Gamma}(\theta)$ of corresponding
system (9.22), (9.29).  Moreover the solutions of system (10.60) appear
exactly by $\langle\check{\Gamma}(\theta)\rangle$!

Not going into details let us remark that properties of Theorem 7 for
$Z_{\Psi}\bigl({\stackrel{\circ}{\Gamma}}(\theta), \Phi^{\ast}(\theta)\bigr)$
written in the form (10.14) taking account of the formula (10.16) follow from
the next transformations of invariance for integrand written by means of
the 1st order on $\theta $ system  $6(n+3m)$ ODE
\renewcommand{\theequation}{\arabic{section}.\arabic{equation}\alph{lyter}}
\begin{eqnarray}
\setcounter{lyter}{1}
{} & {} & \hspace{-1em}
\displaystyle\frac{d_r\hat{\Gamma}^{p}(\theta)}{d\theta\phantom{xxxx}}=
\left(\hat{\Gamma}^{p}(\theta),
S_{H}\bigl(\hat{\Gamma}(\theta),{\stackrel{\circ}{
\hat{\Gamma}}}(\theta),\hbar\bigr) +
G_A\Bigl(\Phi(\theta),\Phi^{\ast}(\theta)-{\Phi'}^{\ast}(\theta)\Bigr)
\lambda^A(\theta)\right)^{(\hat{\Gamma})}_{\theta},
\\
\setcounter{equation}{66}
\setcounter{lyter}{2}
{} & {} & \hspace{-1em}
\displaystyle\frac{d_r\Phi^{\ast}_A(\theta)}{d\theta\phantom{xxxx}}=
\displaystyle\frac{d_r{\stackrel{\circ}{\Phi}}{}'^{\ast}_A(\theta)}{d
\theta\phantom{xxxxx}}=
\displaystyle\frac{d_r\lambda^{A}(\theta)}{d\theta\phantom{xxxx}}=
\displaystyle\frac{d_r\pi^{A}(\theta)}{d\theta\phantom{xxxx}}=0\;. 
\end{eqnarray}
where $\hat{\Gamma}^{p}(\theta)\equiv (\Phi^{A}(\theta), {\Phi'}^{\ast}_{A
}(\theta))$ and
$S_{H}\bigl(\hat{\Gamma}(\theta), {\stackrel{\circ}{\hat{\Gamma}}}(\theta),
\hbar\bigr)$
satisfies to the hypotheses of Theorem 4 (for
$S_{H}\bigl(\hat{\Gamma}(\theta), \hbar\bigr)$ $\in
C^{k}(T^{\ast}_{odd}\hat{\cal M}_{ext})$ to the conditions of Theorem 3).
Moreover
$S_{H}\bigl(\hat{\Gamma}(\theta)$, ${\stackrel{\circ}{
\hat{\Gamma}}}(\theta), \hbar\bigr)$ may satisfy to Eqs.(6.11) or (8.5)
nevertheless the same system (10.66) induces the invariance of integrand in
(10.14) with respect to corresponding to (10.66) change of variables.
Note that even the subsystem (10.66a) is not solvable.

By another important system of $6(n+3m)$ ODE of the 1st order on $\theta$
with the same variables, that in (10.66) and playing the same role as one
(10.66) it appears the system having following nonzero right-hand side
part
\begin{eqnarray}
\setcounter{lyter}{1}
{} & {} & \hspace{-1em}
\displaystyle\frac{d_r\Phi^A(\theta)}{d\theta\phantom{xxxx}}= \left(
\Phi^A(\theta),
G_B\Bigl(\Phi(\theta),\Phi^{\ast}(\theta)-{\Phi'}^{\ast}(\theta)\Bigr)
\lambda^B(\theta)
\right)^{(\hat{\Gamma})}_{\theta} = - \lambda^A(\theta)\;, \\
\setcounter{equation}{67}
\setcounter{lyter}{2}
{} & {} & \hspace{-1em}
\displaystyle\frac{d_r{\Phi'}^{\ast}_A(\theta)}{d\theta\phantom{xxxxx}}=
\left({\Phi'}^{\ast}_A(\theta),
S_{H}\bigl(\hat{\Gamma}(\theta),{\stackrel{\circ}{\hat{\Gamma}}}(\theta),
\hbar\bigr)\right)^{(\hat{\Gamma})}_{\theta} = (-1)^{\varepsilon_A +1}
\displaystyle\frac{\partial
S_{H}\bigl(\hat{\Gamma}(\theta),{\stackrel{\circ}{\hat{\Gamma}}}(\theta),
\hbar\bigr)}{\partial\Phi^A(\theta)\phantom{xxxxxxxx}}. 
\end{eqnarray}
This system does not satisfy to solvability condition but coincides
one-to-one with BRST transformations in BV method for
$Z\bigl({\stackrel{\circ}{\Gamma}}(\theta), \Phi^{\ast}(\theta), {\cal
J}^{\ast}\bigr)$ in form (10.14) taking the type (10.52) formula into
consideration for $\theta = {\stackrel{\circ}{\Gamma}}(
\theta) = 0$, ${\cal J}^{\ast}_{A}(0) = J_{A}$.
\section{Component Formulation of GSQM}
\setcounter{equation}{0}

The component (on $\theta$) formulation of GSTF constructed in papers [1,2]
for objects and relations defined on ${\cal M}_{cl}$, $T_{odd}{\cal M}_{cl}$,
$T^{\ast}_{odd}{\cal M}_{cl}$, $T_{odd}(T^{\ast}_{odd}{\cal M}_{cl})$ is
directly transferred without any modifications on quantities and operations
determined on ${\cal M}_{s}$, $T_{odd}{\cal M}_{s}$,
$T^{\ast}_{odd}{\cal M}_{s}$, $T_{odd}(T^{\ast}_{odd}{\cal M}_{s})$,
$s\in C_0$ (2.27), (2.30) introduced in present work basically in Sec.II.
For completeness of presentation let us point out the following component on
$\theta $ representation for supervariables $J^{\ast}_{A_s}(\theta)$,
$\lambda^{A_s}_{s}(\theta)$, $\pi^{A_s}_{s}(\theta)$ (10.4), (10.15)
\renewcommand{\theequation}{\arabic{section}.\arabic{equation}}
\begin{eqnarray}
{\cal J}^{\ast}_{A_s}(\theta)=J^{\ast}_{0{}A_s} + J^{\ast}_{1{}A_s}\theta,\
\lambda^{A_s}_s(\theta)=\lambda^{A_s}_{0{}s} + \lambda^{A_s}_{1{}s}\theta,\
\pi^{A_s}_{s}(\theta) = \pi^{A_s}_{0{}s} + \pi^{A_s}_{1{}s}\theta,\ s\in
C_0\;. 
\end{eqnarray}
\section{Reduction to BV Method and to Superfield \protect \\
Quantization  from Paper [3]}
\setcounter{equation}{0}

GSQM in the Lagrangian formalism contains as the particular cases the BV
method [4] and the Lagrangian superfield method [3] for gauge theories
quantization.
\subsection{Connection with BV Method}

In order to pass to BV method inside of GSQM it
is necessary to restrict $C^{k}(T^{\ast}_{odd}{\cal M}_{ext}
\times\{\theta\})$
being used for description of GThGT (including the II class ones)
with $S_{H}^{\Psi}(\Gamma(\theta),\hbar)$ or superalgebra $D^{k}_{ext}$
being used for formal GThGT with
$S_{H}^{\Psi}\bigl(\Gamma(\theta),{\stackrel{\circ}{\Gamma}}(\theta),
\hbar\bigr)$ to $C^{k}(P_{0}(T^{\ast}_{odd}{\cal M}_{ext}))$. In the
restricted superalgebra  all possible products of $P_{0}\Gamma^{p
}(\theta)$ form the basis and all superfunctions are defined for $\theta
=0$.  It is the same superalgebra appears for
${\stackrel{\circ}{\Gamma}}(\theta)$ = $0$ by maximal
subsuperalgebra in $D^{k}_{ext}$ being invariant one with respect to
involution $*$ (2.42). Thus  the BV method quantities are singled out
in the invariant way as superfunctions from $C^{k}(T^{\ast}_{odd}{\cal
M}_{ext}\times\{\theta\})$ being invariant with respect to $*$.
Elements of $C^{k}(P_{0}(T^{\ast}_{odd}{\cal M}_{ext}))$ are
represented in the form
\begin{eqnarray}
{\cal F}\bigl(\Gamma_s(\theta),{\stackrel{\circ}{\Gamma}}_s(\theta),
\theta\bigr)_{\mid C^{k}(P_{0}(T^{\ast}_{odd}{\cal M}_{ext}))}
 = {\cal F}(P_0\Gamma_s(\theta),0,0)\equiv \overline{\cal
F}(\Gamma_{0{}s}),\;\Gamma^p_s(\theta)=\Gamma^p_{0{}s} + \Gamma^p_{1{}s}\theta
\end{eqnarray}
and appear by functionals in the usual sense for $s=ext$, i.e. by mappings
from $P_{0}(T^{\ast}_{odd}{\cal M}_{ext})$ into ${\bf K}$ (${\bf R}$ or
${\bf C}$) being used in BV method.

From the 1st and 2nd orders operators  and various kinds of antibrackets
consider only ${\stackrel{\circ}{U}}_{0}(\theta)$,
${\stackrel{\circ}{V}}_{0}(\theta)$, ${\stackrel{\circ}{W}}_{0}(\theta)$ and
following ones acting nontrivially on $C^{k}(P_0(T^{\ast}_{odd}{\cal M}_{
ext}))$
\begin{eqnarray}
\Delta^{ext}_{00}(\theta)\equiv \Delta_{BV},\
\left(\overline{\cal F}(\Gamma_{0}),\overline{\cal J}(\Gamma_{0})\right)^{
ext}_{00}\equiv
\left(\overline{\cal F}(\Gamma_{0}),\overline{\cal J}(\Gamma_{0})\right)_{BV}
.  
\end{eqnarray}
Projectors $P_{a}(\theta)$  take the values under restriction onto
$P_0(T^{\ast}_{odd}{\cal M}_{ext})$
\begin{eqnarray}
P_{1}(\theta)=0,\ P_{0}(\theta)=1\;. 
\end{eqnarray}
The global symmetry supergroup $J$ and superspace ${\cal M}$ are restricted by
$\bar{J}$ and $\tilde{\cal M}$ respectively.
The superalgebras $\Lambda_1(\theta;{\bf K})$,
$\Lambda_{D\vert Nc + 1}(z^{a}, \theta;{\bf K})$, $\tilde{\Lambda}_{D\vert Nc
+ 1}(z^{a}, \theta;{\bf K})$ pass  into  ${\bf K}$,
$\Lambda_{D\vert Nc}(z^{a};{\bf K})$, $\tilde{\Lambda}_{D\vert
Nc}(z^{a};{\bf K})$  correspondingly. All
superfunctionals of the form (2.40) vanish identically. The concepts
of works [1,2] about Lagrangian and Hamiltonian formalism for GSTF with
respect to  GThGT and GThST in fact are reduced to standard definition of
gauge theory of special type for $\theta=0$ appearing by the irreducible
general gauge theory in the usual sense [4,7,8].
Moreover the so-called HCLF and being equal to them
HCHF
\begin{eqnarray}
\Theta^H_{\imath}(P_0{\cal A}(\theta)) = - \frac{\delta_l\overline{\cal
S}_0(A)}{\delta A^{\imath}\phantom{xxx}}=0  
\end{eqnarray}
are the ordinary equations of motion for gauge theory being described by
functional $\overline{\cal S}_0(A)$
given on $P_{0}{\cal M}_{cl}$ appearing by potential term for
$\overline{S}_{L}(A,\lambda)$ (2.13) or for
$\overline{S}_{H}(A,A^{*})$  (4.4).

It is not possible to pass from Lagrangian formalism of description for
restricted GThGT on $C^{k}(P_{0}(T_{odd}{\cal M}_{ext}))$ into Hamiltonian
one by means of Legendre transform of functional $\overline{\cal S}_0$ with
respect to $\lambda^{\imath}$ $\equiv$ ${\stackrel{\ \circ}{\cal
A}}{}^{\imath}(\theta)(-1)^{\varepsilon_{\imath} + 1}$ in view of absence of
the latter fields.  As consequence of this fact it is impossible to
introduce the usual antifields $A^{*}_{\imath}$ according to above-mentioned
scheme for equivalent formulation of GThGT in Hamiltonian formalism as well.
However, if it is not formally to assume ${\stackrel{\ \circ}{\cal
A}}{}^{\imath}(\theta) = 0$ then the Legendre transform of
$\overline{S}_{L}(A,\lambda)$ on $C^{k}(P_{0}(T_{odd}{\cal M}_{ext}))$ with
respect to $\lambda^{\imath}$ is possible.

The following functionals taking Existence Theorems 1,2 for $\hbar=0$
into account
\begin{eqnarray}
\overline{S}_{BV}(\Gamma_{0{}min}) = P_0(\theta)S_{H{}min}(\Gamma_{min}(
\theta)),\
\overline{S}_{BV}(\Gamma_{0{}ext}) = P_0(\theta)S_{H{}ext}(\Gamma_{ext}(\theta))
\end{eqnarray}
appear by the solutions for classical master equations [4,7] of BV method
written by means of antibracket (12.2) in
$C^k(P_{0}(T^{*}_{odd}{\cal M}_{s}))$, $s=min,ext$ respectively.
On the other hand BV method is based on equation of the form (6.11) for
$\hbar \ne 0$ and $\theta=0$.  In the last case
$\overline{S}_{BV}(\Gamma_{0{}ext},\hbar)$ = $S_{BV}(P_{0}\Gamma_{ext}(
\theta))
$ (5.15) is not the restriction of the corresponding superfunction
determining the II class GThGT. Theorem 3 is valid for
$\overline{S}_{BV}(\Gamma_{0{}s},\hbar)$ under replacement of Eq.(6.2)
onto Eq.(6.11)  respectively
\begin{eqnarray}
\textstyle\frac{1}{2}\left(\overline{S}_{BV}(\Gamma_{0{}s},\hbar),
\overline{S}_{BV}(\Gamma_{0{}s},\hbar)\right)_{BV}=\imath\hbar\Delta_{BV}
\overline{S}_{BV}(\Gamma_{0{}s},\hbar), \ s=min,ext\,.
\end{eqnarray}
The results of Sec.X are reduced to corresponding relations for standard
version of the Lagrangian  quantization (for
${\stackrel{\circ}{\Gamma}}(\theta)=0$ if the formal GThGT with
$S_{H}\bigl(\Gamma(\theta),{\stackrel{\circ}{\Gamma}}(\theta),\hbar\bigr)$
have been made use) for $\theta=0$  and under following identifications
\renewcommand{\theequation}{\arabic{section}.\arabic{equation}\alph{lyter}}
\begin{eqnarray}
\setcounter{lyter}{1}
{} & {} &
d\Phi(\theta)_{\mid\theta=0}\equiv P_0(\theta)d\Phi(\theta) = d\phi,\
d\Gamma(\theta)_{\mid\theta=0}\equiv P_0(\theta)d\Gamma(\theta)= d\Gamma_0
\;, \\
\setcounter{equation}{7}
\setcounter{lyter}{2}
{} & {} &
P_0(\theta){\cal J}^{\ast}_A(\theta) = J^{\ast}_{0{}A},\
J_A = - {\stackrel{\circ}{\Phi}}{}^{\ast}_A(\theta),\
P_0(\theta)\lambda^A(\theta) = \lambda^A_0 \equiv \lambda^A\;, 
\end{eqnarray}
where $J^{\ast}_{0{}A}$, $J_{A}$ and $\lambda^{A}$ are the usual sources to
$P_0(\theta)$-components of Green's functions and auxiliary Lagrangian
multipliers introducing the gauge in BV method  respectively.

FI (10.2a) appears by usually defined [8], in fulfilling of the mentioned
correspondences and restrictions, and its equivalent form (10.14) coincides
after integration with respect to $\pi_0^A$ and $d{\stackrel{\;\circ}{\phi
}}{}'^{\ast}_{0{}A}$ (in omitting the normalization constant
${\rm sdet}\left\|\hbar\delta^A{}_B\right\|$) with generating functional of
Green's functions in BV method. Representations for $\overline{Z}$ (10.2a),
(10.14) have the form respectively
\begin{eqnarray}
\setcounter{lyter}{1}
{} & {} & \hspace{-3em}
Z\bigl({
\stackrel{\circ}{\Gamma}}(\theta),\Phi^{\ast}(\theta),{\cal J}^{\ast}(\theta)
\bigr)_{\mid{\stackrel{\circ}{\Gamma}}(\theta)=\theta=0}\equiv
\overline{Z}(\phi^{\ast},J),\
\overline{\Psi}(\phi) \equiv P_0(\theta)\Psi(\Phi(\theta)),\
J^{\ast}_{0{}A} = J_A\,, \\
\setcounter{equation}{8}
\setcounter{lyter}{2}
{} & {} & \hspace{-3em}
\overline{Z}(\phi^{\ast},J) = \displaystyle\int d\phi\,{\rm exp}\left\{
\displaystyle\frac{\imath}{\hbar}\left(\overline{S}_H\left(\phi,\phi^{\ast} +
\displaystyle\frac{\delta\overline{\Psi}}{\delta\phi\phantom{x}},\hbar\right)
+ J_A\phi^A\right)\right\},
\\
\setcounter{equation}{8}
\setcounter{lyter}{3}
{} & {} & \hspace{-3em}
\overline{Z}(\phi^{\ast},J) = \displaystyle\int d\phi\,d\lambda\,
d{\phi'}^{\ast}{\rm exp}
\left\{
\displaystyle\frac{\imath}{\hbar}\left(\overline{S}_H\left(\phi,{\phi'}^{\ast
},\hbar\right) + \left({\phi'}^{\ast}_A - \phi^{\ast}_A -
\displaystyle\frac{\delta\overline{\Psi}}{\delta\phi^A}\right)\lambda^A
+ J_A\phi^A\right)\right\}. 
\end{eqnarray}
The 2nd summand in the integrand (12.8c) with account of (12.7b), (2.41a) and
results of Sec.XI can be rewritten as follows
\renewcommand{\theequation}{\arabic{section}.\arabic{equation}}
\begin{eqnarray}
\left({\phi'}^{\ast}_A - \phi^{\ast}_A -
\displaystyle\frac{\delta\overline{\Psi}}{\delta\phi^A}\right)\lambda^A =
{\stackrel{\circ}{U}}_0\left(
({\phi'}^{\ast}_A - \phi^{\ast}_A)\phi^A -
\overline{\Psi}(\phi)\right).  
\end{eqnarray}
Ward identities for generating functionals $\overline{Z}(\phi^{\ast},J)$,
$\overline{W}(\phi^{\ast},J)$, $\overline{\mbox{\boldmath$\Gamma$}}(
\langle\phi\rangle,\phi^{\ast})$ (two
latters are defined analogously to $\overline{Z}$ (12.8a)) will be written in
the form
\begin{eqnarray}
J_A\frac{\delta\overline{Q}(\phi^{\ast},J)}{\delta\phi^{\ast}_A\phantom{xxxx}}
\equiv -{\stackrel{\circ}{V}}_0\overline{Q}(\phi^{\ast},J) = 0,\ Q\in
\{\overline{Z},\overline{W}\},\
\left(\overline{\mbox{\boldmath$\Gamma$}}(\langle\phi\rangle,\phi^{\ast}),
\overline{\mbox{\boldmath$\Gamma$}}(\langle\phi\rangle,\phi^{\ast})\right)^{(
\langle\phi\rangle,\phi^{\ast})}_{BV}=0\,. 
\end{eqnarray}
The corresponding to BV method system (10.50) for $\overline{Z}(\phi^{
\ast},J)$
(12.8b) and system (10.67), (10.66b) for $\overline{Z}(\phi^{\ast},J)$ in the
form (12.8c) will pass to standard BRST transformations, under following
formal identification
\begin{eqnarray}
P_0(\theta)\delta_{\mu}\Gamma^{p_s}_{s}(\theta){\hspace{-0.5em}
\phantom{\bigl)}}_{\mid\tilde{\Gamma}(\theta)} =
P_0(\theta)\left(\frac{d_r{\Gamma}^{p_s}_{s}(\theta)}{d\theta\phantom{xxxx}
}\right){\hspace{-0.5em}\phantom{\Bigr)}}_{\mid\tilde{\Gamma}(\theta)}\mu
= \delta_{\mu}{\Gamma}^{p_s}_{0{}s},\ s=ext\,,
\end{eqnarray}
for $\overline{Z}(\phi^{\ast},J)$ given in both forms (12.8b) and (12.8c)
respectively
\begin{eqnarray}
{} & {} &
\delta_{\mu}\phi^A = \left(\phi^A,\overline{S}_H\left(\phi,\phi^{\ast} +
\displaystyle\frac{\delta\overline{\Psi}}{\delta\phi\phantom{x}},\hbar\right)
\right)_{BV}\mu,\ \ \delta_{\mu}\phi^{\ast}_A = 0\;; \\ 
{} & {} &\delta_{\mu}\phi^A = - \lambda^A\mu,\ \ \delta_{\mu}\lambda^A =
\delta_{\mu}\phi^{\ast}_A = 0,\ \
\delta_{\mu}\phi'^{\ast}_A = \left(\phi'^{\ast}_A,\overline{S}_H\left(
\phi,{\phi'}^{\ast},\hbar\right)\right)^{(\phi,{\phi'}^{\ast})}_{BV}\mu\;.
\end{eqnarray}
\subsection{Connection with  Lagrangian Superfield Quantization [3]}

It is easy to determine the correlation of GSQM with quantities and relations
of the  Lagrangian quantization superfield variant [3]  by means of the
formulae from corresponding sections devoted to component formulation of
GSTF [1,2] and remarks from Sec.XI.

Namely the operators $U$, $V$, $\Delta$ from method [3] being denoted further
with index "LMR" are connected with ${\stackrel{\circ}{U}}_0(\theta)$,
${\stackrel{\circ}{V}}_0(\theta)$, $\Delta_{00}(\theta)$ (2.41) by
means of the formulae
\renewcommand{\theequation}{\arabic{section}.\arabic{equation}\alph{lyter}}
\begin{eqnarray}
\setcounter{lyter}{1}
{} & {} &
U_{LMR}=
{\stackrel{\circ}{\Phi}}{}^A(\theta)
\displaystyle\frac{d\phantom{}
}{d\theta}\displaystyle\frac{\delta_l \phantom{xxxx}}{\delta\Phi^A(\theta)}
(-1)^{\varepsilon_A} = (-1)^{\varepsilon_A+1}
\lambda^A\displaystyle\frac{\delta_l \phantom{x}}{\delta\phi^A}=
{\stackrel{\circ}{U}}_0(\theta)
\;, \\
\setcounter{equation}{14}
\setcounter{lyter}{2}
{} & {} &
V_{LMR}=
{\stackrel{\circ}{\Phi}}{}^{\ast}_A(\theta)
\displaystyle\frac{d\phantom{}
}{d\theta}\displaystyle\frac{\delta\phantom{xxxx}}{\delta\Phi^{\ast}_A(
\theta)}(-1)^{\varepsilon_A+1} = - J_A\displaystyle\frac{\delta\phantom{xx
}}{\delta\phi^{\ast}_A}= {\stackrel{\circ}{V}}_0(\theta)\;, \\
\setcounter{equation}{14}
\setcounter{lyter}{3}
{} & {} & \Delta_{LMR}=
\displaystyle\frac{d\phantom{}
}{d\theta}\displaystyle\frac{\delta_l \phantom{xxx}}{\delta\Phi^A(\theta)}
\displaystyle\frac{d\phantom{}
}{d\theta}\displaystyle\frac{\delta\phantom{xxxx}}{\delta\Phi^{\ast}_A(
\theta)}(-1)^{\varepsilon_A+1} = \Delta_{00}(\theta)=\Delta_{BV}\;. 
\end{eqnarray}
These equalities are valid in acting on superfunctions from $D^k_{ext}$ and
superfunctionals of the form (2.40).

Functionals "${}^{\ast}F$" which had been made use in the method  [3]
appear in the framework of GSQM terminology by $P_0(\theta)$-components of
superfunctions from $D^k_{ext}$. It means that the representation from
Sec.VIII of Ref.[2] of the form being dictated by formula (8.1) [2] with
$F_{H{}ext}= \overline{F}_{H{}ext} = 0$ is true for those functionals. To be
more precisely there exists for any ${}^{\ast}F[\Gamma]$ from the method
[3] the  such corresponding  superfunction ${\cal F}(\theta)\in
D^k_{ext}$ that the following identities hold
\begin{eqnarray}
 {}^{\ast}F[\Gamma] \equiv {}^{\ast}F[\phi,\lambda,\phi^{\ast},J]=
P_0(\theta){\cal F}\bigl({\Gamma}(\theta),{\stackrel{\circ}{\Gamma}}(\theta),
\theta\bigr) = {\cal F}\bigl(P_0{\Gamma}(\theta),{\stackrel{\circ}{\Gamma}}(
\theta),0\bigr) \equiv \overline{\cal F}(\Gamma_0,\Gamma_1)\;, 
\end{eqnarray}
conforming with (12.1) but without restriction
${\stackrel{\circ}{\Gamma}}(\theta)=0$.

For antibracket $\left(\ ,\ \right)_{LMR}$ [3] we have (from formula (8.10)
of Sec.VIII Ref.[2])
\renewcommand{\theequation}{\arabic{section}.\arabic{equation}}
\begin{eqnarray}
{} &
\left({}^{\ast}F[\Gamma],{}^{\ast}G[\Gamma]\right)_{LMR}=
\left(
P_0(\theta){\cal F}\bigl({\Gamma}(\theta),{\stackrel{\circ}{\Gamma}}(\theta),
\theta\bigr),
P_0(\theta){\cal J}\bigl({\Gamma}(\theta),{\stackrel{\circ}{\Gamma}}(\theta),
\theta\bigr)\right)_{00}\equiv {} & \nonumber\\
{} & \equiv\left(\overline{\cal F}(\Gamma_0,\Gamma_1),\overline{\cal
J}(\Gamma_0, \Gamma_1)\right)_{00},\
{}^{\ast}F[\Gamma] = \overline{\cal F}(\Gamma_0,\Gamma_1),
{}^{\ast}G[\Gamma] = \overline{\cal J}(\Gamma_0,\Gamma_1). {} & 
\end{eqnarray}
In fact the main object of the method [3] being by functional
${}^{\ast}S[\Phi,\Phi^{\ast}]$ is connected with
$S_{H}\bigl(\Gamma(\theta),{\stackrel{\circ}{\Gamma}}(\theta),\hbar\bigr)$
from GSQM, satisfying to BV similar equation (8.5), by means of relation
conforming with (12.15),  with transformation rule for operators
${\stackrel{\circ}{U}}_0(\theta)$, ${\stackrel{\circ}{V}}_0(\theta)$ (2.50),
(2.51) and with expressions (5.16), (5.17) (in addition see the formulae of
Sec.VII from paper [2])
\begin{eqnarray}
{}^{\ast}S[\Phi,\Phi^{\ast}]=
P_0(\theta)\left(S_H\bigl({\Gamma}(\theta),{\stackrel{\circ}{\Gamma}}(\theta),
\hbar\bigr) - {\stackrel{\circ}{\Phi}}{}^{\ast}_A(\theta)\Phi^A(\theta)
\right) = \overline{S}_H(\Gamma_0,\Gamma_1,\hbar) + J_A\phi^A\;.
\end{eqnarray}
Really ${}^{\ast}S[\Phi,\Phi^{\ast}]$ satisfies by virtue of above expressions
(2.50), (2.51) and (12.14) as well to
generating equation of the method [3]
\begin{eqnarray}
\textstyle\frac{1}{2}\left({}^{\ast}S[\Gamma],
{}^{\ast}S[\Gamma]\right)_{LMR} + V_{LMR}{}^{\ast}S[\Gamma]=
\imath\hbar\Delta_{LMR}{}^{\ast}S[\Gamma]\;.  
\end{eqnarray}

FIs (10.2a), (10.14) for $Z\bigl({\stackrel{\circ}{\Gamma}}(\theta),
\Phi^{\ast}(\theta), {\cal J}^{\ast}(\theta)\bigr)$ are not in general
reduced to the generating functional of Green's functions
${}^{\ast}Z[\Phi^{\ast}]$ from paper [3] using the representation for
${}^{\ast}Z[\Phi^{\ast}]$ of the form (10.14). Really write
 $P_0(\theta)Z\bigl({\stackrel{\circ}{\Gamma}}(\theta)$,
$\Phi^{\ast}(\theta), {\cal J}^{\ast}(\theta)\bigr)$ in the form (10.14)
under condition that
gauge is imposed with help of GNA (7.1)
\begin{eqnarray}
{} & \overline{Z}\left(\lambda,J,\phi^{\ast},J_0^{\ast}\right)=
\displaystyle\int d\phi\,d\lambda'_0\,d{\phi'}^{\ast}
{\rm exp}\Bigl\{
\displaystyle\frac{\imath}{\hbar}\Bigl(\overline{S}_H\left(\phi,{\phi'}^{\ast
},\lambda,J,\hbar\right) + {} & \nonumber \\
{} & \left({\phi'}^{\ast}_A - \phi^{\ast}_A -
\displaystyle\frac{\delta\overline{\Psi}(\phi,\lambda)}{\delta\phi^A
\phantom{xxxx}}\right){\lambda}'^A_0 + J^{\ast}_{0{}A}\phi^A\Bigr)\Bigr\}\;.
{} &
\end{eqnarray}
In this case one can regard that the
superfunction $\Psi(\theta)$ from $D^k_{ext}$ has the form  $\Psi(\theta)$
$\equiv$ $\Psi\bigl(\Phi(\theta), {\stackrel{\circ}{\Phi}}(\theta)\bigr)$
without loss of the Eq.(8.5) validity.
In obtaining of (12.19) it was integrated over $\pi_0^A$,
${\stackrel{\;\circ}{\phi}}{}'^{\ast}_{0{}A}$ in the FI (10.14) in omitting
of ${\rm sdet}\left\|\hbar\delta^A{}_B\right\|$.
Relationships (12.7) and
being analogous ones for $d\lambda(\theta)$, $d\pi(\theta)$,
$d{\stackrel{\;\circ}{\phi}}{}'^{\ast}(\theta)$ were taken into account
together with
rule (12.15) for $\overline{\Psi}(\phi,\lambda)$.

The comparison of FI (12.19) with component formulation for ${}^{\ast
}Z[\Phi^{\ast}]$ from Ref.[3] having the following form  with regard for
formula (12.17)
\begin{eqnarray}
{} & {}^{\ast}Z[\Phi^{\ast}] = \overline{Z}(\phi^{\ast},J)=
\displaystyle\int d\phi\,d\lambda\,d{\phi'}^{\ast}
{\rm exp}\Bigl\{
\displaystyle\frac{\imath}{\hbar}\Bigl(\overline{S}_H\left(\phi,\lambda,
{\phi'}^{\ast},J',\hbar\right)_{\mid J'=0} + {} & \nonumber \\
{} & \left({\phi'}^{\ast}_A - \phi^{\ast}_A -
\displaystyle\frac{\delta\overline{\Psi}(\phi,\lambda)}{\delta\phi^A
\phantom{xxxx}}\right)\lambda^A + J^{\ast}_A\phi^A\Bigr)\Bigr\}
{} &
\end{eqnarray}
shows that the very strong constraints must be fulfilled in order to the
right-hand sides of the expressions (12.19) and (12.20) would coincide.
Namely it is necessary to fulfill the identities
\begin{eqnarray}
\frac{\delta\overline{Z}\left(\lambda,J,\phi^{\ast},J_0^{\ast}\right)}{
\delta\lambda^A\phantom{xxxxxxxxx}} =
\frac{\delta\overline{Z}\left(\lambda,J,\phi^{\ast},J_0^{\ast}\right)}{
\delta J_A\phantom{xxxxxxxxx}} = 0,\ J^{\ast}_{0{}A} = J_A\,,
\end{eqnarray}
the such that the first of them are equivalent to the following ones
\begin{eqnarray}
\frac{\delta\overline{S}_H\phantom{}}{\delta\lambda^A} =
\frac{\delta\overline{S}_H\phantom{}}{\delta J_A} =
\frac{\delta\overline{\Psi}\phantom{x}}{\delta\lambda^A} = 0\,.
\end{eqnarray}

The corresponding BRST transformations for
$\overline{Z}(\lambda,J,\phi^{\ast},J_0^{\ast})$ have the form (12.13) under
change of $\overline{S}_H(\phi, \phi'^{\ast}, \hbar)$ onto
$\overline{S}_H(\phi,\lambda,{\phi'}^{\ast},J,\hbar)$
that with accuracy up to sign coincides with invariance transformations for
${}^{\ast}Z[\Phi^{\ast}]$ written in [3] with superfunction
${}^{\ast}S[\Phi,\Phi'^{\ast}]$ defined in (12.17).
All other relations for $\overline{Z}(\lambda,J,\phi^{\ast},J_0^{\ast})$,
$\overline{W}(\lambda,J,\phi^{\ast},J_0^{\ast})$,
$\overline{\mbox{\boldmath$\Gamma$}}(\lambda,J,\langle\phi\rangle,\phi^{
\ast})$ obtained
in Subsec.XII.1, namely, the Ward identities have the same form as in
(12.10).

The definition of generating functional of Green's functions by means of
relations (12.20) but written
in "superfield form" in terminology of the Ref.[3]
\begin{eqnarray}
{} & {}^{\ast}Z[\Phi^{\ast}] =
\displaystyle\int \tilde{d}\Phi(\theta)\,\tilde{d}{\Phi'}^{\ast}(\theta)
\rho({\Phi'}^{\ast}){\rm exp}\Bigl\{
\displaystyle\frac{\imath}{\hbar}\Bigl({}^{\ast}{S}[\Phi,{\Phi'}^{\ast}]
- U_{LMR}{}^{\ast}\Psi[\Phi] + {} & \nonumber \\
{} & \displaystyle\int d\theta\left({\Phi'}^{\ast}_A(\theta) - \Phi^{\ast
}_A(\theta)\right)\Phi^A(\theta)\Bigr)\Bigr\},\ \rho({\Phi'}^{\ast}) =
\delta\left(\displaystyle\int d\theta{\Phi'}_A^{\ast}(\theta)\right),\
\tilde{d}\Gamma(\theta)\equiv d\Gamma_0 d\Gamma_1 {} &
\end{eqnarray}
although  coincides with FI of the BV method (12.8c) on the hypersurface
defined by equations ${\stackrel{\circ}{\Gamma}}(\theta)=0$ but out of
the last equations leads to the functional dependence in a non-parametric way
upon $\Gamma_1^p$ component fields. It means, in particular, that the
presence of $\lambda^A$ in functional ${}^{\ast}\Psi[\Phi]$ =
$\overline{\Psi}(\phi,\lambda)$ leads to the method for imposing of the
gauge being differred both from the phase AT (9.1) and from GNA (9.10) that
diverges
with ideology of interpretation of transition from one Lagrangian surface  to
another in the sense of statements from Theorems 5,6. Furthermore this fact
appears by obstacle to introduction in superfield form the generating
functional of vertex Green's functions {\boldmath$\Gamma$} in the framework
of the method [3].
\section{GSTF Models in  GSQM}
\setcounter{equation}{0}

Continue the investigation of the models from papers [1,2] in the context of
the quantum theory construction in the framework of GSQM in the Lagrangian
formalism by means of FI.
\subsection{Massive Complex Scalar Superfield Models}

It had been shown in papers [1,2] that given GSTF models appear by
nondegenerate ThST (theory of special type), i.e. nongauge one. Therefore
for construction of the quantum
description in terms of generating functionals of Green's functions
there are not necessity to extend $T^{\ast}_{odd}{\cal M}_{cl}$.

Instead of two cases for the free model and for one with interaction let us
consider the latter ThST which includes the former one.
It is sufficient to realize the 1st stage of the construction procedure for
quantum theory with respect to that I class ThST (see Sec.IV). The model is
described by superfunction
$S_{H{}M}(\Gamma(\theta))$ on $T^{\ast}_{odd}{\cal M}_{cl}$ parametrized by
complex scalar  superfields $\varphi(x,\theta)$, $\overline{\varphi}(x,
\theta)$ $\in$ $\tilde{\Lambda}_{4\mid 0+1}(x^{\mu},\theta;{\bf C})$ given
on ${\cal M} = {\bf R}^{1,3} \times \tilde{P}$ [1] with zero ghost
number and complex scalar\footnote{the sign of complex conjugation for
quantity $g(x,\theta)$: $\overline{g}(x,\theta)$ in this subsection is
differred both from the sign of special involution ${\ast}$ in Sec.II being
denoted as $\overline{g(x,\theta)}$ and from the sign of superantifield
"${\ast}$"} superantifields $(\varphi(x,\theta))^{\ast}$,
$(\overline{\varphi}{}(x,\theta))^{\ast}$ $\in$ $\tilde{\Lambda}_{4\mid
0+1}(x^{\mu},\theta;{\bf C})$ [2] with ghost number being equal to $(-1)$.
Superfunction $S_{H{}M}(\Gamma(\theta))$ has the form [2]
\begin{eqnarray}
S_{H{}M}(\Gamma(\theta)) & = &
T\bigl((\varphi(\theta))^{\ast},(\overline{\varphi}(\theta))^{\ast}\bigr) +
S_{0{}M}(\varphi(\theta), \overline{\varphi}(\theta)),\
S_{0{}M}(\theta)= S_{0}(\theta) - V(\theta)\;,   \\ 
T(\theta) & \equiv & T\bigl((\varphi(
\theta))^{\ast},(\overline{\varphi}(\theta))^{\ast}\bigr) =
\int d^4x \frac{1}{
\imath}(\varphi(x,\theta))^{\ast}(\overline{\varphi}(x,\theta))^{\ast}
\equiv \int d^4x {\cal L}_{\rm
kin}^{\ast}(x,\theta), \\ 
S_0(\theta) & \equiv &
S_{0}\bigl(\varphi(\theta),\overline{\varphi}(\theta)\bigr)
= \int d^4x (\partial_{\mu}
\overline{\varphi}\partial^{\mu}\varphi - m^2\overline{\varphi}\varphi)(
x,\theta)
\equiv \int d^4x {\cal L}_0 (x,\theta)\;, \\
V(\theta) & \equiv & V\bigl(\varphi(\theta), \overline{\varphi}(\theta)\bigr)
= \int d^4x\Bigl(\textstyle\frac{\mu}{3}
\overline{\varphi}\varphi(\overline{\varphi} + \varphi) + \textstyle\frac{
\lambda}{2}(\overline{\varphi}\varphi)^2 + \ldots\Bigr)(x,\theta)\;, \\
{} & {} &(\varphi(x,\theta))^{\ast} =
(\varphi_1(x,\theta))^{\ast} - \imath(\varphi_2(x,\theta))^{
\ast}= (\varphi(x))^{\ast} -\theta J_{\varphi}(x)\;, \nonumber \\
{} & {} & \varphi(x,\theta)=(\varphi_1 + \imath\varphi_2)(x,\theta) =
\varphi(x) + \lambda(x)\theta\,.
\end{eqnarray}
The ghost number distribution for kinetic $T(\theta)$ and potential $S_{0{}M}(
\theta)$ parts of $S_{H{}M}(\Gamma(\theta))$ reads as follows
\begin{eqnarray}
{\rm gh}(T(\theta),S_{0{}M}(\theta)) = (-2,0)\;.  
\end{eqnarray}
Eq.(4.3a) is trivially fulfilled in the given example whereas the
superfunction $S_{0{}M}(\theta)$ = ${\cal S}_{0{}M}(\theta)$ defining the
II class GThST with being simply obtained restricted GHS (4.5) [2]
\renewcommand{\theequation}{\arabic{section}.\arabic{equation}\alph{lyter}}
\begin{eqnarray}
\setcounter{lyter}{1}
{} & {\stackrel{\circ}{\varphi}}(x,\theta) =
{\stackrel{\circ}{\overline{\varphi}}}(x,\theta) =0\;, {} & \nonumber \\
{} &  ({\stackrel{\circ}{\varphi}}(x,\theta))^{
\ast} = (\Box + m^2+
\frac{\mu}{3}(\overline{\varphi}+2\varphi)(x,\theta)
+\lambda(\overline{\varphi}\varphi)(x,\theta)+\ldots)
\overline{\varphi}(x,\theta)\;, {} &\nonumber \\
{} & ({\stackrel{\circ}{\overline{\varphi}}}(x,\theta))^{\ast}=
(\Box + m^2 +
\frac{\mu}{3}(2\overline{\varphi}+\varphi)(x,\theta)
+\lambda(\overline{\varphi}\varphi)(x,\theta)+\ldots)
{\varphi}(x,\theta)\;, {} & \\
\setcounter{lyter}{2}
\setcounter{equation}{7}
{} & \Theta_{\varphi{}M}^H(x,\theta) = -
\displaystyle\frac{\partial
{\cal S}_{0{}M}(\theta)}{\partial{\varphi}(x,\theta)} =
(\Box + m^2 + \textstyle\frac{\mu}{3}(\overline{\varphi}+2\varphi)(x,\theta)
+\lambda(\overline{\varphi}\varphi)(x,\theta)+\ldots)
\overline{\varphi}(x,\theta) = 0\;, {} & \nonumber \\
{} & \Theta_{\overline{\varphi}{}M}^H(x,\theta) =
\overline{\Theta}_{\varphi{}M}^H(x,\theta)=0\,.
\end{eqnarray}
appears by solution for Eq.(4.3b).

Because the supermatrix of the 2nd partial superfield derivatives of
${\cal S}_{0{}M}(\theta)$ with respect to ${\varphi}(x,\theta)$,
$\overline{\varphi}(y,\theta)$ is nondegenerate [1] then it is possible to
construct the FI  (10.2a) or (10.14).
Generating functional of Green's functions $Z_M(\theta)$ for interacting
complex
scalar superfield does not depend upon superantifields and additional
supervariables from ${\cal M}_{add}$. $Z_M(\theta)$ is explicitly calculated
according to the formula (10.11) as FI of quasi-Gaussian superfunction (10.10)
\renewcommand{\theequation}{\arabic{section}.\arabic{equation}\alph{lyter}}
\begin{eqnarray}
\setcounter{lyter}{1}
{} & Z_M\bigl({\cal J}^{\ast}_{\varphi}(\theta),{\cal J}^{\ast}_{\overline{
\varphi}}(\theta)\bigr) =
\displaystyle\int d\varphi(\theta)d\overline{\varphi}(\theta)
{\rm exp}\left\{
\displaystyle\frac{\imath}{\hbar}\left({\cal S}_{0{
}M}(\varphi(\theta), \overline{\varphi}(
\theta)) + {\cal J}^{\ast}_{\varphi}(\theta)\varphi(\theta) +
{\cal J}^{\ast}_{\overline{\varphi}}(\theta)\overline{
\varphi}(\theta)\right)\right\} {} &
\nonumber \\
{} & = {\rm det}^{-1}\left\|\hbar^{-1}(\Box+ m^2)\delta(x-y)\right\|
{\rm exp}\left\{-\displaystyle\frac{\imath}{\hbar}V\Bigl(
\displaystyle\frac{\partial_l\phantom{xxxxxx}}{\partial(\frac{\imath}{\hbar}{
\cal J}^{\ast}_{\varphi}(\theta))},
\displaystyle\frac{\partial_l\phantom{xxxxxx}}{\partial(\frac{\imath}{\hbar}
{\cal J}^{\ast}_{\overline{\varphi}}(\theta))}\Bigr)
\right\}\times {} & \nonumber \\
{} & {\rm exp}\left\{-
\displaystyle\frac{\imath}{\hbar}{\cal J}^{\ast}_{\varphi}(\theta)
{\cal D}(\theta){\cal J}^{\ast}_{\overline{\varphi}}(\theta)
\right\}, {} & \\
\setcounter{equation}{8}
\setcounter{lyter}{2}
{} &
{\cal J}^{\ast}_{\varphi}(\theta){\cal D}(\theta){\cal J}^{\ast}_{
\overline{\varphi}}(\theta) =
\displaystyle\int d^4xd^4y{\cal J}^{\ast}_{\varphi}(x,\theta){\cal
D}(x,y,\theta)
{\cal J}^{\ast}_{\overline{\varphi}}(y,\theta) \;,{} &  \nonumber \\
{} & {\cal D}(x,y,\theta) = {\cal D}(x-y) = - \displaystyle\int
\displaystyle\frac{d^4p}{(2\pi)^4}
\displaystyle\frac{e^{-\imath p(x-y)}}{p^2 -m^2 +\imath\varepsilon}
,\ \varepsilon \to +0,\ p^2=p^{\mu}p_{\mu}\;. {} & 
\end{eqnarray}
The functional determinant in (13.8a) being by a constant normalization
factor usually is ignored in the formula writing in the expression for
$Z_M(\theta)$. Superfunction ${\cal D}(x,y,\theta)$
appears by the Green's function in Feynman representation for Klein-Gordon
operator.

The supesources ${\cal J}^{\ast}_{\varphi}(x,\theta)$, ${\cal J}^{
\ast}_{\overline{\varphi}}(x,\theta)$ are the complex
scalars from $\tilde{\Lambda}_{4\mid 0+1}(x^{\mu},\theta;{\bf C})$ possessing
by the properties
\renewcommand{\theequation}{\arabic{section}.\arabic{equation}}
\begin{eqnarray}
\overline{{\cal J}^{\ast}_{\varphi}}(x,\theta) = {\cal J}^{\ast}_{
\overline{\varphi}}(x,\theta) = {\cal J}^{\ast}_{0{}\overline{\varphi}}(x) -
\theta{\cal J}^{\ast}_{1{}\overline{\varphi}}(x),\
({\rm gh},\varepsilon_P,\varepsilon_{\Pi},\varepsilon){\cal J}^{\ast}_{
\varphi}(x,\theta)=(0,0,0,0)\;. 
\end{eqnarray}
making under identification ${\cal J}^{\ast}_{0{}\varphi}(x) =
{\cal J}_{\varphi}(x)$ the restricted $P_0(\theta)$-component $P_0(\theta
)Z_M(\theta)$ by coinciding with standard generating functional of Green's
functions for complex scalar field $\varphi(x) = P_0(\theta)\varphi(x,
\theta)$.

The invariance transformations of the BRST type for $Z_M(\theta)$ (13.8a)
corresponding to the system (10.50) are trivial
\begin{eqnarray}
{\stackrel{\circ}{\varphi}}(x,\theta) =
{\stackrel{\circ}{\overline{\varphi}}}(x,\theta) =
({\stackrel{\circ}{\varphi}}(x,\theta))^{\ast} =
({\stackrel{\circ}{\overline{\varphi}}}(x,\theta))^{\ast} = 0\;. 
\end{eqnarray}
Formulae (13.1)--(13.10) for $V(\theta)=0$ correspond to the free model
quantum description.
\subsection{Massive Spinor Superfield of Spin $\frac{1}{2}$ Models}

From Ref.[1,2] it follows that given GSTF models appear by singular
nondegenerate, i.e. nongauge ones.
In such case it is sufficient to realize the 1st stage of the construction
procedure for quantum theory with respect to the I class ThSTs in question
in the initial $T^{\ast}_{odd}{\cal M}_{cl}$.

The interacting model is described by superfunction
$S^{(1)}_{H{}M}(\Gamma(\theta))$ on $T^{\ast}_{odd}{\cal M}_{cl}$ parametrized
by superfields $\Psi(x,\theta)$, $\overline{\Psi}(x,\theta)$ and
superantifields $\Psi^{\ast}(x,\theta)$,
$\overline{\Psi}{}^{\ast}(x,\theta)$ from
$\tilde{\Lambda}_{4\mid 0+1}(x^{\mu},\theta;{\bf C})$ with ghost number being
equal to 0 and $(-1)$ respectively.
These supervariables appear by Dirac bispinors [1,2]. Superfunction
$S^{(1)}_{H{}M}(\Gamma(\theta))$ has the form [2]
\begin{eqnarray}
{} & {} & \hspace{-2em} S^{(1)}_{H{}M}(\Gamma(\theta)) =
T\bigl({\Psi}^{\ast}(\theta), \overline{\Psi}{}^{\ast}(\theta)\bigr) +
S_{0{}M}(\Psi(\theta),\overline{\Psi}(\theta)),\
S^{(1)}_{0{}M}(\theta)=S^{(1)}_0(\theta)- V^{(1)}(\theta)\;, \\ 
{} & {} & \hspace{-2em}
T^{(1)}(\theta) \equiv T\bigl({\Psi}^{\ast}(\theta),
\overline{\Psi}{}^{\ast}(\theta)\bigr)
= \displaystyle\int d^4x {\Psi}^{\ast}(x,\theta)\overline{\Psi}{}^{\ast}(x,
\theta) \equiv \displaystyle\int d^4x {\cal L}_{\rm kin}^{\ast{}(1)}(x,
\theta)\;,
\\ 
{} & {} & \hspace{-2em}
S^{(1)}_0(\theta) \equiv S_0(\Psi(\theta),\overline{\Psi}(\theta)) =
\displaystyle\int d^4x \overline{\Psi}(x,\theta)(\imath\Gamma^{\mu}\partial_{
\mu} - m){\Psi}(x,\theta)\equiv \displaystyle\int d^4x {\cal L}^{(1)}_0(
x,\theta)
\;, \\ 
{} & {} & \hspace{-2em}
V^{(1)}(\theta) \equiv  V^{(1)}\bigl(\Psi(\theta),
\overline{\Psi}(\theta)\bigr) = \displaystyle\int d^4x\Bigl[\textstyle\frac{
\lambda_1}{2}(\overline{\Psi}{}\Psi)^2 + \textstyle\frac{\lambda_2}{2}(
\overline{\Psi}\Gamma^{\mu}\Psi)(\overline{\Psi}\Gamma_{\mu}\Psi)
\Bigr](x,\theta)\;, \\
{} & {} & \hspace{-2em}
{\Psi}(x,\theta)=\left({\psi}_{\alpha}(x,\theta),{\chi}^{\dot{\alpha}}(x,
\theta)\right)^T,\
{\Psi}^{\ast}(x,\theta)=\left({\psi}^{\ast\alpha}(x,\theta),{\chi}^{\ast}_{
\dot{\alpha}}(x,\theta)\right).  
\end{eqnarray}
Kinetic $T^{(1)}(\theta)$ and potential $S^{(1)}_{0{}M}(\theta)$ parts of
$S^{(1)}_{H{}M}(\Gamma(\theta))$ possess by the same  ghost numbers values
as the corresponding superfunctions in (13.6).

Eq.(4.3a) is trivially fulfilled for the model in question whereas the
superfunction $S^{(1)}_{0{}M}(\theta)$ = ${\cal S}^{(1)}_{0{}M}(\theta)$
defines the II class GThST  providing the fulfilment of the Eq.(4.3b) with
being simply obtained restricted GHS (4.5).
That GHS has the representation
\renewcommand{\theequation}{\arabic{section}.\arabic{equation}\alph{lyter}}
\begin{eqnarray}
\setcounter{lyter}{1}
{} & {\stackrel{\circ}{\Psi}}(x,\theta) =0,\
{\stackrel{\circ}{\overline{\Psi}}}(x,\theta) =0,\ {}& \nonumber \\
{}& \displaystyle\frac{d_r \Psi^{\ast}(x,\theta)}{d\theta
\phantom{xxxxxx}} = -
\left[\left(\imath\partial_{\mu}
 + \lambda_2(\overline{\Psi}\Gamma^{\mu}\Psi)\right)
\overline{\Psi}(x,\theta)\Gamma^{\mu} + \left(m +
\lambda_1(\overline{\Psi}{\Psi})\right)\overline{\Psi}(x,\theta)\right]\;,
{} &  \nonumber \\
{} & \displaystyle\frac{d_r \overline{\Psi}{}^{\ast}(x,\theta)}{d\theta
\phantom{xxxxxx}} =
- \left(\imath\Gamma^{\mu}\partial_{\mu} - m-
\lambda_1(\overline{\Psi}{\Psi})-
\lambda_2(\overline{\Psi}\Gamma^{\mu}\Psi)\Gamma^{\mu}
\right){\Psi}(x,\theta)\;,
{} & \\
\setcounter{lyter}{2}
\setcounter{equation}{16}
{} & \hspace{-1em}\Theta_{\Psi{}M}^{H}(x,\theta)\hspace{-0.2em}=
\hspace{-0.2em} - \displaystyle\frac{\partial_l
{\cal S}^{(1)}_{0{}M}(\theta)}{\partial{\Psi}(x,\theta)\phantom{x}}\hspace{-0.2em} =\hspace{-0.2em} -
\hspace{-0.1em}\left[\hspace{-0.1em}\left(\imath\partial_{\mu}
 + \lambda_2(\overline{\Psi}\Gamma^{\mu}\Psi)
\right)\overline{\Psi}(x,\theta)\Gamma^{\mu} + \hspace{-0.1em}
\left(m + \lambda_1(\overline{\Psi}{\Psi})\hspace{-0.1em}\right)
\overline{\Psi}(x,\theta)\right]\hspace{-0.1em} = 0, {} & \nonumber \\
{} &  \hspace{-1em}\Theta_{\overline{\Psi}{}M}^H(x,\theta) = -
\displaystyle\frac{\partial_l
{\cal S}^{(1)}_{0{}M}(\theta)}{\partial \overline{\Psi}(x,\theta)\phantom{x}}
= - \hspace{-0.2em}\left(\imath\Gamma^{\mu}\partial_{\mu} - m
-\lambda_1(\overline{\Psi}{\Psi})-
\lambda_2(\overline{\Psi}\Gamma^{\mu}\Psi)\Gamma^{\mu}
\right)\hspace{-0.2em}{\Psi}(x,\theta)\hspace{-0.1em} =\hspace{-0.1em} 0.
{} &  
\end{eqnarray}
Since the supermatrix of the 2nd partial superfield derivatives of
${\cal S}^{(1)}_{0{}M}(\theta)$ with respect to
$\Psi(x,\theta)$, $\overline{\Psi}(y,\theta)$ is nondegenerate
almost everywhere in ${\cal M}_{cl}$ [1] it is possible to
construct the FI (10.2a) or (10.14) in the framework of GSQM in the
Lagrangian formalism.
Generating functional of Green's functions $Z^{(1)}_M(\theta)$ for
interacting spinor superfield does not depend upon superantifields
and additional
superfields from ${\cal M}_{add}$. $Z^{(1)}_M(\theta)$ is explicitly
calculated according to formula (10.11) as FI of quasi-Gaussian superfunction
(10.10)
\renewcommand{\theequation}{\arabic{section}.\arabic{equation}\alph{lyter}}
\begin{eqnarray}
\setcounter{lyter}{1}
{} & Z^{(1)}_M\hspace{-0.2em}\left({\cal J}^{\ast}(\theta),
\overline{\cal J}{}^{\ast}(\theta)\right)
\hspace{-0.1em}= \hspace{-0.1em}\displaystyle\int \hspace{-0.2em}d\Psi(
\theta) d\overline{\Psi}(\theta)\hspace{-0.1em}{\rm exp}\left\{\hspace{-0.1em}
\displaystyle\frac{\imath}{\hbar}\left({\cal S}_{0{}M}(\Psi(\theta),
\overline{\Psi}(\theta)) + {\cal J}^{\ast}(\theta)\Psi(\theta) +
\overline{\Psi}(\theta)
\overline{\cal J}{}^{\ast}(\theta)\hspace{-0.1em} \right)\hspace{-0.1em}
\right\} {} & \nonumber \\
{} & = {\rm det}\left\|\hbar^{-1}(\imath\Gamma^{\mu}\partial_{\mu} -
m)\delta(x-y)
\right\|{\rm exp}\left\{-\displaystyle\frac{\imath}{\hbar}V\Bigl(
\displaystyle\frac{\partial_l\phantom{xxxxxx}}{\partial(\frac{\imath}{\hbar}{
\cal J}^{\ast}(\theta))},
\displaystyle\frac{\partial_l\phantom{xxxxxx}}{\partial(\frac{\imath}{\hbar}
\overline{\cal J}{}^{\ast}(\theta))}\Bigr)
\right\}\times {} & \nonumber \\
{} & {\rm exp}\left\{-
\displaystyle\frac{\imath}{\hbar}{\cal J}^{\ast}(\theta)
S(\theta)\overline{\cal J}{}^{\ast}(\theta)\right\}, {} &
\\
\setcounter{equation}{17}
\setcounter{lyter}{2}
{} &
{\cal J}^{\ast}(\theta) S(\theta)\overline{\cal J}{}^{\ast}(\theta)=
\displaystyle\int d^4xd^4y{\cal J}^{\ast}(x,\theta) S(x,y,\theta)\overline{
\cal J}{}^{\ast}(y,\theta)\;,{} & \\
\setcounter{equation}{17}
\setcounter{lyter}{3}
{} & S(x,y,\theta) = S(x-y) =
\displaystyle\int \displaystyle\frac{d^4p}{(2\pi)^4}
\displaystyle\frac{\Gamma^{\mu}p_{\mu} + m}{p^2 - m^2 +\imath\varepsilon}
e^{-\imath p(x-y)},\ \varepsilon \to +0\;. 
\end{eqnarray}
The functional determinant in (13.17a) is usually considered
as a normalization constant  and therefore is omitted.
Superfunction $S(x,y,\theta)$ appears by Green's function for Dirac operator.

Supersources ${\cal J}^{\ast}(x,\theta)$, $\overline{\cal J}{}^{\ast}(x,
\theta)$ to $\Psi(x,\theta)$, $\overline{\Psi}(x,\theta)$
are the Dirac bispinors possessing the properties
\renewcommand{\theequation}{\arabic{section}.\arabic{equation}}
\begin{eqnarray}
\overline{\cal J}{}^{\ast}(x,\theta)= \Gamma^0\left({\cal J}^{\ast}\right)^{
+}(x,\theta) = \overline{{\cal J}^{\ast}}(x,\theta),\
({\rm gh},\varepsilon_P,\varepsilon_{\Pi},\varepsilon)\overline{\cal J}{
}^{
\ast}(x,\theta)=(0,0,1,1)\;, 
\end{eqnarray}
where bar, for instance, over  $\overline{{\cal J}^{\ast}}(x,\theta)$ appears
in this subsection by Dirac conjugation in contrast to its use in
Subsec.XIII.1.

The concluding remarks of the preceding subsection together with
connection for free model quantum description and with the
trivial system corresponding to BRST type transformations are valid in this
case with corresponding modifications.
\subsection{Free Vector Superfield Models}

Superfield models of massless and massive vector superfields
${\cal A}^{\mu}(x,\theta) \in \tilde{\Lambda}_{D\mid 0+1}(x^{\mu},\theta;
{\bf C})$ given on ${\cal M}$ = ${\bf R}^{1,D-1}\times\tilde{P}$, $x^{\mu}$
$\in$ ${\bf R}^{1,D-1}$ had been considered in details  in Lagrangian [1]
and Hamiltonian [2] formulations of GSTF.

Let us begin from the case of massless theory. It had been shown that the
model in question appears by the I class GThST with GGTST
\begin{eqnarray}
{\cal R}^{\mu}(x,y)=\partial^{\mu}\delta(x-y),\ \mu=0,1,\ldots,D-1\;.
\end{eqnarray}
The only superfunction ${S}_0({\cal A}^{\mu}(\theta))$ below is invariant
under GTST according to [1,2].

In Hamiltonian formulation the model is described for $D=2k$, $k \in
\mbox{\boldmath$N$}$ by superfunction
$S^{(2)}_H(\Gamma(\theta))$ defined on $T^{\ast}_{odd}{\cal M}_{cl}$
parametrized
by superfields ${\cal A}^{\mu}(x,\theta)$ with zero ghost number and by
superantifields ${\cal A}^{\ast}_{\mu}(x,\theta)$ with ghost number being
equal to $-1$. $S^{(2)}_H(\Gamma(\theta))$ has the form in this case [2]
\begin{eqnarray}
{} & {} & \hspace{-1.5em}
S^{(2)}_H(\Gamma(\theta)) = T({\cal A}^{\ast}_{\mu}(\theta)) +
{S}_0({\cal A}^{\mu}(\theta))\;,
\\ 
{} & {} & \hspace{-1.5em}
T({\cal A}^{\ast}_{\mu}(\theta)) = \displaystyle\int d^Dx \textstyle\frac{1}{
2}(\varepsilon^{-1})^{\mu\nu}{\cal A}^{\ast}_{\mu}(x,\theta){\cal A}^{\ast}_{
\nu}(x,\theta)\equiv {\cal L}^{\ast{}(2)}_{\rm kin}(x,
\theta),\ (\varepsilon^{-1})^{\mu\nu}=-(\varepsilon^{-1})^{\nu\mu},
 \\ 
{} & {} & \hspace{-1.5em}
 S_0({\cal A}^{\mu}(\theta)) = \displaystyle\int d^Dx {\cal L
}^{(2)}_0({\cal A}^{\mu}(x,\theta), \partial_{\nu}{\cal A}^{\mu}(x,\theta))
= - \textstyle\frac{1}{4}\displaystyle\int d^Dx F_{\mu\nu}(x,\theta)F^{
\mu\nu}(x,\theta)\;.
\end{eqnarray}
For $D=2k+1$, $k \in \mbox{\boldmath$N$}$ it is possible  to
realize the program of Secs.II,IV for transition to the quantum
description as
well but starting directly from Lagrangian formulation of the model with
$S_L\bigl({\cal A}^{\mu}(\theta), {\stackrel{\ \circ}{\cal
A}}{}^{\mu}(\theta)\bigr)$ given on $T_{odd}{\cal M}_{cl}$ = $\{\bigl({\cal
A}^{\mu}(x,\theta)$, ${\stackrel{\ \circ}{\cal A}}{}^{\mu}(x,\theta)\bigr)\}$
[1] and having the form
\begin{eqnarray}
{} & {} & \hspace{-1.5em}S_{L}^{(2)}(\theta) \equiv
S_L\bigl({\cal A}^{\mu}(\theta), {\stackrel{\ \circ}{\cal
A}}{}^{\mu}(\theta)\bigr)
= T\bigl({\stackrel{\ \circ}{\cal A}}{}^{\mu}(\theta)\bigr) -
S_{0}\bigl({{\cal A}}^{\mu}(\theta)\bigr)\;, \\ 
{} & {} & \hspace{-1.5em}T\Bigl({\stackrel{\ \circ}{\cal
A}}{}^{\mu}(\theta)\Bigr) = \displaystyle\int d^Dx {\cal L}_{\rm kin}^{(2)}
(x,\theta) =\displaystyle\int
d^D x \textstyle\frac{1}{2}\varepsilon_{\mu\nu} {\stackrel{\ \circ}{\cal
A}}{}^{\nu}(x,\theta){\stackrel{\ \circ}{\cal A}}{}^{\mu}(x,\theta),\
(\varepsilon^{-1})^{\mu\nu}\varepsilon_{\nu\rho}=\delta^{\mu}{}_{\rho}\;.
\end{eqnarray}
The distribution of ghost number values for
$T({\cal A}^{\ast}_{\mu}(\theta))$,
$T\bigl({\stackrel{\ \circ}{\cal A}}{}^{\mu}(\theta)\bigr)$,
${S}_0({\cal A}^{\mu}(\theta))$ are given by the expression
\begin{eqnarray}
{\rm gh}\left(T({\cal A}^{\ast}_{\mu}(\theta)),T\bigl({\stackrel{\ \circ}{
\cal A}}{}^{\mu}(\theta)\bigr),{S}_0({\cal A}^{\mu}(\theta))\right)=
(-2,2,0)\;.  
\end{eqnarray}
The 1st stage of the procedure from Sec.IV for construction the II class
GThST is fulfilled by virtue  of equation (4.3a) triviality and choice as
the solution for Eq.(4.3b) [for $D=2k+1$ it is necessary to replace instead of
$S^{(2)}_H(\Gamma(\theta))$ the superfunction $S^{(2)}_L(\theta)$ in (4.3b)]
of the superfunction $S_0({\cal A}^{\mu}(\theta))$ = ${\cal S}_0({\cal A}^{
\mu}(\theta))$ taking account of (13.25).
Restricted GHS (4.5) has the form in this case
\renewcommand{\theequation}{\arabic{section}.\arabic{equation}\alph{lyter}}
\begin{eqnarray}
\setcounter{lyter}{1}
{} & {\stackrel{\ \circ}{\cal A}}{}^{\mu}(x,\theta) = 0,\
\displaystyle\frac{d_r{\cal A}^{\ast}_{\mu}(x,\theta)}{d\theta\phantom{xxxx
xx}} = - (\Box\eta_{\mu\nu} - \partial_{\nu}\partial_{\mu}){\cal A}^{\nu}(x,
\theta)\;, {} &
\\
\setcounter{equation}{26}
\setcounter{lyter}{2}
{} &
\Theta_{\mu}^H(x,\theta) =
- (\Box\eta_{\mu\nu} - \partial_{\nu}\partial_{
\mu}){\cal A}^{\nu}(x,\theta) = 0\;. {} &
\end{eqnarray}
By virtue of existence of the nontrivial generator of GTST (13.19) it follows
that this theory
is the irreducible one with Abelian gauge algebra (${\cal F}_0{}^{\gamma}_{
\alpha\beta}(\theta)$ = ${\cal M}_0{}^{\imath\jmath}_{\alpha\beta}(\theta)$
= $0$ in (4.38)).

It is necessary in correspondence with Sec.II and prescription (2.22) to
associate with GGTST (13.19) in every point $(x,\theta)$ the  superfields
$C(x,\theta)$, $\overline{C}(x,\theta)$, $B(x,\theta)$ and their 3
superantifields $C^{\ast}(x,\theta)$, $\overline{C}{}^{\ast}(x,\theta)$,
$B^{\ast}(x,\theta)$ having the following table of ${\rm gh}$,
$\varepsilon_P$, $\varepsilon_{\Pi}$, $\varepsilon$ gradings
\renewcommand{\theequation}{\arabic{section}.\arabic{equation}}
\begin{eqnarray}
\begin{array}{lccccccccl}
{} & {\cal A}^{\nu}(x,\theta) & {\cal A}^{\ast}_{\nu}(x,\theta) &
C(x,\theta) & \overline{C}(x,\theta) & B(x,\theta) & C^{\ast}(x,\theta) &
\overline{C}{}^{\ast}(x,\theta) & B^{\ast}(x,\theta) & {}\\
{\rm gh} & 0 & -1 & 1 & -1 & 0 & -2  & 0 & -1 & {} \\
\varepsilon_P &  0 & 1 & 1 & 1 & 0 & 0 & 0 & 1 & {} \\
\varepsilon_{\Pi} & 0 &0 &0 &0 &0 &0 &0 &0 & {} \\
\varepsilon &  0 & 1 & 1 & 1 & 0 & 0 & 0 & 1 & \hspace{-1em} .
\end{array}
\end{eqnarray}
The 1st order on $\theta$ system from 4 ODE  (2.21) corresponding to GTST
[1,2]
\begin{eqnarray}
\delta{\cal A}^{\mu}(x,\theta)= \int d^Dy(\partial^{\mu}\delta(x-y))\xi(y,
\theta) = \partial^{\mu}\xi(x,\theta)  
\end{eqnarray}
has the form
\begin{eqnarray}
{\stackrel{\ \circ}{\cal A}}{}^{\mu}(x,\theta) = -\partial^{\mu
}C(x,\theta)\;, 
\end{eqnarray}
and is transformed to system of the form (4.10)
\renewcommand{\theequation}{\arabic{section}.\arabic{equation}\alph{lyter}}
\begin{eqnarray}
\setcounter{lyter}{1}
{} & {} &
{\stackrel{\ \circ}{\cal A}}{}^{\mu}(x,\theta) = - \left({\cal A}^{\mu}(x,
\theta),S_1({\cal A}^{\ast}_{\mu}(\theta),C(\theta))\right)^{({\cal
A}^{\mu},{\cal A}^{\ast}_{\mu})}_{\theta}, \\
\setcounter{equation}{30}
\setcounter{lyter}{2}
{} & {} &
S_1({\cal A}^{\ast}_{\mu}(\theta),C(\theta)) = \displaystyle\int d^Dx
{\cal A}^{\ast}_{\mu}(x,\theta)\partial^{\mu}C(x,\theta)\;. 
\end{eqnarray}
The 2nd stage of the above-mentioned procedure permits with allowance made
for
commutativity of the gauge algebra to write down the unified system
containing (13.26a) and (13.30a) already in form of HS
in $T^{\ast}_{odd}{\cal M}_{min}$ = $\bigl\{({\cal A}^{\mu}(x,\theta)$,
$C(x,\theta)$, ${\cal A}^{\ast}_{\mu}(x,\theta)$,
$C^{\ast}(x,\theta))\bigr\}$, i.e. to realize simultaneously the third stage
of the algorithm in question
\begin{eqnarray}
\setcounter{lyter}{1}
{} & {} &
{\stackrel{\ \circ}{\cal A}}{}^{\mu}(x,\theta)
= -\partial^{\mu}C(x,\theta)= - \left({\cal A}^{\mu}(x,
\theta),S_{[1]}(\theta)\right)^{(\Gamma_{min})}_{\theta}\;, \\
\setcounter{equation}{31}
\setcounter{lyter}{2}
{} & {} &
{\stackrel{\ \circ}{\cal A}}{}^{\ast}_{\mu}(x,\theta) =
- (\Box\eta_{\mu\nu} - \partial_{\nu}\partial_{\mu}){\cal A}^{\nu}(x,\theta)=
  \left({\cal A}^{\ast}_{\mu}(x,
\theta),S_{[1]}(\theta)\right)^{(\Gamma_{min})}_{\theta}\;,  \\
\setcounter{equation}{31}
\setcounter{lyter}{3}
{} & {} &
{\stackrel{\,\circ}{C}}(x,\theta) =
\left(C(x,\theta),
S_{[1]}(\theta)\right)^{(\Gamma_{min})}_{\theta}=0\;,
\\ 
\setcounter{equation}{31}
\setcounter{lyter}{4}
{} & {} &
{\stackrel{\,\circ}{C}}{}^{\ast}(x,\theta) =
\partial^{\mu}{\cal A}^{\ast}_{\mu}(x,\theta)= -\left(C^{\ast}(x,\theta),
S_{[1]}(\theta)\right)^{(\Gamma_{min})}_{\theta}  
\end{eqnarray}
with superfunction $S_{[1]}(\theta)$
\renewcommand{\theequation}{\arabic{section}.\arabic{equation}}
\begin{eqnarray}
S_{[1]}(\theta)=S_{[1]}({\cal A}^{\mu}(\theta),
{\cal A}^{\ast}_{\mu}(\theta),C(\theta))= {\cal S}_0({\cal A}^{\mu}(\theta))+
S_1({\cal A}^{\ast}_{\mu}(\theta),C(\theta))\,.  
\end{eqnarray}
$S_{[1]}(\theta)$ satisfies both to master equation (4.30) and to its
consequence (4.33) with
use of the continued antibracket $(\ ,\ )^{cl}_{\theta}$ and operator
$\Delta^{cl}(\theta)$
[2] from $T^{\ast}_{odd}{\cal M}_{cl}$ to $T^{\ast}_{odd}{\cal M}_{min}$
\begin{eqnarray}
{} &
\left({\cal F}(\theta),
{\cal G}(\theta)\right)^{(\Gamma_{min})}_{\theta}
= \displaystyle\int d^D x \Biggl(
\frac{\partial {\cal F}(\Gamma(\theta))}{\partial {\cal A}^{\mu}(x,
\theta)}
\frac{\partial {\cal G}(\Gamma(\theta))}{\partial {\cal A}^{\ast}_{\mu}(x,
\theta)} +
\frac{\partial {\cal F}(\Gamma(\theta))}{\partial C(x,\theta)\phantom{x}}
\frac{\partial {\cal G}(\Gamma(\theta))}{\partial C^{\ast}(x,\theta)} -
{} & \nonumber \\
{} &
\displaystyle\frac{\partial_r {\cal F}(\Gamma(\theta))}{\partial
{\cal A}^{\ast}_{\mu}(x,\theta)\phantom{x}}
\displaystyle\frac{\partial_l {\cal G}(\Gamma(\theta))}{\partial
{\cal A}^{\mu}(x,\theta)\phantom{x}}-
\displaystyle\frac{\partial_r {\cal F}(\Gamma(\theta))}{\partial C^{\ast
}(x,\theta)\phantom{x}}
\displaystyle\frac{\partial_l {\cal G}(\Gamma(\theta))}{\partial
C(x,\theta)\phantom{x}}\Biggr),\ {\cal F}(\theta), {\cal G}(\theta)\in
C^k(T^{\ast}_{odd}{\cal M}_{min})\;, &{}
 \\ 
{} & \Delta^{min}(\theta) =
\displaystyle\int d^D x\left[
\frac{\partial_l \phantom{xxxxx}}{\partial{\cal A}^{\mu}(x,\theta)}
\frac{\partial\phantom{xxxxxx}}{\partial{\cal A}^{\ast}_{\mu}(x,\theta)}-
\frac{\partial_l \phantom{xxxxx}}{\partial C(x,\theta)}
\frac{\partial \phantom{xxxxx}}{\partial C^{\ast}(x,\theta)}
\right]. {} &
\end{eqnarray}
Therefore, HS (13.31) is solvable and $S_{[1]}(\theta)$ appears by its
integral. Superfunction $S_{ext}(\theta)$
\begin{eqnarray}
S_{ext}(\theta) = S_{[1]}(\theta) + \int d^Dx \overline{C}{}^{\ast}(x,\theta)
B(x,\theta)  
\end{eqnarray}
is the solution for master equation (5.2) in $T^{\ast}_{odd}{\cal M}_{ext}$
and its consequence of the type (4.33) with corresponding antibracket and
operator $\Delta^{ext}(\theta)$
\renewcommand{\theequation}{\arabic{section}.\arabic{equation}\alph{lyter}}
\begin{eqnarray}
\setcounter{lyter}{1}
{} &
\left({\cal F}(\theta),
{\cal G}(\theta)\right)^{(\Gamma_{ext})}_{\theta} =
\left({\cal F}(\theta),
{\cal G}(\theta)\right)^{(\Gamma_{min})}_{\theta} +
\displaystyle\int d^Dx  \Biggl(
\frac{\partial {\cal F}(\Gamma(\theta))}{\partial \overline{C}(x,\theta)
\phantom{x}}
\frac{\partial {\cal G}(\Gamma(\theta))}{\partial \overline{C}{}^{\ast
}(x,\theta)} +
{} & \nonumber \\
{} &
\displaystyle\frac{\partial {\cal F}(\Gamma(\theta))}{\partial B(x,\theta)
\phantom{x}}
\displaystyle\frac{\partial {\cal G}(\Gamma(\theta))}{\partial B^{
\ast}(x,\theta)}
- \displaystyle\frac{\partial_r {\cal F}(\Gamma(\theta))}{\partial
\overline{C}{}^{\ast}(x,\theta)\phantom{x}}
\displaystyle\frac{\partial_l {\cal G}(\Gamma(\theta))}{\partial
\overline{C}(x,\theta)\phantom{x}}
- \displaystyle\frac{\partial_r {\cal F}(\Gamma(\theta))}{\partial
B^{\ast}(x,\theta)\phantom{x}}
\displaystyle\frac{\partial_l {\cal G}(\Gamma(\theta))}{\partial
B(x,\theta)\phantom{x}}\Biggr), {} &
\\ 
\setcounter{equation}{36}
\setcounter{lyter}{2}
{} & \Delta^{ext}(\theta)=\Delta^{min}(\theta) +
\displaystyle\int d^D x\left[ -
\frac{\partial_l \phantom{xxxxx}}{\partial \overline{C}(x,\theta)}
\frac{\partial \phantom{xxxxxx}}{\partial \overline{C}{}^{\ast}(x,\theta)} +
\frac{\partial_l \phantom{xxxxx}}{\partial B(x,\theta)}
\frac{\partial \phantom{xxxxxx}}{\partial B^{\ast}(x,\theta)}
\right]. {} &
\end{eqnarray}
$S_{ext}(\theta)$ can be chosen as the solution for analogous quantum
equations (6.2), (6.9). This superfunction satisfies to BV similar equation
(6.11) as well so that the difference between the forms of the mentioned
generating equations (for GSQM and for direct superfield generalization of
the BV master equation) is not essential in that example.

To construct the $S^{\Psi}_{ext}(\theta)$ let us perform the linear phase AT
(9.1) with gauge fermion being quadratic with respect to superfields according
to general formulae (9.17), (9.33)
\renewcommand{\theequation}{\arabic{section}.\arabic{equation}\alph{lyter}}
\begin{eqnarray}
\Psi({\cal A}^{\mu}(\theta),\overline{C}(\theta),B(\theta))=
\int d^Dx \overline{C}(x,\theta)\left[\partial_{\mu}n^{\mu}{}_{\nu}
{\cal A}^{\nu}(x,\theta) + \frac{\alpha_{0}}{2}B(x,\theta)\right],
\end{eqnarray}
with arbitrary matrix $\left\|n^{\mu}{}_{\nu}\right\|$ not depending upon
${\cal A}^{\mu}(x,\theta)$. Corresponding $S^{\Psi}_{ext}(\theta)$ has the
form
\begin{eqnarray}
\hspace{-0.5em}S^{\Psi}_{ext}(\theta)\hspace{-0.2em}=\hspace{-0.2em}
S_{ext}(\theta)\hspace{-0.1em} + \hspace{-0.2em}
\int \hspace{-0.3em} d^Dx \hspace{-0.1em}\left[B(x,\theta)\partial_{\mu}
n^{\mu}{}_{\nu}
{\cal A}^{\nu}(x,\theta)\hspace{-0.1em} + \hspace{-0.1em}\frac{\alpha_{0}}{2}
B^2(x,\theta)\hspace{-0.1em} -\hspace{-0.1em}
\partial_{\mu}\overline{C}(x,\theta)n^{\mu}{}_{\nu}\partial^{\nu}C(x,\theta)
\hspace{-0.1em}\right].  
\end{eqnarray}
Let us write the only physical interesting generating functional of Green's
functions for free massless vector superfield in the form (10.2a) with help of
the last  expression
\begin{eqnarray}
{} &
Z^{(2)}_0\bigl({\cal A}^{\ast}_{\mu}(\theta),{\cal J}^{\ast}_{\mu}(\theta),
{\cal J}^{\ast}_{C}(\theta), {\cal J}^{\ast}_{\overline{C}}(\theta),
{\cal J}^{\ast}_{
B}(\theta)\bigr)= \displaystyle\int d{\cal A}(\theta)\,dC(\theta)\,
d\overline{C}(\theta)\,dB(\theta){\rm exp}\Bigl\{\displaystyle\frac{\imath}{
\hbar}\Bigl(S^{\Psi}_{ext}(\theta)+ {} & \nonumber  \\
{} & \hspace{-1em}
\displaystyle\int d^Dx\bigl[{\cal J}^{\ast}_{\mu}{\cal A}^{\mu
} + {\cal J}^{\ast}_{C}C + {\cal J}_{\overline{C}}^{\ast}\overline{C}  +
{\cal J}^{\ast}_{B}B\bigr](x,\theta)\Bigr)\Bigr\}, & {} 
\end{eqnarray}
with following table of gradings for supersources
\begin{eqnarray}
\begin{array}{lccccl}
{} & {\cal J}^{\ast}_{\nu}(x,\theta) &
{\cal J}^{\ast}_{C}(x,\theta) & {\cal J}_{\overline{C}}^{\ast}(x,\theta) &
{\cal J}^{\ast}_B(x,\theta) & {}\\
{\rm gh} & 0 & -1 & 1 & 0 & {} \\
\varepsilon_P &  0 & 1 & 1 & 0  & {} \\
\varepsilon_{\Pi} & 0 &0 &0 &0 & {} \\
\varepsilon &  0 & 1 & 1 &  0 & \hspace{-1.5em}.
\end{array}
\end{eqnarray}
$Z^{(2)}_0(\theta)$ (13.39) is the FI of Gaussian superfunction
$S^{\Psi}_{ext}(\theta)$ and is explicitly calculated according to formula
(10.6). Let us calculate  this FI in the so-called Lorentz gauge
corresponding  to $n^{\mu}{}_{\nu} = \delta^{\mu}{}_{\nu}$, $\alpha_0=1$
\renewcommand{\theequation}{\arabic{section}.\arabic{equation}\alph{lyter}}
\begin{eqnarray}
\setcounter{lyter}{1}
{} & {} & \hspace{-2.5em}
Z^{(2)}_0\bigl({\cal A}^{\ast}_{\mu}(\theta),{\cal J}^{\ast}_{\mu}(\theta),
{\cal J}^{\ast}_{C}(\theta), {\cal J}_{\overline{C}}^{\ast}(\theta),{\cal
J}^{\ast}_{B}(\theta)\bigr)=
{\rm det}\left\|\hbar^{-1}\Box\delta(x-y)\right\|
{\rm det}^{-\frac{1}{2}}\left\|\hbar^{-1}\eta_{\mu\nu}\Box\delta(x-y)\right\|
 \nonumber \\
{} & {} & \hspace{-2.0em}
\times{\rm exp}\left\{-
\displaystyle\frac{\imath}{2\hbar}{{\cal J}^{\ast}_{B}}^2(\theta)\right\}
{\rm exp}\left\{- \displaystyle\frac{\imath}{\hbar}
{\cal J}_{\overline{C}}^{\ast}(\theta)Sc(\theta){\cal J}^{\ast}_{C}(\theta)
\right\}{\rm exp}\left\{-
\displaystyle\frac{\imath}{2\hbar}{\cal J}^{\ast}_{\mu}(\theta)\Delta^{\mu
\nu}(\theta){\cal J}^{\ast}_{\nu}(\theta)\right\}, \\
\setcounter{equation}{41}
\setcounter{lyter}{2}
{} & {} & \hspace{-2.5em}
{\cal J}^{\ast}_{\overline{C}}(\theta)Sc(\theta){\cal J}^{\ast
}(\theta)\hspace{-0.2em}  =  \hspace{-0.2em}
\displaystyle\int d^Dx d^Dy {\cal J}_{\overline{C}}^{\ast}(x,\theta)Sc(
x,y,\theta){\cal J}_C^{\ast}(y,\theta),\;{{\cal J}^{\ast}_{B}}^2(\theta)
\hspace{-0.2em} = \hspace{-0.3em}
\displaystyle\int \hspace{-0.2em} d^Dx{{\cal J}^{\ast}_{B}}^2(x,\theta)\,,
\nonumber \\
{} & {} & \hspace{-2.5em}
{\cal J}^{\ast}_{\mu}(\theta)\Delta^{\mu\nu}(\theta){\cal J}^{\ast}_{\nu
}(\theta)= \displaystyle\int d^Dx d^Dy
{\cal J}^{\ast}_{\mu}(x,\theta)\Delta^{\mu\nu}(x,y,\theta){\cal J}^{\ast}_{\nu
}(y,\theta)
\;, \\
\setcounter{equation}{41}
\setcounter{lyter}{3}
{} & {} & \hspace{-2.5em}
Sc(x,y,\theta) =
\displaystyle\int \displaystyle\frac{d^Dp}{(2\pi)^D}
\displaystyle\frac{e^{-\imath p(x-y)}}{p^2 +\imath\varepsilon},\
\Delta^{\mu\nu}(x,y,\theta) = -
\displaystyle\int \displaystyle\frac{d^Dp}{(2\pi)^D}
\displaystyle\frac{e^{-\imath p(x-y)}}{p^2 +\imath\varepsilon}\eta_{\mu\nu},\
\varepsilon \to +0 \;.
\end{eqnarray}
Considering all functional determinants in (13.41a) as the normalization
constants usually one does not write them in (13.41a). The superfunctions
$Sc(x,y,\theta) = Sc(x-y)$, $\Delta^{\mu\nu}(x,y,\theta) = \Delta^{\mu\nu}(x
-y)$ (13.41c) are the Green's
functions in Feynman representation for ghost superfields $C(x,\theta)$,
$\overline{C}(x,\theta)$ and for massless vector superfields
${\cal A}^{\mu}(x,\theta)$ (for $D=4$ $\Delta^{\mu\nu}(x,y,\theta)$ is the
Green's function for free electromagnetic superfield in Lorentz gauge).

Corresponding to Theorem 7 invariance transformations for $Z^{(2)}_0(
\theta)$ are based on the type (9.22) system  having the form in this case
\begin{eqnarray}
\setcounter{lyter}{1}
{} & {} &
{\stackrel{\ \circ}{\cal A}}{}^{\mu}(x,\theta)
= -\partial^{\mu}C(x,\theta),\
{\stackrel{\,\circ}{C}}(x,\theta) = {\stackrel{\,\circ}{B}}(x,\theta) = 0,\
{\stackrel{\,\circ}{\overline{C}}}(x,\theta) = B(x,\theta)\;, \\
\setcounter{equation}{42}
\setcounter{lyter}{2}
{} & {} &
{\stackrel{\ \circ}{\cal A}}{}^{\ast}_{\mu}(x,\theta) =
- (\Box\eta_{\mu\nu} - \partial_{\nu}\partial_{\mu}){\cal A}^{\nu}(x,\theta)
 + \partial^{\mu}B(x,\theta),\
{\stackrel{\,\circ}{C}}{}^{\ast}(x,\theta) =
(\partial^{\mu}{\cal A}^{\ast}_{\mu} - \Box\overline{C})(x,\theta)\;,
\nonumber \\
{} & {} &
{\stackrel{\,\circ}{\overline{C}}}{}^{\ast}(x,\theta) = \Box C(x,\theta),\
{\stackrel{\,\circ}{B}}{}^{\ast}(x,\theta) = - (\partial_{\mu}{\cal A}^{\mu} -
\alpha_0B)(x,\theta)\;.
\end{eqnarray}
The ODE system being used in BV method are obtained from (13.42) by means of
vanishing of the right-hand sides in Eqs.(13.42b). Corresponding to
Eqs.(13.42) invariance transformations for $Z_0^{(2)}(\theta)$ with
$\check{\Gamma}(\theta)$ being by an integral curve for this system have the
form
\begin{eqnarray}
\setcounter{lyter}{1}
{} & {} & \hspace{-2.5em}
\delta{\cal A}^{\mu}(x,\theta)_{\mid\check{\Gamma}(\theta)}
\hspace{-0.2em}=\hspace{-0.1em} \partial^{\mu}C(x,\theta)\mu,\
\delta{C}(x,\theta)_{\mid\check{\Gamma}(\theta)} \hspace{-0.2em}=
\hspace{-0.1em}\delta{B}(x,\theta)_{
\mid\check{\Gamma}(\theta)} \hspace{-0.2em}=\hspace{-0.1em} 0,\
\delta{\overline{C}}(x,\theta)_{\mid\check{\Gamma}(\theta)}\hspace{-0.2em} =
\hspace{-0.1em}B(x,\theta)\mu,
\\
\setcounter{equation}{43}
\setcounter{lyter}{2}
{} & {} & \hspace{-2.5em}
\delta{\cal A}^{\ast}_{\mu}(x,\theta)_{\mid\check{\Gamma}(\theta)} =
\bigl[\partial^{\mu}B -(\Box\eta_{\mu\nu} - \partial_{\nu}\partial_{\mu}){\cal
A}^{\nu}\bigr](x,\theta)\mu,\
\delta C^{\ast}(x,\theta)_{\mid\check{\Gamma}(\theta)} =
(\Box\overline{C} - \partial^{\mu}{\cal A}^{\ast}_{\mu})(x,\theta)\mu\;,
\nonumber \\
{} & {} & \hspace{-2.5em}
\delta\overline{C}{}^{\ast}(x,\theta)_{\mid\check{\Gamma}(\theta)} = -
\Box C(x,\theta)\mu,\
\delta B^{\ast}(x,\theta)_{\mid\check{\Gamma}(\theta)} =  (\alpha_0B-
\partial_{\mu}{\cal A}^{\mu})(x,\theta)\mu\;.
\end{eqnarray}
The superfield BRST transformations being used in BV method for
$\theta = 0$ follow from (13.43) in vanishing of the right-hand sides in the
expressions (13.43b).

Ward identity for $Z^{(2)}_0(\theta)$ and component formulation are
obtained in the obvious way according to Secs.X,XI.

For construction of the generating functional of Green's functions
$Z^{(2)}_{0{}m}(\theta)$ in the case of free massive vector superfield ${\cal
A}^{\mu}(
x,\theta)$ theory as the singular nondegenerate ThST described in Lagrangian
[1] and Hamiltonian [2] formulations for GSTF it is sufficient to fulfill the
only 1st step of the procedure from Sec.IV.

As far as the $\theta$-superfield Proca  model is nongauge one then there are
not any gauge transformations and therefore the formulae (13.19),
(13.28)--(13.38) are not valid in question. The given ThST formulation is
defined by the formulae (13.20)--(13.22) for Hamiltonian formalism and
(13.23), (13.24) for Lagrangian one in which it is necessary to replace the
superfunctions $S_H^{(2)}(\theta)$, $S_L^{(2)}(\theta)$, $S_0({\cal
A}^{\mu}(\theta))$ onto $S_{H{}m}^{(2)}(\theta)$,
$S_{L{}m}^{(2)}(\theta)$, $S_{0{}m}(\theta)$ of the form
[1,2]
\begin{eqnarray}
{} &  S_{H{}m}^{(2)}(\theta)
= T\bigl({\cal A}^{\ast}_{\mu}(\theta)\bigr) +
S_{0{}m}\bigl({\cal  A}^{\mu}(\theta)\bigr)\;,
 & {}  \\ 
{} &  S_{L{}m}^{(2)}(\theta)
= T\Bigl({\stackrel{\ \circ}{\cal A}}{}^{\mu}(\theta)\Bigr) -
S_{0{}m}\bigl({\cal  A}^{\mu}(\theta)\bigr)\;,
 & {}  \\ 
{} & S_{0{}m}\bigl({\cal  A}^{\mu}(\theta)\bigr) =
\displaystyle\int d^Dx {\cal
L}_{0{}m}^{(2)}(x,\theta) = \displaystyle\int d^Dx \Bigl(-
\textstyle\frac{1}{4}
F_{\mu\nu}F^{\mu\nu} + \textstyle\frac{m^2}{2}{\cal A}_{\mu}{\cal A}^{
\mu}\Bigr)(x,\theta)\,.& {} 
\end{eqnarray}
The distribution of ghost number values of the form (13.25) remains valid in
this case so that the superfunction $S_{0{}m}(\theta)$ = ${\cal S}_{0{}
m}(\theta)$ is the solution for Eq.(4.3b).
The restricted GHS (4.5) has the form in question
\renewcommand{\theequation}{\arabic{section}.\arabic{equation}\alph{lyter}}
\begin{eqnarray}
\setcounter{lyter}{1}
{} & {\stackrel{\ \circ}{\cal A}}{}^{\mu}(x,\theta) = 0,\
{\stackrel{\ \circ}{\cal A}}{}^{\ast}_{\mu}(x,\theta)= - \Bigl((\Box+m^2)
\eta_{\mu\nu} - \partial_{\mu}\partial_{\nu}\Bigr)
{\cal A}^{\nu}(x,\theta)\;, {} &
\\
\setcounter{equation}{47}
\setcounter{lyter}{2}
{} &
\Theta_{m{}\mu}^H(x,\theta) =
- \Bigl((\Box+ m^2)\eta_{\mu\nu} - \partial_{\mu}\partial_{\nu}\Bigr){
\cal A}^{\nu}(x,\theta) = 0\,. {} &
\end{eqnarray}
In view of the nongauge character of the model it is not required the
introduction of the superfields $C(x,\theta)$, $\overline{C}(x,\theta)$,
$B(x,\theta)$ and their superantifields. Therefore from the table (13.27) it
is necessary to retain the only gradings for ${\cal A}^{\mu}(x,\theta)$,
${\cal A}^{\ast}_{\mu}(x,\theta)$.
The remarks for quantization of the model for odd and even values of
dimension $D$ being used in massless theory remains valid in this case.

To calculate the generating functional of Green's functions
$Z^{(2)}_{0{}m}(\theta)$ for free massive vector superfield in the form
(10.2a) not being dependent from the superantifields
\renewcommand{\theequation}{\arabic{section}.\arabic{equation}}
\begin{eqnarray}
Z^{(2)}_{0{}m}(\theta)  \equiv
Z^{(2)}_{0{}m}\bigl({\cal J}^{\ast}_{\mu}(\theta)\bigr)= \int d{\cal A}(
\theta){\rm exp}\left\{\frac{\imath}{\hbar}\Bigl({\cal S}_{0{}m}\bigl({\cal
A}^{\mu}(\theta)\bigr) + \int d^Dx({\cal J}^{\ast}_{\mu}{\cal A}^{
\mu})(x,\theta)\Bigr)\right\} 
\end{eqnarray}
with corresponding supersources ${\cal J}^{\ast}_{\mu}(x,\theta)$ having the
properties as in (13.40) it is sufficient to make use the formula (10.6) for
FI of Gaussian superfunction. The final result has the form
\begin{eqnarray}
{} & \hspace{-2.2em}
Z^{(2)}_{0{}m}\bigl({\cal J}^{\ast}_{\mu}(\theta)\bigr)\hspace{-0.2em}=
\hspace{-0.2em}
{\rm det}^{-\frac{1}{2}}\hspace{-0.1em}\left\|\hbar^{-1}\Bigl(\eta_{\mu\nu}(
\Box+ m^2)-\partial_{\mu}
\partial_{\nu}\Bigr)\delta(x-y)\right\|\hspace{-0.1em}
{\rm exp}\left\{\hspace{-0.2em}-
\displaystyle\frac{\imath}{2\hbar}{\cal J}^{\ast}_{\mu}(\theta)\Delta^{\mu
\nu}_m(\theta){\cal J}^{\ast}_{\nu}(\theta)\hspace{-0.1em}\right\}\hspace{
-0.1em},{}& \nonumber
\\
{} & \hspace{-2.0em}
\Delta^{\mu\nu}_m(x,y,\theta)\hspace{-0.1em} = \hspace{-0.1em}
\Delta^{\mu\nu}_m(x-y) \hspace{-0.1em}=\hspace{-0.1em} -\hspace{-0.1em}
\displaystyle\int \hspace{-0.2em}\displaystyle\frac{d^Dp}{(2\pi)^D}\Bigl(
\eta_{\mu\nu}\hspace{-0.1em} - \hspace{-0.1em}
\frac{p_{\mu}p_{\nu}}{m^2}\Bigr)
\displaystyle\frac{e^{-\imath p(x-y)}}{p^2 -m^2 +\imath\varepsilon}
\eta_{\mu\nu},\; \varepsilon \to +0.{}&
\end{eqnarray}
The functional  determinant in (13.49)  may be omitted if to consider it as
the normalization constant. Superfunction $\Delta^{\mu\nu}_m(x,y,\theta)$
appears by causal Green's function for massive vector superfield with
allowance made for condition $\partial_{\mu}{\cal A}^{\mu}(x,\theta)=0$ [1].
The superfield BRST transformations for the superfunction (13.48) is trivial.

Restriction of the considered quantum superfield models  to ordinary ones
from gauge fields theory is established by means of involution $*$ or by
simple setting $\theta=0$ in all relations and quantities of this section
taking  of Sec.XI and Subsec.XII.1 into account.
\section{Conclusion}

On the basis of GSTF [1,2] the direct rules of GSQM for superfield (on
$\theta$) gauge theories in the Lagrangian (in the usual sense) formalism,
i.e.
for the GThGT and GThST in terminology of papers [1,2] are constructed. To
this end the new interpretation for extension of supermanifold
${\cal M}_{cl}$ to ${\cal M}_{min}$, ${\cal M}_{(min,\overline{C})}$,
${\cal M}_{ext}$ is proposed together with explicit group-theoretic
construction of the ghost number in question.

The procedure for construction of the II class GThGT defined on
$T^{\ast}_{odd}{\cal M}_{min}$ and next on $T^{\ast}_{odd}{\cal M}_{ext}$
starting from the original I class GThGT (GThST) defined on
$T_{odd}{\cal M}_{cl}$, $T^{\ast}_{odd}{\cal M}_{cl}$ is successively
formulated in the framework of GSTF and
mathematical means developed in Ref.[1,2] in the form of exact algorithm.

The
series of the Existence Theorems in explicit formulation both for the
generating
equations of GSQM and for the BV similar master equations are formulated. The
deformation of the obtained GThGT into the supermanifolds
$T_{odd}(T^{\ast}_{odd}{\cal M}_s)$, $s\in\{min$, $(min,\overline{C})$,
$ext\}$ in powers of ${\stackrel{\circ}{\Gamma}}(\theta)$ is realized. The
crucial role of the HS and HS type systems is demonstrated for superfield
interpretation of BRST similar transformations.

The arbitrariness in solutions of above-mentioned generating equations is
described by means of operations of GNA and GTA. It was shown the
anticanonical transformations
and GNA play principal
role in the covariant formulation of quantum theory in constructing of
the generating functionals of Green's functions.

The procedure to impose the gauge, being interpreted as a choice of
different Lagrangian surfaces on which the Hesse supermatrix of superfunction
$S_H(\Gamma(\theta),\hbar)$ is nondegenerate, is realized by means of ATs and
GNA.

The definition of functional integral of superfunctions in the framework of
perturbation theory is introduced for GSQM. With its help the main GSQM
object being by superfunction $Z(\Phi^{\ast}(\theta),{\cal J}^{\ast}(\theta))$
is written  by means of FI.

The nontrivial properties of that superfunction are studied in details. They
contain the new ones. The additional interpretation
for Ward identities is obtained as the invariance of
$Z(\Phi^{\ast}(\theta),{\cal J}^{\ast}(\theta))$ with respect to translation
transformation relative to
odd time $\theta$ on an arbitrary parameter $\mu\in{}^1\Lambda_1(\theta)$
along integral curve of solvable HS built with respect to quantum gauge fixed
action $S_H^{\Psi}(\Gamma(\theta),\hbar)$. It should be noted  the last
transformation is the special anticanonical transformation in GSQM with
one's generator ${\rm ad}_{S_H^{\Psi}(\theta,\hbar)}$ and unit
Berezinian.

For the first time it is introduced in the superfield form the effective
action {\boldmath$\Gamma$}$(\langle\Phi(\theta)\rangle,\Phi^{\ast}(\theta))$
for which the Ward identities repeat the master equations for
$S_H^{\Psi}(\theta,\hbar)$ in GSQM. This fact makes by possible the
construction of
HS with respect to {\boldmath$\Gamma$}$(\theta)$ and obtaining the quantum
BRST similar transformations in space parametrized by average supervariables
$\langle\Gamma^p(\theta)\rangle$. The analogous objects were obtained and
considered in fulfilling of BV similar generating equations and for deformed
GThGT with respect to  ${\stackrel{\circ}{\Gamma}}{}^p(\theta)$ as well.

Component formulation of GSQM was proposed. The inclusion of the BV method
into
GSQM is established. Detailed connection with superfield method of the
Lagrangian quantization from Ref.[3] is obtained with simultaneous
description of its problem moments. The examples of the GSTF models for
free and interacting scalar, spinor superfields and for free massless and
massive  vector superfields
suggested in [1,2] have been developed in context of the demonstration of
the general rules for quantization  constructed in Secs.II-XI.

\noindent
{\bf Acknowledgments:}
Author is very grateful to Mishchuk B. for discussion of some results of the
present paper.
\vspace{1ex}
\begin{center}
{\large{\bf References}}
\end{center}
\begin{enumerate}
\item A.A. Reshetnyak, General Superfield Quantization Method. I. General
Superfield Theory of Fields: Lagrangian Formalism, hep-th/0210207.
\item A.A. Reshetnyak, General Superfield Quantization Method. II. General
Superfield Theory of Fields: Hamiltonian Formalism, hep-th/0303262.
\item P.M. Lavrov, P.Yu. Moshin and A.A. Reshetnyak, Mod. Phys. Lett. A10
(1995) 2687; JETP Lett. 62 (1995) 780.
\item I.A. Batalin and G.A. Vilkovisky,
Phys. Lett. B 102 (1981) 27.
\item C.N. Yang and R. Mills, Phys. Rev. 96 (1954) 191; R. Utiyama, Phys. Rev.
101 (1956) 1597.
\item I.A. Batalin and G.A. Vilkovisky,  Phys. Rev. D 28 (1983) 2567.
\item I.A. Batalin and G.A. Vilkovisky, J. Math. Phys. 26 (1985) 172.
\item D.M. Gitman and I.V. Tyutin,
Quantization of Fields with Constraints (Springer-Verlag, Berlin and
Heidelberg, 1990).
\item A.A. Slavnov, Teor. Mat. Fiz. 22 (1975) 177.
\end{enumerate}
\end{document}